\documentclass{jpsj3}
\title{Fermi Surface Properties, Metamgnetic Transition and Quantum Phase Transition of CeRu$_2$Si$_2$ and Its Alloys Probed by the dHvA Effect\\--Fermi Surface Studies on the Itinerant and Localized Dichotomy of $f$ Electron Nature}
\author{Haruyoshi Aoki$^{1}$, Noriaki Kimura$^{1}$ and Taichi Terashima$^{2}$}

\inst{$^{1}$Graduate School of Science and Center for Low Temperature Science, Tohoku University, Sendai, Miyagi 980-8578
 \\$^{2}$National Institute for Materials Science, Tsukuba, Ibaraki 305-0003}

\abst{This article describes the Fermi surface properties of  CeRu$_2$Si$_2$ and its alloy systems CeRu$_2$(Si$_x$Ge$_{1-x}$)$_2$ and Ce$_x$La$_{1-x}$Ru$_2$Si$_2$ studied by the de Haas - van Alphen (dHvA) effect.  We pay particular attention to how the Fermi surface properties and the f electron state change with magnetic properties, in particular how they change associated with metamagnetic transition and quantum phase transition.  After summarizing the important physical properties of  CeRu$_2$Si$_2$, we present the magnetic phase diagrams of  CeRu$_2$(Si$_x$Ge$_{1-x}$)$_2$ and Ce$_x$La$_{1-x}$Ru$_2$Si$_2$ as a function of temperature, magnetic field and concentration $x$.  From the characteristic features of the magnetic phase diagram, we argue that the ferromagnetic interaction in addition to the anitferromagnetic interaction and the Kondo effect is responsible for the magnetic properties and that the metamagnetic transitions in these systems are relevant to the ferromagnetic interaction.  We summarize the Fermi surface properties of  CeRu$_2$Si$_2$ in fields below the metamagnetic transition where the f electron state is now well understood theoretically as well as experimentally.  We present experimental results in fields above the metamagnetic transitions in CeRu$_2$(Si$_x$Ge$_{1-x}$)$_2$ and Ce$_x$La$_{1-x}$Ru$_2$Si$_2$ as well as  CeRu$_2$Si$_2$ to show that the Fermi surface properties above the metamagnetic transitions are significantly different from those below in many important aspects.  We argue that the Fermi surface properties above the metamagnetic transitions are not appropriately described in terms of either itinerant or localized f electron.  The experimental results in fields below the metamagnetic transitions in CeRu$_2$(Si$_x$Ge$_{1-x}$)$_2$ and Ce$_x$La$_{1-x}$Ru$_2$Si$_2$ are presented to discuss the f electron state in the ground state.  The Fermi surface properties of dilute Kondo alloys of Ce$_x$La$_{1-x}$Ru$_2$Si$_2$ have been revealed as a function of Ce concentration and temperature.  We show that the f electron state can be regarded as itinerant in the ground state together with the definition of the term ``itinerant'' in this case.  The Fermi surface properties are measured also in high concentration alloys of CeRu$_2$(Si$_x$Ge$_{1-x}$)$_2$ and Ce$_x$La$_{1-x}$Ru$_2$Si$_2$ as a function of $x$.  With the help of the angle resolved photoemisson spectroscopy studies, we show that the f electron nature does not change at the quantum phase transition between the paramagnetic and antiferromagnetic phases.  However, the picture for the f electron state may be ambiguous and depend on which property one considers in the magnetic states of these systems.  The ambiguity and confusion of the f electron state may come from the inherent dual nature of the f electron and we would like to point out that it is sometimes misleading and may not be fruitful to discriminate the f electron state either as itinerant or localized without any clear definition for the terms `` itinerant''  and ``localized''.     }

\begin{document}

\maketitle

\section{Introduction}
In the past few decades, a large number  of studies have been performed to understand the intriguing properties of the strongly correlated f electron system.    The low energy phenomenon like heavy Fermion, superconductivity,  quantum phase transition and so on have been the main issues of this field.  Therefore, the Fermi surface properties  are essential information to reveal the mechanism of the properties.    The de Haas - van Alphen (dHvA) effect has been the most powerful method to investigate the Fermi surface properties and has been applied to the studies of various strongly correlated f electron compounds.

In this article we present the Fermi surface properties of the strongly correlated f electron system probed by the dHvA effect.  We report and discuss mostly the results of CeRu$_2$Si$_2$ and its alloys.   They have been studied via various experimental methods and their physical properties are most well known among the heavy Fermion compounds.   In this respect, CeRu$_2$Si$_2$ is a most interesting compound whose Fermi surface properties are worth to be clarified thoroughly.  

The magnetic phase diagrams of  CeRu$_2$Si$_2$ and its alloys CeRu$_2$(Si$_x$Ge$_{1-x}$)$_2$ and Ce$_x$La$_{1-x}$Ru$_2$Si$_2$ are presented as a function of temperature, magnetic field and concentration of alloying element.  In the former system the ferromagnetic state of CeRu$_2$Ge$_2$ evolves to an antiferromagnetic state and then to the heavy Fermion state of CeRu$_2$Si$_2$ as a function of Si concentration. In the latter system, the normal metal state of LaRu$_2$Si$_2$ evolves to an antiferromagnetic state and then to the heavy Fermion state of CeRu$_2$Si$_2$ with alloying of Ce.  Both systems exhibit metamagnetic transition or crossover which is often observed in the heavy Fermion compounds but whose mechanism is still controversial.  We report and discuss how the f electron state as well as the Fermi surface properties changes associated with the quantum phase transition or metamagnetic transition.  

Particular attention is paid to the issue whether the f electron of the system can be regarded as localized or itinerant.  In the early stage of studies on the transition metals, the same issue was discussed.  For the strongly correlated f electron system this issue is still controversial and is revived with the progress of the studies on the quantum phase transition. 

The volume of the Fermi surface is conventionally used to categorize the f electron state into either ``itinerant" and ``localized".  Depending on whether the f electron is counted as a conduction electron or not, the term ``large Fermi surface" or ``small Fermi surface" is also used. In the paramagnetic ground state, when the measured volume of the Fermi surface is large, the term ``itinerant" can be used for the f electron state without ambiguity owing to the Luttinger theorem and the physics behind the term ``itinerant" is well understood owing to the studies in the past few decades.  For example, no one claims that the term ``itinerant" is  inconsistent with the observation that the valence of Ce is close to 3+ but not 4+.  In the present study, such an example is presented for the state of CeRu$_2$Si$_2$ in fields below the metamagnetic transition field.  Such a state or a compound is referred as the ``itinerant f electron system in the paramagnetic ground state" in this article.  On the other hand, when the measured volume of the Fermi surface is small and the value of the magnetic moment is well explained by the crystal electric field scheme, it seems to be widely accepted that the f electron state can be expressed as ``localized". Such an example in the present case is CeRu$_2$Ge$_2$.  In this article, we refer such a compound as the ``magnetic localized f electron system" to avoid confusion in the descriptions for the f electron state.  We also use the terms ``large Fermi surface" or ``small Fermi surface" for the alloys of CeRu$_2$Si$_2$ depending on whether the f electron of each Ce is counted as a conduction electron or not.
 
However, we think that in some cases the term ``localized" seems to be used without clear consensus among the researchers about what are meant physically by the term \cite{Matsumoto08,Harima11}.  We argue in this article that even though the observed volume of the Fermi surface is the same as or close to that of the small Fermi surface, in some cases it is not appropriate to use the term ``localized" for the f electron state.  The f electron state in fields above the metamagnetic transition field, in the ground state of the dilute Kondo alloy and presumably in the antiferromagnetic states of the alloys CeRu$_2$(Si$_x$Ge$_{1-x}$)$_2$ and Ce$_x$La$_{1-x}$Ru$_2$Si$_2$ presented in this article are such cases.  In these cases, some of the observed features appear to be similar to those observed in the ``itinerant f electron system in the paramagnetic ground state", although the other features appear to be similar to those observed in the ``magnetic localized f electron system".  Because of this confusion, we would like to describe the properties as they are observed and do not rush into the discrimination of either itinerant or localized.  

On one hand CeRu$_2$Si$_2$ and its alloys are interesting as strongly correlated f electron systems, but on the other hand,  they are also very interesting from the point of view of the dHvA effect measurements.  It is not very difficult to grow their single crystals whose quality and size are enough for the dHvA effect measurements.  They exhibit various interesting behaviors which cannot be observed in the dHvA effect measurements of normal metals.  We would like to highlight also some peculiar features observed by the dHvA effect measurements in the strongly correlated f electron system.  Although the dHvA effect is a powerful tool, high magnetic fields are indispensable for the measurements. The high magnetic field could affect the Fermi surface properties of the strongly correlated f electron system whose important interaction is the Kondo effect.  Then, the Fermi surface properties measured by the dHvA effect may not be the same as those in the ground state.  Since CeRu$_2$Si$_2$ and its alloys have very strong Ising character, it is possible to study the magnetic field effect on the Fermi surface properties, comparing  the Fermi surface properties with fields parallel and perpendicular to the easy axis.

$\S$2 describes the physical properties and magnetic phase diagrams of CeRu$_2$Si$_2$ and its alloys  CeRu$_2$(Si$_x$Ge$_{1-x}$)$_2$ and Ce$_x$La$_{1-x}$Ru$_2$Si$_2$ as a function of concentration $x$ of alloying element, temperature, and magnetic field with particular attention to the metamagnetic transitions.  An interpretation of the magnetic properties and magnetic phase diagram is also proposed.

$\S$3 reports the Fermi surface properties of CeRu$_2$Si$_2$ in fields below and above the metamagnetic transition field together with those above the metamagnetic transition fields in CeRu$_2$(Si$_x$Ge$_{1-x}$)$_2$ and Ce$_x$La$_{1-x}$Ru$_2$Si$_2$.  We present how the Fermi surface properties change associated with the metamagnetic transition in CeRu$_2$Si$_2$ and discuss how the f electron state above the metamagnetic transition is described.  

$\S$4 presents the Fermi surface properties measured with fields in the (001) plane which can be assumed to be substantially the same as those of the ground state in the present systems.  We argue how the f electron state changes with magnetic properties, in particular with the quantum phase transition.  

$\S$5 is a remark on how we should describe properly the f electron state  of the strongly correlated f electron system.   

Since this article describes mostly the results of the dHvA measurements, we give a rather lengthy appendix so that readers who are not familiar with the dHvA measurements can understand the statements related with the dHvA effect measurements and analyses.  We also describe the problems of analysis and some peculiar observations in the strongly correlated f electron system. 

\section{Physical Properties and Magnetic Phase Diagrams of CeRu$_2$Si$_2$ and Its alloys}
In $\S$2.1 we summarize the important physical properties of CeRu$_2$Si$_2$ for later descriptions and discussions on the magnetic properties and Fermi surface properties.  In $\S$2.2.1 and  $\S$2.2.2, we present the magnetic phase diagrams of CeRu$_2$(Si$_x$Ge$_{1-x}$)$_2$ and Ce$_x$La$_{1-x}$Ru$_2$Si$_2$ as a function of temperature and concentration $x$, respectively.  $\S$2.2.3 describes the transport properties of CeRu$_2$(Si$_x$Ge$_{1-x}$)$_2$ and Ce$_x$La$_{1-x}$Ru$_2$Si$_2$ to derive important relations among the magnetic transitions and the Fermi surface properties reflected in the transport properties.  In $\S$2.3.1,  we describe the metamagnetic transitions and the phase diagrams under magnetic fields in CeRu$_2$(Si$_x$Ge$_{1-x}$)$_2$ and Ce$_x$La$_{1-x}$Ru$_2$Si$_2$.  From these observations we conjecture the relation between the metamagnetic transition and the magnetism in $\S$2.3.2.  In $\S$2.4 we propose an interpretation of the origins of the magnetic phase diagrams.  The present article is mostly concerned with the Fermi surface properties and their related properties.  For more broad aspects of the physics in CeRu$_2$Si$_2$ and its alloys, we refer the reader to the review papers and the textbook by J. Flouquet\cite{Flouquet95,Flouquet02,Flouquet04,Flouquet05,Flouquet10,Flouquet07} 

\subsection{Physical properties of CeRu$_2$Si$_2$}
\begin{figure}[htbp]
\begin{center}
\includegraphics[width=0.3\linewidth]{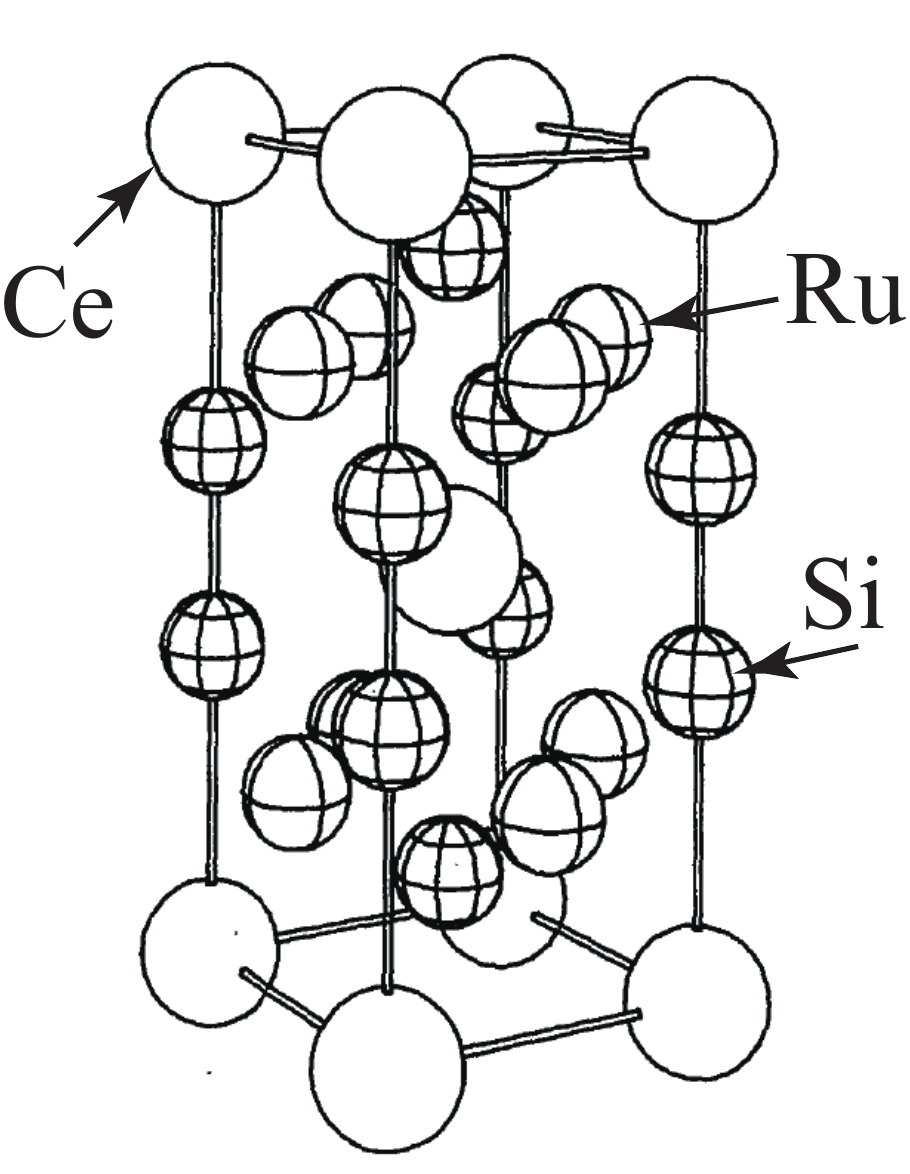}
\end{center}
\caption{Crystal structure of CeRu$_2$Si$_2$.}
\label{fig:90025Fig1}
\end{figure}

 CeRu$_2$Si$_2$ crystallizes in the ThCr$_2$Si$_2$ structure as shown in Fig. \ref{fig:90025Fig1}\cite{Yamagami92}.  The ground state of the crystalline electric field is the $\Gamma_7$ Kramers doublet $\alpha|5/2, \pm5/2 \rangle + \beta|5/2, \mp3/2\rangle$ with $\alpha^2+\beta^2 =1$ and $\alpha$ being close to 1.0.  It is separated from the excited state by 220 - 363 K\cite{Besnus85,Boucherle01,Willers12}.  If we do not consider the magnetism of a tiny moment\cite{Amato94} or the properties under very small magnetic fields\cite{Takahashi03},  the ground state can be assumed to be paramagnetic.  However, neutron scattering experiments\cite{Mignod88,Sato99,Kadowaki04} report that antiferromagnetic fluctuations with incommensurate wave vectors of  $\mbox{\boldmath${k}$}_{1}$ = [0.3, 0, 0],  $\mbox{\boldmath${k}$}_{2}$ = [0.3, 0.3, 0], and $\mbox{\boldmath${k}$}_{3}$ = [0, 0, 0.35] are present.   The Kondo temperature $T_{\rm K}$ is determined to be 24 K from the specific heat measurements\cite{Besnus85}. Electronic specific heat coefficient $\gamma$ is measured to be 320-360 mJ/mol$\cdot$K$^2$\cite{Besnus85,Fisher91}.   The resistivity shows the Fermi liquid behavior and the coefficient $A$ of the $T^2$ term in the temperature dependence of resistivity nearly satisfies the Kadowaki-Woods relation with the proportionality constant of $1\times10^{-5}\mu\Omega$cm(Kmol/mJ)$^2$\cite{Kadowaki86,Tsujii05}. 

\begin{figure}[htbp]
\begin{center}
\includegraphics[width=0.6\linewidth]{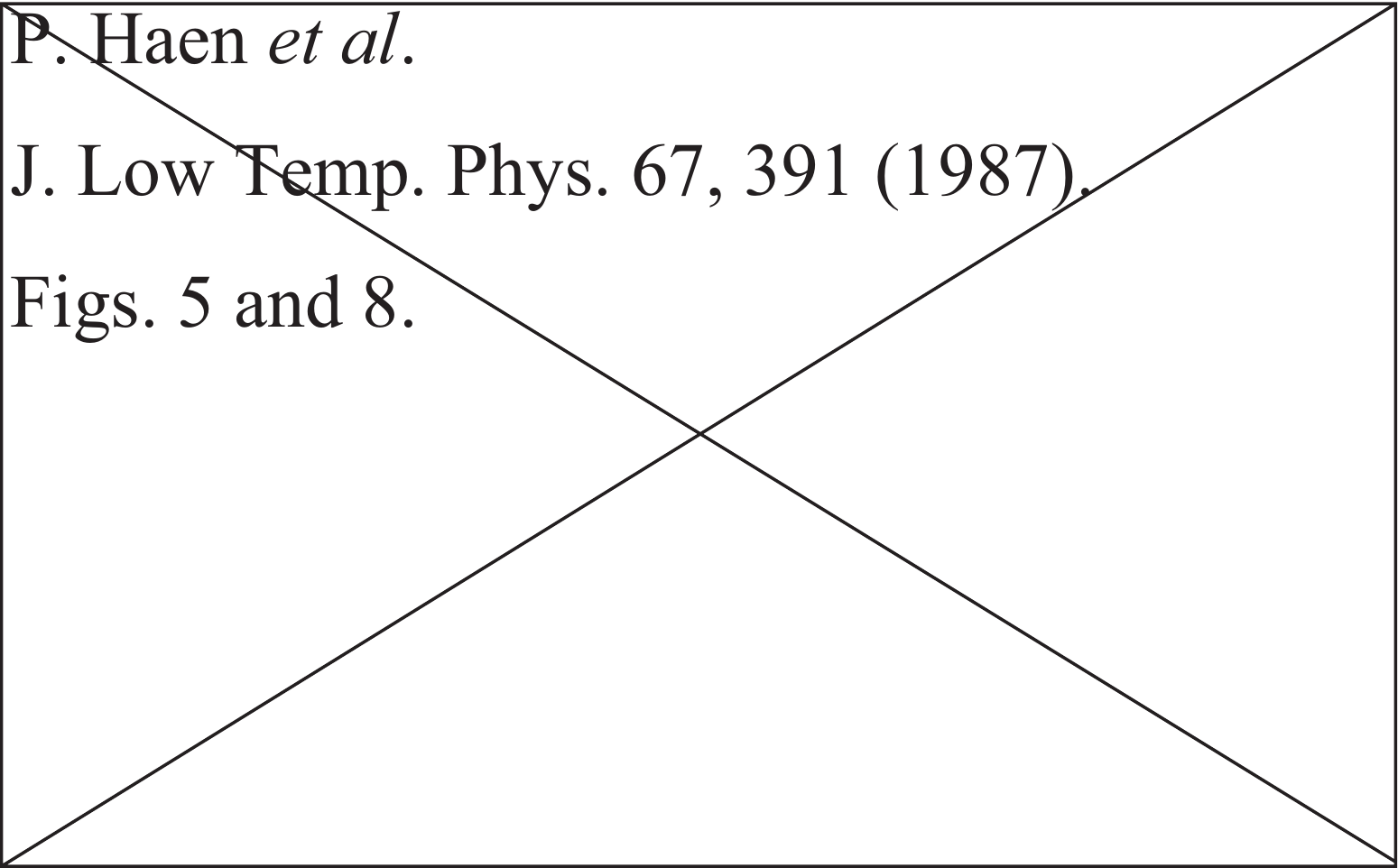}
\end{center}
\caption{(a)Temperature variations of magnetic susceptibility with fields parallel and perpendicular to the  [001] direction (c axis)\cite{Haen87}. (b) Magnetization as a function of applied magnetic field at various temperatures\cite{Haen87}.  The field is applied parallel to the c axis. }
\label{fig:90025Fig2}
\end{figure}

 It has a large magnetic anisotropy\cite{Haen87}.  Figure \ref{fig:90025Fig2} shows the magnetic susceptibility as a function of temperature. When a magnetic field is applied perpendicular to the [001] direction (c axis), the susceptibility is small and gradually increases with decreasing temperature.  On the other hand, the susceptibility is much enhanced for magnetic fields parallel to the [001] direction and has a broad maximum as a function of temperature. Hereafter, we denote the temperature for the maximum as $T_{\rm m}$.  The magnetization increases largely with magnetic field and then around 7.7 T exhibits a rapid increase (metamagnetic transition) as shown in Fig. \ref{fig:90025Fig2}(b)\cite{Haen87}.  We denote the metamagnetic transition field as $H_{\rm m}$.  In this article we also use the term ``polarized state'' for the state where a large magnetic moment is induced.  The increase of the magnetization around 7.7 T becomes broad with increasing temperature and is not obvious above the temperature comparable to $T_{\rm K}$.  With decreasing temperature, the increase becomes sharper but is reported to have a finite width at lowest temperatures\cite{Sakakibara95}.  This metamagnetic transition is thought to be crossover rather than first order transition.  For the fields in the (001) plane, the magnetization increases gradually with magnetic field and shows no metamagnetic behavior at least up to the fields available in laboratories. 

The volume magnetostriction $\Delta V/V$ increases by about 0.5$\times$10$^{-3}$ at the fields close to $H_{\rm m}$,  jumps by about 0.5 $\times$ 10$^{-3}$ across $H_{\rm m}$ and then continues to increase with increasing field.\cite{Paulsen90}    The increase of $\Delta V/V$ is about 0.8 $\times$ 10$^{-3}$ from $H_{\rm m}$ to 12 T.  The compressibility below $H_{\rm m}$ is reported to be 0.95 $\times$10$^{-3}$/kbar\cite{Lacerda89}. Then the volume increase below $H_{\rm m}$ is assumed to correspond to the application of negative pressure of about 0.5 kbar.  The valence of Ce is measured by X-ray absorption spectroscopy to be 3.053 at zero magnetic field, and decreases with increasing field.  The decrease of the valence becomes larger at fields around and above $H_{\rm m}$ and is about 0.005 from 8 T to 18 T\cite{Matsuda12a}.  This observation indicates that the f electron becomes more localized at higher fields.  The f electron state above $H_{\rm m}$ is further discussed in sections 3.2.2, 3.3 and 3.4.  The change in the valence is also correlated with the volume expansion.  The value of the valence change is smaller than those observed upon the valence transition. For example it is from 3.03 (at 0.15 GPa) to 3.19 (at 2 GPa) upon the $\gamma-\alpha$ transition in Ce\cite{Rueff06} and from 2.84 to 2.96 accompanied with the metamagnetic transition in YbInCu$_4$\cite{Matsuda07}.  Correspondingly, the volume changes upon these valence transitions are larger than that in CeRu$_2$Si$_2$ by one order of magnitudes.  

As described in the following sections, the metamagnetic transition in CeRu$_2$Si$_2$ accompanies many anomalous features of the Fermi surface properties such as the effective mass enhancement, a drastic change in the Fermi surface properties and so on which are sometimes observed at the quantum critical point.  To our understanding, the mechanism of the metamagnetic transition as well as that of the anomalous behavior around the metamagnetic transition is still very controversial and there have been many proposals for the mechanism\cite{Ohkawa89,Ohkawa92,Ono98,Ohara99,Watanabe00,Satoh01,Miyake06,Bauer09,Ohara09,Weickert10,Aoki11,Bercx12,Howczak12,Kubo13a,Kubo13b}.  The Fermi surface properties in fields above the metamagnetic transition may be the key feature to understand the mechanism but even the interpretation of the observation for the Fermi surface properties seems to be still unsettled as described in $\S$ 3.4.

\subsection{Magnetic phase diagrams of CeRu$_2$(Si$_x$Ge$_{1-x}$)$_2$ and Ce$_x$La$_{1-x}$Ru$_2$Si$_2$}

\subsubsection{Magnetic phase diagram of CeRu$_2$(Si$_x$Ge$_{1-x}$)$_2$}
\begin{figure}[htbp]
\begin{center}
\includegraphics[width=0.6\linewidth]{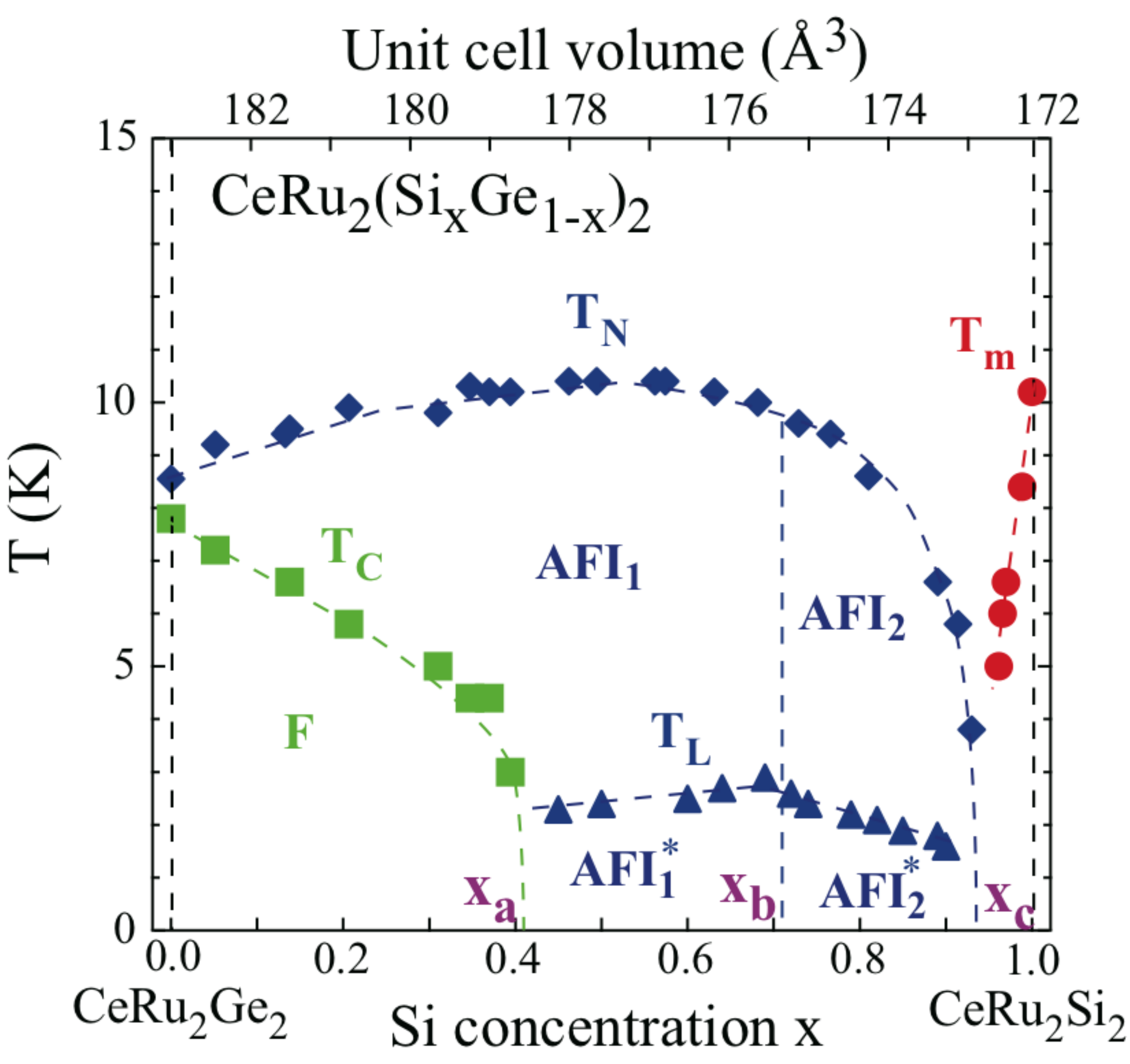}
\end{center}
\caption{(Color on line) Magnetic phase diagram of CeRu$_2$(Si$_{x}$Ge$_{1-x}$)$_2$ plotted on temperature vs. Si concentration $x$ (bottom axis) or unit cell volume (top axis) plane\cite{Matsumoto11}.  $T_c$ and $T_N$ are the ferromagnetic and antiferromagnetic transition temperatures, respectively.  $T_L$ is the transition temperature between two antiferromagnetic phases.  $T_m$ is the temperature where the susceptibility vs. temperature curve becomes maximum.  The broken lines are guides to the eye. }
\label{fig:90025Fig3}
\end{figure}

Figure \ref{fig:90025Fig3} shows the magnetic phase diagram of CeRu$_2$(Si$_x$Ge$_{1-x}$)$_2$ as a function of temperature and Si concentration $x$\cite{Matsumoto11}.  This system has the ThCr$_2$Si$_2$ structure irrespective of the concentration $x$.  The upper abscissa shows the unit cell volume of the crystal.  By replacing Ge by Si, the unit cell volume decreases or chemical pressure is applied. 

CeRu$_2$Ge$_2$ orders antiferromagnetically at 8.5 K and then ferromagnetically at 7.5 K.  With increasing Si concentration, the antiferromagnetic transition temperature $T_{\rm N}$ first increases and becomes maximum around $x$ = 0.6.  Then, it decreases and vanishes around $x_c$ = 0.935.  The ground state changes from ferromagnetic to antiferromagnetic at $x_a$ = 0.42.  The transition at $x_a$ seems to be first order, although that is not proved.  In the antiferromagnetic phase, another magnetic transition takes place at $T_{\rm L}$ with decreasing temperature.  This transition cannot be detected by magnetic susceptibility and magnetization measurements, but can be detected by specific heat, thermal expansion and transport measurements\cite{Haen99}.  Since the value of $T_{\rm L}$ at $x_a$ is approximately the same as that of the ferromagnetic transition temperature $T_{\rm c}$ at $x_a$, the magnitude of $T_{\rm L}$ seems to be determined by the competition between antiferromagnetic and ferromagnetic interactions.  The magnetic phase diagram in the antiferromagnetic state seems to change qualitatively around $x_b$ = 0.71 as shown later in Fig. \ref{fig:90025Fig10}.  It is also noted that $T_{\rm L}$ becomes maximum at $x_b$, although the maximum of $T_{\rm N}$ resides at the lower concentration side. The magnetic structure have not been revealed thoroughly, although the main propagation vector is thought to be $\mbox{\boldmath${q}$}=(0.31,0,0)$ in the ground states of antiferromagnetic phases\cite{Mignot90,Mignot91}.  The transition at $T_{\rm L}$ has probably the same character as that observed in Ce$_x$La$_{1-x}$Ru$_2$Si$_2$ which is described in the following section.  It is reported that a squaring up of the spin density modulation may take place at $T_{\rm L}$\cite{Mignot90,Mignot91}.  

The lengths of the $a$ and $c$ axes decrease about 2\% from CeRu$_2$Ge$_2$ to CeRu$_2$Si$_2$, which gives rises to the compression of the volume by about 6\%.  The $c/a$ ratio changes from 2.353 of CeRu$_2$Ge$_2$ to 2.335 of CeRu$_2$Si$_2$, indicating that the compression along the $c$ axis is slightly larger than that along the $a$ axis. The magnetic phase diagram of CeRu$_2$Ge$_2$ is studied as a function of pressure using transport measurements.  It is reported that the magnetic phase diagram as a function of pressure and temperature is similar to that in Fig. \ref{fig:90025Fig3} and that the CeRu$_2$Ge$_2$ becomes paramagnetic around 7.8 - 8.7 GPa\cite{Wilhelm99,Wilhelm04}.  We may assume that the substitution of Ge by Si gives nearly the same effect on the magnetic properties as that of pressure.  We may also assume that the substitution does not affect the Fermi surface properties significantly except for the chemical pressure effect, because the electronic structures of Si and Ge are similar and the contribution to the electronic structure at the Fermi surface from Si or Ge can be assumed to be small\cite{Runge95}. We also plot the temperature $T_{\rm m}$ in Fig. \ref{fig:90025Fig3}.  The pressure studies on CeRu$_2$Si$_2$ shows that $T_{\rm m}$ further increases with pressure\cite{Mignot88}. 

\subsubsection{Magnetic phase diagram of Ce$_x$La$_{1-x}$Ru$_2$Si$_2$}
\begin{figure}[htbp]
\begin{center}
\includegraphics[width=0.5\linewidth]{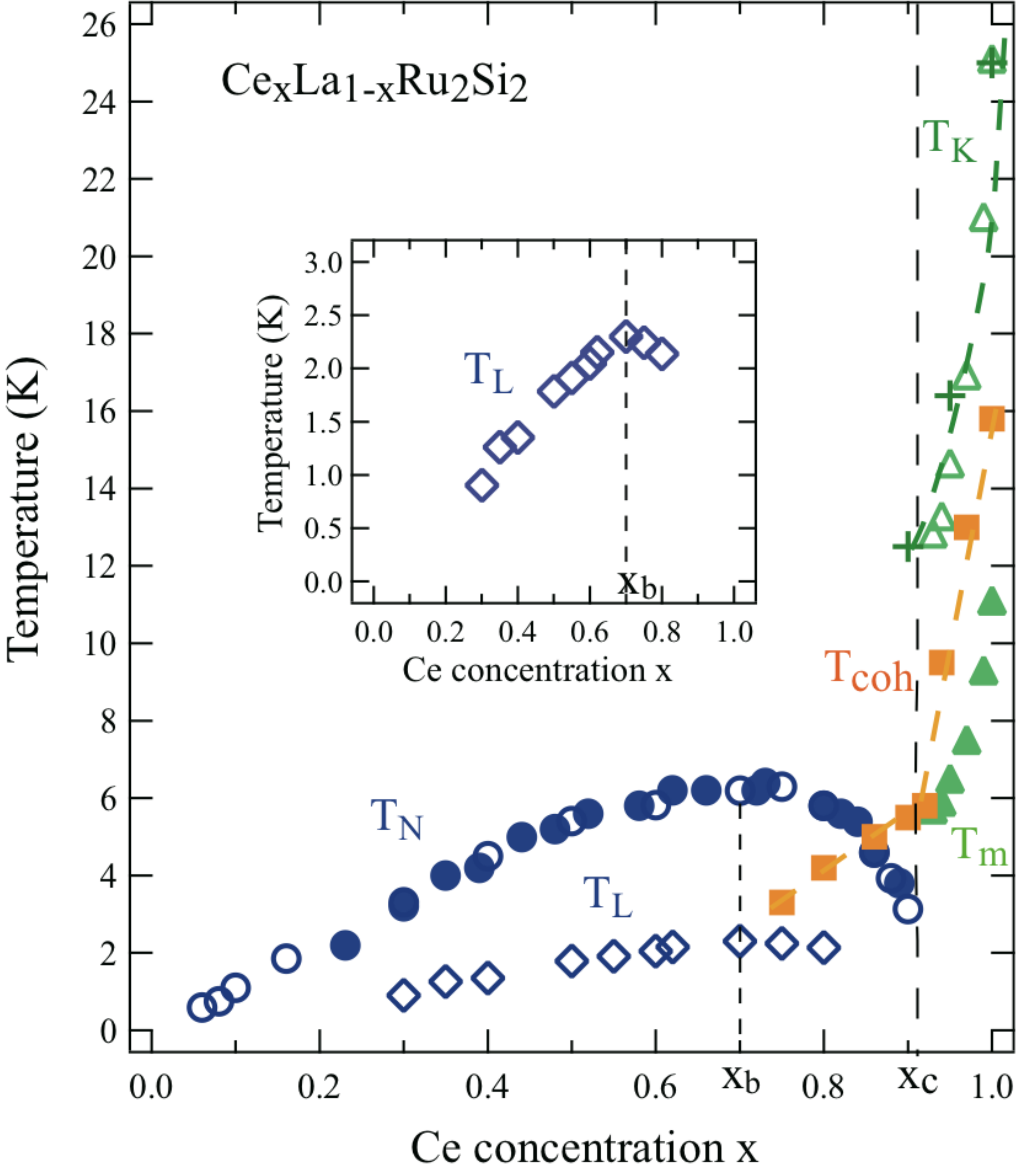}
\end{center}
\caption{(Color online) Magnetic phase diagram of Ce$_x$La$_{1-x}$Ru$_2$Si$_2$ as a function of temperature and Ce concentration\cite{Shimizu12}.  $T_{\rm N}$ (circle) and $T_{\rm L}$ (diamond) are antiferromagnetic transition temperatures.  The closed and open symbols denote the transition temperatures determined by magnetic and resistivity measurements, respectively. $T_{\rm m}$ (closed triangle) is the temperature where the susceptibility vs. temperature curve becomes maximum.  $T_{\rm K}$ is the Kondo temperature determined by specific heat measurement\cite{Fisher91} (cross) or is derived from the value of $T_{\rm m}$ (open triangle).  $T_{\rm coh}$ is the coherence temperature derived from the maximum of the Hall coefficient vs. temperature curve.  See also text.  The broken lines are guides to the eye.  The inset shows the enlarged view for $T_{\rm L}$ to show that $T_{\rm L}$ becomes maximum around $x_b$.}
\label{fig:90025Fig4}
\end{figure}

Figure \ref{fig:90025Fig4} shows the magnetic phase diagram of Ce$_x$La$_{1-x}$Ru$_2$Si$_2$ as a function of temperature and Ce concentration $x$\cite{Shimizu12}.  This system has also the ThCr$_2$Si$_2$ structure.  The unit cell volume of LaRu$_2$Si$_2$ is  176.4 ${\rm \AA}^3$\cite{Pearson91}and that of CeRu$_2$Si$_2$ is 171.3${\rm \AA}^3$\cite{Severing89}.   The unit cell volume of the alloy changes almost linearly with $x$.  Since the ionic size of Ce is smaller than that of La, the substitution of La by Ce gives chemical pressure effect in addition to the effect that the inter-site interaction develops among the Ce atoms.  In low Ce concentration samples, no magnetic order develops and the dilute Kondo behavior is observed as described later in $\S$2.2.3 and $\S$4.1.1.  With increasing $x$ anitiferromagnetic order develops and $T_{\rm N}$ increases with $x$ and has a maximum at around $x$ = 0.75.  Then it decreases with $x$ and disappears at $x_c$ = 0.91.   $T_{\rm L}$ becomes maximum around $x_b = 0.7$.  The main propagation vector in the ground state is the same $\mbox{\boldmath${q}$}=(0.31,0,0)$ as that in CeRu$_2$(Si$_x$Ge$_{1-x}$)$_2$.  The magnetic phase diagram for the antiferromagnetic ground state is similar to that of CeRu$_2$(Si$_x$Ge$_{1-x}$)$_2$, probably because for high Ce concentration samples the chemical pressure effect rather than the inter-site effect determines the magnetic properties effectively. 

We also include $T_{\rm m}$, $T_{\rm K}$ and $T_{\rm coh}$ in Fig. \ref{fig:90025Fig4}.  The plus symbol (+) for $T_{\rm K}$ denotes the value determined by the specific heat measurements\cite{Fisher91}, while the open triangle denotes the value derived from the value of $T_{\rm m}$ by assuming that $T_{\rm m}$ is proportional to $T_{\rm K} $ and that the proportional constant for the samples with $x < 1.0$ is the same as that of CeRu$_2$Si$_2$.  Both the data points lie well on a single curve indicating that the proportionality between $T_{\rm m}$ and $T_{\rm K}$ holds well in the paramagnetic state with $x > x_c$.  $T_{\rm coh}$ is the coherence temperature as described below.  

\begin{figure}[htbp]
\begin{center}
\includegraphics[width=0.6\linewidth]{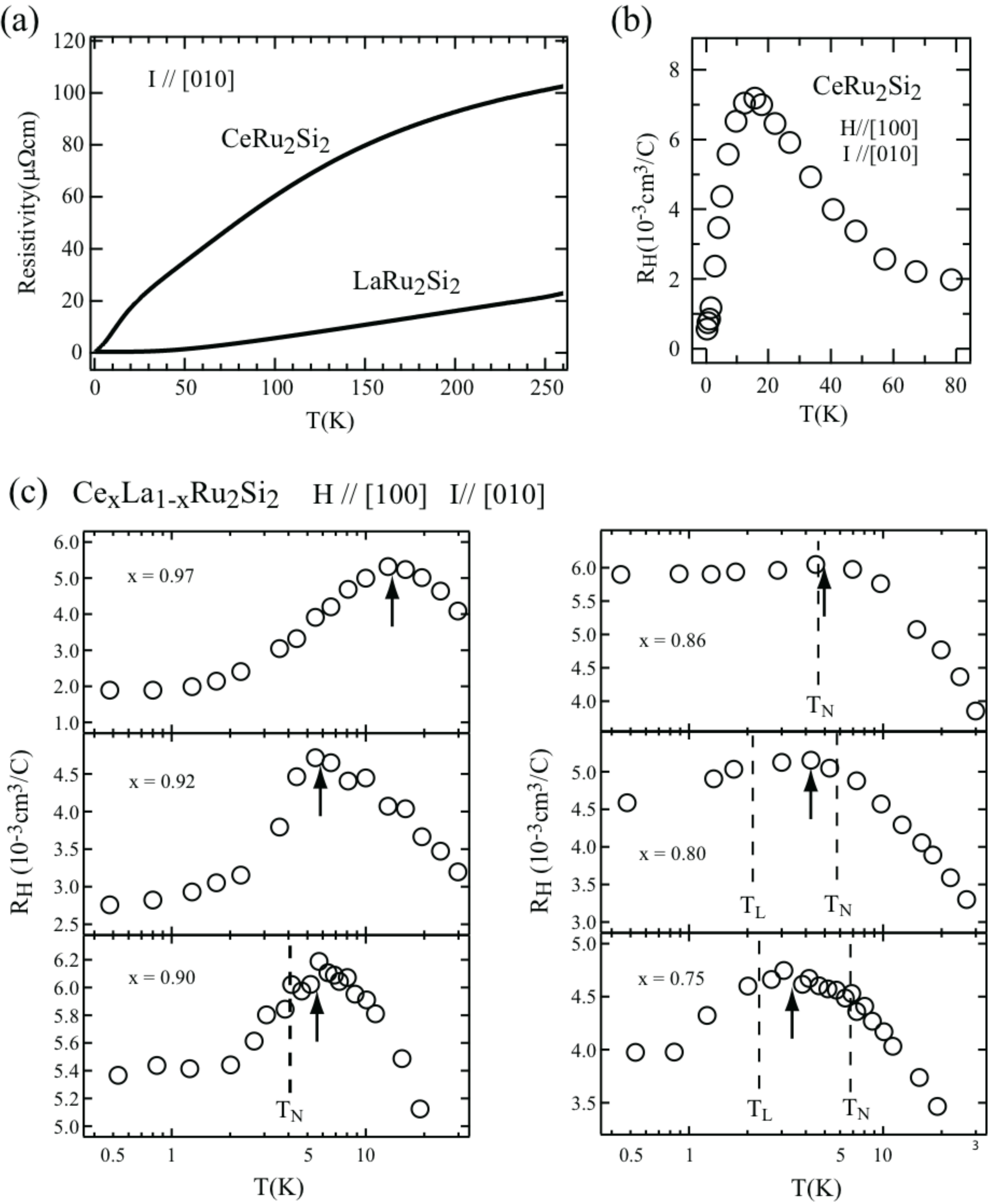}
\end{center}
\caption{(a) Resistivities of CeRu$_2$Si$_2$ and LaRu$_2$Si$_2$ as a function of temperature.  (b) Variation of the Hall coefficient as a function of temperature in CeRu$_2$Si$_2$.  (c)Temperature variation of the Hall coefficient for some selected Ce$_x$La$_{1-x}$Ru$_2$Si$_2$ samples.  The arrows indicate the positions of maximum and the broken lines indicate the antiferromagnetic transition temperatures\cite{Shimizu12}.}
\label{fig:90025Fig5}
\end{figure}

Figure \ref{fig:90025Fig5}(a)  shows the temperature variations of resistivity in CeRu$_2$Si$_2$ and LaRu$_2$Si$_2$.  The resistivity of CeRu$_2$Si$_2$ has a convex structure and starts to decrease more rapidly from about 20 K which is attributed to the formation of  coherence.  The onset of the coherence in a heavy Fermion compound can be  more obviously observed as a peak in the Hall coefficient vs. temperature curve \cite{Hadzic86,Fert87,Lapierre87} as shown in Fig. \ref{fig:90025Fig5}(b) for CeRu$_2$Si$_2$.  A similar peak can be also observed for $x < x_c$ when a magnetic field is applied in the (001) plane as shown in Fig. \ref{fig:90025Fig5}(c). We plot the temperature of the peak as $T_{\rm coh}$  in Fig. \ref{fig:90025Fig4}.   If we compare the unit cell volume of LaRu$_2$Si$_2$ with those of CeRu$_2$(Si$_x$Ge$_{1-x}$)$_2$, we note that the CeRu$_2$(Si$_x$Ge$_{1-x}$)$_2$ sample with the volume of LaRu$_2$Si$_2$ is near $x_b$ in the AFI$_1^\ast$ phase.  It is therefore reasonable that no ferromagnetic order develops in Ce$_x$La$_{1-x}$Ru$_2$Si$_2$ with increasing $x$ or decreasing volume.  

\subsubsection{Transport properties of CeRu$_2$(Si$_x$Ge$_{1-x}$)$_2$ and Ce$_x$La$_{1-x}$Ru$_2$Si$_2$}

\begin{figure}[htbp]
\begin{center}
\includegraphics[width=0.5\linewidth]{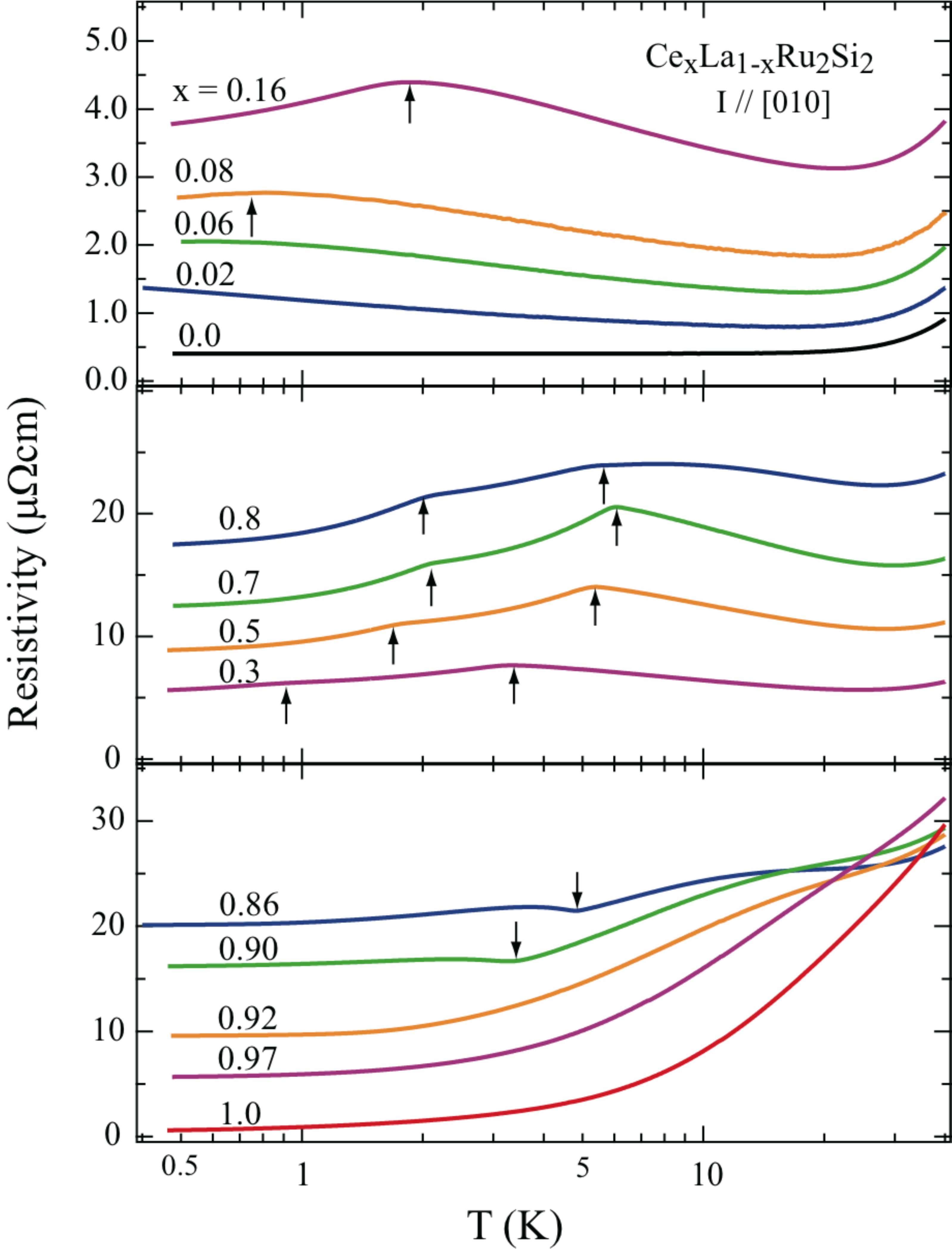}
\end{center}
\caption{(Color online) Temperature variations of resistivity at low temperatures for some selected Ce$_x$La$_{1-x}$Ru$_2$Si$_2$ samples\cite{Shimizu12}.  The arrow indicates the peak or dip position corresponding to the onset of short or long range antiferromagnetic order.}
\label{fig:90025Fig6}
\end{figure}

Figure \ref{fig:90025Fig6} shows the resistivity vs temperature curves for various concentration samples ranging from $x $= 0.0 to 1.0 in Ce$_x$La$_{1-x}$Ru$_2$Si$_2$.  In dilute Ce concentration samples, a typical behavior of the dilute Kondo alloy can be observed. That is, the resistivity increases with decreasing temperature proportionally to $- \log T$ and becomes nearly constant at lowest temperatures as shown for the sample with $x $ = 0.02 on the top panel of Fig. \ref{fig:90025Fig6} and later in Fig. \ref{fig:90025Fig32}. With further increase in Ce concentration, decrease of resistivity is observed at lowest temperatures indicating that a magnetic order develops, while the increase of resistivity proportional to $- \log T$ is observed at higher temperatures.  With further increase in Ce concentration another drop of resistivity is observed at a lower temperature below $T_{\rm N}$ as shown on the middle panel of Fig. \ref{fig:90025Fig6}.  The temperature corresponds to $T_{\rm L}$ in Fig. \ref{fig:90025Fig4}.   In the higher Ce concentration samples, the resistivity does not increase with decreasing temperature and the transition can be detected by observation of a dip rather than the hump as shown on the bottom panel of Fig. \ref{fig:90025Fig6}.    Considering that the increase in the resistivity upon the transition can be observed typically at the SDW transition, the change in the behavior of resistivity upon the transition may imply that the f electron becomes more itinerant at $T_{\rm N}$ compared with the lower concentration samples.  When $T_{\rm coh} > T_{\rm N}$,  the resistivity has a dip upon the antiferromagnetic transition, while it has a hump when $T_{\rm coh} < T_{\rm N}$ as noted from Figs. \ref{fig:90025Fig4}, \ref{fig:90025Fig5} and \ref{fig:90025Fig6}.  Therefore, $T_{\rm coh}$ for $x < x_c$ may be also interpreted as a measure of the temperature where the coherence starts to develop.  

\begin{figure}[htbp]
\begin{center}
\includegraphics[width=0.5\linewidth]{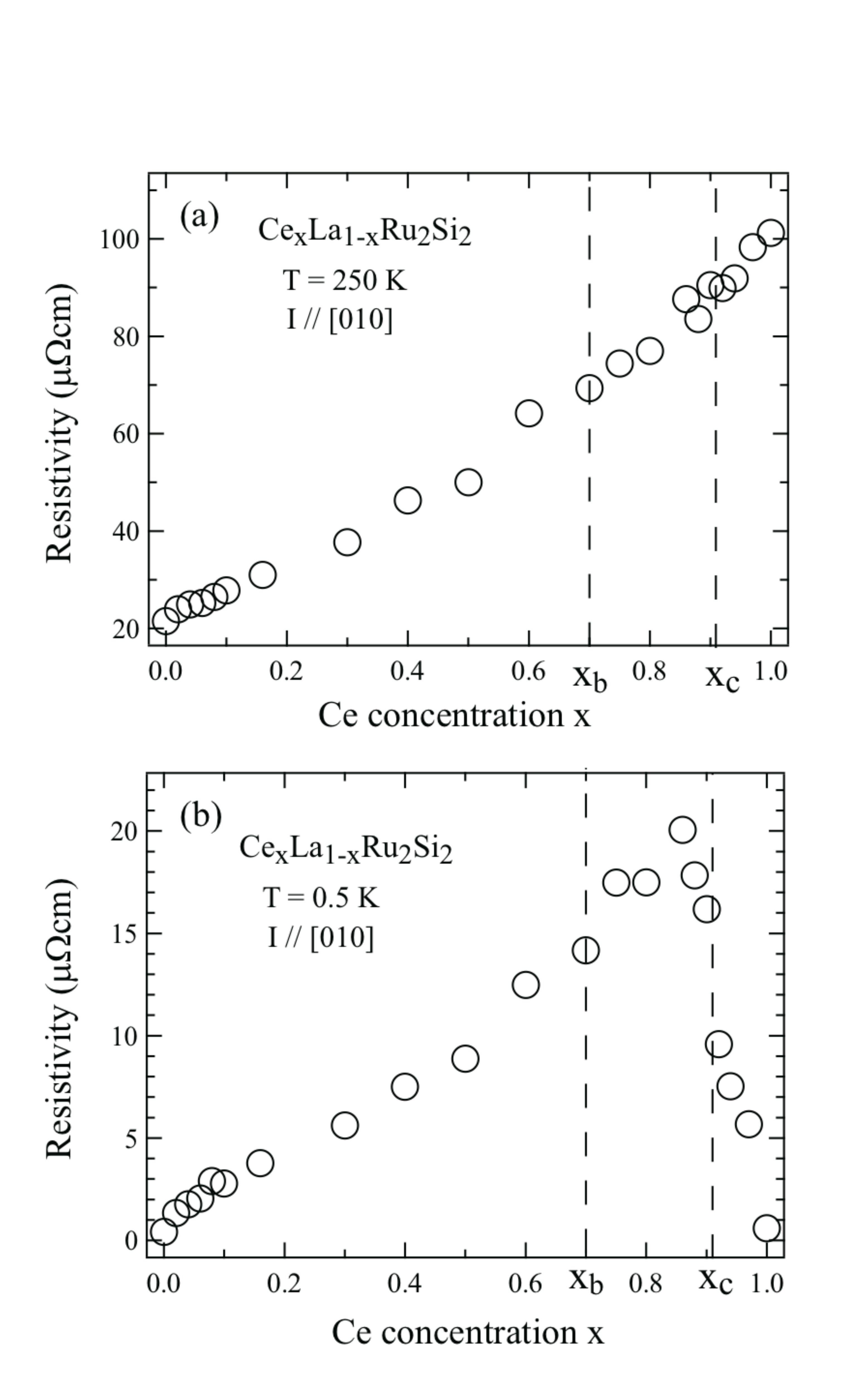}
\end{center}
\caption{Resistivities (a) at 250 K and (b) at 0.5 K as a function of Ce concentration\cite{Shimizu12}.}
\label{fig:90025Fig7}
\end{figure}

When $T_{\rm coh}$ is approximately equal to $T_{\rm N}$, it is found that the transport properties like residual resistivity, magnetoresistivity and Hall resistivity are anomalously enhanced\cite{Shimizu12}.  We show in Fig. \ref{fig:90025Fig7}(a) and (b) the resistivities of Ce$_x$La$_{1-x}$Ru$_2$Si$_2$ at 250 K and 0.5 K as a function of Ce concentration.  We may assume that the resistivity at 0.5 K is almost the same as the residual resistivity. The resistivity at 250 K is proportional to the Ce concentration while the residual resistivity has a maximum in the antiferromagnetic state. The maximum of the residual resistivity in a normal metal alloy is normally expected to appear around $x = 0.5$, which is known as the Nordheim law.  The similar anomalous enhancements of the residual resistivity in the antiferromagnetic state are also observed for CeCu$_5$Au\cite{Wilhelm01} and YbNi$_2$Ge$_2$\cite{Knebel01}.  The observation is attributed to an enhanced  charge fluctuation and resultant enhanced impurity scattering due to the competition between the antiferromagnetic interaction and the Kondo effect when $T_{\rm N} \approx T_{\rm K}$\cite{Hattori10}.   It is noted that the $T_{\rm coh}$ line seems to meet the maximum of $T_L$ around $x_b$ and $T_{\rm L}$ is suppressed with increasing $x$ for $x > x_b$.  For the samples with $x >x_c$, the resistivity decreases monotonically with decreasing temperature indicating that no magnetic order is present. 

\begin{figure}[htbp]
\begin{center}
\includegraphics[width=0.5\linewidth]{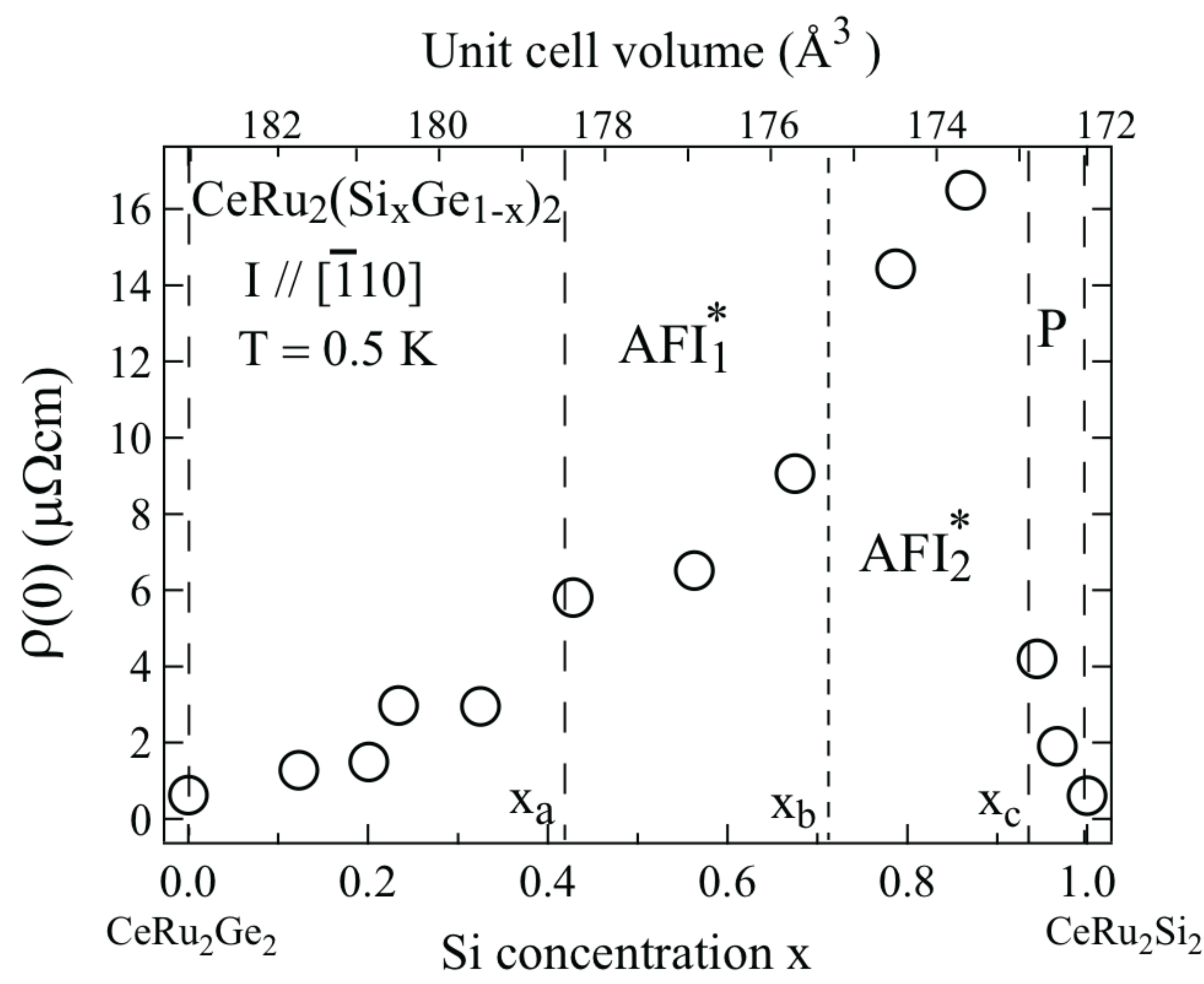}
\end{center}
\caption{Resistivities at 0.5 K as a function of Si concentration\cite{Matsumoto11}.}
\label{fig:90025Fig8}
\end{figure}

We also show the residual resistivity of CeRu$_2$(Si$_x$Ge$_{1-x}$)$_2$ in Fig. \ref{fig:90025Fig8} plotted against Si concentration.  It is likely that the maximum exists in the antiferromagnetic state. The resistivity as a function of temperature shows the similar behavior to those observed in Ce$_x$La$_{1-x}$Ru$_2$Si$_2$, i.e. the resistivity shows a hump at $T_{\rm N}$ in the antiferromagnetic state for $x < x_b$, but shows a dip for $x$ which is close to $x_c$.  The anomalous enhancements of transport properties similar to those found in Ce$_x$La$_{1-x}$Ru$_2$Si$_2$ are also observed near $x_c$ in the antiferromagnetic state\cite{Matsumoto11}. 

It is noted in passing that the behavior of residual resistivity as well as the phase diagram as a function of the concentration of alloying element differs depending on the alloy system.  For example, in Ce$_x$La$_{1-x}$Cu$_6$,  it is reported that LaCu$_6$ evolves to CeCu$_6$ without magnetic order and the residual resistivity has a maximum around $x = 0.5$\cite{Sumiyama86}.   The difference from the Ce$_x$La$_{1-x}$Ru$_2$Si$_2$ system may come from the following fact.  The crystal structure of LaCu$_6$ is monoclinic, while that of  CeCu$_6$ is orthorhombic at room temperature.  However, the distortion from orthorhombic is very small and the crystal structure of CeCu$_6$ becomes monoclinic at low temperatures. Moreover, the difference between the unit cell volumes of LaCu$_6$ and CeCu$_6$ is smaller than 0.1 \% \cite{Asano85}.

From the phase diagrams of CeRu$_2$(Si$_x$Ge$_{1-x}$)$_2$ and Ce$_x$La$_{1-x}$Ru$_2$Si$_2$, it is often stated that CeRu$_2$Si$_2$ is situated close to the quantum critical point.  While the quantum phase transition at $x_c$ in Ce$_x$La$_{1-x}$Ru$_2$Si$_2$ could be weakly first order\cite{Knafo09}, the quantum critical behavior around $x_c$ is successfully discussed on the basis of conventional scenario of quantum critical point, mostly in terms of the self consistent renormalization (SCR) theory\cite{Moriya95,Moriya03}. We will not go into the details of the discussions but refer the reader to the references\cite{Kambe96,Kadowaki04,Lohneysen07,Knafo09}.

\subsection{Metamagnetic transitions in CeRu$_2$(Si$_x$Ge$_{1-x}$)$_2$ and Ce$_x$La$_{1-x}$Ru$_2$Si$_2$}

\subsubsection{Metamagnetic transitions and magnetic phase diagram}
\begin{figure}[htbp]
\begin{center}
\includegraphics[width=0.6\linewidth]{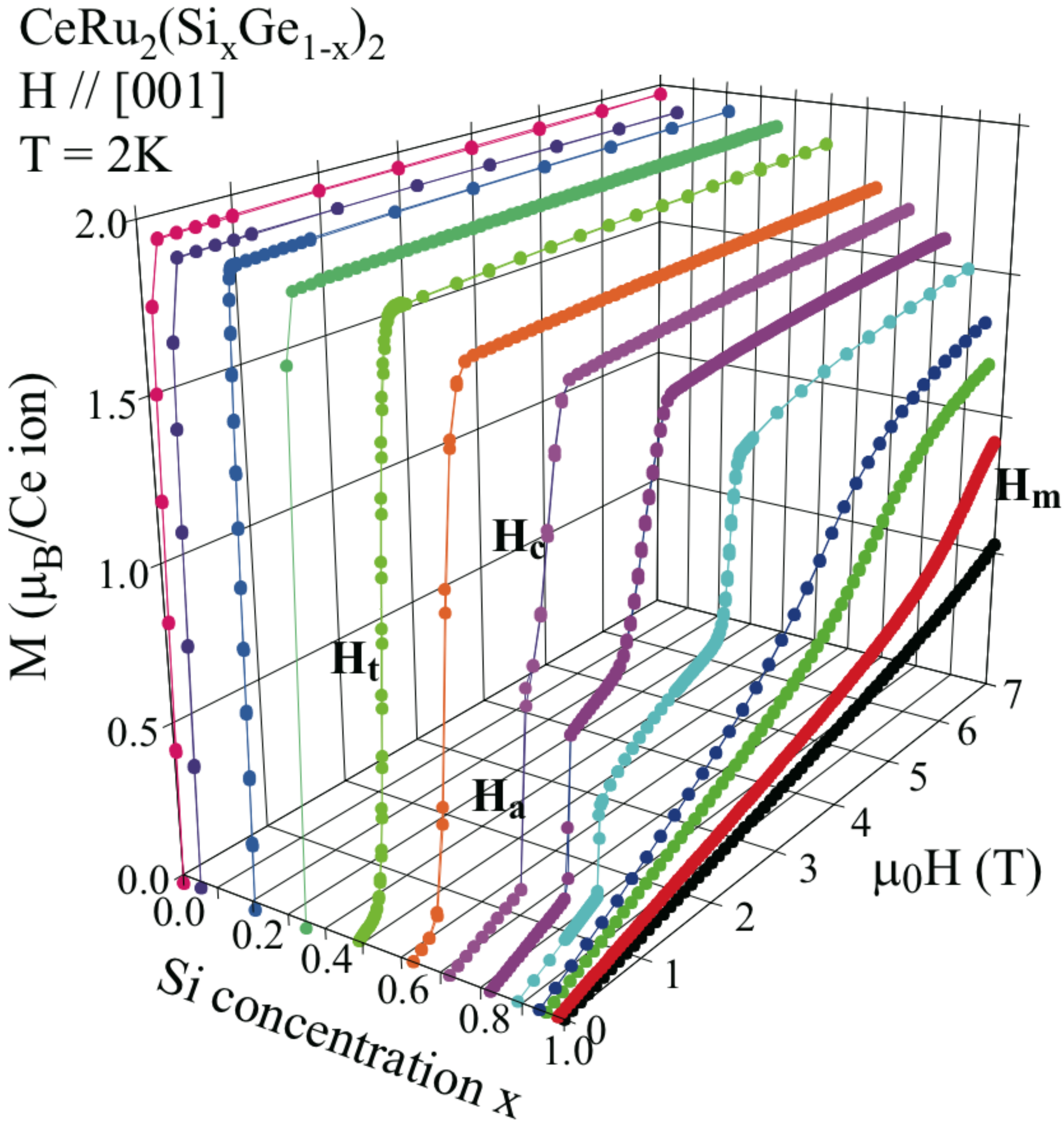}
\end{center}
\caption{(Color on line) Magnetization curves of  CeRu$_2$(Si$_{1-x}$Ge$_x$)$_2$ samples plotted as a function of magnetic field applied parallel to the [001] direction at 2 K.  Each curve includes both the data points measured with increasing and decreasing magnetic fields between 0 and 7 T\cite{Matsumoto11}. }
\label{fig:90025Fig9}
\end{figure}

Both CeRu$_2$(Si$_x$Ge$_{1-x}$)$_2$ and Ce$_x$La$_{1-x}$Ru$_2$Si$_2$ have the strong magnetic anisotropy. When a magnetic field is applied along the [001] direction, metamagnetic transition takes place as observed in CeRu$_2$Si$_2$.  On the other hand, when a magnetic field is applied in the (001) plane, the magnetization increases gradually with field and no metamagnetic transition has been observed. 
To demonstrate an overall metamagnetic behavior at low temperatures, we show in Fig. \ref{fig:90025Fig9} the DC magnetization of CeRu$_2$(Si$_x$Ge$_{1-x}$)$_2$ samples as a function of magnetic field for fields parallel to the [001] direction\cite{Matsumoto11}.  Magnetic fields are applied up to 7 T at 2 K.  Each magnetization curve includes both the data points in increasing and decreasing field processes. For $x$ = 0, the sample is in the ferromagnetic phase at 2 K.  With application of a small magnetic field the magnetization increases abruptly and becomes nearly constant.  The value of magnetization is about 1.95 $\mu_B$/Ce at 7 T.  The value is close to the value of 2.14 $\mu_B$/Ce expected for the pure $|J = \pm 5/2>$ state.   For the samples in the ferromagnetic ground states ($x$ = 0.07, 0.20 and 0.35), the feature of magnetization curve is similar to that of $x$ = 0, but the value of magnetization at 7 T slightly decreases with increasing $x$.   For $x$ = 0.48 and 0.62,  the sample is in the antiferromagnetic phase AFI$^\ast_1$.  The magnetization first increases rather rapidly with magnetic field and then exhibits a sharp increase at around a field denoted by $H_{\rm t}$.  Then it becomes nearly constant.  The sharp increase around $H_{\rm t}$ has weak but appreciable hysteresis which is not very obvious in the figure. 

\begin{figure}[htbp]
\begin{center}
\includegraphics[width=0.6\linewidth]{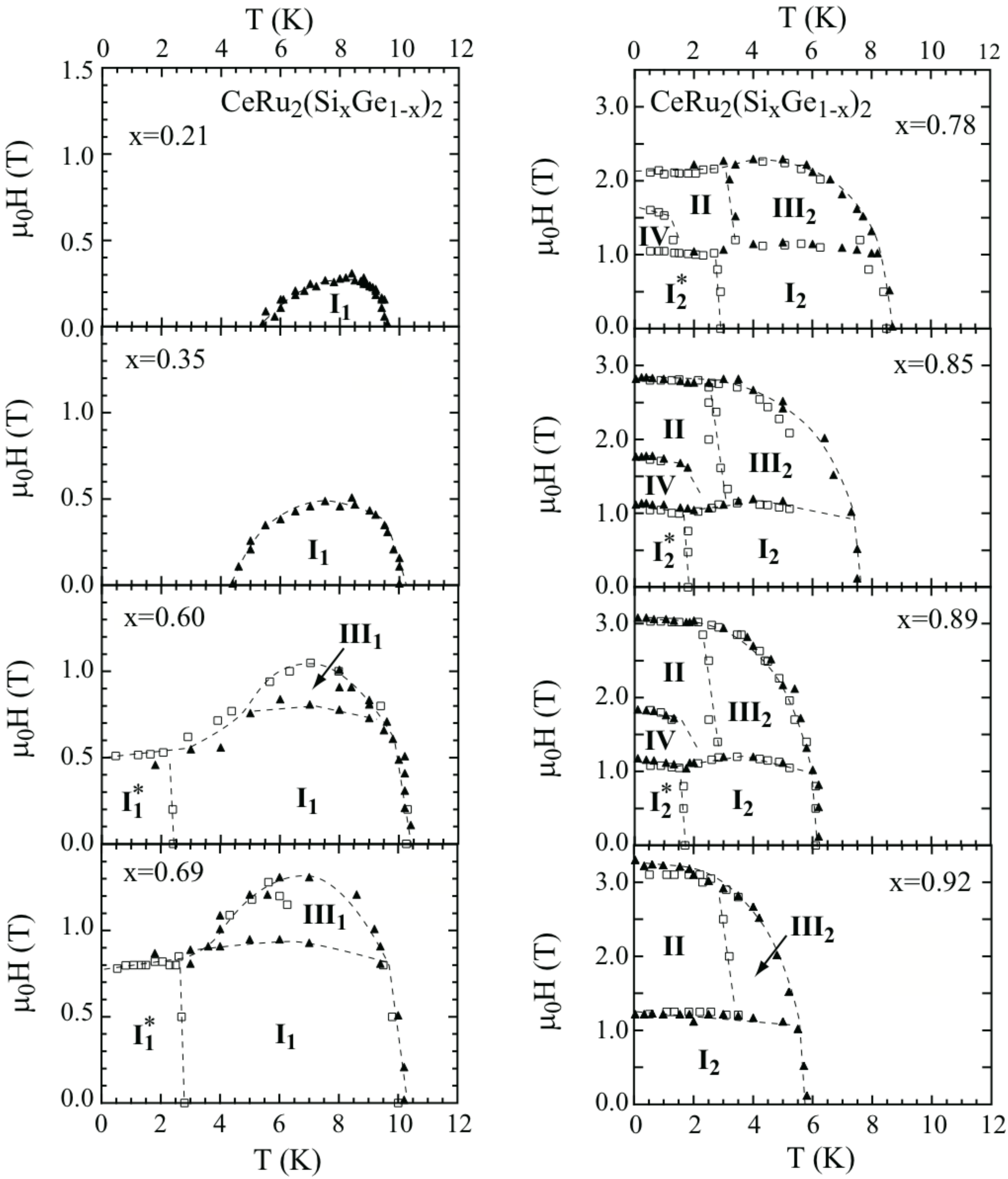}
\end{center}
\caption{Magnetic phase diagrams of CeRu$_2$(Si$_{1-x}$Ge$_x$)$_2$ as a function of  magnetic field and temperature for some selected samples with $x$ = 0.21, 0.35, 0.60, 0.69, 0.78, 0.85, 0.89 and 0.92\cite{Matsumoto11}.  Magnetic fields are applied parallel to the [001] direction.  The closed triangles denote the data obtained from magnetic measurements and the open squares denote the data obtained from the resistivity measurements. The names of the phases are denoted after Haen\cite{Haen99} except for phase IV.}
\label{fig:90025Fig10}
\end{figure}

\begin{figure}[htbp]
\begin{center}
\includegraphics[width=0.6\linewidth]{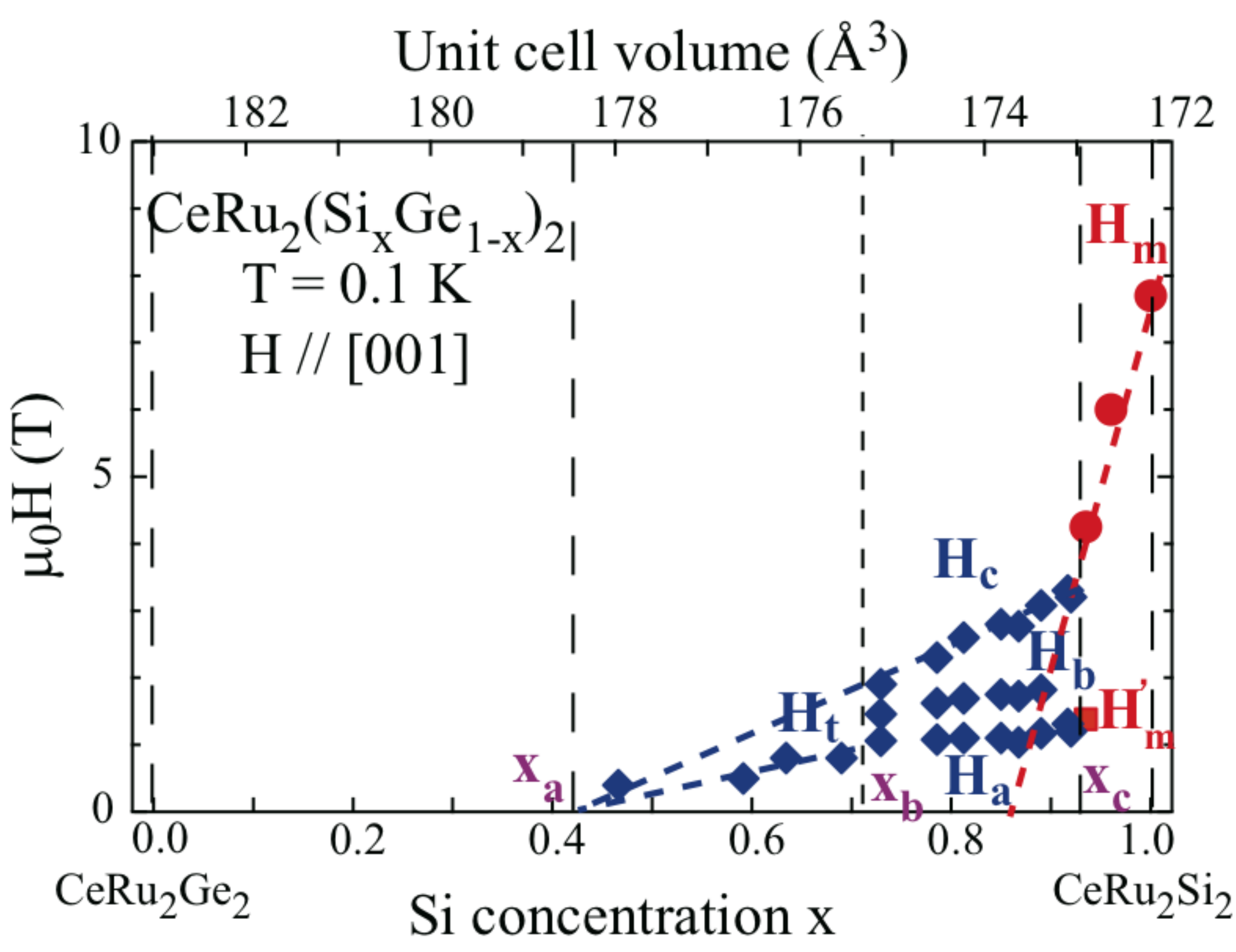}
\end{center}
\caption{(Color on line) Magnetic phase diagram of CeRu$_2$(Si$_x$Ge$_{1-x}$)$_2$ plotted on magnetic field vs. Si concentration $x$ (bottom axis) or unit cell volume (top axis) plane\cite{Matsumoto11}.  $H_{\rm a}$, $H_{\rm b}$ and $H_{\rm c}$ are first order metamagnetic transition fields. $H_{\rm m}$ and $H_{\rm m}^{\prime}$ are metamagnetic crossover fields.  The broken lines are the guides to the eye.}
\label{fig:90025Fig11}
\end{figure}

For $x$ = 0.72, 0.82 and 0.89, the sample is near the boarder between the AFI$^\ast_2$ and AFI$_2$ phases or in the AFI$_2$ phase at 2 K as noted from Fig. \ref{fig:90025Fig3}.  The magnetization rapidly increases around the two magnetic fields denoted by $H_{\rm a}$ and $H_{\rm c}$. Each magnetization curve has hystereses around $H_{\rm a}$ and $H_{\rm c}$ as well as the fields in between.  $H_{\rm a}$ and $H_{\rm c}$ shift to higher fields with increasing $x$.  The magnetization curves for $x $ = 0.82 and 0.89 are rounded between $H_{\rm a}$ and $H_{\rm c}$.  This feature arises from another metamagnetic transition at a field $H_{\rm b}$ between them which appears at lower temperatures below 1 K\cite{Sugi08} as shown in Figs. \ref{fig:90025Fig10} and \ref{fig:90025Fig11}.  

For $x$ = 0.94, 0.96, 0.98 and 1.0, the sample is in the paramagnetic phase at 2 K. The magnetization increases gradually with magnetic field. 
$H_{\rm m}$ shifts to higher fields with increasing $x$ and is above 7 T for $x = 0.98$ and 1.0.    The initial susceptibility increases with decreasing $x$ towards $x_c$ indicating that the density of states at the Fermi energy is enhanced around $x_c$.  It is noted that the magnetization at 7 T decreases continuously with increasing $x$, implying that the f electron state changes continuously with $x$.  The change in the magnetization across the metamagnetic transition also decreases with $x$ implying that the magnetic moment of Ce at the transition decreases with $x$.  It is also noted that the magnetization increases almost linearly with field at low fields, while that above the metamagnetic transition field is not linear.  These observations will be referred when we discuss the electronic structures above the metamagnetic transitions.

The magnetization of  CeRu$_2$Si$_2$  reaches about 1.5 $\mu_B$/Ce at 19 T \cite{Mignot89} and does not seem to saturate. The value is larger than the value of 1.29 $\mu_B$/Ce expected for the pure $|J = \pm 3/2>$ state,  but is smaller than the value of  2.14 $\mu_B$/Ce expected for the pure $|J = \pm 5/2>$ state or that of CeRu$_2$Ge$_2$.  

The transition at $H_{\rm b}$ becomes less obvious upon approaching $x_c$,  i.e. the peak height of AC susceptibility becomes smaller and the width of the peak becomes broader, and it disappears near $x_c$.  The one at $H_{\rm a}$ also becomes less obvious.  However, in addition to the large peak at $H_{\rm m}$  a tiny but still appreciable peak at $H_{\rm m}^\prime$ is observed in the AC susceptibility at low temperatures below 1 K in the  sample with $x = 0.94 > x_c$\cite{Sugi08}.   With increasing temperature the crossover feature at $H_{\rm m}^\prime$ disappears and can not be observed above 1 K.  It is difficult to determine experimentally whether the feature observed at $H_{\rm m}^\prime$ is intrinsic or is due to an inhomogeneity in the sample.  

We show magnetic phase diagrams as a function of temperature and magnetic field for some representative concentration samples in Fig. \ref{fig:90025Fig10}.  For $x < x_b$,  there is only one phase at lowest temperatures, while there are more phases for $x > x_b$.  Upon approaching $x_c$, the number of the phases decreases at lowest temperatures or the phase boundaries become difficult to detect.  The end point of the first order transition or the tricritical point may be present somewhere on the phase boundary between the paramagnetic and antiferromagnetic states\cite{Stryjewski}, but it has not been identified in CeRu$_2$(Si$_{1-x}$Ge$_x$)$_2$ or Ce$_x$La$_{1-x}$Ru$_2$Si$_2$. 

We plot $H_{\rm t}$, $H_{\rm a}$, $H_{\rm b}$, $H_{\rm c}$ and $H_{\rm m}$ determined from the AC susceptibility measurements at about 0.1 K in Fig. \ref {fig:90025Fig11}.  $H_{\rm t}$ increases almost linearly to $x_b$ from $x_a$.  The single step transition changes to the three-step transition around  $x_b$.   The metamagnetic transition at $H_{\rm c}$ becomes weakly first order upon approaching $x_c$ and changes to the crossover at $H_{\rm m}$ almost exactly at $x_c$.  The increasing rate of the metamagnetic transition field with respect to increasing $x$ changes both at $x_b$ and $x_c$.  However, if we extend the slope of $H_{\rm c}$ to the smaller $x$ side,  it seems to meet $x_a$ at zero magnetic field. These features of $H_{\rm t}$, $H_{\rm c}$ and $H_{\rm m}$ may imply that the metamagnetic transition is somehow related with the ferromagnetism of this system.  

\begin{figure}[htbp]
\begin{center}
\includegraphics[width=0.6\linewidth]{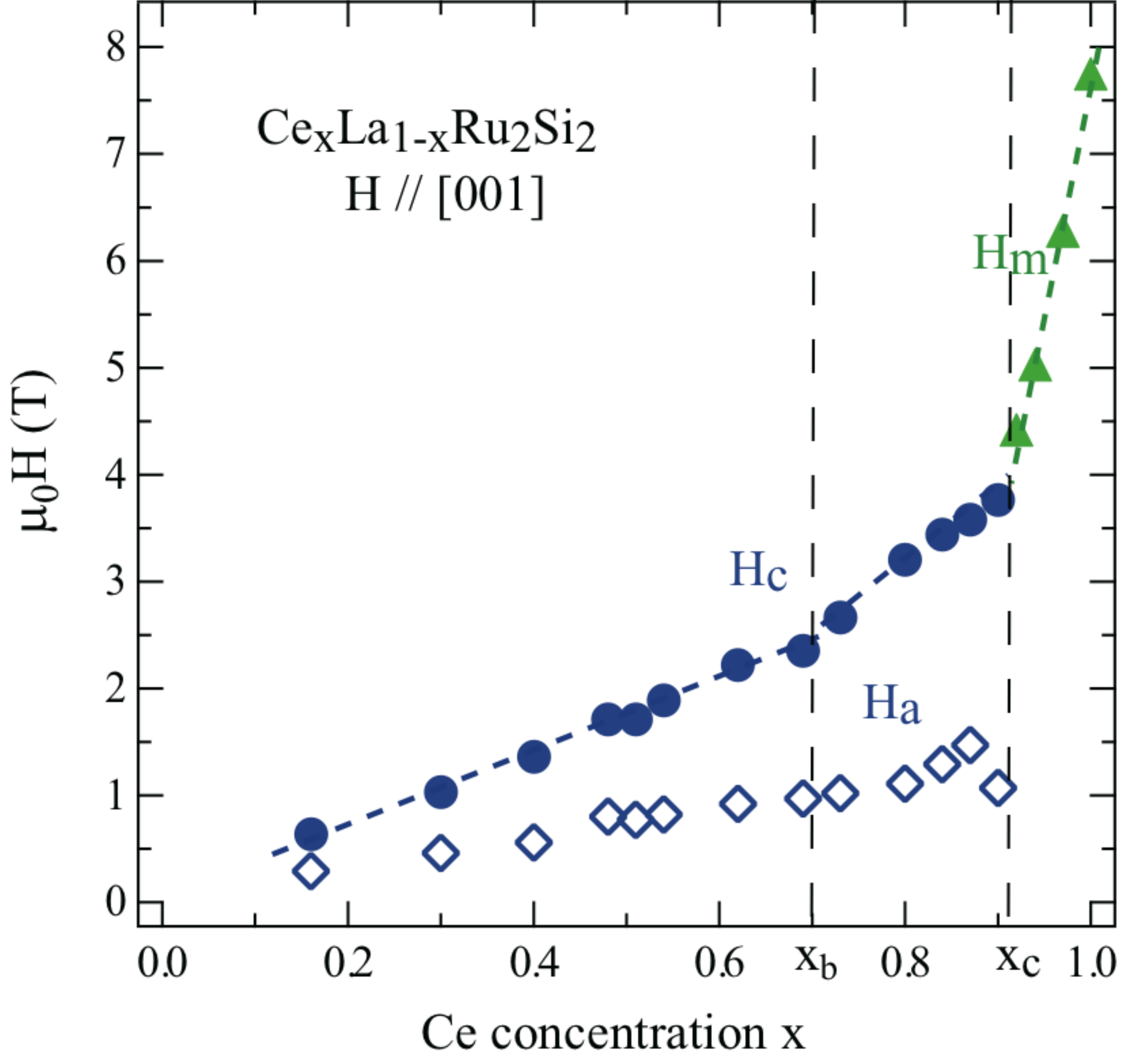}
\end{center}
\caption{(Color on line) Magnetic phase diagram of Ce$_x$La$_{1-x}$Ru$_2$Si$_2$ as a function of magnetic field and Ce concentration\cite{Shimizu12}. $H_{\rm a}$  and $H_{\rm c}$ are first order metamagnetic transition fields. $H_{\rm m}$ is metamagnetic crossover field.  The broken lines are guides to the eye.}
\label{fig:90025Fig12}
\end{figure}

We also plot the metamagnetic transition fields at 0.1 K in Ce$_x$La$_{1-x}$Ru$_2$Si$_2$ as a function of Ce concentration in Fig. \ref{fig:90025Fig12}. The magnetic phase diagrams of Ce$_x$La$_{1-x}$Ru$_2$Si$_2$ as a function of temperature and magnetic field are simpler than those of CeRu$_2$(Si$_x$Ge$_{1-x}$)$_2$ and has  two phases at lowest temperatures\cite{MatsumotoM}.  The first order metamagnetic transition at $H_{\rm c}$  changes to the metamagnetic crossover almost exactly at $x_c$ as in the case of CeRu$_2$(Si$_x$Ge$_{1-x}$)$_2$.  It is also noted that the slope of the $H_{\rm c}$ line changes around $x = x_b (=0.7)$ probably corresponding to the change in the metamagnetic behavior at $x_b$ in CeRu$_2$(Si$_x$Ge$_{1-x}$)$_2$. 

\begin{figure}[htbp]
\begin{center}
\includegraphics[width=0.5\linewidth]{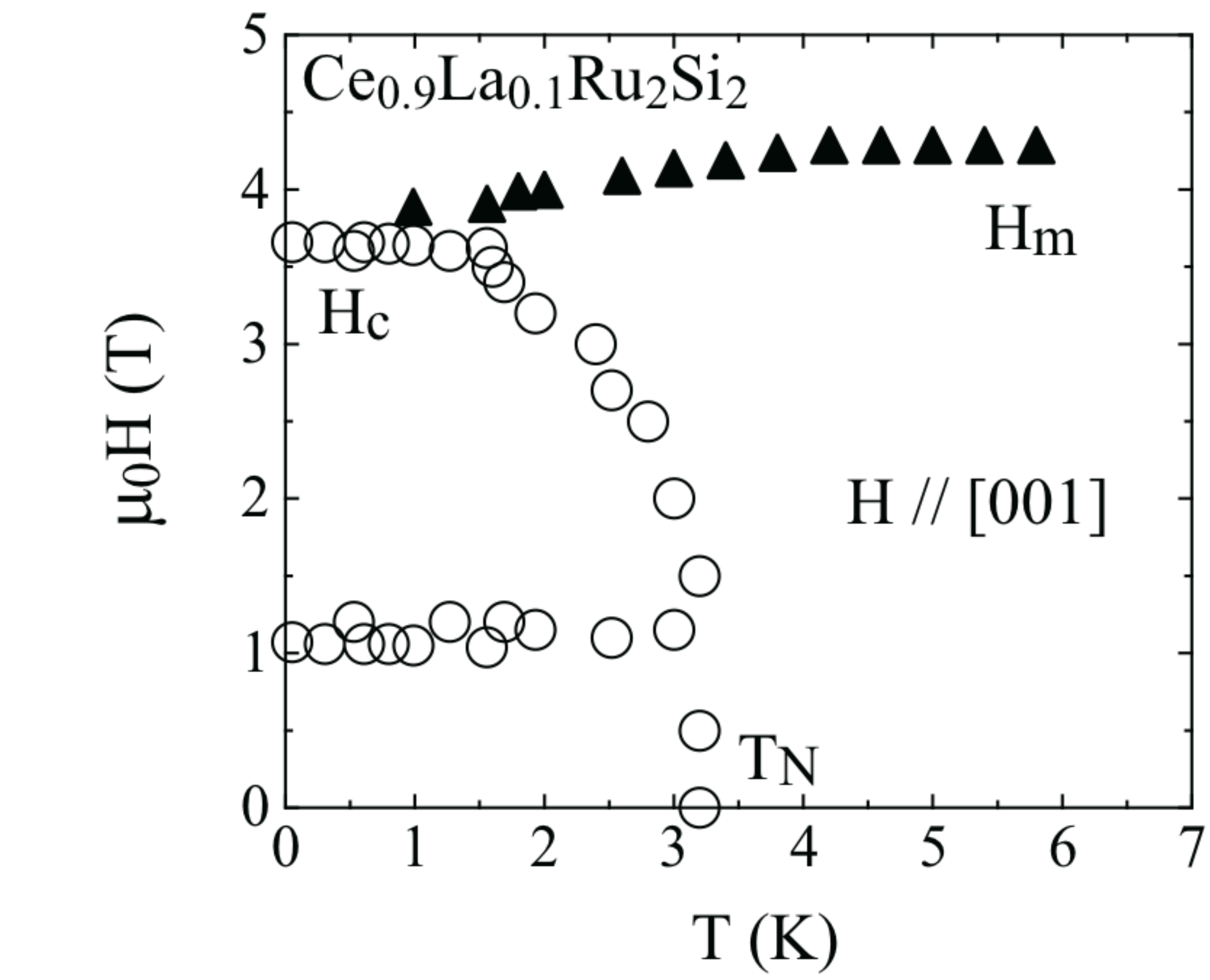}
\end{center}
\caption{Magnetic phase diagram of Ce$_{0.9}$La$_{0.1}$Ru$_2$Si$_2$\cite{MatsumotoM}. }
\label{fig:90025Fig13}
\end{figure}

While at lowest temperatures the first order metamagnetic transition seems to change to the metamagnetic crossover at $x_c$, the metamagnetic crossover remains for $x <x_c$ at finite temperatures.  Figure \ref{fig:90025Fig13} shows the magnetic phase diagram of $x = 0.9$ which is close to $x_c$ but is smaller than $x_c$\cite {MatsumotoM}.   The transition at $T_{\rm L}$ is not obvious for this concentration.  In addition to the phase boundary of the antiferromagnetic state (the open circles),  metamagnetic crossover (closed triangles) is observed at higher temperatures.  At low temperatures, the crossover changes to the first order transition\cite{Haen90,Fisher91,MatsumotoM}.  The separation between the metamagnetic crossover and the phase boundary of the antiferromagnetic state is more clearly demonstrated in  Ce(Ru$_{0.92}$Rh$_{0.08}$)$_2$Si$_2$\cite{Aoki12}.  It is also noted that for $x_b < x < x_c $,  $H_{\rm c}$ increases with increasing $x$ although $T_{\rm N}$ or $T_{\rm L}$  decreases with increasing $x$.  That is, the behavior of the first order metamagnetic transition is different from that in an ordinary  antiferromagnet  where the metamagnetic transition field is larger for the larger $T_{\rm N}$. 
 
\subsubsection{Metamagnetic transitions and magnetism}

\begin{figure}[htbp]
\begin{center}
\includegraphics[width=0.6\linewidth]{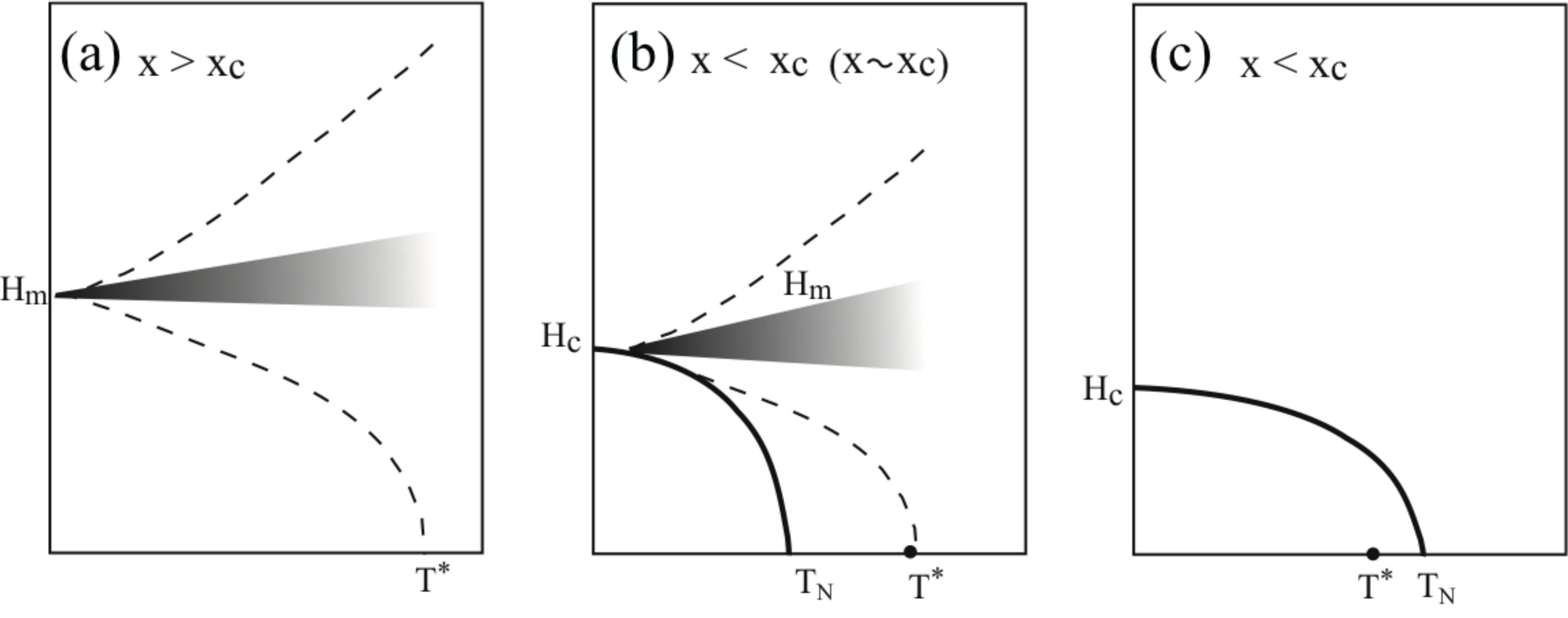}
\end{center}
\caption{(a) Schematic phase diagrams for $x > x_c$, (b) for $x < x_c$ but close to $x_c$, and (c) $x > x_c$.   }
\label{fig:90025Fig14}
\end{figure}

\begin{figure}[htbp]
\begin{center}
\includegraphics[width=0.6\linewidth]{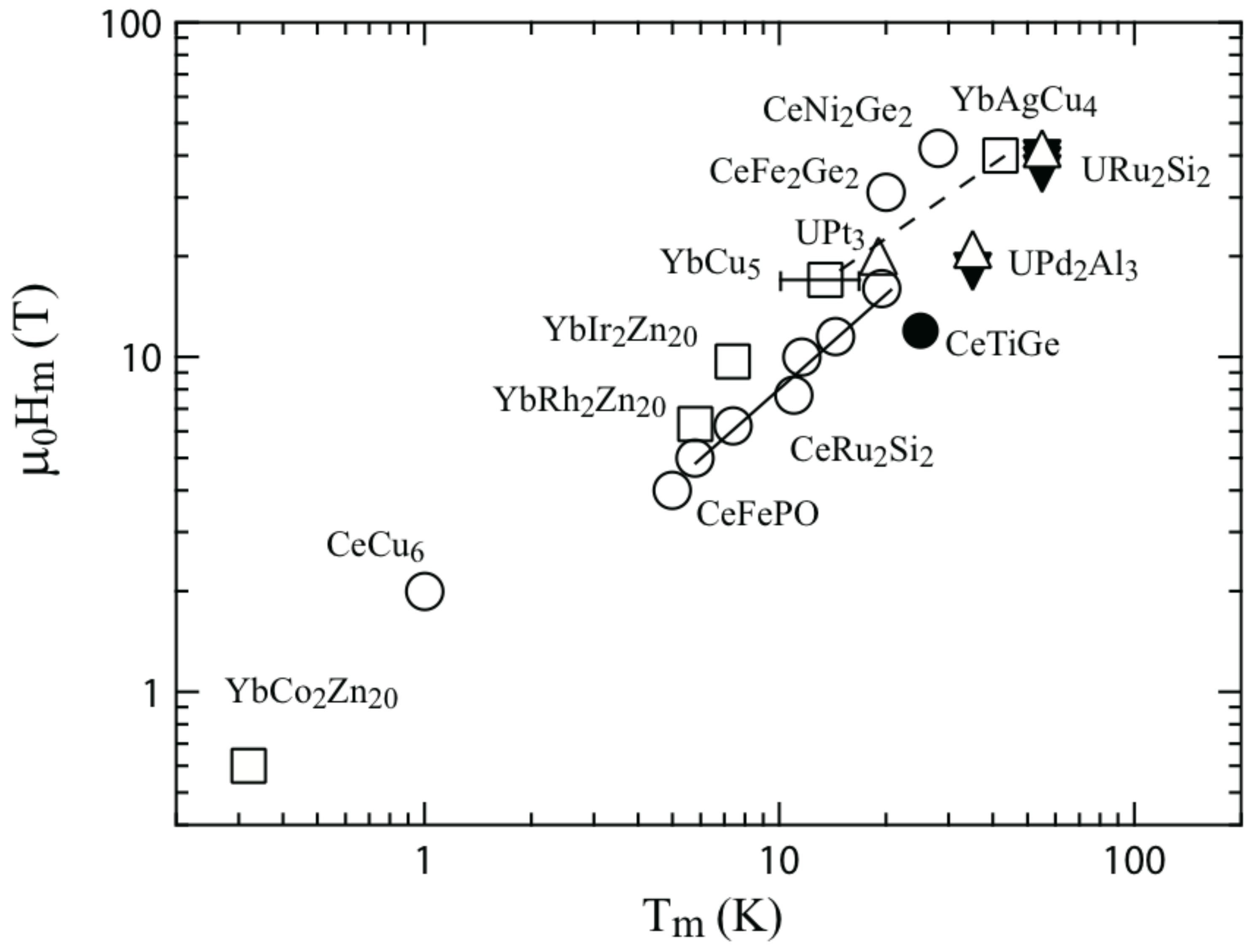}
\end{center}
\caption{$H_{\rm m}$ versus $T_{\rm m}$ of various heavy Fermion compounds. (CeCu$_6$\cite{Onuki87}, CeNi$_2$Ge$_2$\cite{Fukuhara96},  CeFe$_2$Ge$_2$\cite{Sugawara00}, CeRu$_2$Si$_2$ and Ce$_x$La$_{1-x}$Ru$_2$Si$_2$\cite{Mignot88,Aoki01,Matsumoto08}, CeFePO\cite{Kitagawa11}, CeTiGe\cite{Deppe12}, YbAgCu$_{4}$, YbCu$_{5}$, YbAg$_x$Cu$_{5-X}$\cite{Tsujii97b,Tsujii01,Mito12}, YbT$_2$Zn$_{20}$ (T = Co, Rh, Ir)\cite{Torikachvili07,Takeuchi10,Ohya10,Onuki11}, UPt$_3$\cite{Sugiyama99}, URu$_2$Si$_2$\cite{Sugiyama99b}, UPd$_2$Al$_3$\cite{Sugiyama00}), $T_{\rm m}$'s of   CeCu$_6$, CeFe$_2$Ge$_2$ are determined from the temperatures where the susceptibility as a function of temperature starts to saturate. For URu$_2$Si$_2$ and UPd$_2$Al$_3$, the first order transition field from the ordered state is denoted by closed triangle.  In these compounds the cross over field $H_{\rm m}$ like that shown in Fig. \ref{fig:90025Fig13} is observed. They are denoted by the open triangles.  It is reported that CeTiGe does not exhibit magnetic order but exhibits the first order metamagnetic transition.  The solid line for CeRu$_2$Si$_2$ is the guide to the eye for the data points under pressure and those of Ce$_x$La$_{1-x}$Ru$_2$Si$_2$.  The broken line denotes that various data points of YbAg$_x$Cu$_{5-x}$ lie approximately on this line. }
\label{fig:90025Fig15}
\end{figure}

We describe how the heavy Fermion state, antiferromagnetism  and ferromagnetism are related with the metamagneic crossover.  
Figures \ref{fig:90025Fig14}(a), (b) and (c) show  schematic phase diagrams of  Ce$_x$La$_{1-x}$Ru$_2$Si$_2$ as a function of magnetic field and temperature for  $x > x_c$,  $x < x_c $ but $x $ is close to $x_c$,  and $x < x_c$, respectively.    The broken line denotes crossover line and is drawn schematically from the points where thermal expansion coefficient shows anomaly\cite{Holtmeier95} or $C/T$ becomes maximum\cite{Meulen91,YAoki98}.  The temperature for the end point at zero magnetic field of the crossover line is denoted as $T^\ast$.  It is reported to be about 9 K in CeRu$_2$Si$_2$\cite{Holtmeier95} which is comparable to $T_{\rm m}$ or $T_{\rm coh}$ in Fig. \ref{fig:90025Fig4}. We may regard $T^\ast$ as a measure below which a coherent heavy Fermion state is formed.   It is reported that the lower part and the upper part of the enclosed area by the broken line are connected at lowest temperatures\cite{Holtmeier95,Ishida98,Daou06}.  The shape of the enclosed area  is also similar to the area where the resistivity shows the Fermi liquid behavior, although in this case $T^\ast$ is about 0.5 K\cite{Daou06}.  The shaded band shows schematically the behavior of the metamagnetic crossover.  That is,  (1) the metamagnetic transition is very sharp at low temperatures but still has a finite transition width at the lowest temperatures,  and (2) the width becomes broader with increasing temperature and the transition behavior becomes obscured at temperatures comparable to $T^\ast$ or $T_{\rm K}$.   It is reported in CeRu$_2$Si$_2$ and other heavy Fermion compounds \cite{Mignot88,Continentino93}  that  $T_{\rm m}$  as well as $H_{\rm m}$, $1/\sqrt{A}$, $1/\chi_0$ increase with increasing pressure and the proportionality constant among them is nearly the same at all pressures.  Figure \ref{fig:90025Fig15} plots $H_{\rm m}$ against $T_{\rm m}$ observed in several heavy Fermion compounds under positive and negative (chemical) pressures as well as at ambient pressure.   A good correlation is observed between  $H_{\rm m}$ and $T_{\rm m}$.  

Figure\ref{fig:90025Fig14}(b) corresponds to the phase diagram  of Fig. \ref{fig:90025Fig13}.   The solid line denotes the phase boundary of the antiferromagnetic state.  If we assume that $T^\ast$ is represented by $T_{\rm m}$ or $T_{\rm coh}$, we note from Fig. \ref{fig:90025Fig4} that they are larger than $T_{\rm N}$ for $x =0.9$.  This situation is shown in the schematic phase diagram (b).  When $x < x_c$ and $x$ is not close to $x_c$, metamagnetic crossover is not observed and $T_{\rm m}$ or $T_{\rm coh}$ is smaller than $T_{\rm N}$.   This situation is illustrated in  Fig. \ref{fig:90025Fig14}(c).   

From the observations above, we may assume that the metamagnetic crossover is not directly related with the transition from the antiferromagnetic state but is related with the heavy Fermion state whose characteristic energy is reflected in $T^\ast$, $T_{\rm coh}$ or $T_{\rm m}$.   The metamagnetic crossover is likely to take place when the Zeeman split state of the ground state characterized by $T_{\rm m}$ shifts to another state by magnetic field.  

 Next we pay attention to the relation between ferromagnetism and metamagnetism considering the phase diagram of CeRu$_2$(Si$_x$Ge$_{1-x}$)$_2$ system.  In pyrite compounds Co(Si$_{1-x}$Se$_x$)$_2$ and Laves phase compounds Y(Co$_{1-x}$Al$_x$)$_2$, Lu(Co$_{1-x}$Al$_x$)$_2$, first order metamagnetic transition takes place from the paramagnetic ground state.  These compounds are on the verge of ferromagnetism and its mechanism is attributed to the peculiar pseudo gap structure in the density of states near the Fermi level\cite{Yamada93}.  It is reported that there is also a good proportionality between $H_{\rm m}$ and $T_{\rm m}$.  But the proportionality constant $\mu_0H_{\rm m}/T_{\rm m}$ is about 1/2 - 1/3 of that in CeRu$_2$Si$_2$\cite{Sakakibara90}, implying that the energy scale relevant to metamagnetism is different but the mechanism may not be very different.  

We have not determined $T_{\rm coh}$ in CeRu$_2$(Si$_x$Ge$_{1-x}$)$_2$ system.  From the similarity between the magnetic phase diagrams and properties of Ce$_x$La$_{1-x}$Ru$_2$Si$_2$ and CeRu$_2$(Si$_x$Ge$_{1-x}$)$_2$, we may assume that a similar relation to that observed in Ce$_x$La$_{1-x}$Ru$_2$Si$_2$ system holds among $T_{\rm N}$, $T_{\rm L}$ and $T_{\rm coh}$.  That is when $T_{\rm N} \approx T_{\rm coh}$, the residual resistivity becomes maximum implying that the charge fluctuation is enhanced.  $T_{\rm L}$ becomes maximum at $x_b$ when $T_{\rm L} \approx T_{\rm coh}$ implying that $T_{\rm L}$ is suppressed by the interaction whose magnitude is reflected in the value of $T_{\rm coh}$.  Below $T_{\rm coh}$ itinerant antiferromagnetic order may develop. For $x_a < x < x_b$,  the interaction corresponding to $T_{\rm coh}$ becomes less significant and the value of $T_{\rm L}$ seems to be mostly determined as a consequence of competition between ferromagnetic and antiferromagnetic interactions.
 
\begin{figure}[htbp]
\begin{center}
\includegraphics[width=0.6\linewidth]{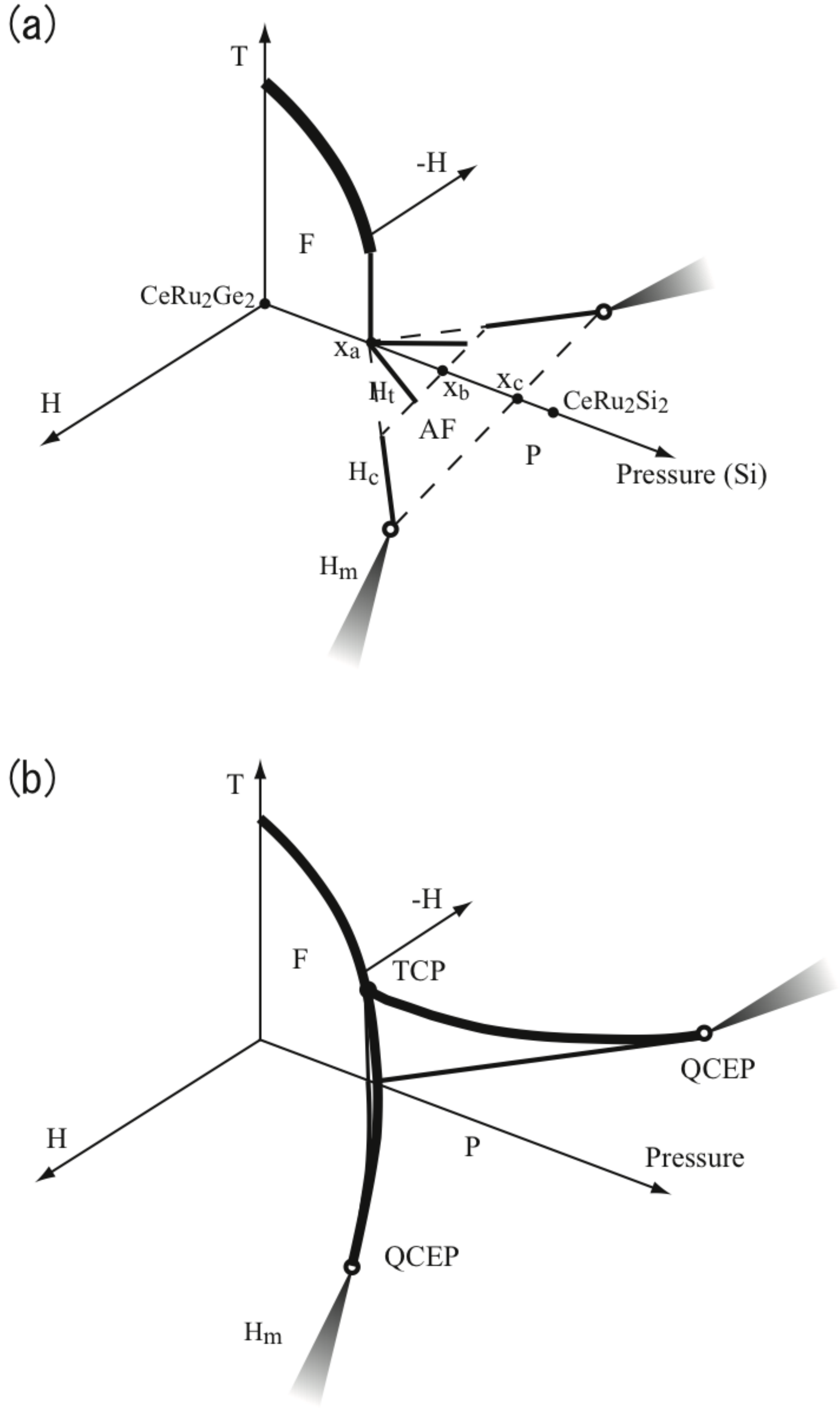}
\end{center}
\caption{Schematic phase diagrams of  (a) CeRu$_2$(Si$_x$Ge$_{1-x}$)$_2$ and (b) a weak ferromagnet.  Thick solid lines, thin solid lines and shaded bands denote the second order transition lines, the first order transition lines and the metamagnetic crossovers, respectively.  In (a) only the phase boundary of the ferromagnetic state and the metamagnetic transition fields are drawn.  In (b) the plane below the critical line TCP-QCEP separating the paramagnetic state from the field induced ferromagnetic state is first order transition planes.}
\label{fig:90025Fig16}
\end{figure}

If we remove the antiferromagnetic state from the phase diagrams of Figs. \ref{fig:90025Fig3}, \ref{fig:90025Fig10} and \ref{fig:90025Fig11}, and only retain the ferromagnetic phase and metamagnetic transitions of $H_{\rm t}$, $H_{\rm c}$ and $H_{\rm m}$ at lowest temperatures, we obtain a schematic $T - P - H$ phase diagram as shown in Fig. \ref{fig:90025Fig16}(a).  $H_{\rm t}$ increases almost linearly from $x_a$.  It is expected that $H_{\rm t}$ goes to zero at $x_a$, but experimentally it is not proved.  Together with the observation in Fig. \ref{fig:90025Fig10} that the metamagnetic transition at the lowest temperatures seems to be the revival of the ferromagnetic state under magnetic fields, this behavior indicates that the competition between antiferromagnetic and ferromagnetic interactions is relevant to the metamagnetic transition from the antiferromagnetic ground state.   The competition is also suggested from the behavior of $T_{\rm L}$ as described before.  

The relation among the ferromagnetic phase, first order metamagnetic transition and metamagnetic crossover somehow reminds us of the phase diagram of itinerant weak ferromagnet\cite{Belitz05,Belitz05b} like UGe$_2$\cite{Aoki11} and ZrZn$_2$\cite{Kabeya12} as shown schematically in Fig. \ref{fig:90025Fig16}(b).  In the case of the weak ferromagnet, the critical line vanishes at the quantum critical endpoint and  the metamagnetic crossover starts from the point.  In CeRu$_2$(Si$_x$Ge$_{1-x}$)$_2$, as stated in $\S$2.3.1 we have not identified the end point of the first order transition or the tricritical point on the phase boundary between the paramagnetic and antiferromagnetic states for each $x$.  Therefore, it is not determined how the critical line disappears, or whether or not the end point of the first order metamagnetic transition line is the quantum critical point.  Although the ground state below the first order metamagnetic transition is antiferromagnetic but not paramagnetic, it might be interesting to note that in ZrZn$_2$ antiferromagnetic fluctuation is reported to be present in the paramagnetic state\cite{Kabeya12}.  The partial similarity might imply that the physics behind are not very different among CeRu$_2$Si$_2$ alloys and the weak ferromagnets \cite{Hoshino13}.      

\subsection{An interpretation of magnetic phase diagram}
The magnetic phase diagram of  CeRu$_2$(Si$_x$Ge$_{1-x}$)$_2$  seems to suggest that the interactions responsible for the magnetic properties are naively composed of three types.  One is the interaction to favor antiferromagnetic order and the interaction could be described by the RKKY type interaction.  The second one is the hybridization or the Kondo effect which favors paramagnetic state or the itinerant behavior of the f electrons.  The behavior of the magnetic phase diagram for $x>x_a$ may be mostly understood in terms of the competition between these two interactions like the Doniach model.  Since for $x<x_a $ the ferromagnetic order is present, another mechanism which gives rise to the ferromagnetic interaction should be present.  The magnetic phase diagrams suggest that due to this mechanism the ground state changes from the antiferromagnetic state to the ferromagnetic state at $x_a$ through possibly first order transition with decreasing chemical pressure or increasing volume.   We speculate that the mechanism is operative irrespective of the range of $x$ and  may be also responsible for the polarized state under magnetic fields as suggested from the behavior of the metamagnetic transition fields.   

Hoshino and Kuramoto successfully explain the magnetic phase diagram of CeRu$_2$(Si$_x$Ge$_{1-x}$)$_2$ under zero magnetic field by using the Kondo-Heisenberg model\cite{Hoshino13}.   By using a similar model, we conjecture that the change of the phase diagram with $x$ comes from the change in the relative strengths of the interactions as follows.  We denote the strengths of the antiferromagnetic interaction,  the ferromagnetic interaction, and the Kondo effect as $I_{\rm AF}$, $I_{\rm F}$ and $I_{\rm K}$.    For $x < x_a$,  the phase diagram may be understood by the competition of the three interactions with the relative strengths of the three interactions being $I_{\rm F} > I_{\rm AF} > I_{\rm K}$.  For $x_a < x < x_b$, $I_{\rm AF} > I_{\rm F} > I_{\rm K}$ and the main competition between $I_{\rm F}$ and $I_{\rm AF}$ would lead the observed behaviors of  $T_{\rm L}$ and the dominant interaction $I_{\rm AF}$ would give rise to the behavior of $T_{\rm N}$.  Since the phase diagram of the antiferromagnetic state for $x_b < x <x_c$ changes from that for $x_a < x < x_b$,  the relative strengths could change to $I_{\rm AF} \ge I_{\rm K} > I_{\rm F}$. The change in the strength of $I_{\rm K}$ with $x$ may be reflected in the experimentally observed values of  $T_{\rm K}$, $T_{\rm m}$ or $T_{\rm coh}$.  For $x > x_c$, $I_{\rm K} > I_{\rm AF} > I_{\rm F}$ and the ground state is paramagnetic.  Since the relation between the magnetic phase diagrams of  CeRu$_2$(Si$_x$Ge$_{1-x}$)$_2$ and Ce$_x$La$_{1-x}$Ru$_2$Si$_2$ is like that shown in $\S$2.2.2,  these three interactions can be assumed to be present in Ce$_x$La$_{1-x}$Ru$_2$Si$_2$ although no ferromagnetic order is present. 

If we assume that the interaction $I_{\rm F}$ gives rise to the polarized state with the help of magnetic field or volume expansion competing with the dominant interaction in each case, the behavior of the metamagnetic field as a function of $x$ may be qualitatively understood.   For $x_a < x < x_b$, $I_{\rm F}$ compete with $I_{\rm AF}$ whose magnitude
is comparable to that of $I_{\rm F}$.  Then, the metamagnetic field is relatively small.  For $x_b < x <x_c$, $I_{\rm F}$ competes with $I_{\rm AF}$ and also $I_{\rm K}$ whose magnitudes becomes larger compared with those for  $x_a < x < x_b$. The metamagnetic field becomes larger irrespective of decreasing $T_{\rm N}$ with $x$. For $x > x_c$, the main competition is with $I_{\rm K}$ whose magnitude is larger than those of  $I_{\rm AF}$ and $I_{\rm K}$ for $x_a < x < x_b$.  Since $I_{\rm K}$ is not the interaction for magnetic order, the metamagnetic behavior could be different from those for $x_a < x < x_c$.  

Although the underlying microscopic model for the inter-site interactions is not clarified, we speculate that the intra atomic interaction $U$ among the f electron could be important for the mechanism to give rise to the ferromagnetic order or the polarized state.  The inelastic neutron scattering experiments on CeRu$_2$Si$_2$ reports that the magnetic correlation changes from antiferromagnetic below $H_{\rm m}$ to ferromagnetic above $H_{\rm m}$\cite{Raymond98,Sato04}. Satoh and Ohkawa report that this observation can be accounted by assuming a characteristic pseudo gap structure of the quasi particle density of states near the Fermi level and that the Kondo collapse mechanism due to volume expansion is also important\cite{Satoh01}.  We conjecture that this model for the switching may be relevant with present interpretation in the sense that the pseudo gap structure with the large density of states is related with ferromagnetic instability and a large value of $U$.  We further discuss the mechanism of the metamagnetic transition and the role of $U$ in $\S$3.4.

In the following $\S$3 and $\S$4, we describe how the f electron state changes depending on $x$ or on magnetic field, and also discuss how we should describe the f electron state.

\section{Fermi Surface Properties of CeRu$_2$Si$_2$ below and above the Metamagnetic Transition} 
In this section we describe how the Fermi surface properties of CeRu$_2$Si$_2$ change associated with the metamagnetic transition to give more insight into the mechanism.  In $\S$3.1 we present the Fermi surface properties observed by the dHvA effect in fields below ($\S$3.1.1) and above ($\S$3.1.2) the metamagnetic transition.    The Fermi surface properties above $H_{\rm m}$ is crucially important to understand the mechanism of the metamagnetic transition.  However, the interpretation of the results seems to be controversial.  Some controversial points are also discussed.  In $\S$3.2, we describe anomalous and interesting features of the Fermi surface properties observed associated with metamagnetic transition.  $\S$3.2.1   describes about the spin dependent effective mass which may be a common feature in a polarized state of the strongly correlated f electron system under magnetic fields.   $\S$3.2.2 demonstrates how the Fermi surface properties change together with the metamagnetic transition.  In $\S$3.3, the Fermi surface properties above the metamagnetic transitions in  Ce$_x$La$_{1-x}$Ru$_2$Si$_2$ and CeRu$_2$(Si$_x$Ge$_{1-x}$)$_2$ are presented to show that the Fermi surface properties above $H_{\rm m}$ in CeRu$_2$Si$_2$ continuously change with $x$ to those of LaRu$_2$Si$_2$ or CeRu$_2$Ge$_2$.  In $\S$3.4,  we summarize the important differences in the Fermi surface properties between below and above $H_{\rm m}$ and discuss models for the Fermi surface properties above $H_{\rm m}$.  We would like to also point out that it may be misleading and may not be fruitful to categorize the f electron state above $H_{\rm m}$ as either itinerant or localized.

\subsection{Fermi surface properties of CeRu$_2$Si$_2$}
\subsubsection{Fermi surface properties of CeRu$_2$Si$_2$ below $H_{\rm m}$}
\begin{figure}[htbp]
\begin{center}
\includegraphics[width=0.6\linewidth]{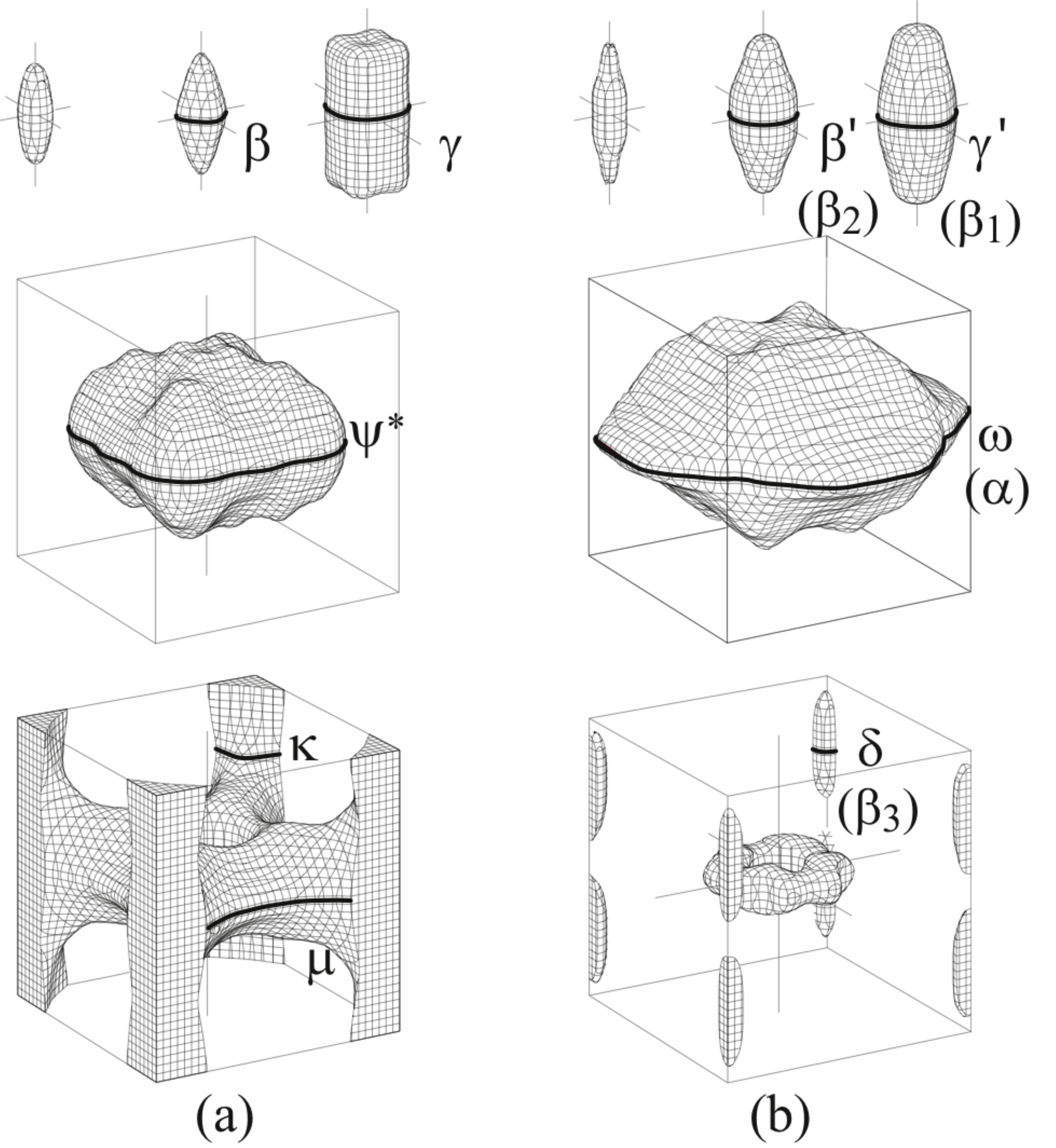}
\end{center}
\caption{(a) Fermi surface of CeRu$_2$Si$_2$\cite{Yamagami93}.  (b) Fermi surface of LaRu$_2$Si$_2$\cite{Yamagami92}.  The orbits which are responsible for the observed dHvA oscillations are also shown on the Fermi surface.  The symbols in the parentheses are those used in the previous paper of LaRu$_2$Si$_2$\cite{Settai95}.}
\label{fig:90025Fig17}
\end{figure}

The Fermi surface properties of CeRu$_2$Si$_2$ below $H_{\rm m}$ seem to be well established now from the dHvA effect studies \cite{Lonzarich88,Onuki92,Aoki92,Aoki93a,Aoki93b,Julian94,Tautz95,Aoki95,Takashita96}, the angle resolved photo emission spectroscopy (ARPES) studies\cite{Denlinger01,Yano08,Okane09}, and the Compton scattering studies\cite{Koizumi11} as well as the band structure calculations \cite{Yamagami93,Runge95,Suzuki10,Sakai13}.  Figure \ref{fig:90025Fig17}(a) shows the Fermi surface of CeRu$_2$Si$_2$ \cite{Yamagami93} obtained from the band structure calculation assuming that the f electron behaves as a conduction electron.  The orbits which correspond to the dHvA oscillations are also shown on the Fermi surface when a magnetic field is applied parallel to the [001] direction.  The orbits when a magnetic field is applied parallel to the [100] direction are shown in Fig. \ref{fig:90025Fig33}.  The top panel shows the hole surfaces centered at the Z point, the middle the large hole surface centered at the Z point and the bottom the electron surface centered at the X point. The experimental results indicate that the Fermi surface of CeRu$_2$Si$_2$ is quite similar to that shown in  Fig. \ref{fig:90025Fig17} except for some small differences stated below. 

\begin{figure}[htbp]
\begin{center}
\includegraphics[width=0.6\linewidth]{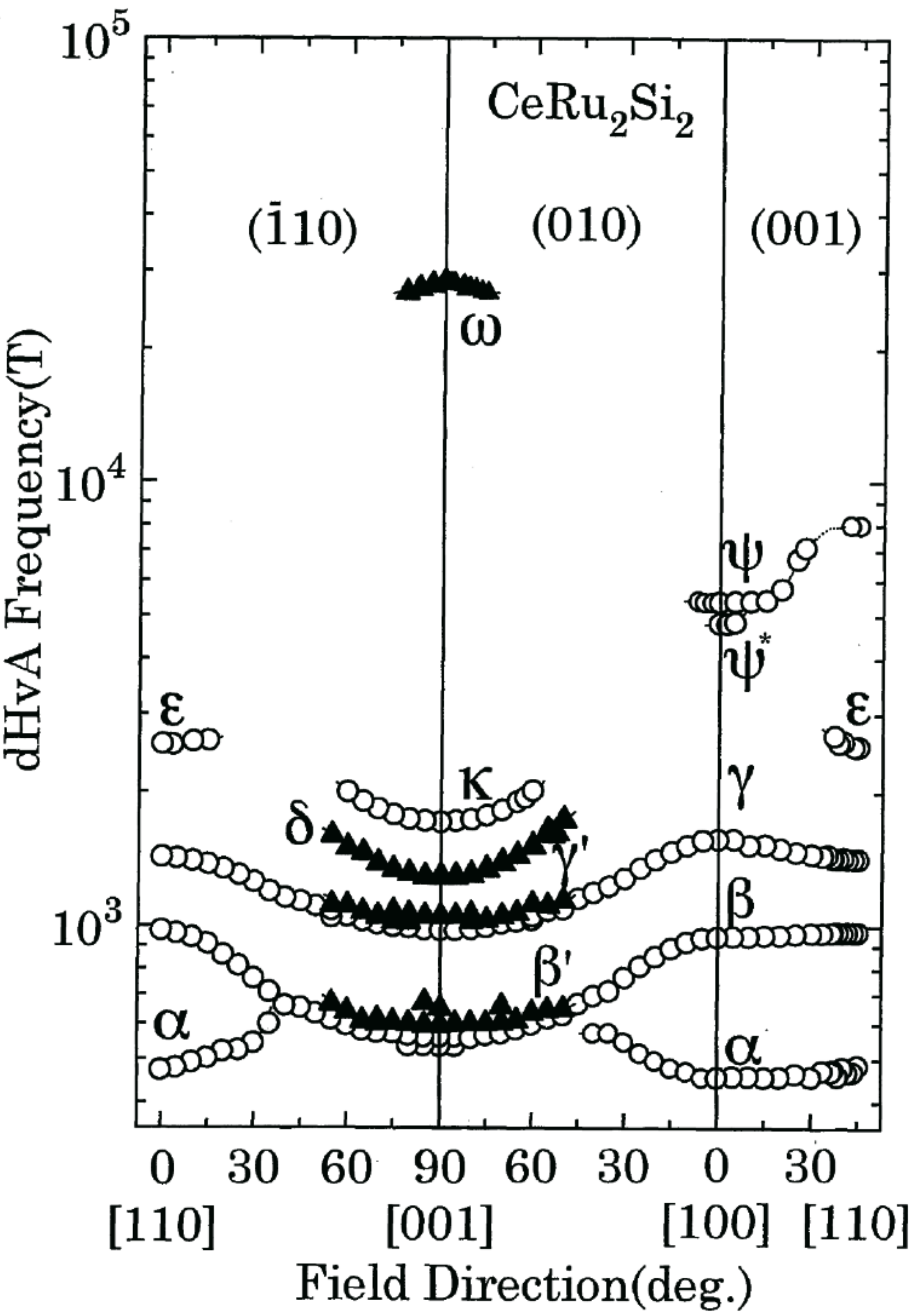}
\end{center}
\caption{Angular dependences of the frequencies observed in fields below $H_{\rm m}$ (open circles) and above $H_{\rm m}$ (closed triangles)\cite{Takashita96}. The central and non-central orbits for the $\psi^\ast$ and $\psi$ oscillations merge with tilting the field direction from [100] to [110].}
\label{fig:90025Fig18}
\end{figure}

The angular dependences of the frequencies are plotted in Fig. \ref{fig:90025Fig18} together with those above $H_{\rm m}$. The $\psi^\ast$ and $\psi$ oscillations are attributed to the central and non-central orbits on the large hole surface. The effective masses of the oscillations are of the order of 100 $m_0$.  Here $m_0$ is the electron rest mass.  The $\psi^\ast$ and $\psi$ oscillations become unobservable with tilting the field direction from the (001) plane.  This is probably because the effective mass and the curvature factor ($\S$A.1) become larger for these field directions.  The $\kappa$ oscillation\cite{Aoki93b} is attributed to the orbit on the arm of the electron surface extending along the [001] direction\cite{Tautz95,Takashita96} and the $\alpha$ oscillation to that extending to the [100] direction. The $\varepsilon$ oscillation is possibly attributed to an orbit on the electron surface.  The effective masses of these oscillations are of the order of 10 - 20 $m_0$.  The $\beta$ and $\gamma$ oscillations arise from the small hole surfaces on the top panel.  The values of the effective masses of these oscillations are 1-2 $m_0$. 

The oscillation which is probably attributed to the $\mu$ orbit in Fig. \ref{fig:90025Fig17}(a) can be observed only under high pressures and the effective mass is measured to be 34 $m_0$ at 6 kbar\cite{Takashita98,Aoki01}. The $\mu$ oscillation has probably a large curvature factor ($\S$A.1) as noted from the shape of the Fermi surface as well as large effective mass.  With increasing pressure the effective mass decreases and the field range available for measurements increases because $H_{\rm m}$ increases. Consequently, the signal amplitude increases in the measurements at high pressures, if we do not consider the experimental difficulties at high pressures ($\S$A.1).  The frequency of the $\mu$ oscillation increases with pressure.  The frequency at ambient pressure is estimated to be about 2.69$\times10^4$ T by extending the observed frequency as a function of pressure to ambient pressure\cite{Aoki01}.  From the pressure dependence of the effective mass of the $\kappa$ oscillation which arises from the same electron surface, the effective mass at ambient pressure is estimated to be 50 $m_0$\cite{Aoki01}.  Its angular dependence has not been measured.  However, the frequency is expected to increase with tilting angle from the [001] direction from the shape of the Fermi surface.  The small hole surface on the left hand side of the top panel is now confirmed to be absent from various measurements and more recent band structure calculations\cite{Runge95,Suzuki10,Sakai13}.  The values of the frequencies and effective masses in high symmetry directions are summarized in Table \ref{tab:dHvAbelowHm}.  

\begin{table}[t]
\caption{Frequencies and effective masses of the dHvA oscillations observed below $H_{\rm m}$.  The - symbol denotes that the oscillation is not present judging from the Fermi surface. The $\ast$ symbol denotes that the signal could not be observed although the oscillation is expected to be present.  $^{1)}$The data taken from the paper by Tautz et al.\cite{Tautz95}}
\label{tab:dHvAbelowHm}
\begin{center}
\begin{tabular}{c|ccc} \hline\hline
 
Oscillation & [001] & [110] & [100]\\ 
\hline   
$\alpha$ & - & $4.84\times10^2$ T & $4.54\times10^2$  T\\ 
      & - & 12.3 $m_0^{1)}$ & 15 $m_0$\\ 
\hline   
$\beta$& $5.36 \times10^2$ T (at 3 T) & $9.7\times10^2 $ T & $9.43\times10^2 T$\\
 & 1.5 $m_0$ (at 5.4 T) & 1.8 $m_0^{1)}$ & 1.5 $m_0^{1)}$\\
\hline   
$\gamma$ & $9.8 \times10^2$ T (at 5.4 T) & $1.42\times10^3$ T& $1.57\times10^3 $ T  \\ 
  & 1.6 $m_0$ (at 5.4 T) & 2.3 $m_0^{1)}$ & 2.8 $m_0^{1)}$ \\ 
\hline   
$\varepsilon$ & - & 2.54$\times10^3$ T & - \\ 
 & - & 20 $m_0^{1)}$ & - \\ 
\hline   
$\kappa$ & $1.65\times10^3$ T (at 6 T) & - & - \\
 & 11  $m_0$ (at 6.2 T) & - & - \\  
\hline   
$\psi^\ast$ & $\ast$ & $8\times10^3$ T& $4.84\times10^3$  T\\ 
 &$\ast$ & 140 $m_0^{1)}$  & 120 $m_0$ \\ 
\hline   
 $\psi$ & - & - &  $5.42\times10^3$  T\\ 
 & - & - &  120 $m_0$ \\
\hline
\hline   
 $\mu$ & 2.69 $\times 10^4$ T(at ambient P) & - &  - \\ 
  & 2.84 $\times 10^4$ T (at 6kbar) & - &  - \\
   & 50 $m_0$ (at ambient P) & - &  - \\ 
  & 34 $m_0$ (at 6kbar) & - &  - \\ 
\hline 
\end{tabular}
\end{center}
\end{table}

\begin{table}[t]
\caption{Frequencies and effective masses above $H_{\rm m}$}
\label{tab:dHvAaboveHm}
\begin{center}
\begin{tabular}{c|c} \hline\hline
 
Oscillation & [001] \\ 
\hline   
$\beta^\prime$ & $6.13 \times10^2$ T (at 13.1 T) \\
 & 1.0 $m_0$ (at 10.5 T) \\
\hline   
$\gamma^\prime$ & $1.04 \times10^3$ T (at 12.5 T)  \\ 
  & 1.1 $m_0$ (at 12.5 T)  \\ 
\hline   
$\delta$ & $1.24 \times10^3$ T (at 12.6 T)   \\ 
 &3.6 $m_0$ (at 12.6 T) \\ 

\hline   
$\omega$ &$2.82 \times10^4$ T (15.65 - 15.9 T)  \\ 
 & 8.2 $m_0$ (15.65 - 15.9 T)  \\ 
\hline

\end{tabular}
\end{center}
\end{table}

The band structure calculations show that the f content is rich on the large hole surface, moderate on the multiply connected electron surface, and poor on the ellipsoidal hole surfaces\cite{Yamagami93,Runge95,Suzuki10,Sakai13}. The observed value of the effective mass in CeRu$_2$Si$_2$ approximately corresponds to the amount of the calculated f content \cite{Aoki93b,Tautz95}. 

As a reference we also show the Fermi surface of LaRu$_2$Si$_2$ obtained by the band structure calculation \cite{Yamagami92}. The Fermi surface of LaRu$_2$Si$_2$ has been revealed by the dHvA effect measurements\cite{Settai95} and is found to be quite similar to that shown in Fig. \ref{fig:90025Fig17}(b) except that there is no small hole surface on the left hand side of the top panel\cite{Matsumoto08}.  The extremal orbits responsible for the dHvA oscillations are shown on the Fermi surface.  The effective masses of LaRu$_2$Si$_2$ are about $1-2 m_0$ on the large hole surface, about $1 m_0$ on the electron surfaces and about $0.5 m_0$ on the small hole surfaces. If we compare the Fermi surface of LaRu$_2$Si$_2$ with that of CeRu$_2$Si$_2$, we note that the volume of the large hole surface in LaRu$_2$Si$_2$ is larger than that in CeRu$_2$Si$_2$ while the volume of the electron surface in  LaRu$_2$Si$_2$ is smaller than that in CeRu$_2$Si$_2$. There is no large difference in the volumes between the small hole surfaces.  The difference in the total volume of the Fermi surface comes from the fact that the number of the conduction electrons per primitive unit cell is smaller by one in LaRu$_2$Si$_2$ than in CeRu$_2$Si$_2$.

Thus, the dHvA experimental results are consistent with the picture that the f electron in CeRu$_2$Si$_2$ contributes to form the Fermi surface or the Fermi surface is large, although the valence of Ce is close to +3 as described in $\S$2.1.  The results are also consistent with the theoretical picture developed for the heavy Fermion compounds\cite{Shiba90,Shibata99,Oshikawa00,Otsuki09}.  

In the paramagnetic ground state like that $H<H_{\rm m}$ in CeRu$_2$Si$_2$, the term ``itinerant" can be used without ambiguity from the observation of the ``large Fermi surface" or the two terms ``itinerant" and ``large Fermi surface" can be used equivalently.  However, even when the observed volume of the Fermi surface is the same as or close to the volume of the small Fermi surface, the term ``localized" may not be appropriate to characterize the f electron state in some cases.  We discuss such examples in the following sections.

We use the Lifshitz - Kosevich (LK) formula ($\S$A.1) for analysis assuming that the LK formula is valid for the strongly correlated f electron system.  When the magnetic fields are applied in the (001) plane or when the fields along the [001] direction are well below $H_{\rm m}$, there seems no experimental observation against the implicit assumptions in the LK formula  ($\S$A.2 and $\S$A.3).   In the following sections, we also present some peculiar features which are not consistent with the implicit assumptions.    

\subsubsection{Fermi surface properties above the metamagnetic transition}
As shown in Fig. \ref{fig:90025Fig18}, we observe four different frequency branches above $H_{\rm m}$.  The frequencies and the effective masses of these oscillations are summarized in Table \ref{tab:dHvAaboveHm}.   As described below the Fermi surface properties depend on spin direction as well as on magnetic field. Then, the listed values are an average value of the two spin directions at a particular field strength well above  $H_{\rm m}$.   Since the metamagnetic field increases with tilting angle $\theta$ from the [001] direction as $1/\cos\theta$, the oscillations above  $H_{\rm m}$ can be observed in limited angular directions around the [001] direction.  The $\beta^\prime$ and $\gamma^\prime$ oscillations have similar frequencies and angular dependences of frequency to those of the $\beta$ and $\gamma$ oscillations below  $H_{\rm m}$, respectively.  The effective masses are slightly smaller than those of the $\beta$ and $\gamma$ oscillations.  The frequency of the $\delta$ oscillation is comparable to that of the $\kappa$ oscillation and its angular dependence indicates that it comes from an approximately cylindrical Fermi surface.  The effective mass is about 3.6 $m_0$ and is much smaller than that of the $\kappa$ oscillation.  The frequency of the $\omega$ oscillation and its angular dependence are similar to those of the $\omega$ oscillation in LaRu$_2$Si$_2$ or CeRu$_2$Ge$_2$.  

The values of the $\omega$ oscillations in LaRu$_2$Si$_2$\cite{Settai95} and CeRu$_2$Ge$_2$\cite{Ikezawa97} are listed in Table \ref{tab:dHvALRSandCRG} for reference.  With fields along the [001] direction only one frequency is observed for the $\omega$ oscillation of CeRu$_2$Ge$_2$, while spin split frequencies are observed with fields in the (001) plane.   Moreover, the effective mass of the $\omega$ oscillation in CeRu$_2$Ge$_2$ is strongly field dependent for fields parallel to the [001] direction, while those with fields in the (001) plane do not depend on field as presented later in $\S$4.2.  Since the curvature factors of the up and down spin Fermi surfaces are not expected to be very different, we suspect that the effective mass of the other oscillation $\omega_{-\sigma}$ may be considerably larger for the [001] direction. 

\begin{table}[t]
\caption{Frequencies and effective masses of the $\omega$ oscillations in LaRu$_2$Si$_2$ \cite{Settai95}and CeRu$_2$Ge$_2$\cite{Ikezawa97}. The dHvA effect measurements on CeRu$_2$Ge$_2$ were performed also by King and Lonzarich\cite{King91}.  However, the results are not consistent with our results. We suspect that their sample alignment might not be good enough.  We refer the reader to our paper \cite{Matsumoto11} where our present understanding of the Fermi surface of CeRu$_2$Ge$_2$ is summarized. $^{\rm a)}$The value of the effective mass changes sensitively also with tilting angle of the field direction from the [001] direction as well as with field strength\cite{IkezawaM}.  The value listed is taken from the measurements of Fig. \ref{fig:90025Fig41}\cite{DoiM}. }
\label{tab:dHvALRSandCRG}
\begin{center}
\begin{tabular}{c|ccc} \hline\hline
LaRu$_2$Si$_2$ &  & & \\  
Oscillation & [001] & [110] & [100]\\ 
\hline
 $\omega$ & $2.72\times10^4$ T &  $1.3\times10^4$ T &  $1.38\times10^4$ T\\ 
      & 2.4 $m_0$ & 1.4 $m_0$ & 1.7 $m_0$\\ 
\hline   
\hline   
CeRu$_2$Ge$_2$ &  &  & \\
\hline
 $\omega_{+\sigma}$ & $2.7\times10^4$ T &  $1.28\times10^4$ T &  not measured\\ 
      & 3.8 $m_0^{\rm a)}$ (at 16 T) & 5.5 $m_0$ & 4.3 $m_0$\\ 
\hline 
$\omega_{-\sigma}$ & -  &  $1.33\times10^4$ T & not measured\\ 
      & - & 6.5 $m_0$ & 4.7 $m_0$ \\   
\hline   

\end{tabular}
\end{center}
\end{table}

We note from Figs. \ref{fig:90025Fig17} and \ref{fig:90025Fig18} that the observed frequencies of the oscillations above $H_{\rm m}$ and their angular dependences are well explained by the Fermi surface of LaRu$_2$Si$_2$.  From this observation we reported that the f electron is localized above $H_{\rm m}$\cite{Aoki92} according to the conventional definition of ``localized''.  Further understanding of the f electron state above $H_{\rm m}$ has been brought about by the subsequent studies as described in the following sections.

The frequency of the $\omega$ oscillation above $H_{\rm m}$ is also similar to that of the $\mu$ oscillation below $H_{\rm m}$. One may be tempted to attribute the $\omega$ oscillation to the orbit responsible for the $\mu$ oscillation\cite{Daou06} and may think that the Fermi surface remains substantially the same above $H_{\rm m}$.  However, as discussed above in $\S$3.1.1, the frequency of the oscillation from the orbit is expected to increase with the tilting angle from the [001] direction, while that of the $\omega$ oscillation decreases with the tilting angle.  

Other auxiliary evidence that the origins of the $\mu$ and $\omega$ are different comes from the pressure effect measurements on the dHvA oscillations\cite{Takashita98,Aoki01}.  The frequencies of the $\beta$, $\beta^\prime$, $\gamma$ and $\gamma^\prime$ oscillations from the hole surfaces decrease with pressure both below and above $H_{\rm m}$, while those of the $\kappa$ and $\delta$ oscillations from the electron surfaces increase with pressure.   The frequency of the $\mu$ oscillation below $H_{\rm m}$ increases with pressure, while that of the $\omega$ oscillation above $H_{\rm m}$ decreases with pressure.  This observation may need an explanation if the origins of the $\mu$ and $\omega$ oscillations are the same.  

The pressure effect study also tells that the electronic structures below and above $H_{\rm m}$ are qualitatively different. The values of the coefficient $A$  and the initial susceptibility $\chi_0$ decrease with pressure below $H_{\rm m}$ and the effective mass of the $\kappa$ oscillation is also found to decrease with pressure.   Although the accurate effective mass measurements were difficult because of the experimental difficulty under pressure ($\S$A.1), the effective mass of the $\omega$ oscillation under pressure is found to be nearly the same or larger than that at ambient pressure.   This observation for the $\omega$ oscillation is also consistent with the observation that the value of the $A$ coefficient and the effective masses above $H_{\rm m}$ increase with chemical pressure in Ce$_x$La$_{1-x}$Ru$_2$Si$_2$ and CeRu$_2$(Si$_x$Ge$_{1-x}$)$_2$ as described in $\S$3.3.  We also present the Fermi surface properties above $H_{\rm m}$ in Ce$_x$La$_{1-x}$Ru$_2$Si$_2$ and CeRu$_2$(Si$_{1-x}$Ge$_x$) in $\S$3.3 to show the Fermi surface properties are qualitatively different from those below $H_{\rm m}$.

It was also reported that the oscillation with frequency of 3.2 kT seemed to be present above $H_{\rm m}$\cite{Tautz95}.   However, we could not find any obvious oscillation other than the third harmonic frequency oscillation ($\S$A.1) of the $\gamma^\prime$ oscillation around 3 kT\cite{Takashita96}.

\subsection{Anomalous features of the Fermi surface properties associated with the metamagnetic transition}

\subsubsection{Spin dependent effective mass}
Before we present the details of the metamagnetic behavior observed by the dHvA effect measurements, we describe that the effective mass depends on spin direction under magnetic fields.  This spin dependence of the effective mass is crucially important for the interpretation of the electronic structure above $H_{\rm m}$.  However, the spin dependence may not be surprising because the most significant interaction of the strongly correlated f electron system is the Kondo effect and the spin dependent scattering is observed in the resistivity\cite{Felsch73,Costi00} and in the dHvA effect \cite{Coleridge70,Alles73} of the dilute Kondo alloys under magnetic fields ($\S$A.3).   If the crystal has the inversion symmetry, the degeneracy corresponding to the spin degeneracy remains even when the spin-orbit interaction is present.  The dHvA oscillations in the strongly correlated f electron systems have anomalous field and temperature dependences which cannot be understood by the LK formula.  However, if we assume that the effective mass depends on the spin direction although the frequencies of the two spin directions are approximately the same ($\S$A.2), the anomalous features of the dHvA oscillations in various strongly correlated f electron systems can be well explained\cite{Aoki93a,Aoki93b,Takashita96,Endo02,Endo03,Sheikin03,Nakayama04,McCollam05,Endo04,Endo05,Endo06,Daou06b}.   The spin dependence has been discussed also theoretically\cite{Spalek90,Spalek06,Otsuki07,Bauer07,Onari08}.

The spin dependence is observed when the f electron is in a polarized state under magnetic fields.  It seems that the spin dependence is more significant when the polarization is larger.  
The effective mass of one spin direction can be by more than a few times as large as that of the other direction.  When the crystal does not have the inversion symmetry, the degeneracy is removed by the spin-orbit interaction.  In this case we observe a pair of frequency branches whose angular dependences are similar, but whose effective masses are considerably different\cite{Iida11}.  

\begin{figure}[htbp]
\begin{center}
\includegraphics[width=0.6\linewidth]{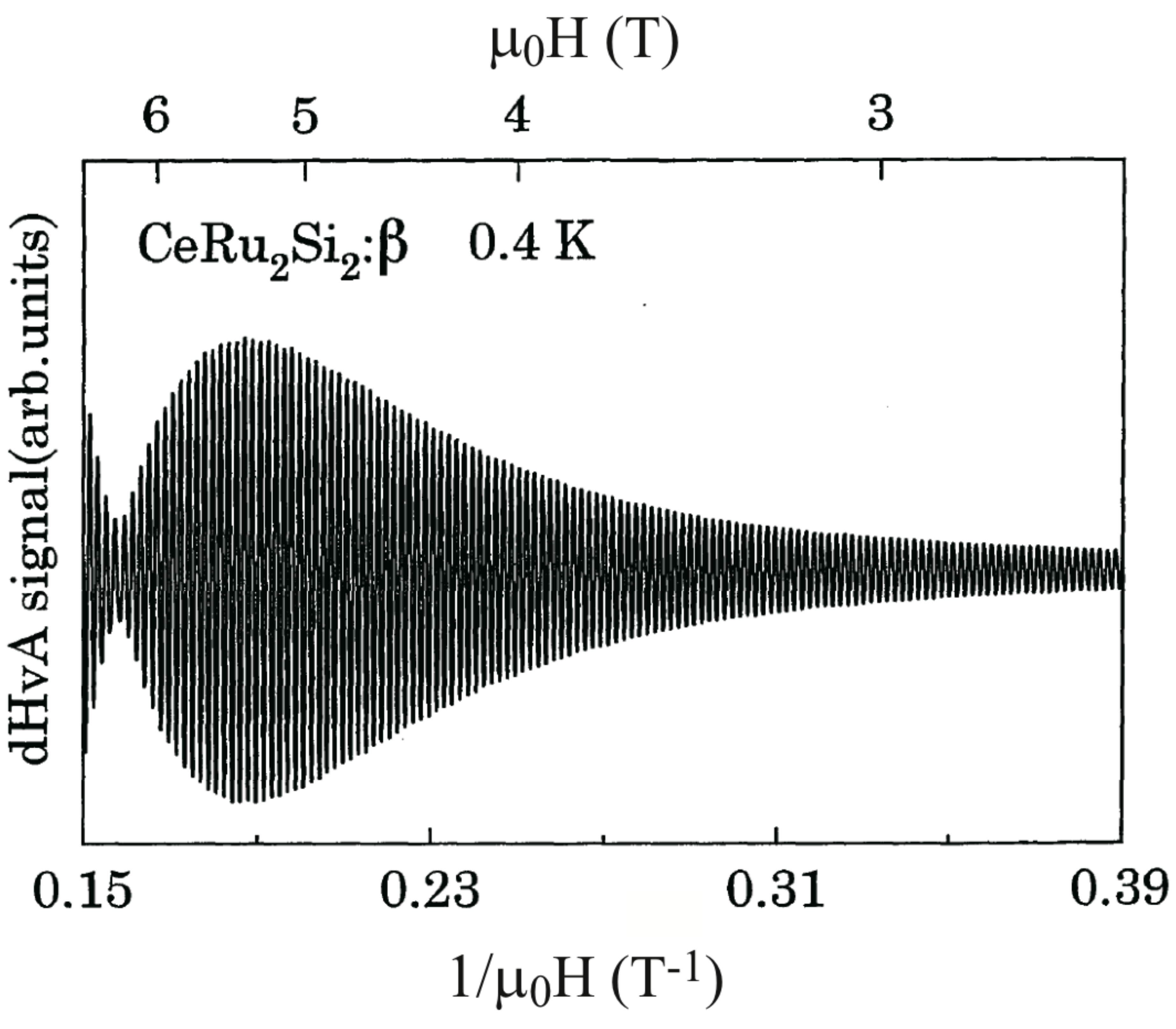}
\end{center}
\caption{The $\beta$ oscillation below $H_{\rm m}$ plotted as a function of magnetic field or inverse field\cite{Takashita96}. }
\label{fig:90025Fig19}
\end{figure}

\begin{figure}[htbp]
\begin{center}
\includegraphics[width=0.6\linewidth]{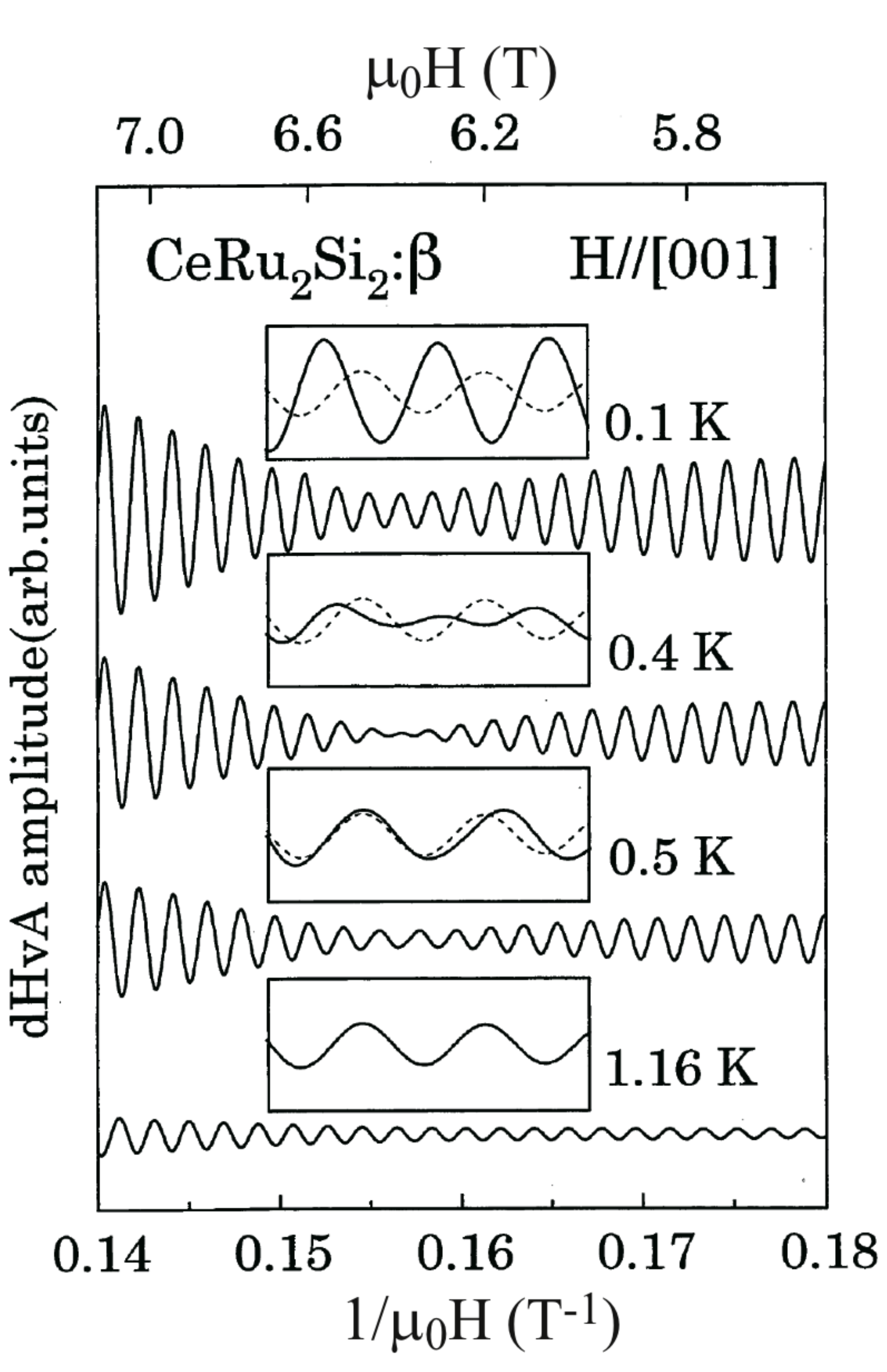}
\end{center}
\caption{$\beta$ oscillations below $H_{\rm m}$ observed at 0.1 K, 0.4 K, 0.5 K and 1.16 K for the fields parallel to the [001] direction\cite{Takashita96}.  The insets shows the expanded views of the waveforms at the minimum positions.  The waveform observed at 1.16 K was shown in each insets by the broken line to compare the frequencies and phases.} 
\label{fig:90025Fig20}
\end{figure}

As a simple example of the spin dependent effective mass, we show the $\beta$ oscillation below $H_{\rm m}$ in CeRu$_2$Si$_2$ at 0.4 K as a function of inverse field in Fig. \ref{fig:90025Fig19}\cite{Takashita96}.  The oscillation amplitude has a minimum as a function of inverse field near $H_{\rm m}$ where the polarization becomes significant.  Such a beat structure is sometimes observed owing to the shape of the Fermi surface or bi-crystal structure\cite{Ogawa79,Endo05}.  If we examine the field and temperature dependences of the oscillation, particularly at the minimum position, we note that the behavior of the oscillation can not be explained by such a mechanism.  Figure \ref{fig:90025Fig20} shows the oscillations near the minimum position at several temperatures.  The amplitude and the phase of the oscillation change anomalously with temperature.   However, the behavior of the oscillation can be successfully explained if we assume that the amplitude and phase of the dHvA oscillation depends on the spin direction\cite{Takashita96}.  

To incorporate the field and spin dependences explicitly, we used the following phenomenological equation to analyze the fundamental freqeuncy oscillation\cite{Takashita96,Endo04,Endo05} ($\S$A.1). 
\begin{eqnarray}
	\tilde M_{fnd} &=& \tilde M_{\uparrow}(H,T)+ 	\tilde 
M_{\downarrow}(H,T)\nonumber\\
    &=& A_{\uparrow}(H,T)
					\sin\left(\frac{2\pi F_{\uparrow}(H)}{H}
							+\xi_{\uparrow}+\xi_0\right) 
			       	+A_{\downarrow}(H,T)
					\sin\left(\frac{2\pi F_{\downarrow}(H)}{H}
							+\xi_{\downarrow}+\xi_0\right) . 
\label{eq:mod_LK0}
\end{eqnarray}
Here, $A_{\sigma}$ and $F_{\sigma}$ denote the amplitude and frequency of the dHvA oscillation from up spin electrons ($\sigma =\uparrow$) or down spin electrons ($\sigma =\downarrow$), respectively.  Since so far no obvious difference in the frequencies of the up and down spin oscillations has been found, the frequencies are assumed to be the same ($\S$A.2).  Here we also assume that the Fermi surface and effective mass change slowly and monotonically in the field range of interest.  We assume that the difference in the phase between the up and down spin oscillations comes from the difference in the effective mass of the up and down spin electrons and the phase $\xi_\sigma(H)$ is given by
\begin{eqnarray}
	\xi_\sigma(H) = \mp\frac{\pi g m_\sigma^\ast(H)}{2m_0}.
\label{eq:phase0}
\end{eqnarray}
Here, $m_\sigma^\ast(H)$ is the effective mass which depend on spin direction and magnetic 
field.  The $-$ and $+$ signs correspond to up and down spin electrons, respectively.  $g$ is the g factor which is assumed to be independent of 
spin direction and magnetic field strength in the field range of interest. If all 
the quantities are independent of spin and magnetic field,  
eq.(\ref{eq:mod_LK0}) reduces to the conventional LK formula ($\S$A.1).
The details of the phenomenological analysis based on these assumptions and the LK formula are given in the appendix of the reference\cite{Endo05}. It is noted that the spin dependent phase or effective mass is necessary to reproduce the behavior of the phase around the minimum position.  

Thus, the effective mass determined by the conventional method ($\S$A.1 and $\S$A.2) is an average value of the effective masses of up and down spin electrons.  The experimental effective mass is likely to be closer to the smaller one because the signal amplitude of the oscillation with smaller effective mass has normally larger amplitude and dominates the temperature dependence of the signal amplitude.  As described below, the effective mass increases with increasing field towards $H_{\rm m}$.  The minimum appears when the average effective mass becomes about 1.5 $m_0$.  If the value of $g$ factor can be assumed to be 2.0, this minimum corresponds to the spin splitting zero condition when the effective masses of both the up and down spin electrons are the same ($\S$A.1).  The effective masses of the $\beta^\prime$, $\gamma$, $\gamma^\prime$, $\kappa$, and $\delta$ oscillations are very likely to depend on the spin direction from the analysis of the behavior of the dHvA oscillations\cite{Takashita96,Daou06b}. For the $\mu$ and $\omega$ oscillations,  it was difficult to judge whether or not the oscillations are spin dependent, mostly because the signal amplitudes are not large enough to examine the details of the behavior ($\S$A.1).   Therefore, following two cases are possible: (1) The effective mass as well as the Dingle temperature is spin independent.  (2) The differences in the effective masses and/or the Dingle temperatures between the up and down spin oscillations are large and consequently only the dominant oscillation of one spin direction can be clearly observed.  This case can not be discriminated experimentally from the former case ($\S$A.1).  The former indicates that the $\mu$ and $\omega$ oscillations with relatively large effective masses are spin independent while the other oscillations with smaller effective masses are spin dependent.  Since the spin dependence most likely originates from the spin dependent Kondo effect under magnetic fields,  we think that the former case is very unlikely.

\subsubsection{Metamagnetic transition probed by the dHvA effect measurements}

\begin{figure}[htbp]
\begin{center}
\includegraphics[width=0.6\linewidth]{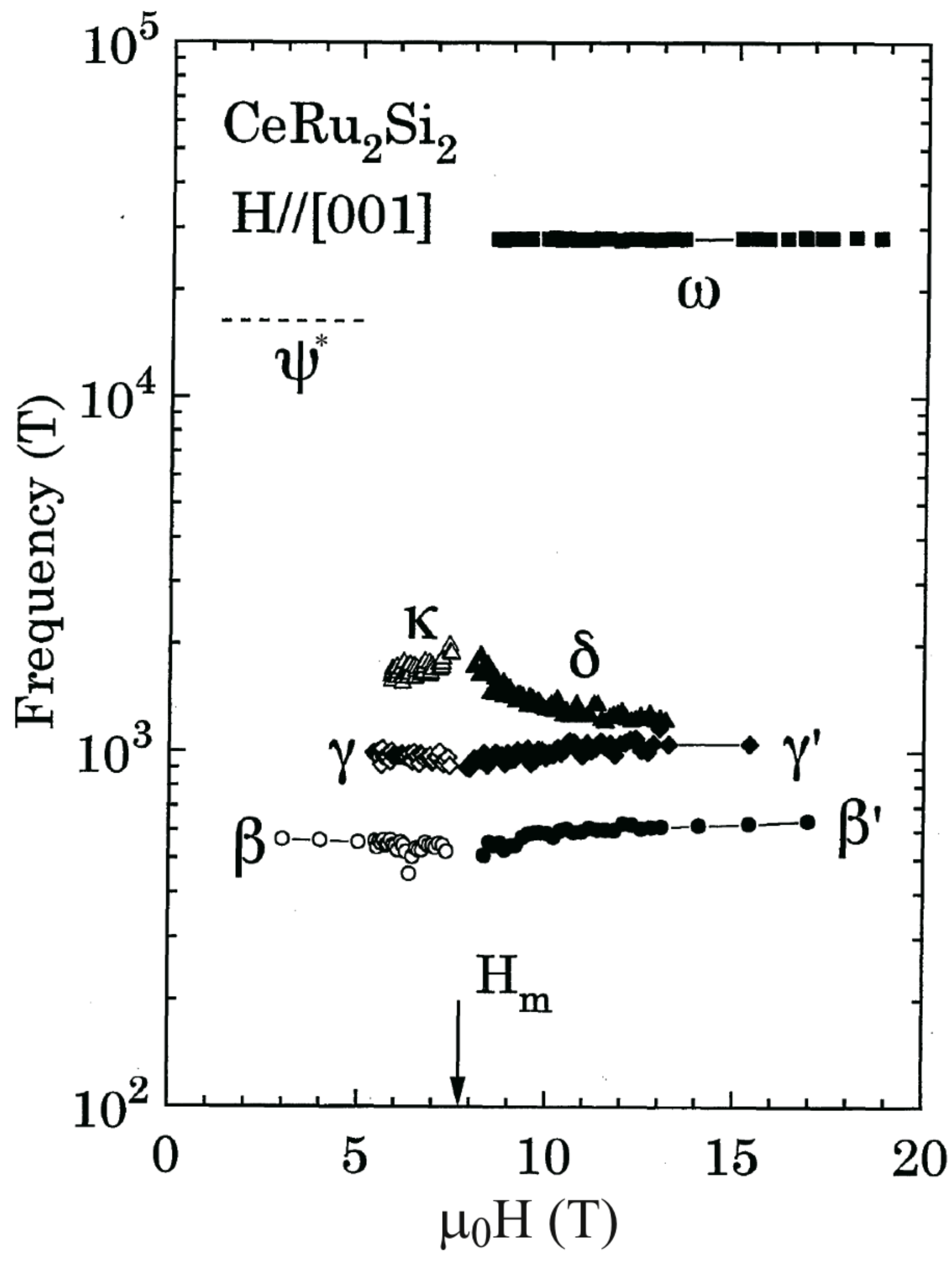}
\end{center}
\caption{dHvA frequencies as a function of magnetic field below and above $H_{\rm m}$ with fields parallel to the [001] direction\cite{Takashita96}. }
\label{fig:90025Fig21}
\end{figure}

\begin{figure}[htbp]
\begin{center}
\includegraphics[width=0.6\linewidth]{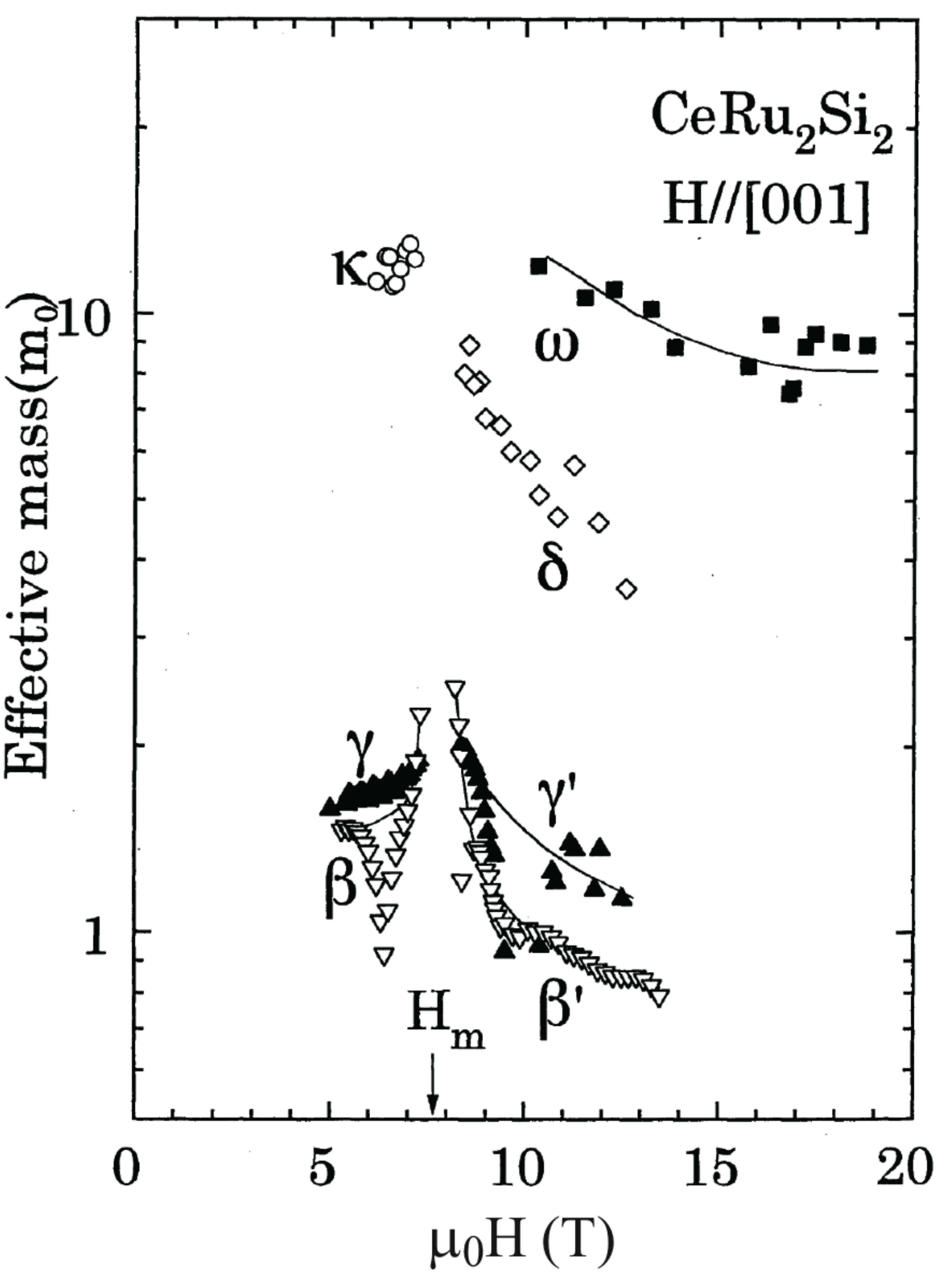}
\end{center}
\caption{Effective masses as a function of magnetic field below and above $H_{\rm m}$ with fields parallel to the [001] direction\cite{Takashita96}. }
\label{fig:90025Fig22}
\end{figure}

Figures \ref{fig:90025Fig21} and \ref{fig:90025Fig22} plot the frequencies and effective masses of the dHvA oscillations as a function of magnetic field, respectively.  They are observed below and above $H_{\rm m}$ for fields parallel to the [001] direction at ambient pressure\cite{Takashita96}. The value of the $\psi^\ast$ oscillation is obtained from the band structure calculation\cite{Yamagami93}. The data points of the effective masses appear to be scattered.  For example, the value of the effective mass of the $\beta$ oscillations below $H_{\rm m}$ has a dip at around the field where the minimum of the $\beta$ oscillation is observed.  Since the amplitude of the $\beta$ oscillation at the minimum position does not increase monotonically with decreasing temperature as shown in Fig. \ref{fig:90025Fig20}, the conventional analysis  gives unrealistic small values at the minimum position ($\S$A.1 and $\S$A.2).   The scatter of the data points comes from the unusual behavior of the dHvA oscillations which are mostly owing to the spin dependent effective mass and scattering.  The effective masses are nearly constant at low fields, and are enhanced around $H_{\rm m}$, then decreases with increasing field.  It is noted that the effective masses of all the oscillations are similarly enhanced around $H_{\rm m}$.  

This behavior of the effective mass is qualitatively consistent with the behavior of the electronic specific heat coefficient under magnetic fields\cite{Meulen91}.  On the other hand, the value of the electronic specific heat coefficient at high fields of about 20 T decreases to about 1/4 of that well below $H_{\rm m}$.  Assuming that the effective masses on the large hole surface are comparable to those of the $\psi$ and $\psi^\ast$ oscillations, i.e. of the order of 100-200 $m_0$ at zero magnetic field, we would expect to observe the dHvA oscillations with the effective mass of several tens of $m_0$ above $H_{\rm m}$.  Since the largest value of the effective mass observed at high fields is 8 $m_0$ of the $\omega$ oscillation, it is suspected that a Fermi surface with heavier mass has not been observed.  

Two interpretations are possible for this observation.\\
\noindent
(I) The band structure above $H_{\rm m}$ is approximately the same as that below $H_{\rm m}$. The missing Fermi surface is the Fermi surface corresponding to the heavy hole surface which may have the effective mass of the order of several tens of $m_0$.  According to one of the interpretations along this line the $\omega$ oscillation arises from the $\mu$ orbit on the electron surface\cite{Daou06}. \\
\noindent
(II) The Fermi surface changes to that similar to the Fermi surface of LaRu$_2$Si$_2$ and the $\omega$ oscillation arises from the large hole surface similar to that in Fig. \ref{fig:90025Fig17}(b)\cite{Aoki93a,Aoki93b,Takashita96, Matsumoto08,Suzuki10}.  The effective mass of the $\omega$ oscillation is spin dependent and the other oscillation with the heavier mass may have the effective mass of the order of several tens of $m_0$.

We discuss this problem again in $\S$3.4 together with the models for the metamagnetic transition and electronic structure above $H_{\rm m}$ 

It would be interesting to see how the enhancement  of the effective mass or electronic specific heat coefficient around $H_{\rm m}$ changes with pressure.  At positive pressure side of CeRu$_2$Si$_2$,  the value of $\sqrt{A}$ at zero magnetic field decreases with increasing pressure and at 5 - 6 kbar becomes about 1/3 as large as that of ambient pressure\cite{Mignot88,Aoki11}.  The values of $A(H)/A(0)$'s at different pressures are also reported to be well expressed almost by a single function as $\phi(H/H_{\rm m}$)\cite{Aoki11}.  It is also reported that at ambient pressure the value of the electronic specific heat coefficient at $H_{\rm m}$  reaches the value at $x_c$ under zero magnetic field\cite{Aoki11}.   

\begin{figure}[t]
\begin{center}
\includegraphics[width=0.6\linewidth]{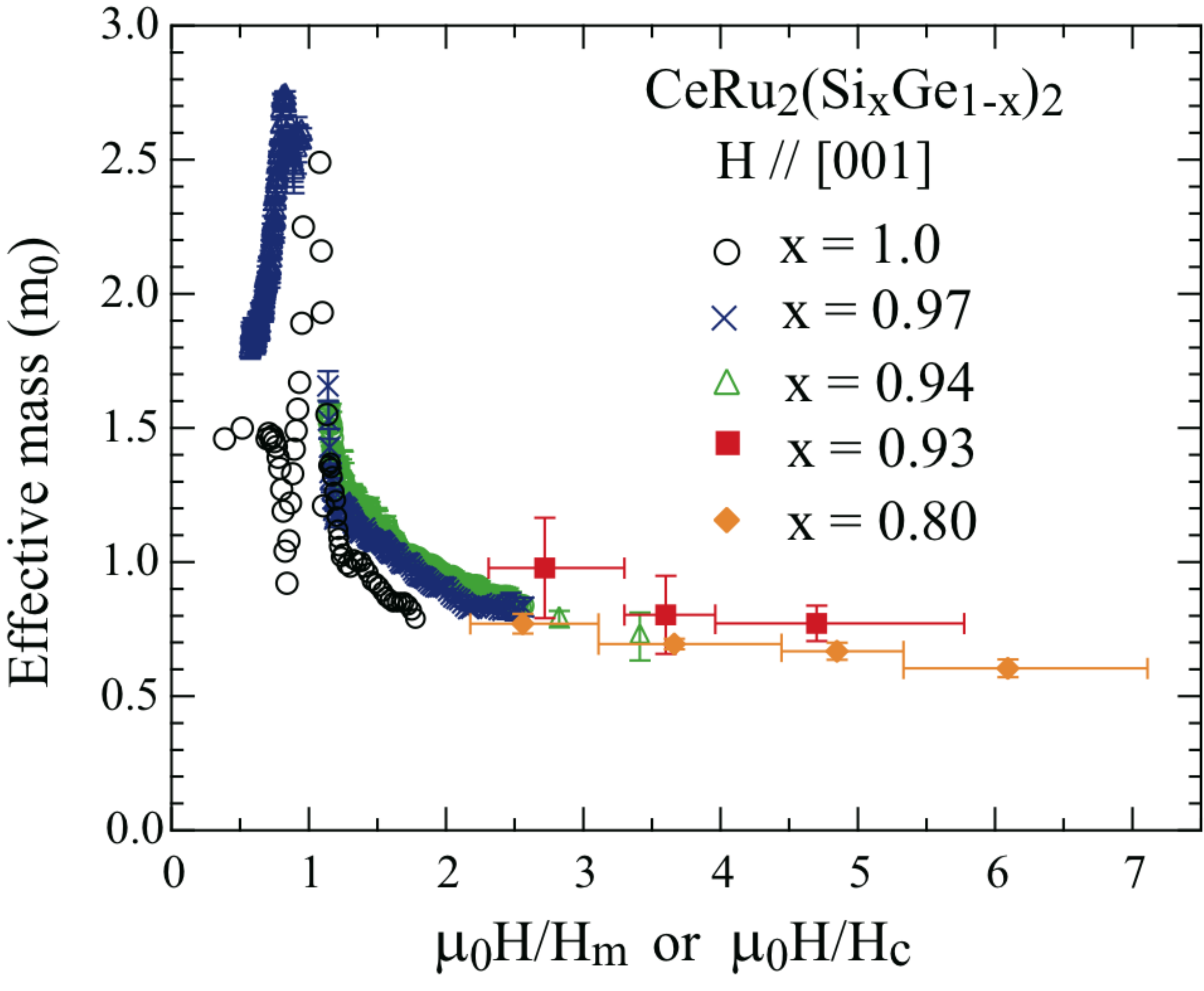}
\end{center}
\caption{(Color on line) Variation of the effective masses of the $\beta$ and $\beta^\prime$ oscillations in CeRu$_2$(Si$_{1-x}$Ge$_x$)$_2$ samples with $x$ = 1.0, 0.97, 0.94, 0.93 and 0.80 as a function of normalized field $\mu_0H/H_{\rm m}$ or  $\mu_0H/H_{\rm c}$\cite{Matsumoto11}. }
\label{fig:90025Fig23}
\end{figure}

At negative pressure side, the value of $A$ coefficient in Ce$_x$La$_{1-x}$Ru$_2$Si$_2$ increases upon approaching $x_c$ and the value near $x_c$ becomes about twice as much as that of CeRu$_2$Si$_2$.  The value of the electronic specific heat coefficient at $x_c$ in Ce$_x$La$_{1-x}$Ru$_2$Si$_2$ is also measured to be about twice as much as that of CeRu$_2$Si$_2$\cite{Fisher91,Aoki11}.  
To see how the effective mass is enhanced at $H_{\rm m}$ at the negative chemical pressure side,  we plot the data for the effective masses of the $\beta$ and $\beta^\prime$ oscillations in CeRu$_2$(Si$_x$Ge$_{1-x}$)$_2$\cite{Sugi08} as a function of $\mu_0H/H_{\rm m}$ or $\mu_0H/H_{\rm c}$ in Fig. \ref{fig:90025Fig23}. For the samples of the paramagnetic ground state ($ x = 1.0, 0.97, 0.94$), the effective mass at $H_{\rm m}$ does not seem to increase much with decreasing $x$ or the degree of the enhancement in the effective mass  at $H_{\rm m}$ seems to decrease. 
The electronic specific heat coefficient (or C/T value at 0.45 K) for $x$ = 0.925 in Ce$_x$La$_{1-x}$Ru$_2$Si$_2$  is reported to be enhanced about 10 \% at $H_{\rm m}$\cite{Aoki12} compared with that at zero magnetic field.  (In this report  $x_c$ is thought to be 0.925.)  From these observations, it is likely that the effective mass or the electronic specific heat coefficient at zero magnetic field increases upon approaching $x_c$, but the enhancement at $H_{\rm m}$  becomes smaller.

For $x<x_c$, there is no systematic study for the enhancement in CeRu$_2$(Si$_x$Ge$_{1-x}$)$_2$, but in Ce$_x$La$_{1-x}$Ru$_2$Si$_2$ it is shown that there is no mass enhancement around $H_{\rm c}$ for the sample $x$ = 0.90\cite{Fisher91,Aoki11}.   On the other hand, for the sample with $x = 0.87$\cite{Fisher91} or 0.90 \cite{Aoki11}, a broad maximum of the electronic specific heat coefficient is observed around 1.5 - 2 T which may correspond to the transition at $H_a$.    The magnetoresistance experiment on CeRu$_2$Ge$_2$ under high pressure reports the variation in the value of the coefficient $A$  as a function of magnetic field\cite{Wilhelm99}.  The value is enhanced around 2 T for samples under pressures of 6.5 and 6.9 GPa which are lower than the critical pressure of about 8.7 GPa.  The enhancement around 2 T was also attributed to an anomaly associated with $H_{\rm a}$.  If the phase diagram under high pressure is the same as that  in Fig. \ref{fig:90025Fig3},  the enhancement could be also attributed to the anomalies associated with $H_{\rm b}$.  It is likely that the effective mass is not enhanced around the first order transition at $H_{\rm c}$.  But they are possibly enhanced around $H_a$ or $H_b$ for $x$ near $x_c$. 

The effective masses above $H_{\rm c}$ decrease with magnetic field and decreasing rate seems to be similar to those above $H_{\rm m}$.  This observation is expected from the result that the electronic structures above the metamagnetic transitions continuously vary across $x_c$ as a function of $x$ as described in $\S$3.3.  

\begin{figure}[htbp]
\begin{center}
\includegraphics[width=0.6\linewidth]{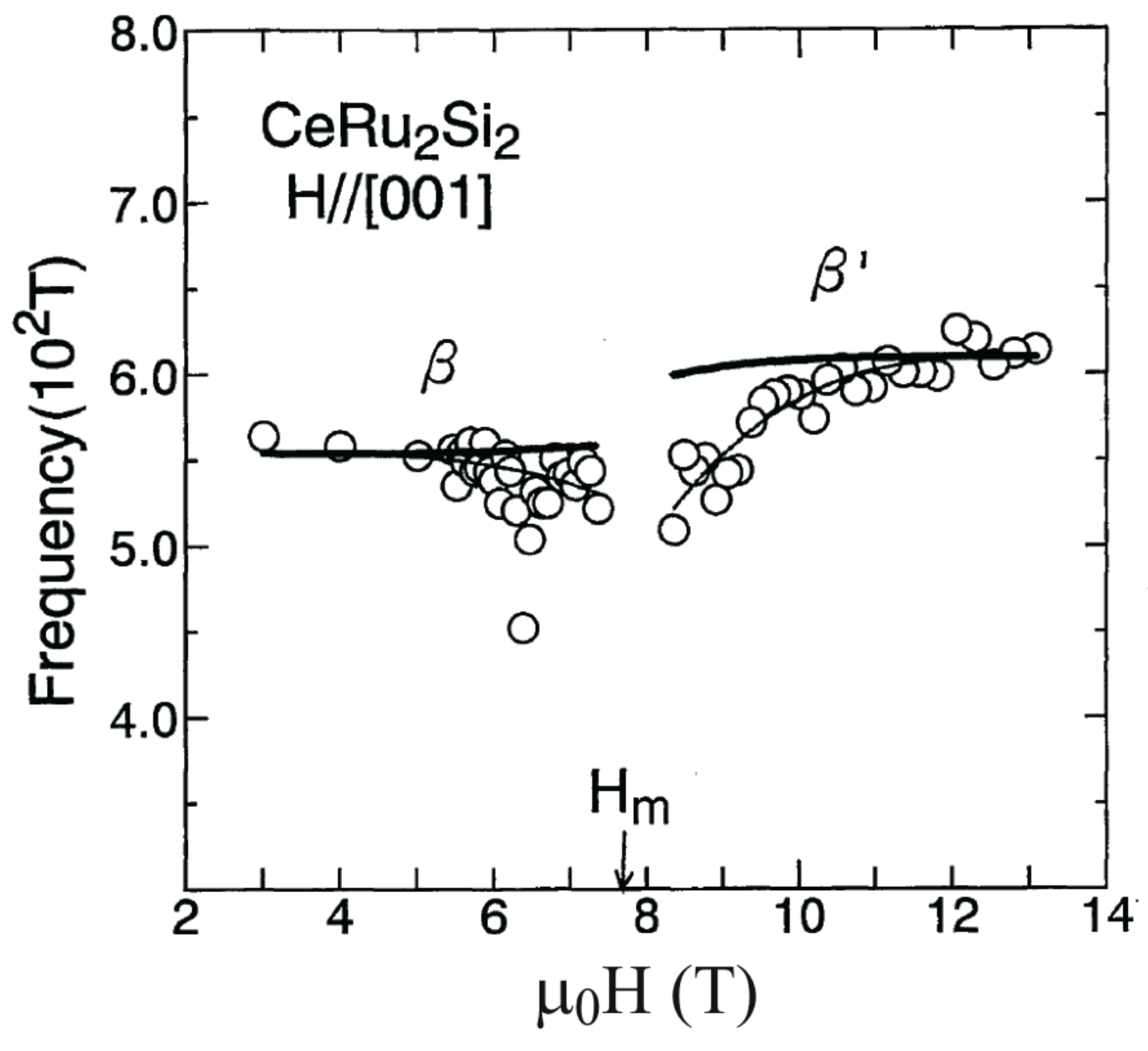}
\end{center}
\caption{Variation of the frequencies of the $\beta$ and $\beta^\prime$ oscillations with field\cite{Takashita96}.  The open circles denote the observed frequency and the thin solid line is a fit to the data points.  The thick solid line is the true frequency calculated from the frequencies of the thin solid line using eq.(\ref{eq:truefrequency}).  }
\label{fig:90025Fig24}
\end{figure}

Next we describe how the Fermi surface changes accompanied with the metamagnetic transition.  Figure \ref{fig:90025Fig24} shows the measured frequencies of the $\beta$ and $\beta^\prime$ oscillations by open circle as a function of magnetic field.  The scatter of the data points mostly comes from the anomalous phase change due to the spin dependent mass as described above in $\S$3.2.1.  The thin solid lines denote the traces which approximately follow averaged values of the frequency.  The frequency appears to change considerably around $H_{\rm m}$.  This observation comes mostly from the effect that the real frequency change with field is magnified owing to the fact that the dHvA frequency is a function of $1/B$ or $1/H$ ($\S$A.1 and $\S$A.2).  The difference in the magnitudes of $B$ and $H$ are assumed to be negligible except for the fields very close to the metamagnetic transition field where the dHvA oscillations could not be observed.  Then, we obtain the relation between the observed frequency $F_{ob}(H)$ and the true frequency $F(H)$ as \cite{Ruitenbeck82,Sigfusson84,Aoki93b}
\begin{eqnarray}
F_{ob}(H)=F(H) - H\frac{dF(H)}{dH}.	
\label{eq:truefrequency}
\end{eqnarray}
By using this equation and by assuming that the observed frequencies at the highest and lowest fields are field independent, we obtain the $F(H)$'s of the $\beta$ and $\beta^\prime$ oscillations from the $F_{ob}(H)$'s. We show them in Fig. \ref{fig:90025Fig24} as thick solid lines.  

We note that both the $F(H)$'s of the $\beta$ and $\beta^\prime$ oscillations slightly increases with increasing field.  The changes are found to be explained by the volume expansion described in $\S$2.1. The pressure effect on the dHvA frequencies have been studied \cite{Takashita98,Aoki01} to derive how much the frequency changes with compression. If we assume that the volume expansion gives rise to the same amount of the frequency change with the reversed sign as the volume contraction does, we can estimate the frequency change near $H_{\rm m}$ from the observed volume expansion  by using the data of the frequency change with pressure and the compressibility.  Although there is no report for the compressibility above $H_{\rm m}$, we may assume that the value is nearly the same as that below $H_{\rm m}$.  By using these values, the magnitudes of the volume magnetostriction can account for the changes in $F(H)$'s of the $\beta$ and $\beta^\prime$ oscillations in Fig. \ref{fig:90025Fig24} almost quantitatively.  However, the difference between the two frequencies of the $\beta$ and $\beta^\prime$ oscillations across $H_{\rm m}$ is too large to be explained simply by the volume expansion.  That is, the frequencies are nearly constant below and above $H_{\rm m}$ except for the effect of the volume expansion, but they change in a very narrow range of fields around $H_{\rm m}$, which is narrower than the transition width of magnetization.  This observation is in accord with the results that the metamagnetic transition takes place with constant volume condition\cite{Matsuhira97}. 
The relation between the changes in the magnetization and the frequency can be more explicitly demonstrated under high pressures where the metamagnetic transition width becomes broader\cite{Mignot88,Mignot89}.  

\begin{figure}[htbp]
\begin{center}
\includegraphics[width=0.6\linewidth]{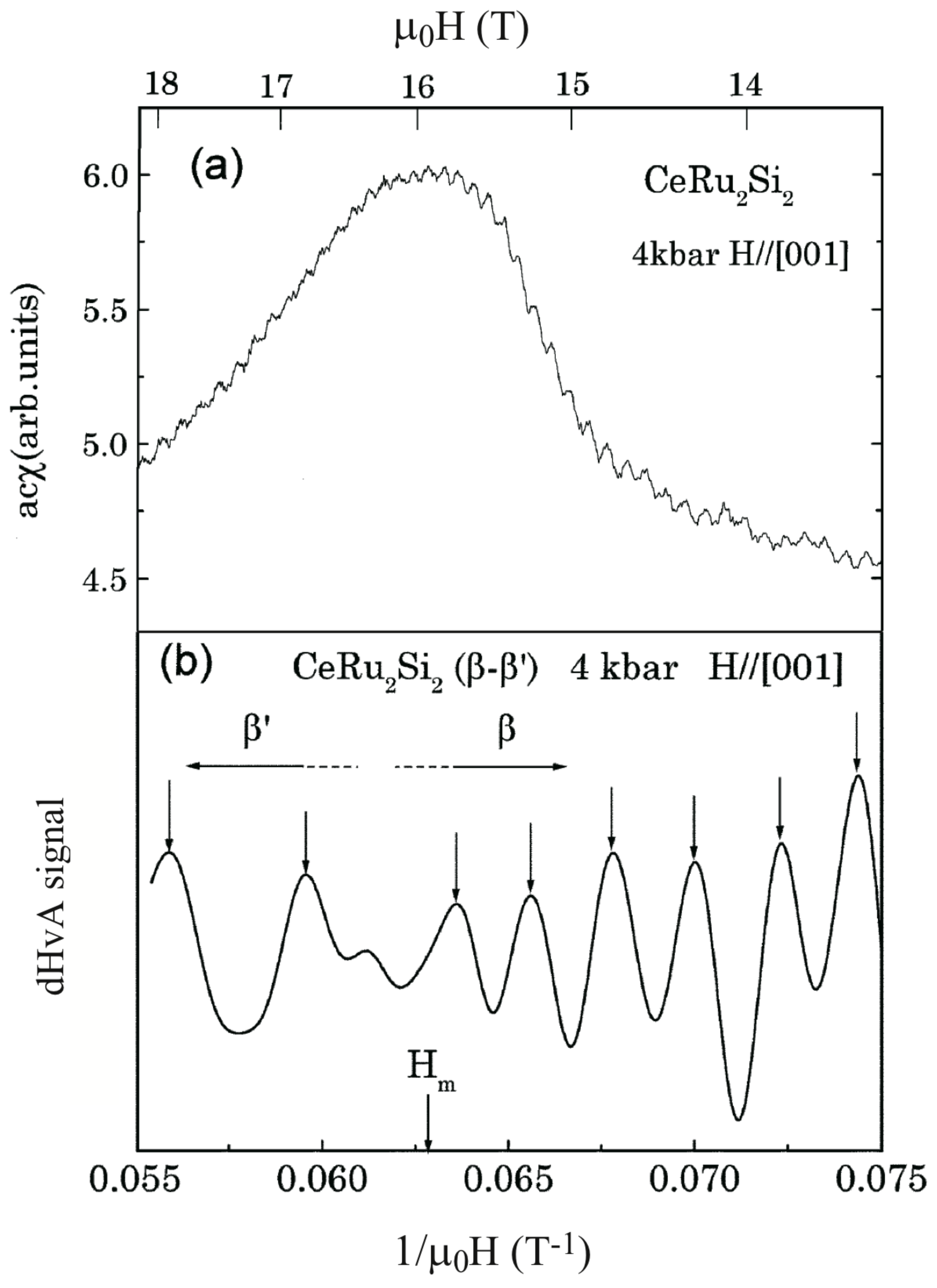}
\end{center}
\caption{ (a) AC susceptibility as a function of magnetic field around $H_{\rm m}$ measured at 4 kbar.  (b) Traces of the $\beta$ and $\beta^\prime$ oscillations plotted as a function of magnetic field around $H_{\rm m}$\cite{Aoki01}.}
\label{fig:90025Fig25}
\end{figure}

\begin{figure}[htbp]
\begin{center}
\includegraphics[width=0.6\linewidth]{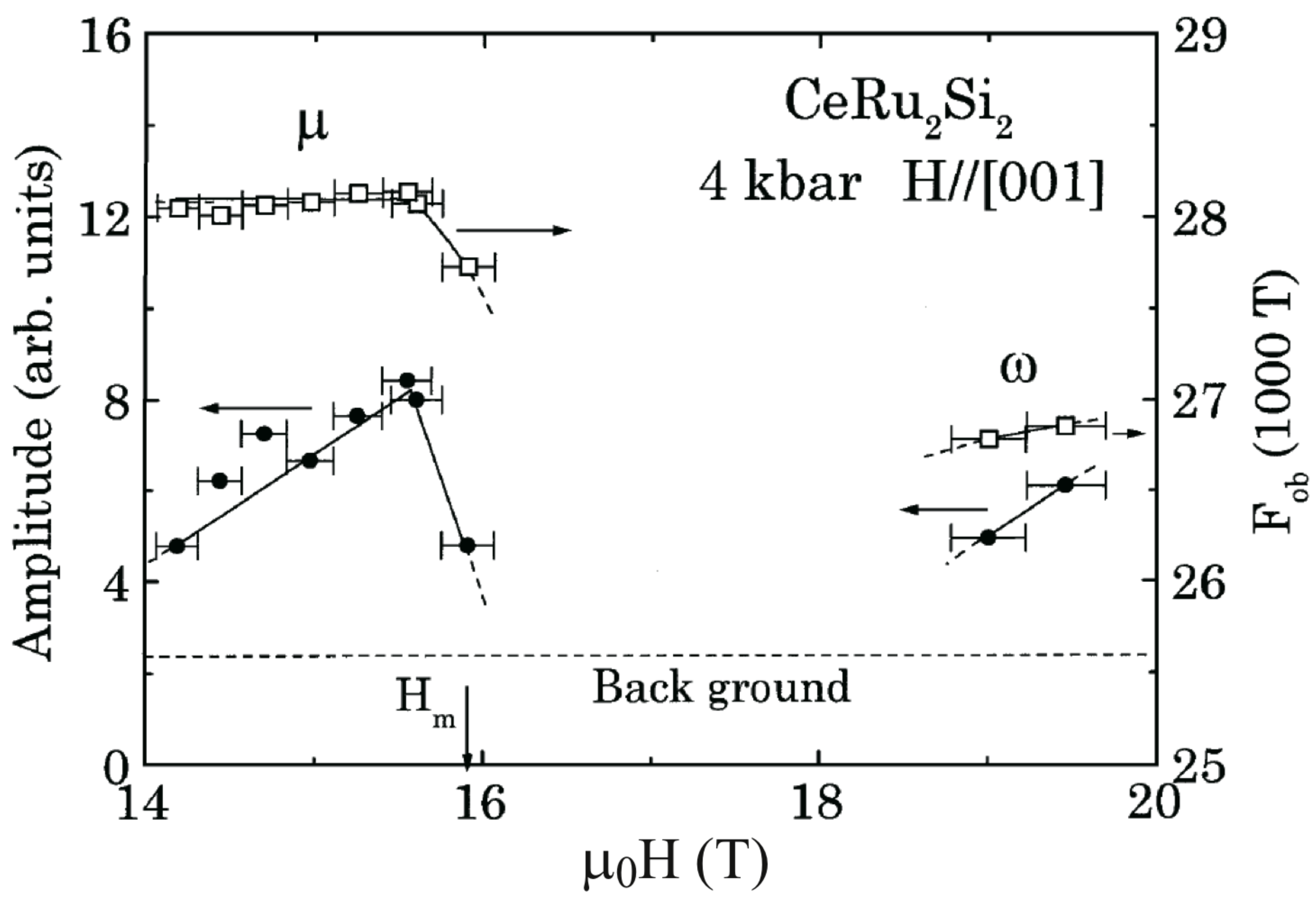}
\end{center}
\caption{Transition behavior of the $\mu$ and $\omega$ oscillations around $H_{\rm m}$.  The open and closed symbols denote the observed frequencies and amplitudes, respectively. }
\label{fig:90025Fig26}
\end{figure}

We show the AC susceptibility as a function of magnetic field around $H_{\rm m}$ observed at 4 kbar in Fig. \ref{fig:90025Fig25}.  The traces of the $\beta$ and $\beta^\prime$ oscillations are picked up by using a low frequency filter to show how they change as a function of magnetic field around $H_{\rm m}$.  Since the $\beta$ and $\beta^\prime$ oscillations have similar frequencies, we can observe the two oscillations with the same filtering condition.  It is noted that the $\beta$ oscillation switches to the $\beta^\prime$ oscillation in much narrower range than the transition width of magnetization.  It can be also shown that the $\gamma$ and $\kappa$ oscillations observed below $H_{\rm m}$ disappear and the $\gamma^\prime$ and $\delta$ oscillations above $H_{\rm m}$ appear in a much shorter range of field than the width of the magnetization change.  This observation is also true for the $\mu$ and $\omega$ oscillations.  Figure \ref{fig:90025Fig26} plots the frequencies and amplitudes of the $\mu$ and $\omega$ oscillations as a function of magnetic field at 4 kbar where both the oscillations can be observed.  The $\mu$ oscillation persists close to $H_{\rm m}$ and then disappears.  On the other hand, the $\omega$ oscillation starts to have the observable amplitude from a little bit above $H_{\rm m}$.  It is noted that the amplitude of the $\omega$ oscillation seems to be small compared with that of the $\mu$ oscillation, considering that the effective mass of the $\omega$ oscillation is 8 $m_0$ while that of the $\mu$ oscillation is 34 $m_0$.  This observation is discussed in the next section.

The fact that the change of the Fermi surface takes place in a much narrower range than the transition width of magnetization indicates that the change in the magnetization around $H_{\rm m}$ may not be explained by a simple picture that the magnetization comes from the difference in the numbers of  the up and down spin conduction electrons or the Zeeman splitting of the conduction bands.

In passing we would like to comment on the metamagnetic behavior in YbIr$_2$Zn$_{20}$\cite{Yoshiuchi09,Takeuchi10}.  As shown in Fig. \ref{fig:90025Fig15} the proportional constant between $H_{\rm m}$ and $T_{\rm m}$ of this compound is nearly the same as those of other heavy compounds. It is reported that no significant changes are observed for the dHvA frequency at $H_{\rm m}$, although similar behaviors of the mass enhancement and the transport properties to those of CeRu$_2$Si$_2$ are observed associated with the metamagnetic transition.   However, the dHvA frequency presented for the $\alpha$ oscillation seems to be the observed frequency $F_{ob}(H)$ in eq.(\ref{eq:truefrequency}), but not the true frequency $F(H)$.  It appears to change significantly and in a complicated way with magnetic field below and above $H_{\rm m}$.  The frequencies below and above $H_{\rm m}$ seem to become nearly the same at $H_{\rm m}$.  This behavior is similar to that of $F_{ob}(H)$ presented for the $\beta$ and $\beta^\prime$ oscillations in Fig. \ref{fig:90025Fig24} and to the other oscillations presented in Fig. \ref{fig:90025Fig21}\cite{Takashita96}.  It might be interesting to obtain the true frequency change together with pressure dependence of the frequency and to confirm whether or not the frequency is constant across $H_{\rm m}$.  

\subsection{Fermi surface properties in fields above the metamagnetic transitions of Ce$_x$La$_{1-x}$Ru$_2$Si$_2$ and CeRu$_2$(Si$_x$Ge$_{1-x}$)$_2$}

\begin{figure}[htbp]
\begin{center}
\includegraphics[width=0.6\linewidth]{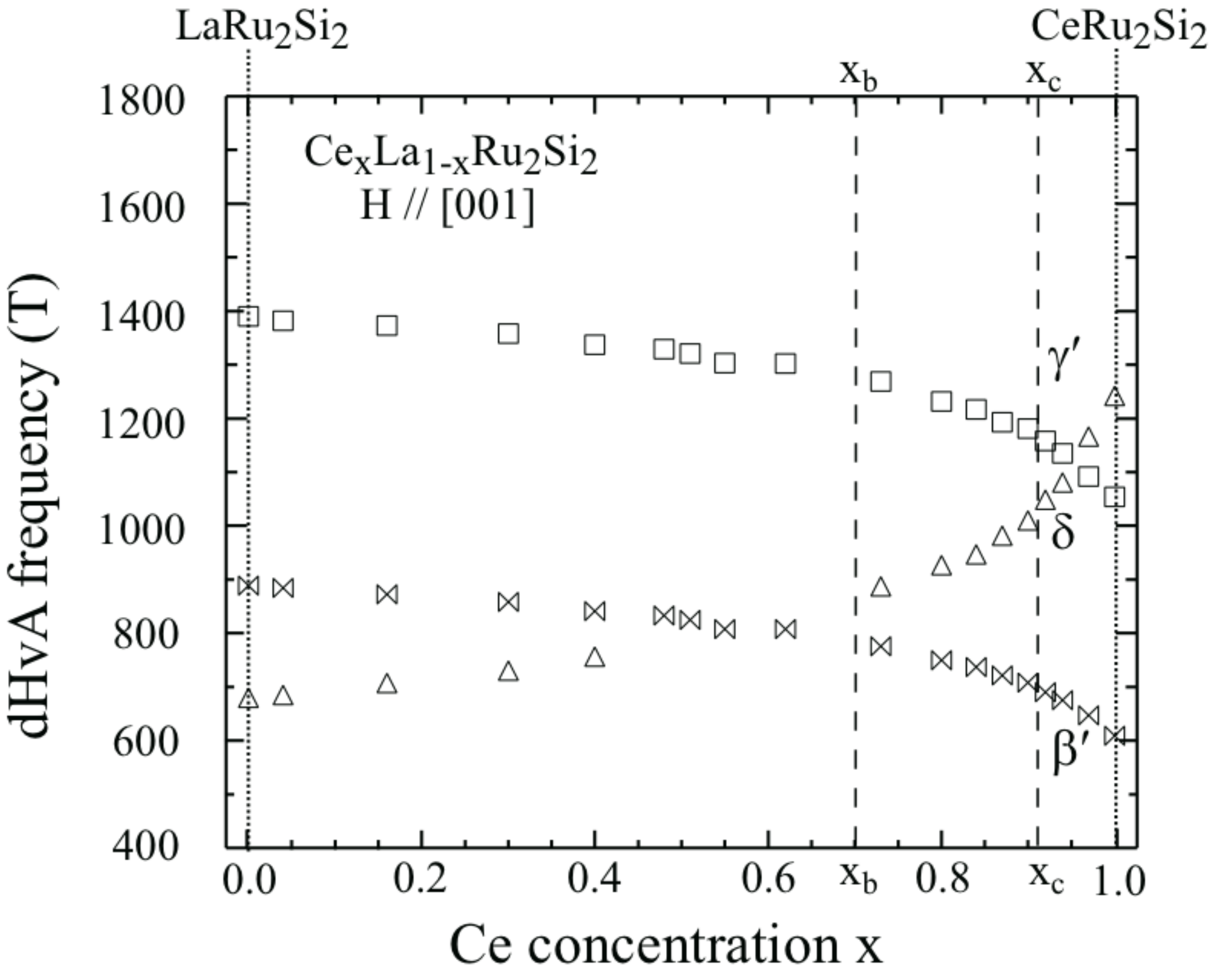}
\end{center}
\caption{Frequencies of the $\beta^\prime$, $\gamma^\prime$ and $\delta$ oscillations for fields above metamagnetic transition fields plotted as a function of Ce concentration\cite{Matsumoto08}.}
\label{fig:90025Fig27}
\end{figure}

\begin{figure}[htbp]
\begin{center}
\includegraphics[width=0.6\linewidth]{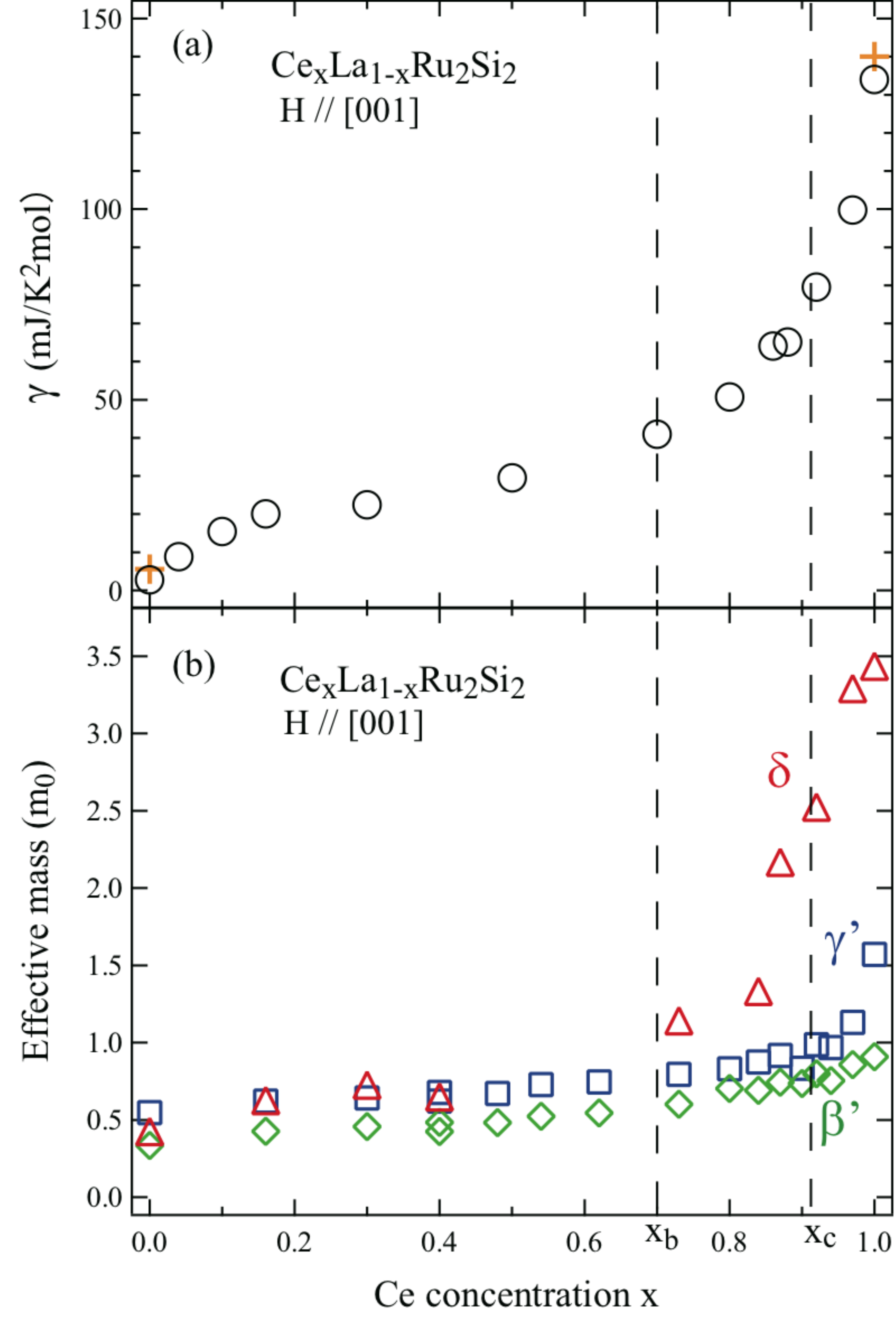}
\end{center}
\caption{(Color on line) (a)Electronic specific heat coefficient derived from the $A$ coefficient of $T^2$ dependence of resistivity at 13 T as a function of Ce concentration (open circle).  The cross symbol denotes the electronic specific heat coefficient determined by the specific heat measurements\cite{Meulen91,Besnus85}.  (b) Effective masses of the $\beta^\prime$, $\gamma^\prime$ and  $\delta$ oscillations above $H_{\rm m}$\cite{Shimizu12}. }
\label{fig:90025Fig28}
\end{figure}

\begin{figure}[htbp]
\begin{center}
\includegraphics[width=0.6\linewidth]{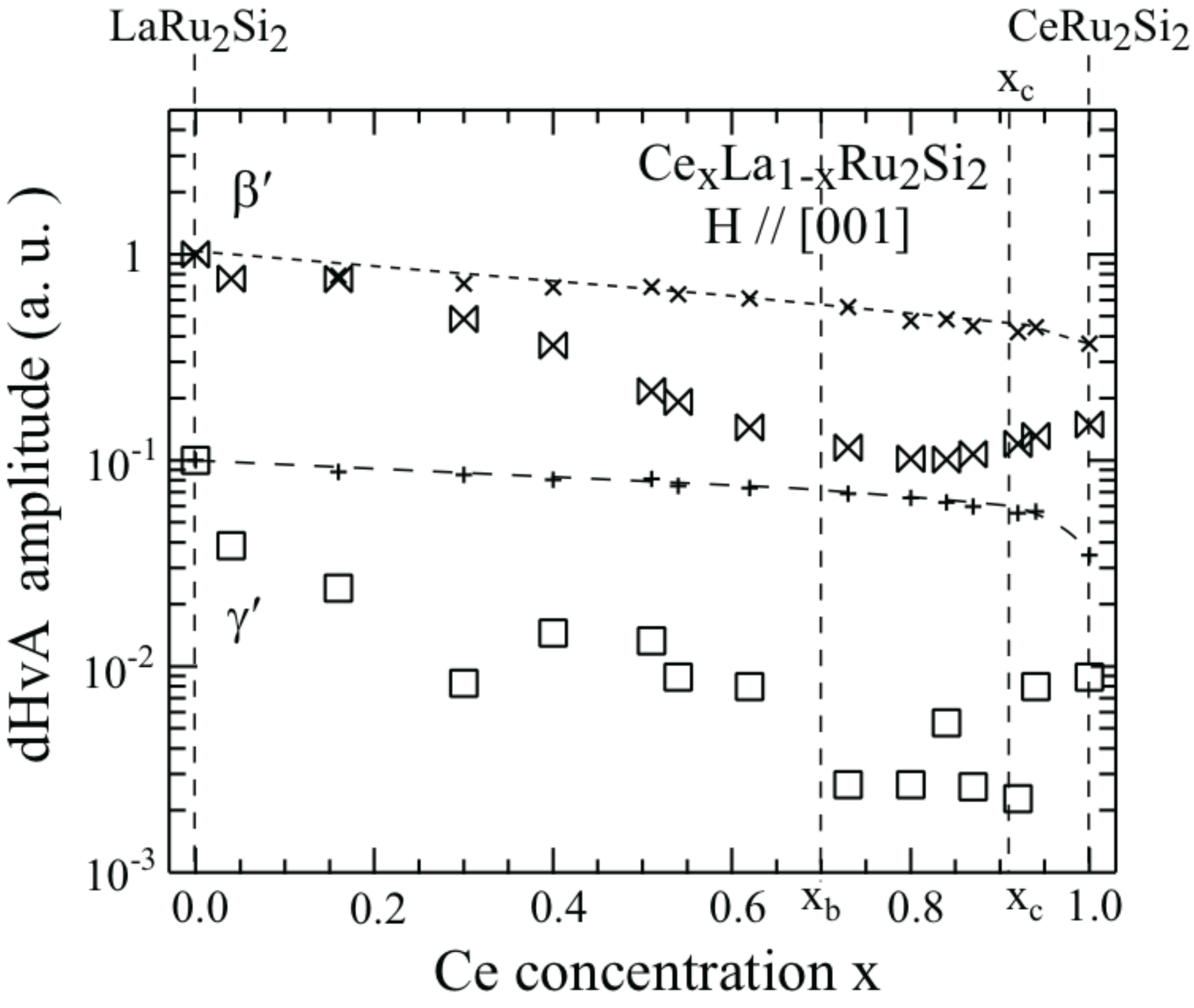}
\end{center}
\caption{Relative signal magnitudes of the $\beta^\prime$ and $\gamma^\prime$ oscillations plotted as a function of Ce concentration\cite{Matsumoto08}. The magnitude of the $\beta^\prime$ oscillation is normalized to 1 in LaRu$_2$Si$_2$ and that of the $\gamma^\prime$ oscillation to 0.1. The $\times$ and + symbols denote the magnitudes calculated by assuming that the oscillation amplitude is determined only by the effective mass and frequency. ($\S$A.1).  The broken lines are guides to the eye.}
\label{fig:90025Fig29}
\end{figure}

To discuss the electronic structure in fields above $H_{\rm m}$, we present the Fermi surface properties in fields observed well above the metamagnetic transition fields in CeRu$_2$Si$_2$ and its alloys.  Figures \ref{fig:90025Fig27},  \ref{fig:90025Fig28} and \ref{fig:90025Fig29} show the frequencies, effective masses and signal amplitudes of the oscillations observed in Ce$_x$La$_{1-x}$Ru$_2$Si$_2$, respectively.   We also include the value of the electronic specific heat coefficient $\gamma$ derived from the measurements of coefficient $A$ by using the Kadowaki-Woods relation\cite{Kadowaki86}.  Three oscillations can be observed for all the concentration range.  The high frequency oscillation $\omega$ can be observed only in CeRu$_2$Si$_2$ or LaRu$_2$Si$_2$, because of the small amplitudes of the oscillations.  The frequencies of the three oscillations change continuously from CeRu$_2$Si$_2$ to LaRu$_2$Si$_2$ within the experimental accuracy.    A similar continuous change is observed in CeRu$_2$(Si$_x$Ge$_{1-x}$)$_2$.  In this system,  the frequency changes continuously for $x > x_a$ and connects to the average of the spin split frequencies for $x < x_a$\cite{Sugi08}. 

The values of the effective mass and the coefficient $A$ also continuously increase with Ce concentration $x$ and are not enhanced at $x_c$.  This behavior is in contrast to that with fields in the (001) plane in which the effective mass is enhanced around $x_c$ ($\S$4).  The effective mass starts to increase more rapidly from around $x_b$.  The behavior of the value of $\gamma$ is similar to those of the effective masses, particularly to that of the $\delta$ oscillation whose f content is relatively large.   The behavior is qualitatively similar to that of $T_{\rm K}$, if $T_{\rm K}$ in Fig. \ref{fig:90025Fig4} can be extended continuously to the lower concentration side and to the value of 1.3 K of a dilute alloy of Ce$_{0.02}$La$_{0.98}$Ru$_2$Si$_2$ ($\S$4.1.1).  The values of the effective mass and $\gamma$ increase with Ce concentration or chemical pressure in the paramagnetic state.  The effective masses of the $\beta^\prime$, $\gamma^\prime$ and $\delta$ oscillations in CeRu$_2$(Si$_x$Ge$_{1-x}$)$_2$ show  similar behaviors as a function of Si concentration $x$\cite{Sugi08}. 
 
Figure \ref{fig:90025Fig29} shows the relative signal amplitudes of the $\beta^\prime$ and $\gamma^\prime$ oscillations to those of LaRu$_2$Si$_2$,  the values of LaRu$_2$Si$_2$ being taken as 1 and $10^{-1}$ for the $\beta^\prime$ and $\gamma^\prime$ oscillations, respectively.  Since the effective mass depends on field, the Dingle temperature cannot be determined by the conventional method ($\S$A.2).  The signal amplitude is obtained by keeping the experimental condition like sample volume and shape the same.  The amplitudes become minimum approximately around $x=0.8$ and are about 1/20 of those of LaRu$_2$Si$_2$.    Considering the case of normal metal alloys ($\S$A.3), the decrease of the amplitude due to alloying is surprisingly small.  This implies that the electronic state of Ce probed by the conduction electrons is similar to that of La.  Such a small reduction in the signal amplitude in alloys can be observed for Ce$_x$La$_{1-x}$B$_6$\cite{Goodrich99,Nakamura06,Endo06} at high magnetic fields or Ce$_x$La$_{1-x}$Sb\cite{Nakanishi01} where the f electron is conventionally thought to be localized, although no one may have confirmed whether the definition of the term ``localized" in these systems is the same among the researchers.  For the details of the Fermi surface properties, we refer the reader to the references\cite{Goodrich99,Nakamura06,Endo06, Nakanishi01} and other references therein.  This small reduction with alloying is also in contrast to the observations for the itinerant f electron system in the paramagnetic ground state where a small amount of impurity degrades the signal unobservable.  

\begin{figure}[htbp]
\begin{center}
\includegraphics[width=0.6\linewidth]{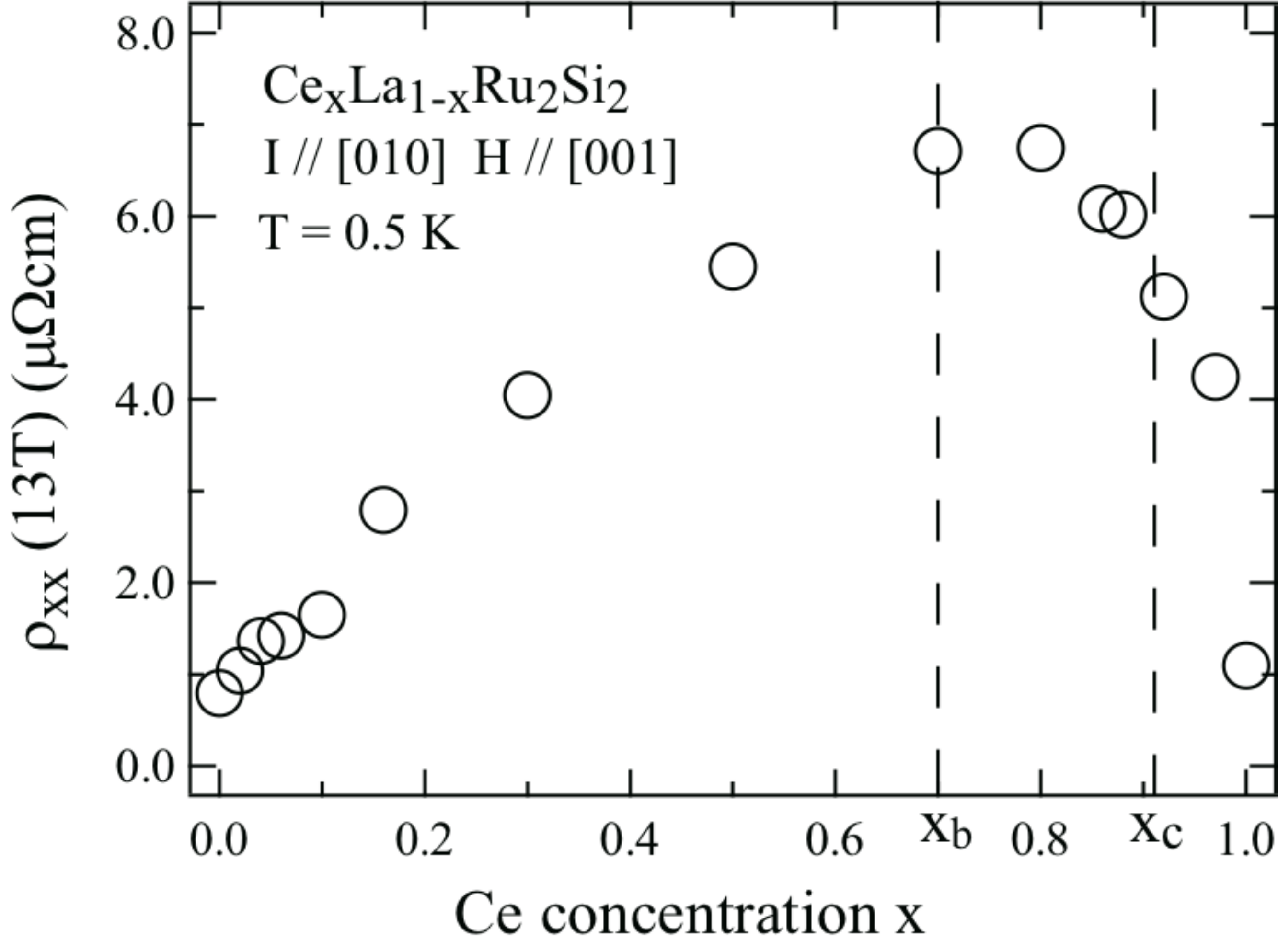}
\end{center}
\caption{Resistivity at 13 T as a function of Ce concentration\cite{Shimizu12}. }
\label{fig:90025Fig30}
\end{figure}

The broken lines show the relative amplitudes calculated by assuming that the amplitudes are determined only by the effective mass and the frequency ($\S$A.1), i.e. the contribution from the scattering is not taken into account.  The difference between the observed amplitudes and the broken lines indicate that the magnitudes of the effective mass and the frequency are not main factors to determine the amplitudes. We show the residual resistivity at 13 T as a function of Ce concentration in Fig. \ref{fig:90025Fig30}.  It takes a maximum around $x = 0.8$ and is in accord with the behavior of the amplitudes.  The same correlation is found between the residual resistivity at zero magnetic field and the signal amplitudes for the case with fields in the (001) plane as demonstrated in $\S$4.  However, it is noted that the position of the maximum at 13 T is different from that in the residual resistivity at zero magnetic field (Fig. \ref{fig:90025Fig7}) and the value at the maximum are smaller by a few times.  If the scattering comes mostly from the charge fluctuation as mentioned in $\S$2.2.3,  the difference may arise from the suppression of the charge fluctuation.  The magnetic field along the easy axis would suppress the magnetic fluctuation and also the charge fluctuation by driving the f electron in a more polarized and localized state. The suppression may also shift the position of the maximum closer to $x=0.5$ where the maximum of the resistivity could reside if the charge and magnetic fluctuations were absent.  The observations for the signal amplitude and resistivity are consistent with the observation for the valence of Ce ($\S$2.1) that the f electron is more localized above $H_{\rm m}$.

In $\S$3.1.2, we have discussed that the Fermi surface above $H_{\rm m}$ is likely to resemble that of LaRu$_2$Si$_2$.  The Fermi surface properties above $H_{\rm m}$ in the alloys are also different from those below the metamagnetic transition fields and in some aspects resemble those of the magnetic localized f electron system.  If the  transition in the volume of the Fermi surface from large to small takes place at a certain concentration, we would observe discontinuous changes in the frequency and the effective mass as well as the transport and magnetic properties at that concentration.  The continuous changes indicate that the electronic structures in fields well above the metamagnetic transitions are continuously connected to that of LaRu$_2$Si$_2$ or CeRu$_2$Ge$_2$. 

\subsection{Models for the electronic structure in fields above the metamagnetic transition}

The observations described in previous sections 3.1, 3.2 and 3.3 indicate that the electronic structure above $H_{\rm m}$ is significantly different from that below $H_{\rm m}$ in many aspects:  \\
\noindent 
(1) The shape of the Fermi surface resemble that of the small Fermi surface, while that below $H_{\rm m}$ is that of the large Fermi surface.\\
 \noindent 
(2) The effective masses are much smaller than those below $H_{\rm m}$ and decreases with magnetic field, while those well below  $H_{\rm m}$ are nearly constant.  \\ 
\noindent 
(3) The effective masses increase with pressure while those below $H_{\rm m}$ decrease with pressure.  \\
\noindent  
(4) The reduction in signal amplitude due to alloying is much smaller than that below $H_{\rm m}$.  \\
\noindent 
(5) The Fermi surface properties above the metamagnetic transitions change smoothly and continuously to those of LaRu$_2$Si$_2$ or CeRu$_2$Ge$_2$. \\
\noindent
(6) The signal amplitudes above $H_{\rm m}$ are found to be systematically smaller than those below $H_{\rm m}$\cite{Takashita96}, although this observation has not been demonstrated in this article except for the case of the $\omega$ oscillation in Fig. \ref{fig:90025Fig26}.  The difference cannot be explained by the change in the parameters used in the LK formula.

 We also note interesting features of the metamagnetic transition:\\ 
\noindent 
(a)The Fermi surface properties change from those below $H_{\rm m}$ to above $H_{\rm m}$ in a much narrower field range than the transition width of magnetization.\\
\noindent 
(b) The effective masses are similarly enhanced irrespective of the characters of the Fermi surface responsible for the oscillation.

\begin{figure}[htbp]
\begin{center}
\includegraphics[width=0.6\linewidth]{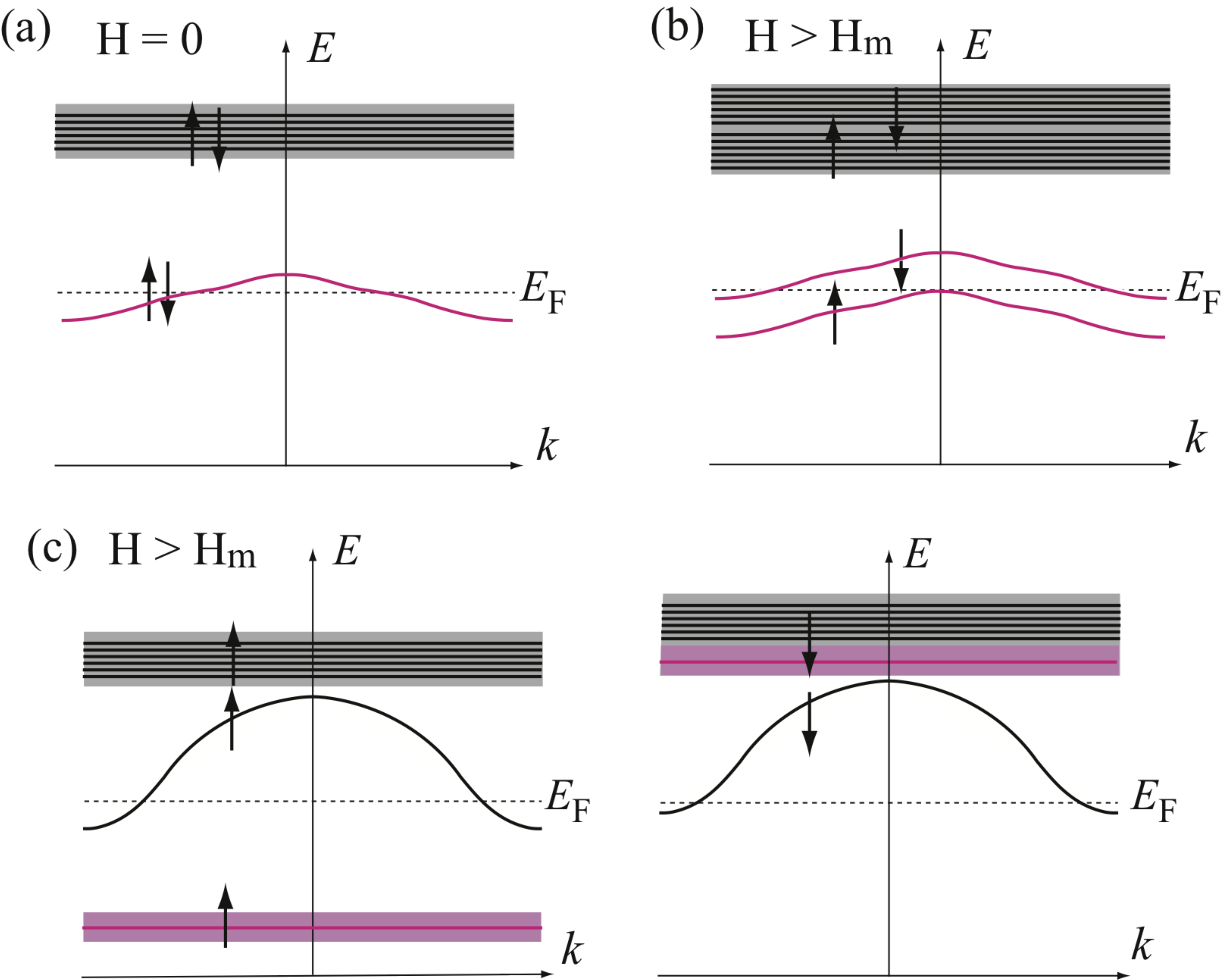}
\end{center}
\caption{(Color on line) (a)Schematic illustration for the band structure of  CeRu$_2$Si$_2$ below $H_{\rm m}$.  The  band crossing the Fermi level $E_{\rm F}$ is the one corresponding to the large hole surface in Fig. \ref{fig:90025Fig17}(a).  (b) Band structure above $H_{\rm m}$ corresponding to the interpretation (I).  (c)Band structures above $H_{\rm m}$ corresponding to the interpretation (II).  See also the text.}
\label{fig:90025Fig31}
\end{figure}

In the following, we discuss the two naive models of the electronic structure above $H_{\rm m}$.   In Fig. \ref{fig:90025Fig31}(a) we show schematically the band which corresponds to the large hole surface of CeRu$_2$Si$_2$\cite{Suzuki10} and assume that the band represents the f electron state of the system.  It is shown as the pink band in Fig. \ref{fig:90025Fig31}(a).  The main component of the band is the $J=5/2$, $J_{\rm z}=\pm5/2$ state of the f electron\cite{Suzuki10}. We denote these states simply as up and down spin states.  Other bands with different f components like $|5/2, \pm3/2\rangle$, $|5/2, \pm1/2\rangle$ and  $J=7/2$ are situated well above the Fermi level.  We denote schematically these bands with the solid lines and gray band. Upon application of magnetic field, the bands split into the up and down spin bands by the Zeeman effect.  The splitting of the $|5/2, \pm5/2\rangle$ band near the Fermi energy gives rise to a large initial susceptibility due to the large density of states at the Fermi level.  The increase in the magnetization at low fields can be accounted for by the increase in the difference between the numbers of the up and down spin conduction electrons. 

For the electronic structure above the metamagnetic transition, two interpretations are possible as described above in $\S$3.2.2.  Figure \ref{fig:90025Fig31}(b) shows the state at the metamagnetic transition schematically according to interpretation (I).  The band structure are approximately the same below and above $H_{\rm m}$\cite{Miyake06,Daou06} and the metamagnetic transition can be qualitatively explained by the Zeeman splitting of the quasiparticle bands.  In one of the interpretation\cite{Daou06} along this line,  the one spin band comes down below the Fermi level and the Fermi surface of this band disappears at the metamagnetic transition.  The remaining Fermi surface may be the missing Fermi surface which is responsible for the difference between the observed effective mass and the electronic specific heat coefficient.  The magnetization can be mostly accounted in terms of  the difference between the numbers of the up and down spin conduction electrons. 

Figure \ref{fig:90025Fig31}(c) shows a schematic illustration of the model for the state above $H_{\rm m}$ according to the interpretation (II)\cite{Matsumoto08,Suzuki10}.  In this interpretation, the interaction $U$ among the f electron\cite{Suzuki10} or $I_F$ plays an important role for the metamagnetic transition together with the Zeeman effect.  Due to the interaction energy $U$, the correlation energy of the system would be smaller when the electrons with the same spin direction occupied the band.  We show in Fig. \ref{fig:90025Fig31}(c) a schematic electronic structure of the polarized state above $H_{\rm m}$.  In the polarized state, the $J=5/2$, $J_{\rm z}=\pm5/2$ band splits largely so that one of the spin band is below the Fermi level and the other is above the Fermi level.  Owing to the lowered correlation energy and larger gain in the Zeeman energy, we speculate that the energy of the state could become lower than the Zeeman split state of the heavy Fermion band at high fields. The magnetic field strength to drive the Zeeman split state into the polarized state is smaller for the ground state with larger $U$ or smaller $T_{\rm K}$ ($T_{\rm m}$) as shown in Fig. \ref{fig:90025Fig15}.  

We think that this model may explain the characteristic observations (1) - (5), (a) and (b).  Since almost one f electron state per Ce is below the Fermi level and is polarized, the hole surface becomes large which is consistent with the observation (1).    The transition may be continuous but may be accelerated cooperatively with the volume expansion making the transition sharp.  Although the f electron becomes more localized,  the hybridization between the conduction electrons and the f electron remains significant and the f electrons form a band.  In this polarized state the magnetization arises mostly from the polarized f electron state below the Fermi level.  Consequently, the magnetization behavior is different from that below $H_{\rm m}$.  The f electron band shifts below the Fermi level in a very short interval of field giving rise to the very sharp change of the Fermi surface, while the magnetization behavior with field may be more gradual because the polarization of the f electron or the  decoupling from the conduction electrons continues to increase with increasing field (a).   Consequently, the f electron content at the Fermi surface or the effective mass gradually decreases with magnetic field and could be significantly spin dependent (2).  On the other hand, the f electron content at the Fermi surface will increase with pressure because of the increase in the hybridization resulting in the increase of the effective mass (3).  Since most of the f content is well below the Fermi level, the potential of Ce which the conduction electrons feel may not be very different from that of La (4). 

It is thought that in a very dilute alloy the f electron of the Kondo singlet state is continuously connected to the localized state in the high magnetic field limit. Although it is not obvious whether the f electron of the itinerant f electron system becomes a localized state without phase transition, the f electron state of the polarized state in Fig. \ref{fig:90025Fig31}(c)  would smoothly and continuously change to the localized f electron state in the high magnetic field limit.  Here we use the term ``localized" by assuming that the f electron in the high magnetic field limit is completely polarized and the volume of the Fermi surface is small.  As demonstrated in $\S$3.3, the polarized state above $H_{\rm m}$ seems also to change continuously and smoothly to the state of LaRu$_2$Si$_2$ or CeRu$_2$Ge$_2$ (5), i.e., to the state of the small Fermi surface or the magnetic localized f electron state.   The interpretation (II) implies that the f content slips off from all the Fermi surfaces and the decreasing ratio of the f content on the each Fermi surface is approximately the same, while the interpretation (I) would naively implies that the effective mass change with field depends on the Fermi surface.   Although we can not mention a proper mechanism for the enhancement of the effective mass around $H_{\rm m}$,  we think that the interpretation (II) is consistent with observation (b).

Thus, the model (II) is likely to explain the Fermi surface properties above $H_{\rm m}$ and the difference from those below $H_{\rm m}$.  

We may regard this state of the f electron as localized from the several characteristic features (1) - (5) mentioned above, because these features are the same or similar to those observed in the systems where the f electron is conventionally thought to be localized.  On the other hand, the induced magnetic moment above $H_{\rm m}$ is smaller or does not match with that expected from the crystal electric field scheme and increases continuously with increasing magnetic field as stated in $\S$2.1. The X-ray absorption spectra measurements under magnetic fields indicate that the $f_0$ satellite intensity is almost unchanged above $H_{\rm m}$\cite{Okane12}. These observations indicate that significant hybridization between the f electron and the conduction electrons remains above $H_{\rm m}$.   In these respects, the f electron state in the polarized state may have a significant itinerant character.  If the transition is crossover, the polarized state above $H_{\rm m}$ should be connected continuously to the itinerant f electron state in the paramagnetic ground state, while the state is likely to change continuously also to the magnetic localized f electron state.  This implies that there may be no clear boundary where the discontinuous change takes place between the small Fermi surface state and the large Fermi surface state along the path from the itinerant f electron state in the paramagnetic ground state to the magnetic localized f electron state through the metamagnetic crossover.   We think that it may be misleading to categorize the f electron state in the polarized state above $H_{\rm m}$ as either localized or itinerant in the present case\cite{Matsumoto08}.  In this article we refer it as `` the f electron state above $H_{\rm m}$", which has the Fermi surface properties described in this $\S$3.     

\section{Fermi Surface Properties in the Ground State of Ce$_x$La$_{1-x}$Ru$_2$Si$_2$ and  CeRu$_2$(Si$_{1-x}$Ge$_x$)$_2$  }
The changes in the Fermi surface properties and the f electron state at the quantum phase transition have attracted much attention in the past decade and a large number of theoretical as well as experimental studies have been performed.  However, the experimental observations (YbRh$_2$Si$_2$\cite{Paschen04,Gegenwart02}, UGe$_2$\cite{Terashima01,Settai02}, CeRhIn$_5$\cite{Shishido05}, CeIn$_3$\cite{Endo04,Settai05}, CeRhSi$_3$\cite{Terashima07}, CeRh$_{1-x}$Co$_x$In$_5$\cite{Goh08}) differ depending on the system and sometimes on the report. Theoretical results also depend on the values of parameters as well as the theoretical model and the methods of calculation used\cite{Si01,Si10,Vojta10,Senthil03,Senthil04,Senthil05,Vojta08,Watanabe07,Watanabe09,Lanata08,Martin08,Martin10,Leo08,Leo08b,Otsuki09,Hoshino10,Hoshino13,Watanabe10,Kubo13b}.   No unified understanding seems to be established on the Fermi surface properties at the quantum phase transition.   

In this section we present the dHvA effect measurements to study how the Fermi surface properties of normal metal LaRu$_2$Si$_2$ or those of ferromagnetic compound CeRu$_2$Ge$_2$ evolve to those of CeRu$_2$Si$_2$ with alloying.  With magnetic fields in the (001) plane we observe the Fermi surface properties below $H_{\rm m}$. We may assume that they are substantially the same as those in the ground state.  We pay particular attention to the f electron state and discuss how it changes with alloying, in particular at quantum phase transitions. We argue that in the present systems there is no drastic change in the f electron state at the quantum phase transition between the paramagnetic and antiferromagnetic sates, but there is a significant change in the f electron state at the quantum phase transition between the ferromagnetic and antiferromagnetic phases.  Since the definitions of the terms of ``localized'' and ``itinerant'' seem to be ambiguous in magnetically ordered states or in disordered alloys, careful descriptions of the f electron state are made to make clear what we mean by the terms in each case.

In $\S$4.1.1 we discuss the f electron state in the ground state of dilute alloys of Ce$_x$La$_{1-x}$Ru$_2$Si$_2$, in particular whether the f electron state can be understood by the itinerant picture or the localized picture. The changes in the Fermi surface properties with temperature in a dilute alloy are also reported.  In $\S$4.1.2,  we present how the Fermi surface properties change with Ce concentration in Ce$_x$La$_{1-x}$Ru$_2$Si$_2$, with particular attention to the issue whether or not the f electron nature or the Fermi surface volume changes at the quantum phase transition in this system.  In $\S$4.2, we present the results in CeRu$_2$(Si$_{1-x}$Ge$_x$)$_2$ to compare with those in Ce$_x$La$_{1-x}$Ru$_2$Si$_2$.
                 
\subsection{Fermi surface properties of Ce$_x$La$_{1-x}$Ru$_2$Si$_2$ with fields in the (001) plane}

\subsubsection{Fermi surface properties of dilute alloys}
\begin{figure}[htbp]
\begin{center}
\includegraphics[width=0.5\linewidth]{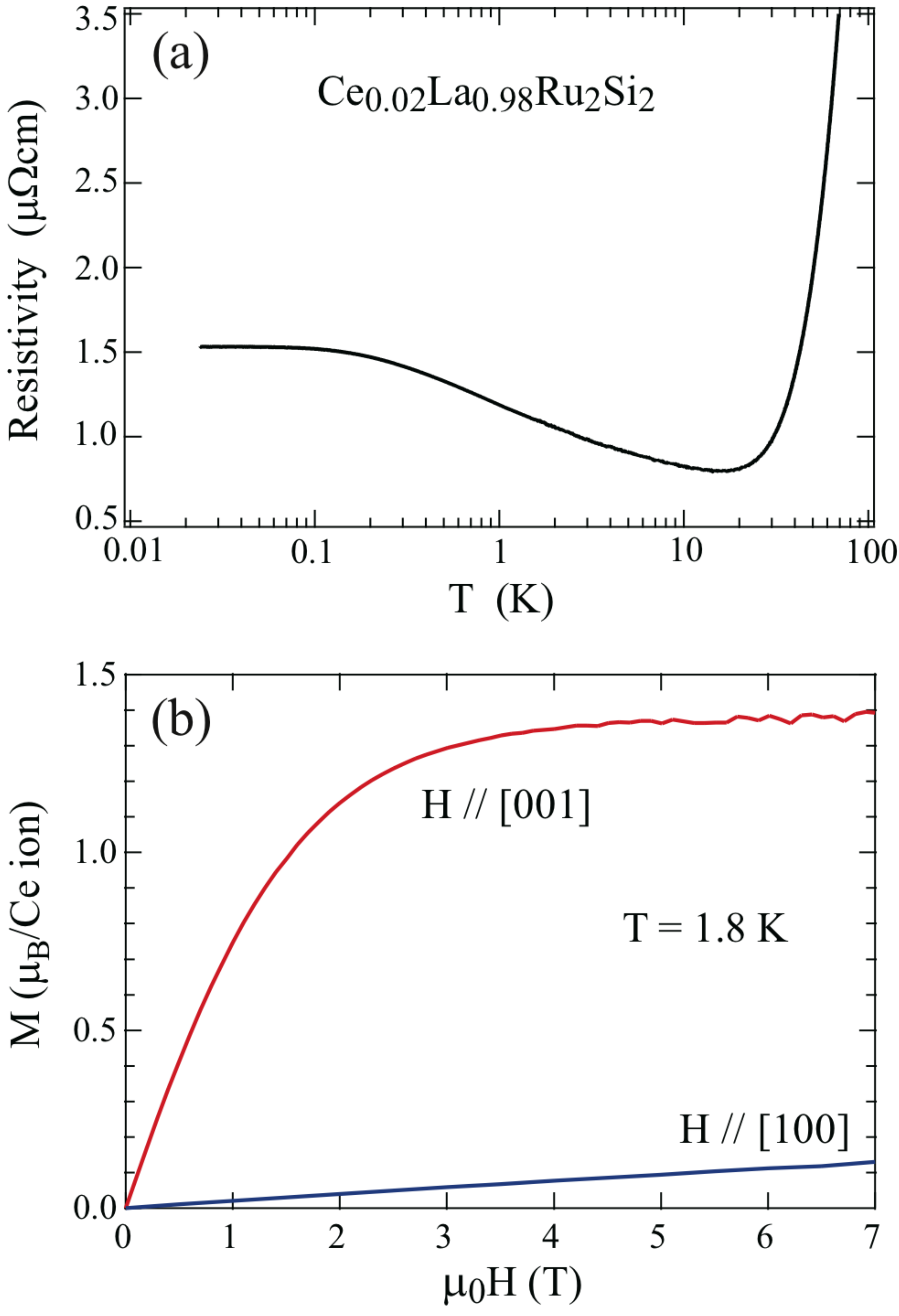}
\end{center}
\caption{(Color on line)(a) Temperature variation of resistivity of Ce$_{0.02}$La$_{0.98}$Ru$_2$Si$_2$.  The resistivity starts to increase from about 20 K and becomes nearly constant below 0.1 K.  (b) Magnetization vs. magnetic field curves for fields parallel to the [001] direction and to the [100] direction\cite{Matsumoto12}. }
\label{fig:90025Fig32}
\end{figure}

Figure \ref{fig:90025Fig32}(a)\cite{Matsumoto12} shows the resistivity of Ce$_{0.02}$La$_{0.98}$Ru$_2$Si$_2$ sample plotted against $\log T$.  The resistivity exhibits the typical dilute Kondo behavior. That is, the resistivity decreases with decreasing temperature and then increases proportionally to $- \log T$.  At the lowest temperatures it becomes nearly constant, indicating that a Kondo singlet state is formed.  The Kondo temperature is estimated to be about 1.3 K from the temperature variation of resistivity.  

Since the ground state of the dilute Kondo alloy evolves to the itinerant f electron state in the paramagnetic ground state of CeRu$_2$Si$_2$ through the antiferromagnetic state, it would be interesting and useful to clarify how the f electron state is described in the dilute Kondo alloy.  At high temperatures the f electron of the dilute Kondo alloy is assumed to behave as a localized f electron.  If we could define the Fermi edge at high temperatures and could measure the Fermi surface volume, the Fermi surface would be small.  An interesting question  may be whether the f electron state is regarded as itinerant or localized in the ground state, or such a distinction may not be useful as discussed in the previous section. 

In the paramagnetic compound the measurements of the Fermi surface volume is quite helpful to judge whether or not the f electron is itinerant as discussed in $\S$3.1. However, in a dilute alloy the measurement of the Fermi surface volume may not be helpful, because as shown in $\S$A.3, the rigid band model may not hold.  In particular, if we use the Kondo model and the phase shift in a particular channel is dominant, then the phase shift could be $\pi/2$.  Consequently, if the equations for normal metal alloys ($\S$A.3) are also valid  for the strongly correlated f electron system, the frequency change or the Fermi surface change will be zero.  

The momentum distribution of a disordered Kondo lattice model is calculated by using the dynamical mean field theory with coherent potential approximation and it is found that the sharp change of the momentum distribution or the Fermi edge does not move with the magnetic impurity concentration\cite{Otsuki10}.  This result could be brought about by using the Kondo model where the charge degrees of freedom is quenched.  Although we might obtain a different result by using the Anderson model, we suspect that the change in the Fermi edge could be smaller than the prediction of the rigid band model for normal metals where the f electron of Ce is assumed to fill the conduction band.   In this subsection we demonstrate that the f electron state may be better described with the term ``itinerant'' rather than the term ``localized", although the Fermi surface volume measured by the dHvA effect is closer to that of the small Fermi surface.   

Even if the measurements of the Fermi surface properties are useful to understand the f electron state, there is an experimental problem how to determine the tiny changes in the Fermi surface properties due to alloying ($\S$A.3).  The resolution of ARPES has been improved remarkably in these ten years, it may be still not enough to determine the tiny change as described in $\S$A.3.  On the other hand, the dHvA effect study needs high magnetic fields for the measurements which may suppress the Kondo effect. Consequently, the Fermi surface properties studied by the dHvA effect may not be that of the ground state. 

The dilute Ce alloys of Ce$_x$La$_{1-x}$Ru$_2$Si$_2$ seem to fortuitously satisfy the suitable condition to study the dilute Kondo alloy.  Fig. \ref{fig:90025Fig32}(b) shows the magnetization of Ce$_{0.02}$La$_{0.98}$Ru$_2$Si$_2$ as a function of magnetic field applied parallel to the easy axis [001] and to the hard axis [100].  As shown in Fig. \ref{fig:90025Fig2}, this system has also the strong magnetic anisotropy.  The magnetic moment direction is aligned along the [001] direction.  The magnetization with fields along the [001] direction increases rapidly with magnetic field and starts to saturate around 3 T, while with fields parallel to the [100] direction the magnetization increases slowly with magnetic field.  The measurements with fields in the (001) plane can be assumed to detect the electronic properties least affected by magnetic field, while the magnetic field component along the [001] direction significantly affects the Kondo effect as shown later and also as demonstrated in the resistivity measurements under magnetic fields\cite{Shimizu12}.  Since the Kondo temperature is as low as 1.3 K, the dHvA study as a function of temperature can also be made across the Kondo temperature.
\\
\noindent
(a) {\it Variation of the Fermi surface properties with Ce concentrations in dilute alloys}

\begin{figure}[htbp]
\begin{center}
\includegraphics[width=0.6\linewidth]{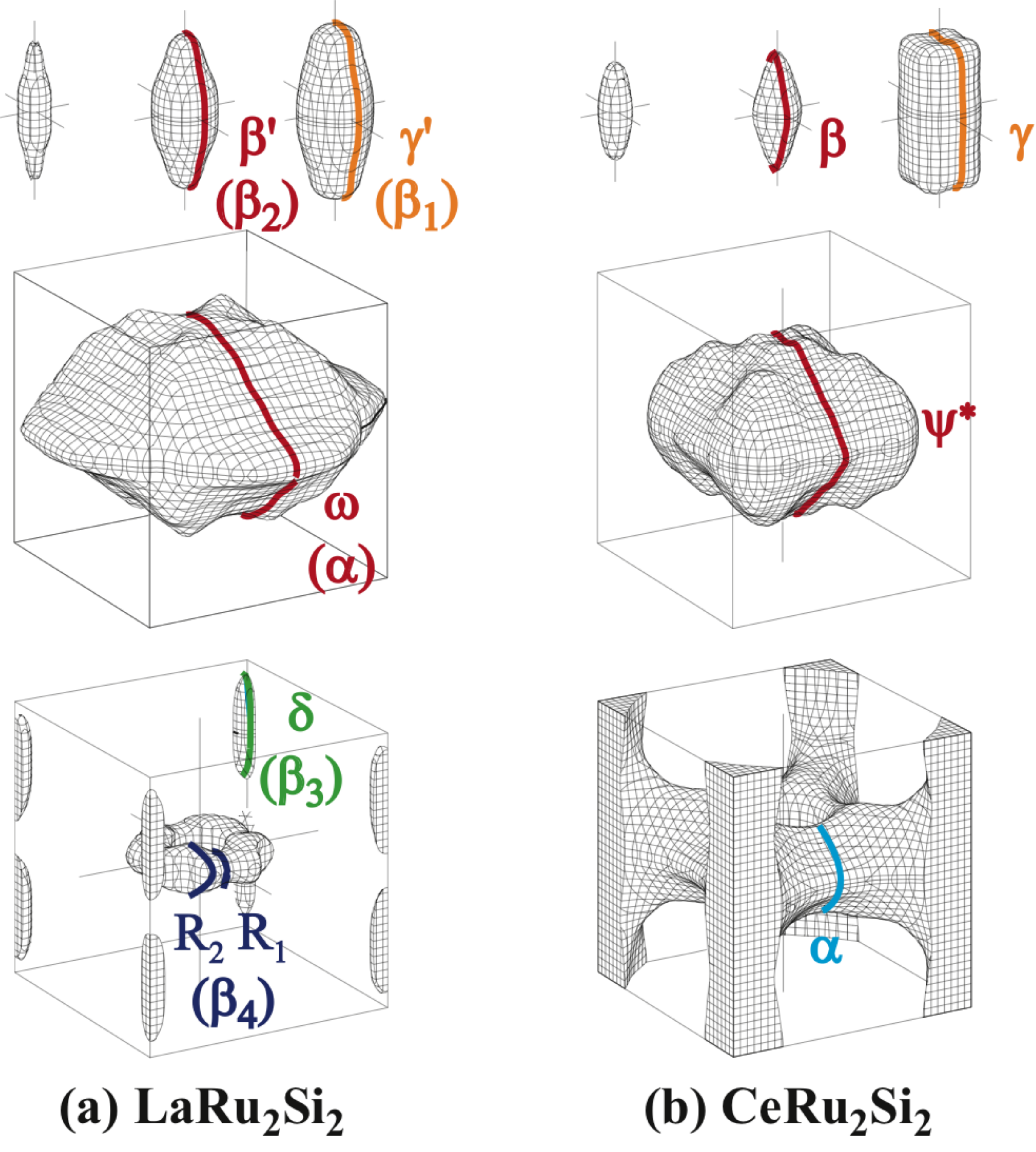}
\end{center}
\caption{(Color on line) Fermi surfaces of  (a) LaRu$_2$Si$_2$ and (b) CeRu$_2$Si$_2$.   The figures indicate the names of the dHvA oscillations and the corresponding orbits when the magnetic field is applied parallel to the [100] direction.  The names used in the previous work on LaRu$_2$Si$_2$ \cite{Settai95} are shown in the parentheses. }
\label{fig:90025Fig33}
\end{figure}

\begin{figure}[htbp]
\begin{center}
\includegraphics[width=0.6\linewidth]{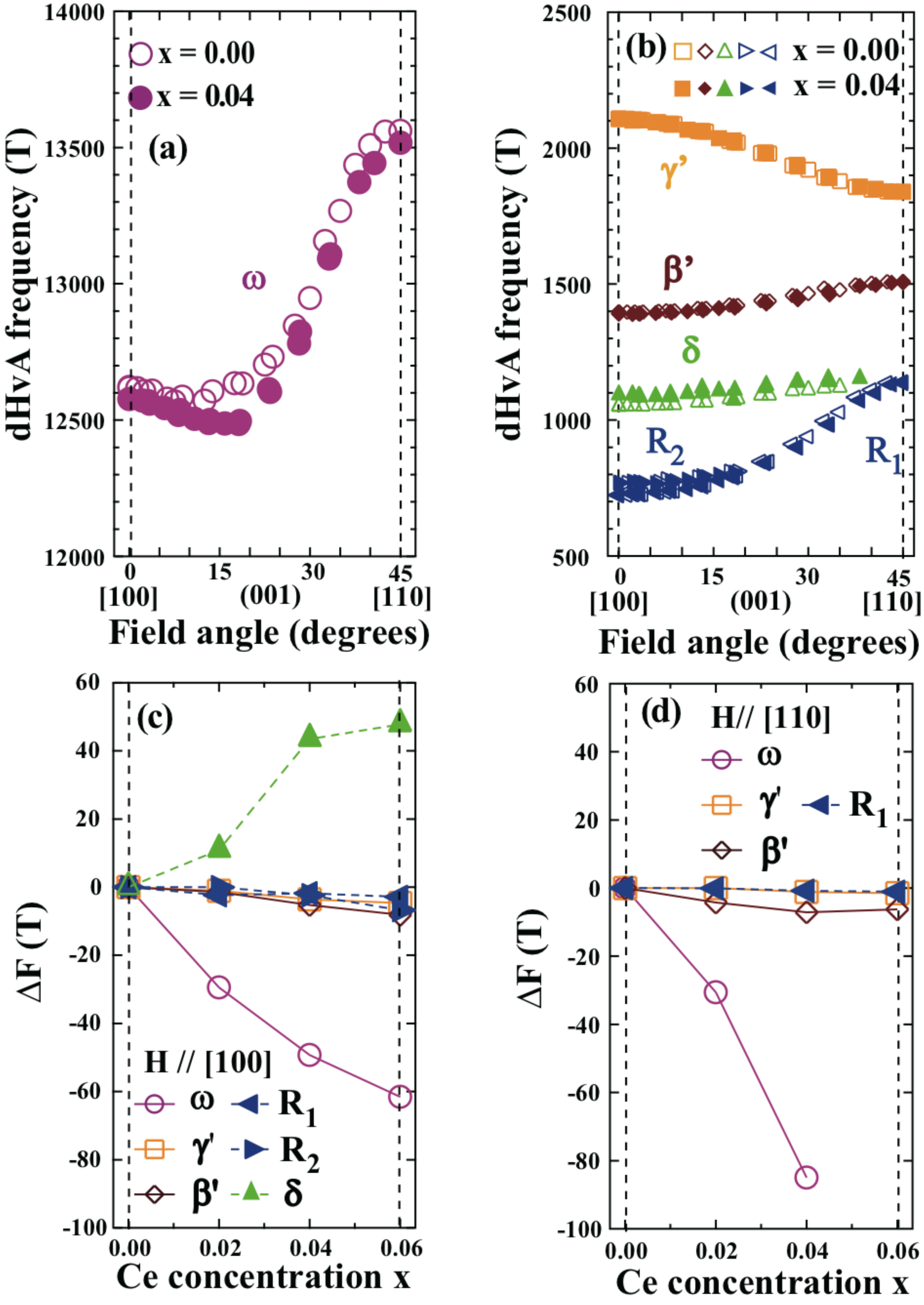}
\end{center}
\caption{(Color on line) Top panels: Angular dependences of the dHvA frequencies in the (001) plane in samples with $x$ = 0.0 and 0.04. (a) $\omega$ oscillation. (b) $\beta'$, $\gamma'$, $\delta$,  $R_1$ and $R_2$ oscillations.
Bottom panels: The change in the frequency $\Delta{F}$  as a function of $x$ with fields parallel to (c) the [100] direction and (d) the [110] direction\cite{Matsumoto10}.  }
\label{fig:90025Fig34}
\end{figure}

Figure \ref{fig:90025Fig33}\cite{Matsumoto12} shows the orbits on the Fermi surfaces of LaRu$_2$Si$_2$ and CeRu$_2$Si$_2$ which are responsible for the dHvA oscillations when a magnetic field is applied parallel to the [100] direction.  The names of the orbit on the small hole surfaces are denoted as $\beta^\prime$ and $\gamma^\prime$ in LaRu$_2$Si$_2$ and as $\beta$ and $\gamma$ in CeRu$_2$Si$_2$.  As described below, it is likely that the small hole surfaces are continuously deformed from LaRu$_2$Si$_2$ to CeRu$_2$Si$_2$.  Here, we use mostly the notation $\beta^\prime$ and $\gamma^\prime$,  because even if the hole surfaces changed discontinuously, it is difficult to tell where they changed from the present experiment.  First we present the changes in the frequencies of the dHvA oscillations in low Ce concentration samples.   The frequency changes are mostly determined from the phase shift of the oscillations ($\S$A.4).  Figures \ref{fig:90025Fig34} (a) and (b) show the angular dependences of the frequencies in the (001) plane for $x$ = 0.0 and 0.04.   In Fig. \ref{fig:90025Fig34} (c) and (d), we plot frequency changes  $\Delta{F} = F - F_0$ as a function of $x$ for the fields parallel to the [100] and [110] directions.  Here, $F$ is the frequency of the alloy and $F_0$ is that of LaRu$_2$Si$_2$.  $\Delta{F}$ of the $\omega$ oscillation decreases considerably, while that of the $\delta$ oscillation increases with $x$.  On the other hand, those of the  $R_1$, $R_2$, $\beta^\prime$ and $\gamma^\prime$ oscillations do not change appreciably.  This observation indicates that the Fermi surface change is very different depending on the Fermi surface sheet.   It is noted from Figs. \ref{fig:90025Fig34} (a), (c) and (d) that the frequency change also depends on the field direction or on the orbit.  

\begin{figure}[t]
\begin{center}
\includegraphics[width=0.6\linewidth]{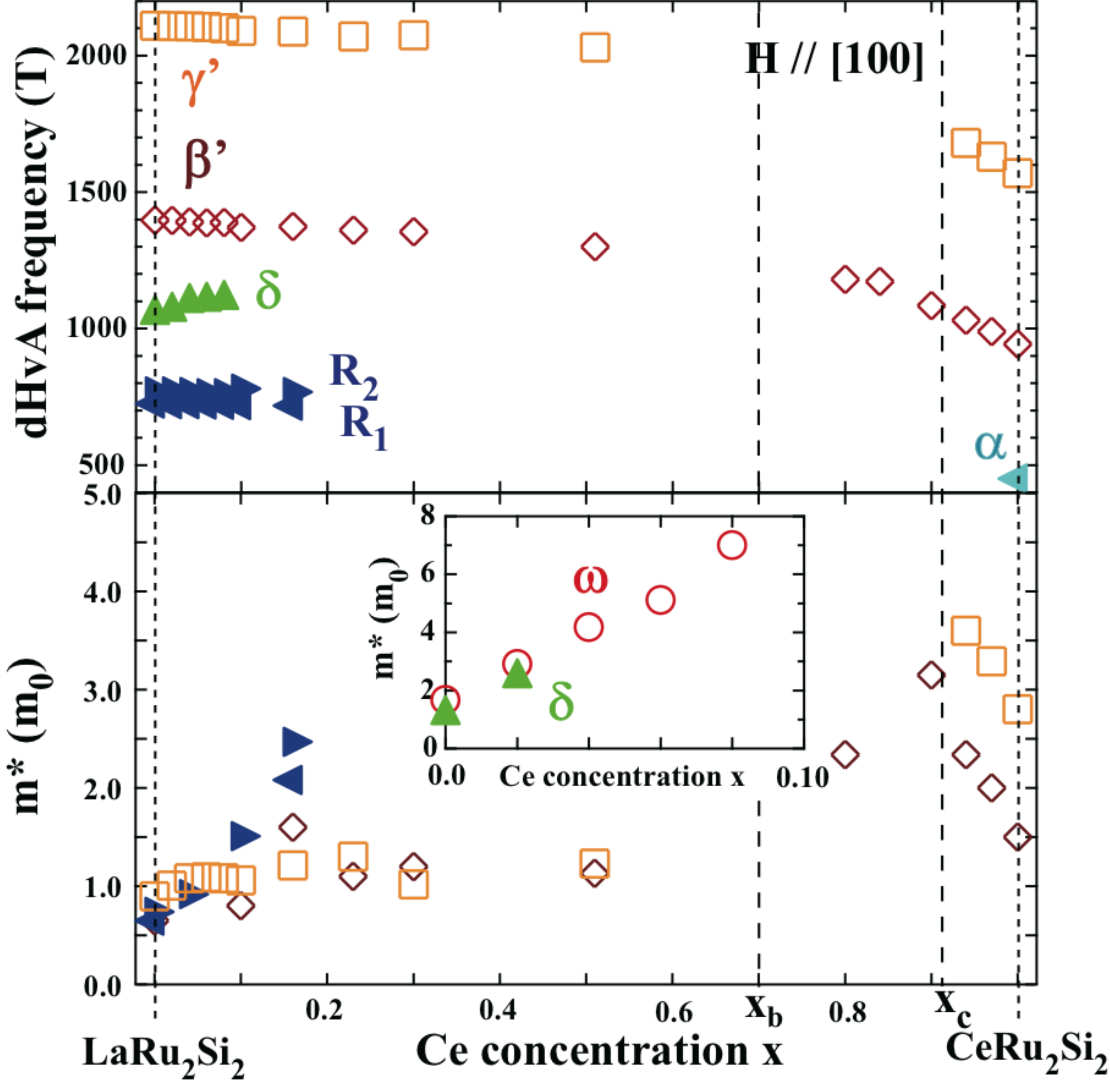}
\end{center}
\caption{(Color on line) Frequencies (upper panel) and effective masses (lower panel) of the $\beta'$, $\gamma'$, $\delta$, $R_1$,  $R_2$ and $\omega$ oscillations as a function of $x$\cite{Matsumoto10}.  The frequency of the $\alpha$ oscillation observed in CeRu$_2$Si$_2$ is also plotted. The magnetic field is applied parallel to the [100] direction.}
\label{fig:90025Fig35}
\end{figure}

We plot the frequencies of the dHvA oscillations as a function of $x$ from $x = 0$ to $x =1.0$ in the upper panel of Fig. \ref{fig:90025Fig35}\cite{Matsumoto10}.  The magnetic field is applied parallel to the [100] direction.  The $\omega$ oscillation can not be observed in samples with $x > 0.08$.   The amplitude of the $\delta$ oscillation also decreases rapidly with $x$.  The $R_1$ and $R_2$ oscillations can be traced up to slightly higher concentration samples.  The $\beta^\prime$  and $\gamma^\prime$ oscillations can be observed over a broad range of $x$.   However, the $\gamma^\prime$ oscillation cannot be observed around $x = 0.8 $.  With fields parallel to the [110] direction, the frequency change of each oscillation with $x$ is similar to that with fields in the [100] direction, but the amplitudes of the oscillations are slightly smaller.  

The lower panel shows the effective masses of the $R_1$,  $R_2$, $\beta^\prime$  and $\gamma^\prime$ oscillations as a function of  $x$.  The inset shows those of the $\omega$ and $\delta$ oscillations in low concentration samples.  The increase in the effective mass with $x$ is very rapid for the $\omega$ and $\delta$ oscillations, while it is moderate or small for the $R_1$, $R_2$, $\beta^\prime$ and $\gamma^\prime$ oscillations. It is noted that the change in the effective mass of the low Ce concentration samples approximately correlates with that in the frequency:  If  $\Delta F$ with $x$ is large, then the increase in the effective mass is large and vice versa.  The effective masses as well as the frequencies change almost linearly with $x$ in low Ce concentration samples.  It is noted that the quasi-linear relation persists to the concentration where the dilute limit approximation may not hold and irrespective of whether or not the signature for the antiferromagnetic transition is present.  The variations of the frequencies and the effective masses in the high concentration alloys are further discussed in $\S$4.1.2.

We examine the anisotropic changes in $\Delta{F}$ and effective mass by assuming that the Fermi surface properties of LaRu$_2$Si$_2$ changes towards those of CeRu$_2$Si$_2$ with alloying Ce.  The ellipsoidal hole surfaces in Figs. \ref{fig:90025Fig33} (a) and (b) are similar in shape and size, while the large hole surfaces on the middle panels are very different in size. The Fermi surface of the $\delta$ oscillation is considerably deformed to make the multiply connected electron surface of CeRu$_2$Si$_2$ and therefore there is no orbit in CeRu$_2$Si$_2$ corresponding to the $\delta$ orbit of LaRu$_2$Si$_2$.  Then, the changes in the frequency are large for the $\omega$ and $\delta$ oscillations, while those for the $\beta^\prime$ and $\gamma^\prime$ oscillations are small.  The changes in the effective mass are large for the $\omega$ and $\delta$ oscillations and small for the $\beta$ and $\gamma$ oscillations corresponding to the difference in the effective masses between LaRu$_2$Si$_2$ and CeRu$_2$Si$_2$.  The changes in the frequencies are small for the $R$ oscillations, while those in the effective masses are moderate.  This is probably because the orbit on the ring Fermi surface can be continuously deformed to that on the multiply connected electron Fermi surface and the difference between the sizes of the orbits is small.  Thus, considering that the effective mass of CeRu$_2$Si$_2$ is correlated with the f content on the Fermi surface, the anisotropic changes in the effective mass and the frequency are likely to be correlated with the $f$ content on the Fermi surface.  Then, a plausible interpretation of these results is that the f electron of Ce in a dilute alloy is incorporated in the sates at the Fermi surface, resulting in the increases in the volume of the Fermi surface and the effective masses.  In this respect, we may assume that the f electron in a dilute alloy of Ce$_x$La$_{1-x}$Ru$_2$Si$_2$ is ``itinerant''.

If we assume that the frequency and effective mass of the large hole surface change linearly and uniformly over the Fermi surface with Ce concentration from LaRu$_2$Si$_2$ to CeRu$_2$Si$_2$, the observed rate of the frequency change is significantly smaller than that expected from the linear change, i.e. about 1/7 for [100] and about 2/5 for [110].  Although it is possible that the Fermi surface changes more largely in the [001] direction, we speculate that this observation may reflect a characteristic feature of a dilute alloy of the strongly correlated f electron system as discussed at the beginning of this section.  On the other hand, the observed rate of the change in the effective mass for [100] is about a half of that expected from the linear change. The value of the effective mass is determined by the conventional mass plot ($\S$A.1 and $\S$A.2) mostly over a temperature range from above 1 K to below 100 mK.  As described below in (b), the value thus determined is smaller than that in the ground state.  If we use the value of the ground state, the observed rate of the change in the effective mass is approximately the same as that obtained by assuming the linear change, i.e. the effective mass seems to be proportional to the Ce concentration.   It is not clear whether or not this observation is accidental, because the intersite interactions could be responsible for the Fermi surface properties in high concentration alloys as well as in CeRu$_2$Si$_2$.   
\\
\noindent
(b) {\it Variation of the Fermi surface properties with temperature}

Since the f electron at high temperatures is thought to behave as a localized electron, then we may observe a change in the Fermi surface properties from high temperatures above $T_{\rm K}$ to low temperatures below $T_{\rm K}$. We present how the temperature variation of the Fermi surface properties of dilute alloys is observed or whether the interpretation that the f electron state changes from localized to itinerant is reasonable or not. No measurements at high temperatures have been performed on the Fermi surface properties in a dilute alloy.  However, we may assume that the Fermi surface will be the same as that of LaRu$_2$Si$_2$, if the Fermi surface can be well defined at high temperatures.  The effective mass may be approximately the same as that of LaRu$_2$Si$_2$.  It is also expected that the transition behavior with temperature is observed more evidently for the oscillations whose f content is larger, and that the transition behavior may not be observed when the magnetic field along the [001] direction is large enough because the magnetic field drives the f electron into a more polarized state where the contribution of the f electron to the Fermi surface is diminished and the volume of the Fermi surface can be assumed to become smaller. 

\begin{figure}[ht]
\begin{center}
\includegraphics[width=0.6\linewidth]{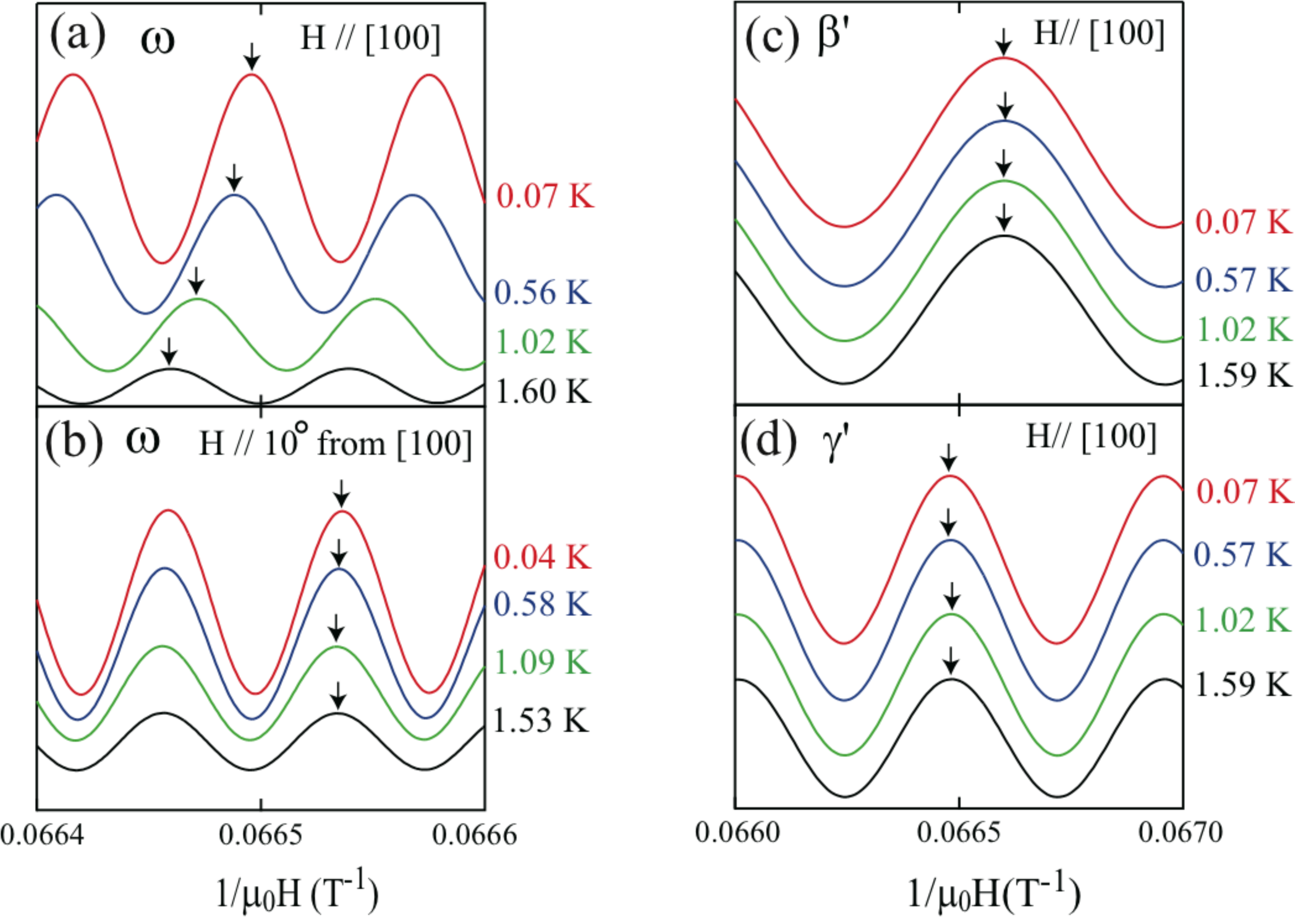}
\end{center}
\caption{(Color online) Traces of the dHvA oscillations in Ce$_{0.02}$La$_{0.98}$Ru$_2$Si$_2$ as a function of inverse field.  (a) $\omega$ oscillation for fields parallel to the [100] direction. (b) $\omega$ oscillation for fields tilted by 10 degrees from the [100] direction towards the [001] direction.  (c) $\beta^\prime$ oscillation for fields parallel to the [100] direction.  (d) $\gamma^\prime$ oscillation for fields parallel to the [100] direction\cite{Matsumoto12}.}
\label{fig:90025Fig36}
\end{figure}

In Fig. \ref{fig:90025Fig36} we show the traces of the dHvA oscillations at several temperatures as a function of reciprocal magnetic field applied parallel to the [100] direction.  Figures \ref{fig:90025Fig36}(a), (c) and (d) show those of the $\omega$, $\beta^\prime$ and $\gamma^\prime$ oscillations at about 15 T, respectively.   The $\delta, R_1$ and $R_2$ oscillations from the electron surface can be observed in this field direction.  However, the signal amplitudes of these oscillations are not strong enough or the frequencies are not enough separated from each other to trace each oscillation individually as a function of magnetic field and temperature.   
To examine the effect of magnetic field component along the [001] axis, we measured the $\omega$ oscillation with fields tilted by 10 degrees from the [100] direction towards the [001] direction.  The magnetic field component along the [001] direction is about 2.8 T when we apply a magnetic field of 16 T.   It is enough to give rise to nearly saturated magnetic moment as noted from Fig. \ref{fig:90025Fig32}(b).  Figure \ref{fig:90025Fig36}(b) shows the traces of the $\omega$ oscillation for this field direction.

The arrows in the figures indicate the positions of the same peak number which is counted from the infinite magnetic field.  The position of the arrow for the $\omega$ oscillation in Fig. \ref{fig:90025Fig36}(a) shifts to the high magnetic field side with increasing temperature indicating that the frequency of the $\omega$ oscillation increases with increasing temperature.  On the other hand, the positions of the arrows in Figs. \ref{fig:90025Fig36}(b), (c) and (d) do not move appreciably indicating that the frequencies are nearly constant. From the shift of the phase, we can evaluate the frequency change ($\S$A.4). 

\begin{figure}[ht]
\begin{center}
\includegraphics[width=0.6\linewidth]{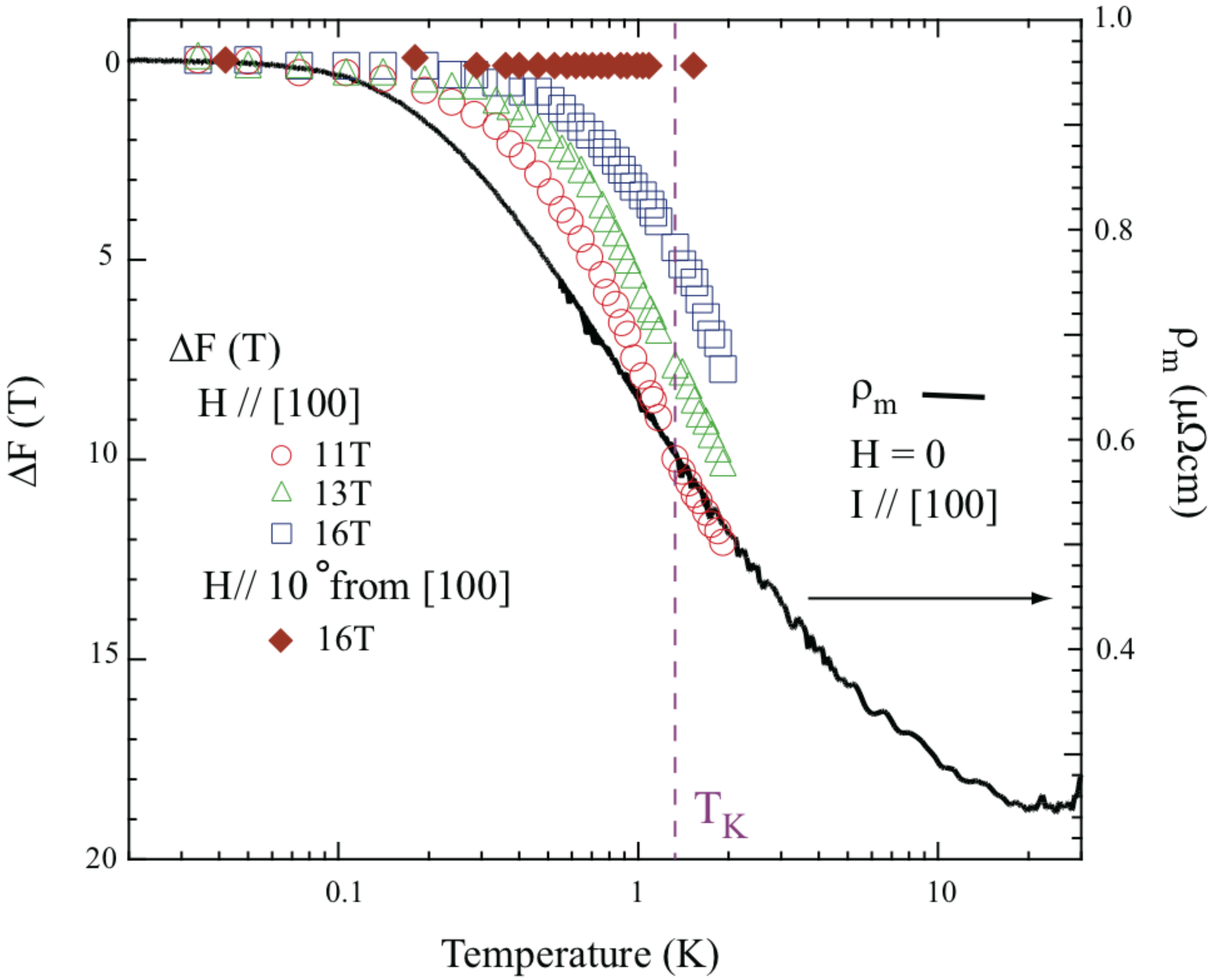}
\end{center}
\caption{(Color online) Frequency changes of the $\omega$ oscillation as a function of temperature at three different fields of 11 T, 13 T and 16 T parallel to the [100] direction and at 16 T with the field direction tilted by 10 degrees from the [100] direction towards the [001] direction.  The magnetic resistivity which is obtained by subtracting the resistivity of LaRu$_2$Si$_2$ from the observed resistivity of Ce$_{0.02}$La$_{0.98}$Ru$_2$Si$_2$ is also plotted to compare the behavior of frequency change with resistivity\cite{Matsumoto12}. }
\label{fig:90025Fig37}
\end{figure}

In Fig. \ref{fig:90025Fig37}, we plot the frequency change of the $\omega$ oscillation with temperature for a few selected magnetic field strengths parallel to the [100] direction and for the fields tilted by 10 degrees from the [100] direction.  Here, $\Delta F$ denotes the frequency change from the lowest temperature. The value of the vertical axis increases toward the downward direction to compare the behavior with that of resistivity.  Since the $\omega$ oscillation arises from the hole surface, the increase in the frequency with increasing temperature indicates that hole surface expands with temperature.  
The magnetic contribution to the resistivity is derived by subtracting the resistivity of LaRu$_2$Si$_2$ from that of Ce$_{0.02}$La$_{0.98}$Ru$_2$Si$_2$ and is plotted in Fig. \ref{fig:90025Fig37}.  It is noted that the temperature variation of the frequency change is qualitatively similar to that of the resistivity.  The frequency increase is of the order of 10 T from the lowest temperatures up to $T_{\rm K}$ and slightly depends on the applied magnetic field strength.  If we assume that the frequency changes similarly to that of the resistivity up to 20 K, the frequency change becomes of the order of 20 T.  This value is comparable to the frequency decrease of about 30 T from LaRu$_2$Si$_2$ to Ce$_{0.02}$La$_{0.98}$Ru$_2$Si$_2$ at low temperatures below $T_{\rm K}$\cite{Matsumoto10}.   

The variation of frequency with temperature for fields parallel to the [100] direction is more moderate for higher applied magnetic fields and is nearly zero for the fields tilted by 10 degrees from the [001] direction.  Considering that the alignment of the sample may have an error of the order of 1 - 2 degrees, the magnetic field effect may be consistently understood by assuming that the magnetic field component along the [001] direction is effective to drive the f electron into a polarized state.  

\begin{figure}[ht]
\begin{center}
\includegraphics[width=0.5\linewidth]{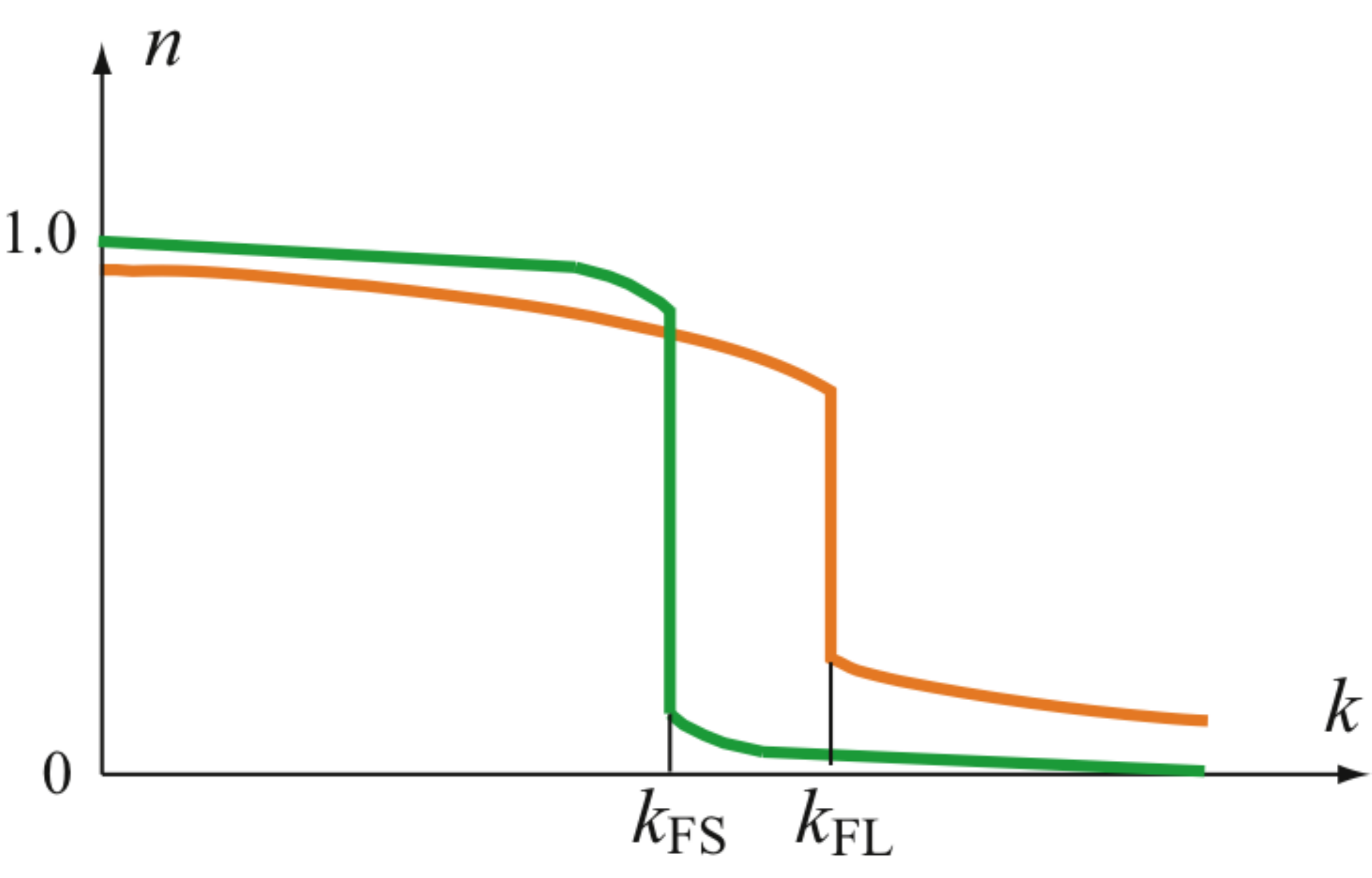}
\end{center}
\caption{(Color online) Schematic illustration of momentum distribution.  $k_{\rm FS}$ and $k_{\rm FL}$ correspond to the Fermi edges at high and low temperatures, respectively.}
\label{fig:90025Fig38}
\end{figure}

The present observation for temperature variation is in contrast with those in the heavy fermion compounds where no observation of an intermediate Fermi surface between those corresponding to the large Fermi surface and the small Fermi surface has been reported from ARPES measurements\cite{Denlinger01,Yano08,Okane09,Koizumi11,Fujimori07}.  It is also reported in UPd$_2$Al$_3$ that the Fermi surface changes rather abruptly between 40 K and 60 K\cite{Fujimori07}.  Then, the continuous change in the observed frequency may not directly correspond to a continuous change of the Fermi edge with temperature. But it could be explained by the two Fermi edges corresponding to the small and large Fermi surfaces and the former one diminishes while the latter one grows with decreasing temperature as expected in the case of compounds\cite{Otsuki09}. We show a schematic illustration for the distribution of the electrons in the momentum space in Fig. \ref{fig:90025Fig38}. In the present case, strictly speaking there will be no well defined Fermi edge because the distribution is broadened due to the impurities or temperature.  But we use the term Fermi surface or Fermi edge for simplicity.  At high temperatures we may assume that the f electron does not contribute to form the Fermi surface and the edge of the Fermi surface stays at $k_{\rm FS}$.  At lowest temperatures we may assume that the f electron contributes to form the Fermi surface whose Fermi edge resides at $k_{\rm FL}$.

One may speculate that the continuous change can be also reproduced by assuming that the observed frequency is an average of the two oscillations with different frequencies of $F_{\rm S}$ and $F_{\rm L}$, and that their relative amplitudes change with temperature and field as follows.

\begin{equation}
g(H,T)\sin (\frac{2\pi F_{\rm L}}{H} + \phi_{\rm L}) + (1-g(H,T))\sin (\frac{2\pi F_{\rm S}}{H} + \phi_{\rm S}) ~.
\label{eq:Ftemp}
\end{equation}
Here, $\phi$ is the phase of the oscillation at $H = \infty$ and $g(H,T)$ is a monotonous function of temperature and magnetic field.  However, this model can not reproduce the observed monotonous change of oscillation amplitude and phase shift with temperature as well as with magnetic field\cite{Matsumoto12}.  Therefore, in the present dilute Kondo alloy, the picture that only one Fermi edge is present and it changes continuously with temperature is more likely than the picture that the two Fermi edges corresponding to the small and large Fermi surfaces are present.

\begin{figure}[ht]
\begin{center}
\includegraphics[width=0.6\linewidth]{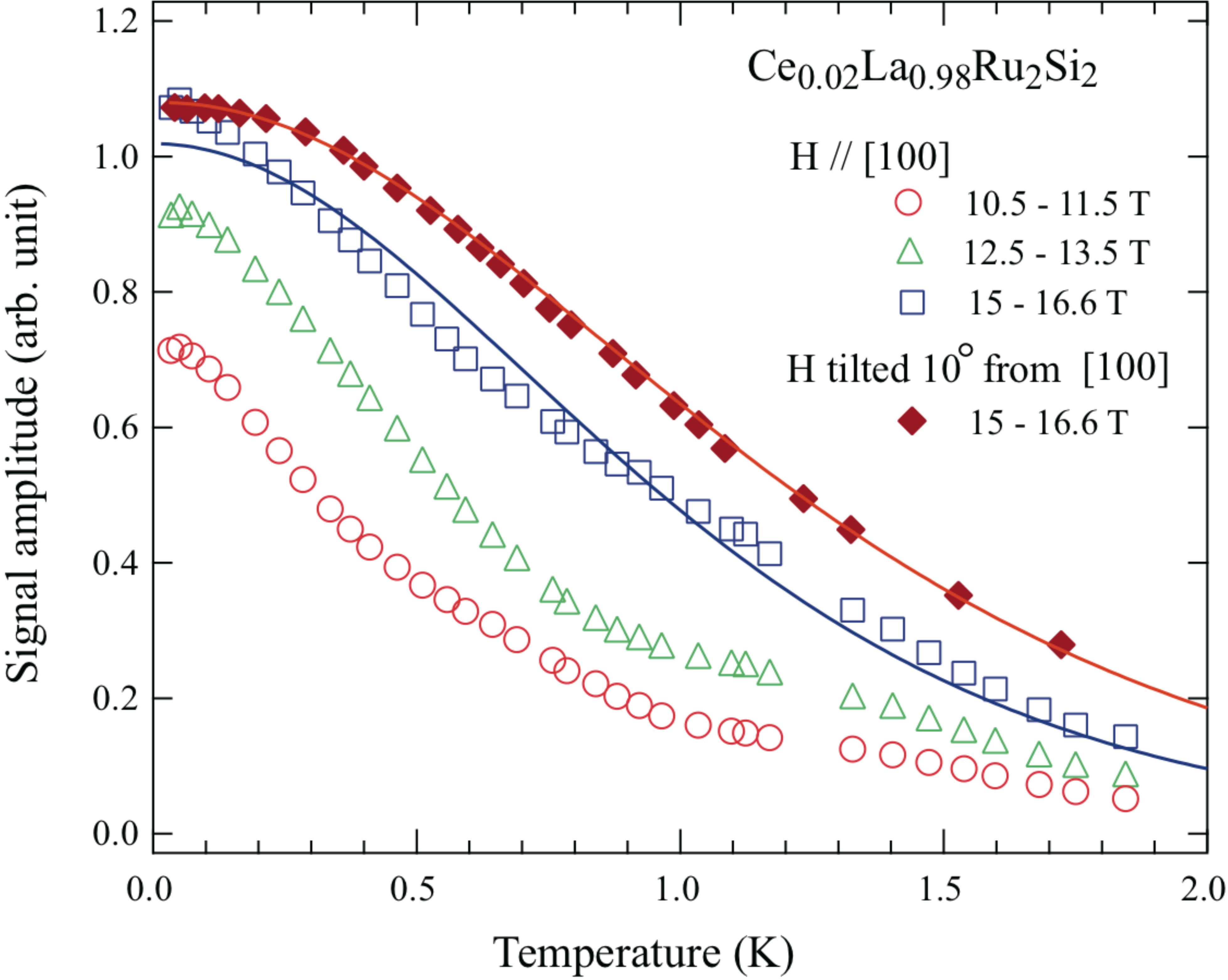}
\end{center}
\caption{(Color online) Temperature dependences of signal amplitudes (mass plots) of the $\omega$ oscillations.  For fields parallel to the [100] direction, the signal amplitudes are obtained from the Fourier analysis with field intervals of 10.5 T - 11.5 T, 12.5 T - 13.5 T and 15 - 16.6 T.  For the fields tilted by 10 degrees from the [100] direction towards the [001] directions, the signal amplitudes obtained from the field interval of 15 - 16.6 T are plotted.  The red and blue lines are fits to the data points which give the values of 2.0 $m_0$ and 2.5 $m_0$, respectively\cite{Matsumoto12}.}
\label{fig:90025Fig39}
\end{figure}

The temperature dependence of the signal amplitude has been found to be explained consistently with the frequency change with temperature.   Figure \ref{fig:90025Fig39} shows the amplitudes of the $\omega$ oscillation as a function of temperature for the fields parallel to the [100] direction and for the fields tilted by 10 degrees from the [100] direction towards the [001] direction.  The amplitude of the oscillation was obtained from the Fourier analysis of the oscillation.  For fields parallel to the [100] direction, we plot three sets of the data points which are obtained from the analysis using three different field ranges, i.e., 10 - 11.5 T, 12.5 - 13.5 T and 15 T - 16.6 T.  For the fields tilted by 10 degrees from the [100] direction, the data for the field range of 15 T - 16.6 T are plotted.  For ordinary cases, the temperature variation of the amplitude is given by eq.(\ref{eq:RT1}) in $\S$A.1 or eq.(\ref{eq:RT2}) in $\S$A.2.  For the fields tilted by 10 degrees from [100] all the data points well lie on a single fitting curve with the effective mass of 2.0 $m_0$. This value is comparable to that of 1.6 $m_0$ in LaRu$_2$Si$_2$ for the same field direction, indicating that for this field direction and strength the Fermi surface properties are nearly the same as those of LaRu$_2$Si$_2$ and do not change with temperature.   On the other hand, for fields parallel to the [100] direction, the data point curve has a structure which cannot be fitted by a single curve given by eq.(\ref{eq:RT1}) or eq.(\ref{eq:RT2}).    However, by assuming that (1) the scattering or the Dingle temperature increases with decreasing temperature as observed in the resistivity in Fig. \ref{fig:90025Fig32} and (2) the effective mass increases with decreasing temperature, the behavior can be explained almost quantitatively\cite{Matsumoto11}.  Then, the effective mass obtained from the curve fitting at the lowest temperatures can be assumed to be that of the ground state.  The value obtained from the fitting is 3.8 $m_0$ which is considerably larger than that of LaRu$_2$Si$_2$ and that obtained from the fitting to the data over a larger temperature range.

The above results indicate that the f electron at high temperatures does not contribute to form the Fermi surface and the effective mass of the conduction electron does not seem to be much enhanced from that of LaRu$_2$Si$_2$. On the other hand,  with decreasing temperature the contribution of the f electron to form the Fermi surface increases and the effective mass of the conduction electron increases.  Thus, the temperature variation of the Fermi surface properties probed by the dHvA effect measurements is consistent with the picture that the f electron is localized at temperatures well above $T_{\rm K}$ and contributes to form the Fermi surface at temperatures well below $T_{\rm K}$.  This is also consistent with the results of the dilute alloys of Ce$_x$La$_{1-x}$Ru$_2$Si$_2$ described in (a) above.  Although the measured volume of the Fermi surface is likely to be closer to that of the small Fermi surface, it may be more appropriate to regard the f electron state in the ground state as itinerant rather than localized.  In this article, we refer this state as ``the itinerant f electron state in the ground state of the dilute Kondo alloy".
 
 A few dHvA effect studies have been performed to investigate the Fermi surface properties in the impurity state of the strongly correlated f electron system. In Ce$_x$La$_{1-x}$B$_6$, the dHvA measurements were performed in dilute alloys down to $x$ = 0.01 to derive the information of the spin dependent Fermi surface properties\cite{Teklu00}.  The dilute Ce alloys of this system exhibit the impurity Kondo behavior\cite{Sato85} but the properties such as the effective mass and magnetoresistance have strong magnetic field dependence\cite{Joss87,Goodrich99,Nakamura06,Endo06}. Therefore, it is not possible to infer the Fermi surface properties at zero or low fields from the measurements of the dHvA effect at high fields. Since the f electron of the terminal compound CeB$_6$ is assumed to be localized because the measured volume of the Fermi surface is small  according to the dHvA effect measurements\cite{Goodrich99,Nakamura06,Endo06}, one would conjecture that the volume of the Fermi surface remains small with alloying or the f electron in dilute alloys is localized. However, from the results in Ce$_x$La$_{1-x}$Ru$_2$Si$_2$ it is also possible that the f electron is regarded as ``itinerant'' in the ground state of the dilute Kondo alloys and possibly in high concentration alloys.  

The dHvA effect measurements in the alloys of Ce$_x$La$_{1-x}$MIn$_5$ with M = Co, Ir, Rh\cite{Harrison04} have been performed by using a pulsed magnet.  It is argued that the f electron is localized at low temperatures in the dilute alloys from the observations of the small Fermi surfaces and the transition from the localized f electron state to the itinerant f electron state at the quantum critical point is discussed.  However, since the dHvA effect measurements use high magnetic fields, the magnetic field effect on the Fermi surface properties may have to be taken into account also in this case.  As discussed for Ce$_x$La$_{1-x}$Ru$_2$Si$_2$ in the case with the strong field component along the [001] direction,  the f electron state might appear differently from that in the ground state in high fields.
 
\subsubsection{Fermi surface properties in high concentration alloys of Ce$_x$La$_{1-x}$Ru$_2$Si$_2$}
We describe how the itinerant f electron state in the ground state of the dilute Kondo alloy evolves into the itinerant f electron state of CeRu$_2$Si$_2$.    

With fields in the (001) plane, the magnetic state can be assumed to change from the Kondo singlet state in a dilute alloy to the anitiferromagnetic state and to the paramagnetic state through $x_c$.  As long as the low frequency oscillations like $\beta^\prime$ and $\gamma^\prime$ are concerned, the Fermi surface does not seem to change abruptly at a certain concentration.  Since the dHvA effect study cannot observe the main Fermi surfaces, the ARPES study is performed on Ce$_{0.84}$La$_{0.16}$Ru$_2$Si$_2$ sample at 10 K which is above $T_{\rm N}$ of 5 K and comparable to $T_{\rm K}$, if $T_{\rm K}$ in Fig. \ref{fig:90025Fig4} is extended to $x<x_c$.   The Fermi surface is found to be quite similar to that of CeRu$_2$Si$_2$ or the f electron of each Ce of the alloy can be counted as a conduction electron\cite{Okane11}.  This is also consistent with the observation that the resistivity vs. temperature curve starts to have a signature reminiscent of the SDW transition from about this concentration.  The Fermi surface could be a nested Fermi surface originated from the large Fermi surface whose measured volume is similar to that of CeRu$_2$Si$_2$.  This interpretation is also consistent with other experiments that the behavior around $x_c$ are well explained by the SCR theory\cite{Kambe96,Kadowaki04,Lohneysen07,Knafo09}.  The effective masses of the $\beta$ and $\gamma$ oscillations are enhanced moderately around $x_c$. The moderate enhancement is consistent with the electronic specific heat coefficient\cite{Fisher91,Kambe96,Aoki11} described in $\S$3.3.2 which is also successfully analyzed by the SCR theory\cite{Kambe96}.  In these respects, the f electron state in the ground state of the antiferromagnetic state near $x_c$ can be denoted as itinerant without ambiguity.

If we assume that the electronic structures above $H_{\rm c}$ are similar to that shown in Fig. \ref{fig:90025Fig31}(c), then with increasing field in the [001] direction the Fermi surface would change at the metamagnetic transitions.   It is possible that the Fermi surface change takes place accompanied with the lower transition at $H_{\rm a}$ (or also $H_{\rm b}$ in the case of CeRu$_2$(Si$_x$Ge$_{1-x}$)$_2$) rather than at $H_{\rm c}$ because the enhancement of the electronic specific heat coefficient or that of the value of the $A$ coefficient is observed at lower transition fields near $x_c$ in the antiferromagnetic state.

Since there is no phase transition from $x = 0$ to $x_c$, we may expect that the f electron state is regarded as itinerant in the sense that the volume of the Fermi surface is larger than that of the small Fermi surface all the way from $x = 0$ to 1.0.  However, this does not mean that the Fermi surface properties change almost linearly with Ce concentration from LaRu$_2$Si$_2$ to CeRu$_2$Si$_2$.  Since we could not observe the oscillations from the large hole surface and the electron surface in high concentration alloys, we infer how the Fermi surface properties change with concentration from the behavior of the low frequency oscillations.

\begin{figure}[t]
\begin{center}
\includegraphics[width=0.5\linewidth]{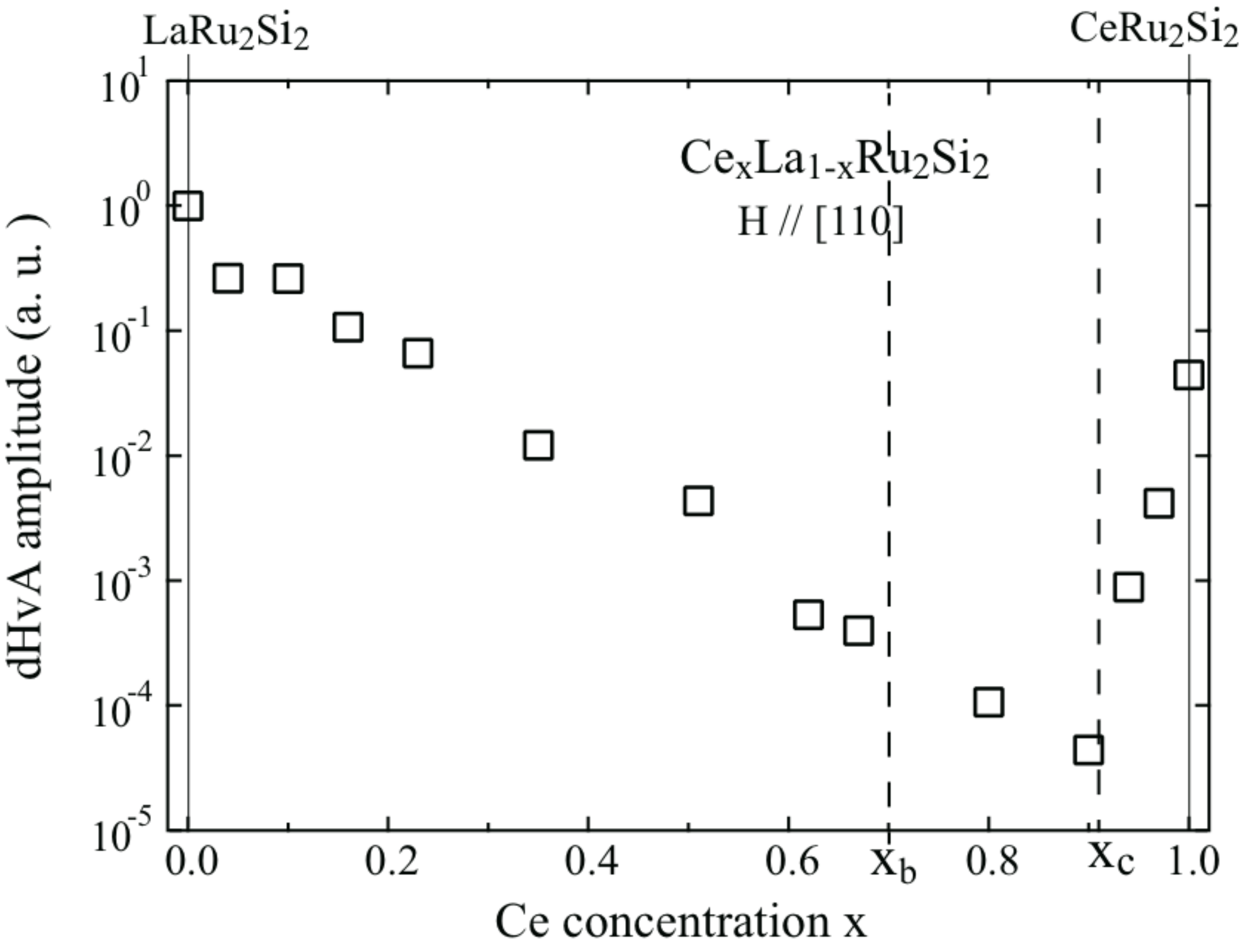}
\end{center}
\caption{Signal amplitude of the $\gamma^\prime$ oscillation as a function of Ce concentration $x$\cite{Matsumoto12b}.  The amplitude is the relative amplitude to that in LaRu$_2$Si$_2$.  The magnetic field is applied parallel to the [110] direction.}
\label{fig:90025Fig40}
\end{figure}

With fields in the [110] direction,  the $\gamma^\prime$ oscillation can be observed in almost all the concentration range.  We show the signal amplitudes normalized to that of LaRu$_2$Si$_2$ in Fig. \ref{fig:90025Fig40}\cite{Matsumoto12b}.   The signal amplitude is diminished so strongly that it can not be explained only in terms of the effective mass or the frequency.  The signal amplitude decreases very quickly with Ce concentration and has a minimum somewhere between $x = 0.8$ and $x_c$.  Since we could not observe the signal between $x$ = 0.8 and 0.9,  we could not determine from this plot where the minimum of the signal amplitude resides. However, the minimum is likely to reside around $x$ = 0.85 where the residual resistivity is found to become maximum as shown in Fig. \ref{fig:90025Fig7}, similarly to the case of the polarized state in $\S$3.3.   It is noted that the reduction of the signal amplitude in the alloy with fields in the (001) plane is much larger than that observed in the polarized state above $H_{\rm m}$ with fields in the [001] direction.  This may imply that the electronic state of Ce probed by the conduction electrons is very different from that of La or from that of Ce above $H_{\rm m}$.  This observation suggests that the f electron state is different from that of the polarized state and is closer to that of the itinerant f electron system rather than that of the magnetic localized f electron system.

In dilute alloys, the Fermi surface properties seem to change almost linearly with Ce concentration.  However, as shown in Fig. \ref{fig:90025Fig35} the frequencies and the effective masses of the $\beta^\prime$ and $\gamma^\prime$ oscillations change very slowly with Ce concentration from about 0.1 to about $x_b = 0.7$, and then more rapidly from about $x_b$.  This observation seems to be correlated with the evolution of magnetic order where $T_{\rm N }$ and $T_{\rm L}$ become maximum around $x_b$.  It also suggests that the Fermi surface properties or the f electron state is correlated with the magnetic order.  To explore the picture of the f electron state in the magnetically ordered state, in the next section we compare the results with those in CeRu$_2$(Si$_x$Ge$_{1-x}$)$_2$.  

\subsection{Fermi surface properties of CeRu$_2$(Si$_x$Ge$_{1-x}$)$_2$ with fields in the (001) plane}

\begin{figure}[t]
\begin{center}
\includegraphics[width=0.6\linewidth]{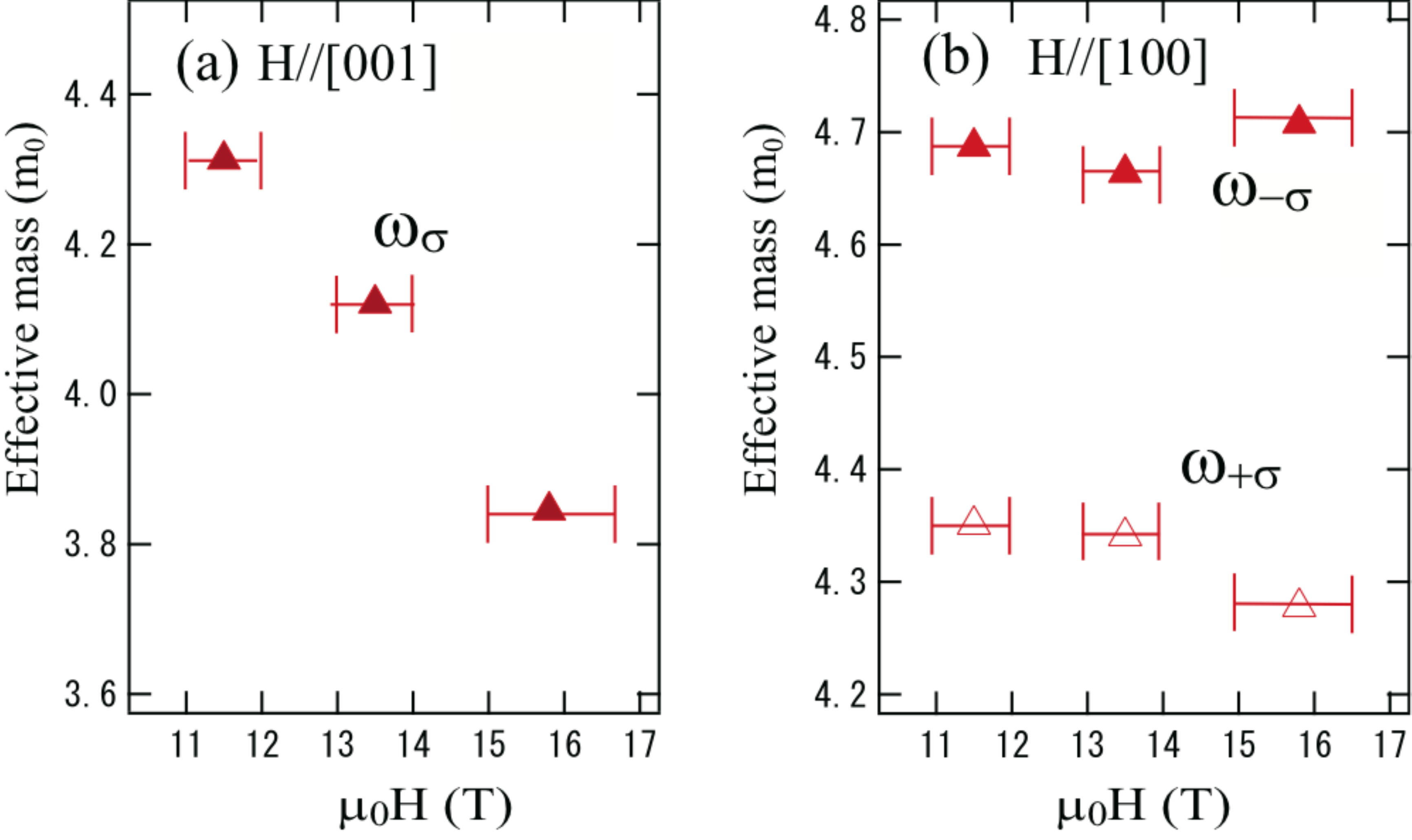}
\end{center}
\caption{(Color on line) Magnetic field dependence of the effective masses (a) with fields parallel to the [001] direction and (b) parallel to the [100] direction\cite{DoiM}.  The value of the effective mass is determined from the Fourier analysis of the oscillations over the field range denoted by the horizontal error bars.  $\sigma$ denotes either of the spin directions.}
\label{fig:90025Fig41}
\end{figure}

We first discuss the electronic structure in the ferromagnetic state.  CeRu$_2$Ge$_2$ has spontaneous magnetic moment of 1.9 $\mu_{\rm B}$ which is comparable to the value of 2.1 $\mu_{\rm B}$ expected for $J=5/2, J_{\rm z}=\pm 5/2$ state.  The electronic specific heat coefficient is measured to be 16.5 mJ/mol$\cdot$ K$^2$\cite{Raymond99}.  The measured Fermi surface is spin dependent owing to the ferromagnetic order but each spin split Fermi surface is similar to that of LaRu$_2$Si$_2$.  Therefore, the localized picture can be applied to the f electron of CeRu$_2$Ge$_2$. We refer the reader to the reference\cite{Matsumoto11} for a more detailed description of the Fermi surface of CeRu$_2$Ge$_2$.  It is noted, however, that this does not mean that there is no significant interaction between the conduction electrons and the f electron.  When a magnetic field is applied parallel to the [001] direction, the magnetic moment increases to 2 $\mu_{\rm B}$ at 20 T\cite{Besnus91}.  The effective mass of the $\omega$ oscillation decreases by 10 - 20 \% from 10 T to 16 T, whereas no appreciable decrease in the effective mass is observed with fields in the (001) plane as shown in Fig. \ref{fig:90025Fig41}\cite{IkezawaM,DoiM}. This observation implies that a significant interaction between the f electron and conduction electrons remains even in the ferromagnetic and localized f electron state.   

In dilute alloys of Ce$_x$La$_{1-x}$Ru$_2$Ge$_2$, a similar dilute Kondo behavior to that in dilute alloys of Ce$_x$La$_{1-x}$Ru$_2$Si$_2$ can be observed in the resistivity and the temperature dependence of the dHvA frequency with fields in the (001) plane.  For small concentration samples less than $x = 0.05$, the effective mass increases with Ce concentration. However, when the ferromagnetic order develops in higher concentration samples, the effective mass first becomes smaller with Ce concentration and then at sufficiently high concentration samples increases with Ce concentration.  It seems that the f electron state in the ground state of the dilute Kondo alloy of Ce$_x$La$_{1-x}$Ru$_2$Ge$_2$ can be regarded as ``itinerant'' like that in the dilute Kondo alloy of Ce$_x$La$_{1-x}$Ru$_2$Si$_2$.  In the case of the ferromagnetic order, the evolution of the magnetic order seems to affect the nature of the f electron considerably, while the Fermi surface is always very similar to that of LaRu$_2$Ge$_2$ or CeRu$_2$Ge$_2$, i.e., the small Fermi surface. We have not made enough study to judge how the itinerant f electron state in the ground state of the dilute Kondo alloy changes to the magnetic localized f electron state with Ce concentration in these alloys\cite{DoiM}. 

\begin{figure}[t]
\begin{center}
\includegraphics[width=0.6\linewidth]{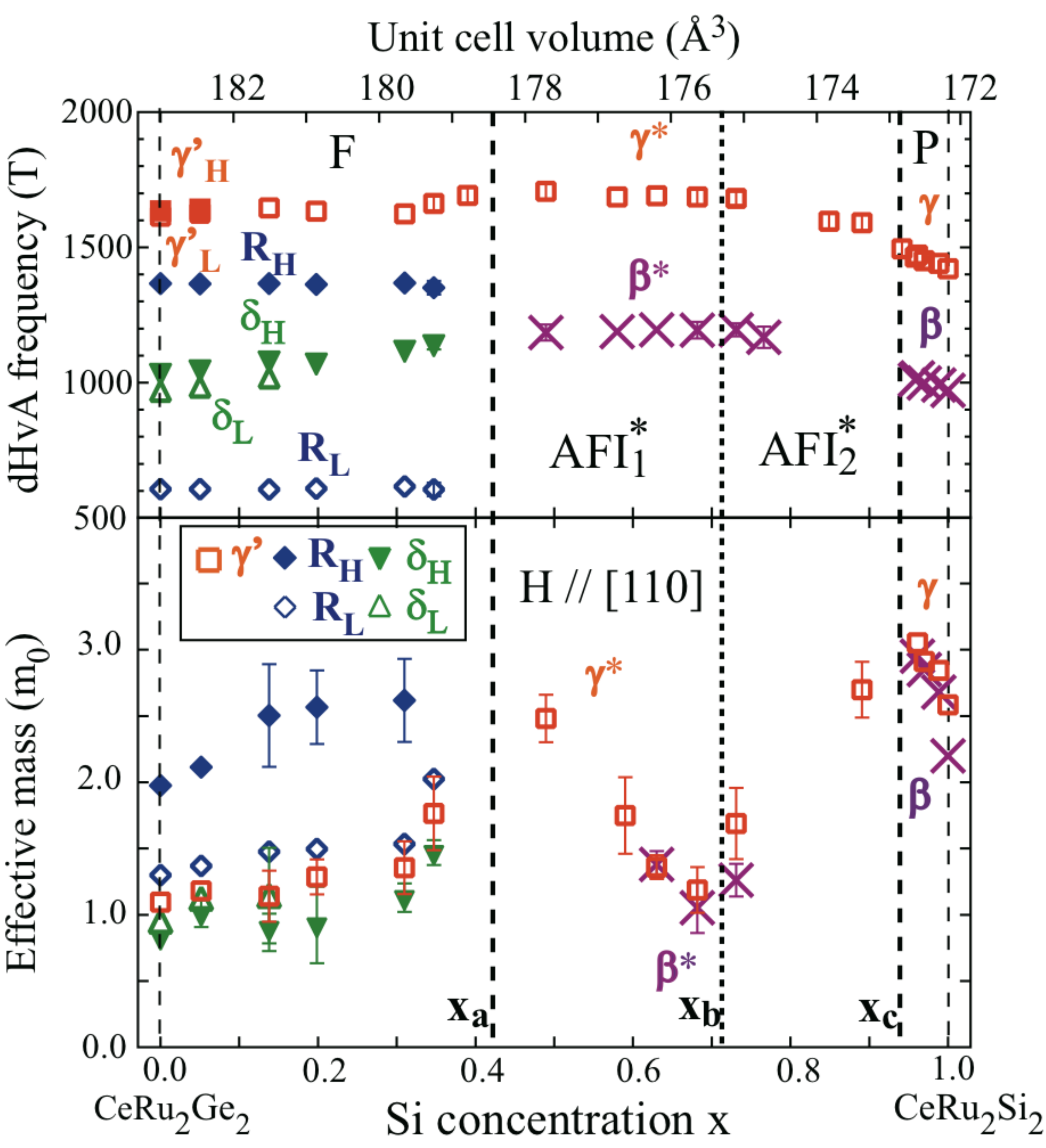}
\end{center}
\caption{(Color on line) DHvA frequencies (upper panel) and effective masses (lower panel) plotted as a function of Si concentration $x$ (bottom axis) or unit cell volume (top axis)\cite{Matsumoto11}.  The open squares denote the frequencies of the $\gamma_L^\prime$, $\gamma^\ast$, $\gamma$  oscillations and the closed one denotes that of  $\gamma_H^\prime$ oscillation.  The crosses denote the frequencies of the $\beta^\ast$ and  $\beta$ oscillations. The open and closed triangles denote the frequencies of the $\delta_L$ and $\delta_H$ oscillations, respectively. The open and closed diamonds denote the frequencies of the $R_L$ and $R_H$ oscillations, respectively.  The symbols for the effective masses are the same as those for the frequencies.}
\label{fig:90025Fig42}
\end{figure}

We plot the dHvA frequencies, effective masses in CeRu$_2$(Si$_x$Ge$_{1-x}$)$_2$ against Si concentration $x$ in Fig. \ref{fig:90025Fig42} for fields parallel to the [110] direction.  The $\gamma^\prime$, $\delta$ and $R$ oscillations can be traced as a function of $x$  in the ferromagnetic phase of  $x < x_a$ and the spin split frequencies are observed for each oscillation.  Both the frequencies of the $R_L$ and $R_H$ oscillations do not change appreciably with $x$ up to $x_a$, while that of $\gamma^\prime$ seems to increase slightly with $x$ and those of $\delta_H$ and $\delta_L$ increase considerably with $x$ towards $x_a$.  The values of the effective masses increase with increasing $x$ towards $x_a$.  With decreasing volume, the magnetization at the same magnetic field strength slightly decreases as demonstrated in Fig. \ref{fig:90025Fig9}.  We suspect that the effective mass increases probably by the increase in the hybridization due to compression as in the case of  the polarized state above $H_{\rm m}$.   

In the antiferromagnetic phase,  the $\gamma^\ast$ and $\beta^\ast$ oscillations are observed.   The measurements of the angular dependences of their frequencies  tell that they are likely to arise from the ellipsoidal Fermi surfaces like those shown in Fig. \ref{fig:90025Fig33}. However, since it is not possible to determine experimentally whether the Fermi surface changes continuously or discontinuously at $x_b$ or $x_c$, we denote them as  $\gamma^\ast$ and $\beta^\ast$ in the antiferromagnetic phase.  The $\gamma^\ast$ oscillation can be observed almost all the concentration range. No signals corresponding to the electron surfaces can be observed. On the other hand, the $\beta^\ast$ oscillation is observed in the antiferromagnetic phase, although the $\beta^\prime$ oscillation can not be observed in the ferromagnetic phase.  In the antiferromagnetic phase between  $x_a$  and $x_b$,  the frequencies of the $\beta^\ast$ and  $\gamma^\ast$ oscillations do not change significantly.  They start to decrease from about $x_b$ with increasing $x$. These behaviors of the frequencies for $x<x_b$ and $x>x_b$ are similar to those in the antiferromagnetic sates of Ce$_x$La$_{1-x}$Ru$_2$Si$_2$ for $x<x_b$ and $x>x_b$, respectively.   In the paramagnetic sate between  $x_c$ and $x = 1.0$,  the $\beta$ and $\gamma$ oscillations are observed except for $x$ close to $x_c$.  Their frequencies decrease rapidly with increasing $x$ as in the case of Ce$_x$La$_{1-x}$Ru$_2$Si$_2$.  

The effective masses of the $\beta^\ast$ and $\gamma^\ast$ oscillations in the AFI$_1^\ast$ phase decrease with increasing $x$ towards $x_b$.  The behavior of the effective mass could be also interpreted in another way as the effective mass is enhanced near $x_a$.   The effective masses of the  $\beta^\ast$ and $\gamma^\ast$ oscillations start to increase from about $x_b$ with increasing $x$ towards $x_c$ making a minimum around $x_b$.  In the paramagnetic phase the effective masses of the $\beta$ and $\gamma$ oscillations decrease rapidly with increasing $x$ making a maximum at $x_c$ as observed in Ce$_x$La$_{1-x}$Ru$_2$Si$_2$.   

The signal amplitude of the $\gamma^\ast$ oscillation in the antiferromagnetic phase decreases to about 10$^{-3}$ of that of  the $\gamma^\prime$ oscillation in the ferromagnetic phase. Around $x$ = 0.8 - 0.9,  the signal intensity is reduced very much and only the signal of the $\gamma^\ast$ oscillation can be observed.  The reduction is related with the enhanced residual resistivity in the antiferromagnetic phase compared with that in the ferromagnetic phase as shown in Fig. \ref{fig:90025Fig8}.  If it arises from the enhanced charge fluctuation as described in $\S$2.2.3, this observation implies that the f electron is more itinerant in the antiferromagnetic phase than in the ferromagnetic phase.  From the measurements of the transport properties and the dHvA effect the electronic structure obviously changes across $x_a$.  However, it is not clarified from these observations whether or not the change is attributed to the transition  of the Fermi surface from the small Fermi surface to the large Fermi surface. 
  
The ARPES study on CeRu$_2$(Si$_{0.82}$Ge$_{0.18}$)$_2$ sample was performed at 20 K and the Fermi surface is found to resemble that of CeRu$_2$Si$_2$\cite{Okane09}.   This observation may indicate that the Fermi surface is large or the f electron is regarded as itinerant near $x_c$ in the antiferromagnetic phase as observed in Ce$_x$La$_{1-x}$Ru$_2$Si$_2$ system.   Although the Fermi surface in the antiferromagnetic ground state has not been revealed, we may assume for $x > x_b$ that the Fermi surface properties and magnetic properties of CeRu$_2$(Si$_x$Ge$_{1-x}$)$_2$ and Ce$_x$La$_{1-x}$Ru$_2$Si$_2$ can be described in terms of itinerant f electron picture.   Therefore, there will be no such a drastic change of the Fermi surface from the small Fermi surface to the large Fermi surface at $x_c$ in the present CeRu$_2$(Si$_x$Ge$_{1-x}$)$_2$ and Ce$_x$La$_{1-x}$Ru$_2$Si$_2$.

\begin{figure}[t]
\begin{center}
\includegraphics[width=0.5\linewidth]{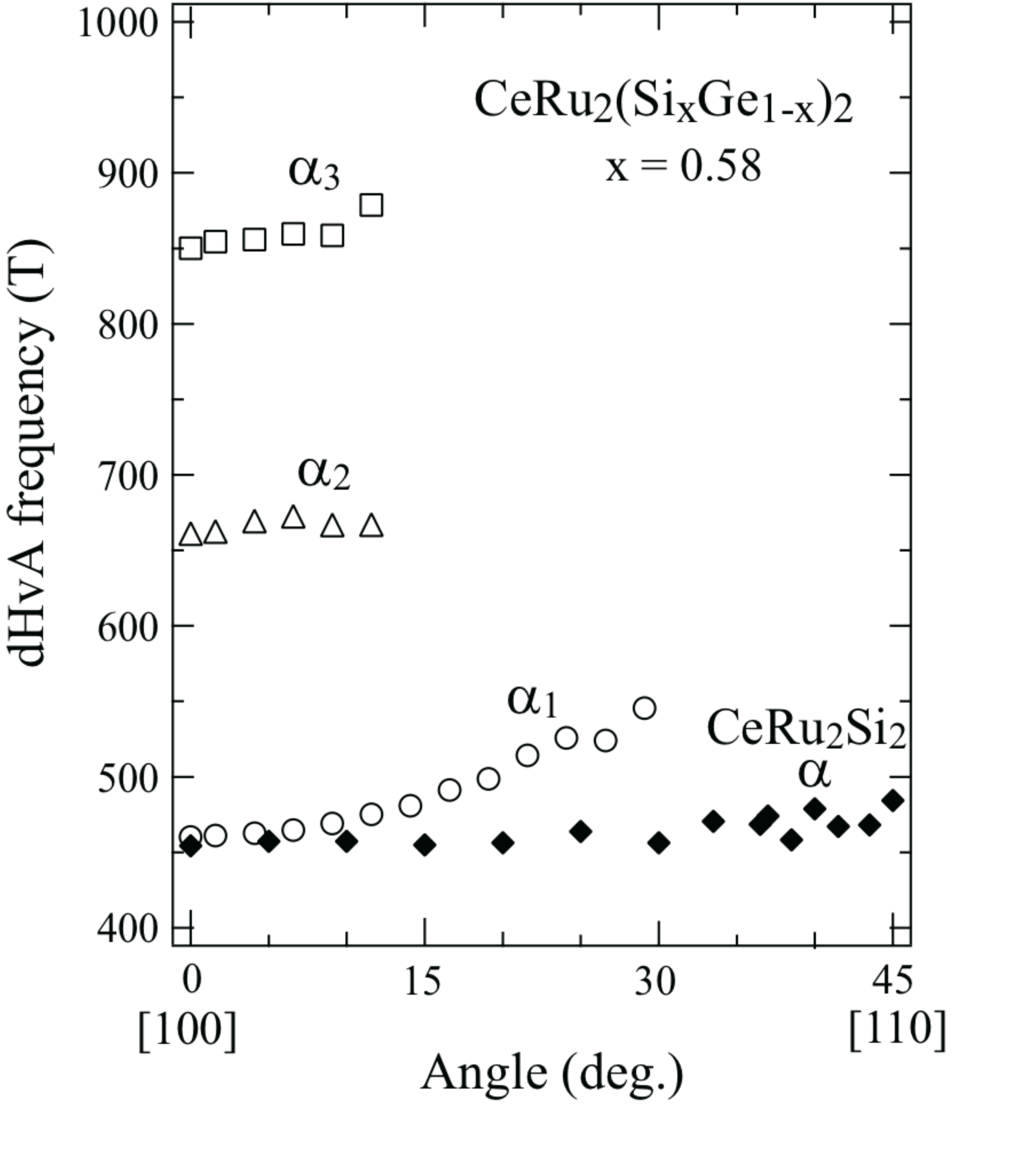}
\end{center}
\caption{Angular variations of the frequencies of the $\alpha_1$, $\alpha_2$ and $\alpha_3$ oscillations in the sample  CeRu$_2$(Si$_{0.58}$Ge$_{0.42}$)$_2$\cite{Matsumoto11}.  That of the $\alpha$ oscillation in  CeRu$_2$Si$_2$ \cite{Takashita96} is also shown.}
\label{fig:90025Fig43}
\end{figure}

It is likely that the f electron has a more localized character for $x_a < x <x_b$ than for $x_b < x <x_c$, judging from the transport and magnetic properties \cite{Matsumoto11} presented in $\S$2.   There may be also competition with the magnetic order as suggested in the case of Ce$_x$La$_{1-x}$Ru$_2$Si$_2$ and Ce$_x$La$_{1-x}$Ru$_2$Ge$_2$.  Although there is no decisive experimental information on the Fermi surface properties for $x_a < x <x_b$, another dHvA frequency with small effective mass observed for $x_a < x <x_b$ might be worth to be examined.  The dHvA signal intensity with fields parallel to the [100] direction is mostly smaller than that with fields parallel to the [110] direction.   However, we could observe rather clear signals of the $\alpha_1$, $\alpha_2$ and $\alpha_3$ oscillations for $x$ around 0.5 - 0.6\cite{Matsumoto11}.  In Fig. \ref{fig:90025Fig43}, we show the angular dependences of the frequencies in the (001) plane for the  $\alpha_1$, $\alpha_2$ and $\alpha_3$ oscillations with  $x$ = 0.58 together with that of the $\alpha$ oscillation in CeRu$_2$Si$_2$\cite{Takashita96}. The difference in the frequencies between  $\alpha_1$ and $\alpha_2$ is the same as that between $\alpha_2$ and $\alpha_3$.  The difference in the effective masses between  $\alpha_1$ and  $\alpha_2$ is also the same as that between $\alpha_2$ and $\alpha_3$.  This observation indicates that the three oscillations are related with each other via the magnetic breakdown effect.  
 
From the angular dependence, the $\alpha_1$ oscillations could arise from the Fermi surface similar to the ring surface in Fig. \ref{fig:90025Fig33}(a) or the multiply connected electron surface in Fig. \ref{fig:90025Fig33}(b).  The frequencies of the $\alpha_1$ and $\alpha$ oscillations are nearly the same for fields parallel to the [100] direction, but the frequency of the $\alpha_1$ oscillation increases more rapidly with the tilting angle from [100] than that of $\alpha$ does.  On the other hand, the average frequency of the spin split frequencies from the ring surface is about 1000 T and is much larger than that of $\alpha_1$.  It seems also difficult to find the magnetic breakdown orbits corresponding to these oscillations from the Fermi surfaces in Figs. \ref{fig:90025Fig33}(a) and (b).  The effective mass of $\alpha_1$ is found to be about 1.5 $m_0$.  On the other hand, the effective masses of the $R$ oscillations are larger than 2$m_0$ near $x_a$ and that of the $\alpha$ oscillation is 13 - 15 $m_0$\cite{Aoki92,Aoki93a,Aoki93b,Tautz95,Takashita96,Aoki01}.  This small value may be also difficult to explain with the orbit on the ring surface or the multiply connected electron surface.  The Fermi surface in the antiferromagnetic state could be deformed by the magnetic Brillouin zone.  Even if the reconstruction due to antiferromagnetic order is taken into account, it may not be easy to find an appropriate orbit for the $\alpha_1$ oscillations from the Fermi surfaces of Figs. \ref{fig:90025Fig33}(a) and (b).   Therefore, the  Fermi surface for $x_a<x<x_b$ might be different from those of Figs. \ref{fig:90025Fig33}(a) and (b). 
 
We think from the features observed in the transport and dHvA measurements that a significant hybridization between the f electron and the conduction electron is present and therefore the f electron state could be more appropriately described with the term ``itinerant'' in the antiferromagnetic state.  On the other hand, the magnetic properties may be better described in terms of the localized moment or the localized f electron for $x_a<x<x_b$.  We speculate that most of the f  electron content lie well below the Fermi energy and that the Fermi surface could be significantly different from that reconstructed from the Fermi surface of the large Fermi surface by the magnetic Brillouin zone.   This state could turn into the magnetic localized f electron state due to the relative increase in the interaction $I_{\rm F}$ of our interpretation without the help of magnetic field.  With application of a small magnetic field, it can turn into the f electron state above $H_{\rm m}$ where the f electron may be regarded as ``itinerant'' as well as ``localized'' as discussed in $\S$3. 

We have argued that the measured volume of the Fermi surface may not be a good criterion to designate the f electron nature as either itinerant or localized in the disordered alloy as well as in the polarized state under high magnetic fields.   In the case of the ordered alloy,  it could be a criterion to characterize the f electron state in most of the cases, but in some cases it may not be an appropriate criterion.   Further experimental studies on the Fermi surface in the antiferromagnetic state are required to understand the f electron state properly in the antiferromagnetic state.         

\section{Concluding Remark}
The f electron state in the strongly correlated f electron system is often categorized as ``itinerant'' or ``localized''.  Whether or not the f electron is itinerant or how we understand the f electron state has been a long standing issue in the field of the strongly correlated f electron system.  However, since the terms of  ``itinerant'' or ``localized'' are used without clear definition, the discussions among reseachers sometimes appear to be confused\cite{Harima11}. 

In this article, we have presented the Fermi surface properties of CeRu$_2$Si$_2$ and its alloys observed by the dHvA effect together with their magnetic properties and magnetic phase diagrams.  We have described these properties as they are observed and they have been examined carefully without looking at only a few properties and rushing into the discrimination of ``itinerant" and ``localized".  Since the dHvA effect measurements use high magnetic fields and the properties of the strongly correlated f electron system are rather sensitive to magnetic field, we utilize the strong anisotropic magnetic properties of these systems to explore the magnetic field effect on the Fermi surface properties or the Fermi surface properties in the ground state.   Although the Fermi surface volume has been conventionally used for the discrimination between ``itinerant" and ``localized" in most of the cases, we have examined whether such a criterion is useful and unambiguous to describe the f electron state properly. 

We have summarized the Fermi surface properties of the paramagnetic heavy Fermion state  well below $H_{\rm m}$ of CeRu$_2$Si$_2$ which are now well explored experimentally and are well understood also theoretically.  We have stressed that in this state the definition for the term ``itinerant'' by the Fermi surface volume is useful and what are implied by the term are unambiguously understood among researchers.  

With the high fields along the [001] direction of the easy axis, we have shown and discussed the anomalous features observed by the dHvA effect measurements like the spin dependent effective mass and the effective mass enhancement accompanied by the metamagnetic transiton in CeRu$_2$Si$_2$.  The metamagnetic transition in CeRu$_2$Si$_2$ probed by the dHvA measurements is discussed to show that the transition may not be understood by the Zeeman splitting of the conduction bands below $H_{\rm m}$.  We have demonstrated that the Fermi surface properties above metamagnetic transitions in the alloys of Ce$_x$La$_{1-x}$Ru$_2$Si$_2$ and CeRu$_2$(Si$_x$Ge$_{1-x}$)$_2$ as well as CeRu$_2$Si$_2$ are considerably different from those below $H_{\rm m}$ of CeRu$_2$Si$_2$.  We have argued that some of the properties like the volume of the Fermi surface are the same or similar to those in the systems whose f electron is conventionally thought to be localized, while some of the properties are similar to those of the itinerant f electron system in the paramagnetic ground state. From the experimental observations we have proposed a schematic model for the electronic structures above the metamagnetic transition fields and have pointed out that in the polarized state above $H_{\rm m}$ of CeRu$_2$Si$_2$ the volume of the Fermi surface does not seem to be a good criterion to characterize the f electron state properly .  

By assuming that we can observe the Fermi surface properties similar to those in the ground state with the fields in the (001) plane of the hard axes, we have explored the evolution of the Fermi surface properties with the content of the alloying element in Ce$_x$La$_{1-x}$Ru$_2$Si$_2$ and CeRu$_2$(Si$_x$Ge$_{1-x}$)$_2$.  In the dilute Kondo alloy of Ce$_x$La$_{1-x}$Ru$_2$Si$_2$, the volume of the Fermi surface is likely to be larger than the volume of the small Fermi surface, but is considerably smaller than that of the large Fermi surface.  We have found that the effective mass is considerably enhanced with alloying Ce, implying that the f electron contributes to the Fermi surface property significantly and that the f electron state may be regarded as ``itinerant" although the measured volume of the Fermi surface is closer to the small Fermi surface.  It is shown that this itinerant f electron state in the ground state of the dilute Kondo alloy is likely to evolve continuously to the itinerant f electron state of antiferromagnetic state near $x_c$ where the magnetic properties can be understood by the SDW model, and then to the itinerant f electron state in the paramagnetic ground state of CeRu$_2$Si$_2$ through $x_c$, indicating that there is no such a drastic change of the Fermi surface from the small Fermi surface to the large Fermi surface at $x_c$.  In disordered alloys the measured Fermi surface volume may not be a good criterion to characterize the f electron state properly because the change in the Fermi surface volume with alloying could be very tiny in the limit of strong correlation.

The localized f electron state of CeRu$_2$Ge$_2$ has been examined by the dHvA effect studies on Ce$_x$La$_{1-x}$Ru$_2$Ge$_2$ as well as on CeRu$_2$Ge$_2$ to reveal the inherent competing interactions in the system.  The localized f electron state of CeRu$_2$Ge$_2$ is found to evolve with increasing Si concentration in CeRu$_2$(Si$_x$Ge$_{1-x}$)$_2$ to the itinerant f electron state of the antiferromagnetic state in $x_b < x <x_c$ and to the itinerant f electron state in CeRu$_2$Si$_2$ through $x_c$.  While it is very likely that a drastic change in the Fermi surface properties takes place at $x_a$, it has not been clarified whether the Fermi surface is large or small for $x_a < x <x_b$ . We have suggested that even in the magnetically ordered state of the compound without disorder, the Fermi surface volume might not be an appropriate criterion to describe the f electron state properly in some cases. 

The ambiguity and the confusion for the terms of  ``itinerant" and ``localized" may arise from the inherent nature of the f electron in the strongly correlated f electron system, i.e., the duality of itinerant and localized or the competition between the hybridization and the intra-atomic coulomb interaction in the Anderson model.  The duality seems to appear differently depending on the situation.  In the paramagnetic ground state, the f electron behaves as a conduction electron, while the excitation from the ground state shows another aspect of the localized electron.  It may not be surprising that with application of a large magnetic field or in the alloys the duality may appear differently from that in the paramagnetic ground state as discussed in this article.    In such a case, the observed Fermi surface volume is one of the features of the system but may not be an appropriate criterion to characterize the f electron state. In other words, we may characterize the f electron state as ``itinerant" and ``localized" by the observed volume of the Fermi surface, but in some cases the observed volume of the Fermi surface appears to be intermediate between those of the large Fermi surface and the small Fermi surface, or other features than the volume of the Fermi surface are different from those of either of the systems whose f electron is conventionally thought to be localized or itinerant.

We feel that the confusion in the experimental results about the quantum phase transition might partially come from improper and insufficient understanding of the f electron state in the magnetically ordered state as well as in a polarized state, where the measurements at high magnetic fields may not reveal the f electron state in the ground state in the strongly correlated f electron system.  Careful measurements with various experimental methods would be necessary to understand properly the f electron state veiled by the duality.   

\section{Acknowledgement}
We thank Prof. Y. Onuki, Prof. S. Uji, Prof. T. Komatsubara, Prof. K. Maezawa, Prof. H. Yamagami, Dr. T. Okane, Dr. M. Takashita, Dr. M. Endo, Dr. T. Isshiki, Dr. Y. Matsumoto, Ms. A. K. Albessard, Mr. H. Ikezawa, Mr. T. Fujiwara, Mr. M. Sugi, Mr. K. Aoki, Mr. Y. Shimizu and Mr. Y. Doi for fruitful collaborations for the studies on CeRu$_2$Si$_2$ and its alloys.  We also thank Prof. H. Harima, Prof. Y. Kuramoto and Prof. J. Otsuki for helpful discussions.  Our studies presented in this article were supported by several grants from Ministry of Education, Culture, Sports, Science and Technology Japan and Japan Society for the Promotion of Science.

\appendix\section{de Haas - van Alphen Effect} 
In $\S$A.1, we describe very briefly how the standard formula for the dHvA effect in normal metals is derived.  The implicit assumptions and problems to use the formula for the strongly correlated f electron system are also described.  For a more detailed description of the dHvA effect, we refer the reader to the review article by A. V. Gold\cite{Gold} or the textbook by D. Shoenberg\cite{Shoenberg}.  In $\S$ A.2, we present some examples which are not described by the standard formula or some peculiar features observed in the strongly correlated f electron system.  Problems to analyze the dHvA effect measurements in the strongly correlated f electron system are also discussed.  In  $\S$ A.3, we briefly summarize the dHvA effect studies on dilute alloys of normal metals to compare them with those in the alloys of the strongly correlate f electron system.  $\S$ A.4 describes the method to determine the small change in the frequency using the phase shift of the oscillation due to alloying, pressure or temperature.

\appendix\subsection{ A brief description of the de Haas - van Alphen effect in normal metals : the LK formula}
\begin{figure}[t]
\begin{center}
\includegraphics[width=0.5\linewidth]{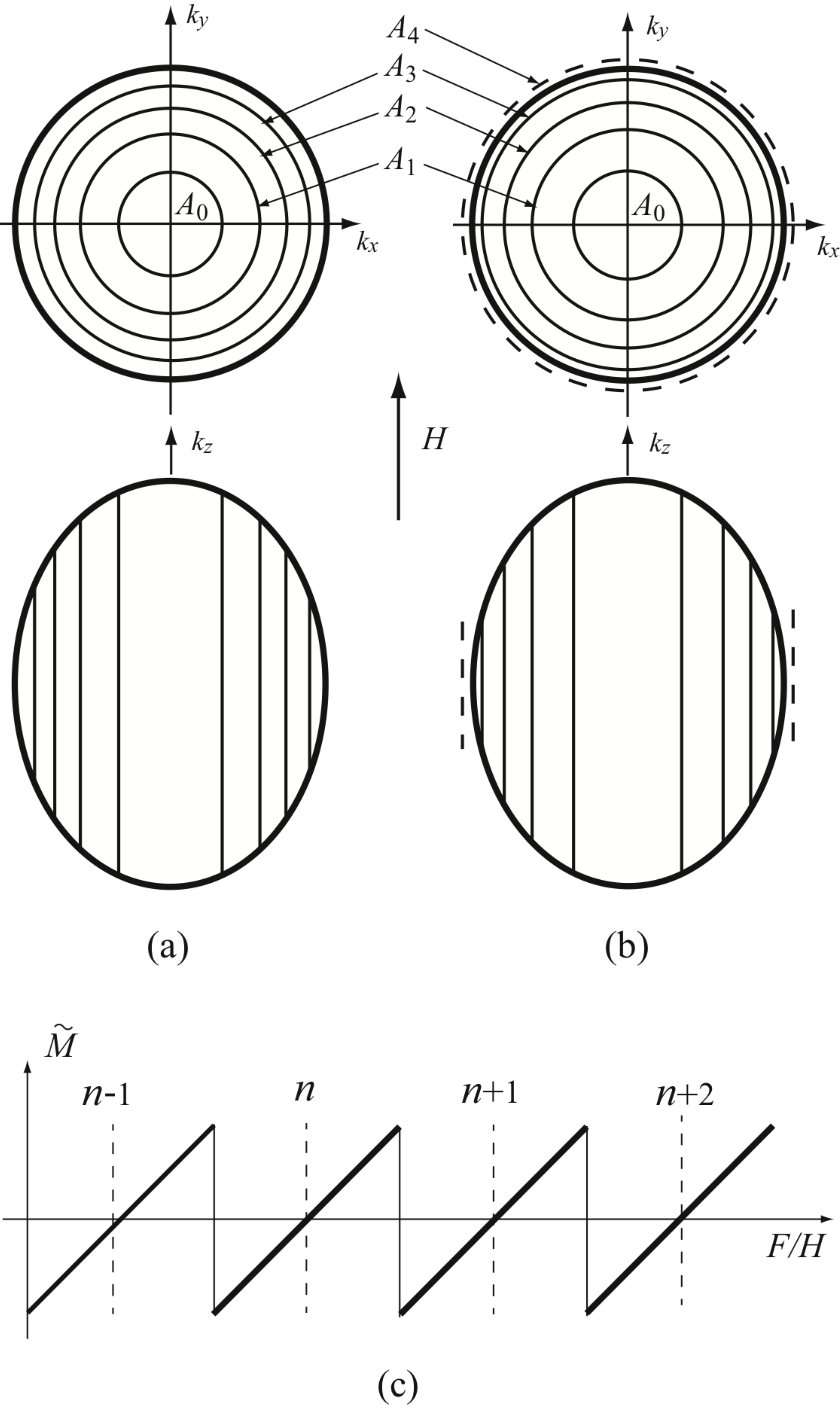}
\end{center}
\caption{(a) and (b):  Schematic illustrations of the Landau tubes for the ellipsoidal Fermi surface when a magnetic field is applied parallel to the long axis of the ellipsoid. The upper figures are cross sections in the $k_x - k_y$ plane and the lower figures are cross sections which include the $k_z$ axis.  In (a)  the highest Landau level matches the Fermi level and in (b) the Fermi level is above the highest Landau level. (c) Saw tooth variation of the magnetization arising from a thin slice of the Fermi surface perpendicular to the $k_z$ axis.}
\label{fig:90025FigA1}
\end{figure}

The dHvA effect, that is the quantum oscillation of magnetization, arises from the quantized cyclotron motion of the conduction electron under magnetic fields. The energy levels in the reciprocal space are also quantized into the so-called Landau levels.  Figure \ref{fig:90025FigA1} (a) and (b) illustrate the Landau levels for the case of an ellipsoidal  Fermi surface centered at the origin of the reciprocal space.  The magnetic field is applied along the long axis of the ellipsoid which is parallel to the $k_z$ axis.  The upper panel shows the cross section of the Fermi surface at the origin in the $k_x - k_y$ plane and the circles show the Landau levels and Fermi surface.  The figure (a) is so drawn that the highest Landau level $n$ matches the Fermi energy or the area $A_n$ in the reciprocal space enclosed by the Landau level $n$ becomes equal to the cross section of the Fermi surface $A_{\rm F}$ at $k_z = 0$.  Since only the motion in the plane perpendicular to the magnetic field is quantized, the quantized levels form the so-called Landau tubes.  The lower panel shows the cross sections in the plane which includes the $k_z$ axis.  The straight lines denote the cross sections of the Landau tubes.  The area in the reciprocal space enclosed by the Landau tube is given by 
\begin{eqnarray}
A_n=2\pi\alpha(n+\gamma). 
\label{eq:A1}
\end{eqnarray}
Here $n$ is an integer, $\alpha$ is given by $\alpha=(eH/\hbar{c})$.   $\gamma$ is the phase factor which is 1/2 for the free electron case and corresponds to the zero point motion of a harmonic oscillator.  In the following, we assume that $\gamma=1/2$.  The spacing between the Landau levels is given by
$\hbar\omega_c$ where $\omega_c$ is the cyclotron frequency which is given by ${eH}/{m_c c}$.  $m_c$ is the cyclotron mass which is defined as  
\begin{eqnarray}
m_c=\frac{\hbar^2}{2\pi}\frac{\partial A}{\partial E} 
\label{eq:A2}
\end{eqnarray}
$A$ is the area in the reciprocal space proportional to the area in real space enclosed by the cyclotron orbit. 

The electron mass determined by the dHvA effect is an average on the cyclotron orbit under magnetic fields but we simply use the term effective mass $m^\ast$ in this article.  Since for the free electron case $\hbar\omega_c$ is 1.16$\times 10^{-3}$ eV at 10 T, $n$ is of the order of a few to ten thousands for a normal metal whose Fermi energy is of the order of a few electron volt.  Only a few Landau levels are shown in Fig. \ref{fig:90025FigA1} for the sake of illustration. 

If we consider a thin slice at a constant $k_z$ with width of $\delta k_z$, each Landau tube incorporates the electrons in the reciprocal space $2\pi\alpha\delta k_z$.   When the magnetic field is increased from that in Fig. \ref{fig:90025FigA1} (a), the energy of the electron at highest $n$ th level in Fig. \ref{fig:90025FigA1} (a) exceeds the Fermi energy, then the electrons on the level is depleted from the level.   Then, the $n-1$ th level becomes the highest energy level.  We show in Fig. \ref{fig:90025FigA1}(b) the case where the highest Landau level is below the Fermi energy and the energy is equal to that at zero magnetic field. With further increase of magnetic field, the $n-1$ th Landau level approaches the Fermi energy and the depletion of the electrons will be repeated.
The change can be found to be periodic in $1/H$ by noting that the magnetic field $H_n$ at which the area $A_n$ enclosed by the highest Landau level $n$ matches the cross sectional area of the Fermi surface $A_{\rm F}$ is given by
\begin{eqnarray}
A_F=A_n(H)=2\pi\frac{eH_n}{\hbar c}(n+\frac{1}{2}).
\label{eq:A4}
\end{eqnarray}
The period $P$ is given by
\begin{eqnarray}
P=\frac{1}{H_{n+1}}-\frac{1}{H_{n}}=\frac{2\pi e}{\hbar c}\frac{1}{A_{\rm F}},
\label{eq:A5}
\end{eqnarray}
and the frequency $F$ by 
\begin{equation}
F=\frac{\hbar c}{2\pi e}A_{\rm F}.
\label{eq:Frequency}
\end{equation}

We have assumed that the Fermi energy does not change with field or the Landau level below the Fermi level is always filled.  If there is only one slice, the number of electrons should be conserved and to conserve it the Fermi energy may oscillate.  However, we have to count the contributions from slices at different $k_z$.  Since $A_{\rm F}$ or the frequency depends on $k_z$,  the magnetic field where the electrons at the highest Landau level are depleted also depends on $k_z$.  Since the value of $n$ is large for a normal metal and there are ordinarily several pieces of the Fermi surface, then the depleted electrons will be transferred to the portion at the Fermi energy of some other slices or on different Fermi surfaces which act as a reservoir.  On the other hand, to fill the unfilled Landau level, the electrons at the Fermi energy of the reservoir will be transferred to the slice.

The difference in the number of the electrons in the slice from that at zero magnetic field is given by
\begin{eqnarray}
2 \pi \alpha \delta k_z \times D_k\times [A_0(H)-(A_{\rm F}-A_n(H))]. 
\label{eq:dHvA1}
\end{eqnarray}
Here $D_k$ is the density of states in the reciprocal space and is $2\Omega/(2\pi)^3$.  $\Omega$ is the volume of the sample.  The mean change in energy of the electrons which are transferred to and from the slice at a general field will be given by  
\begin{eqnarray}
\frac{1}{2}[A_0(H)-(A_{\rm F}-A_n(H))]\bigl( \frac{\partial E}{\partial A}\big)_{E_{\rm F}},
\label{eq:dHvA2}
\end{eqnarray}
in order that the transferred electrons may leave or enter the reservoir at $E_{\rm F}$\cite{Gold}.
The change in the thermodynamic potential of the slice will be given by the product of the two terms, that is, 
\begin{eqnarray}
2 \pi \alpha \delta k_z D_k\times \frac{1}{2}[A_0(H)-(A_{\rm F}-A_n(H))]^2 \bigl( \frac{\partial E}{\partial A}\big)_{E_{\rm F}}.
\label{eq:dHvA3}
\end{eqnarray}
The magnetization $\tilde{M}$ is given by differentiating it by H\cite{Gold,Shoenberg}.
\begin{eqnarray}
\tilde{M} 
=\frac{e\hbar}{c}\frac{A_{\rm F}}{m_c}(\frac{F}{H}-n-1)dk_z D_{\mathbf{k}}.
\label{eq:dHvA4}
\end{eqnarray}
Here, the condition that $n>>1$ is used. The behavior of the magnetization change of the slice becomes a saw tooth wave as a function of magnetic field as illustrated in Fig. \ref{fig:90025FigA1}(c).   The amplitude of the oscillation is inversely proportional to $m_c$ and is proportional to the cross sectional area $A_{\rm F}$ or the frequency.

The magnetic field when the electrons at the highest Landau level are depleted depends on the slice or the phases of the oscillations from the slices are different.  Then the oscillations from the slices are added destructively except for those near the extremal cross sectional area where the oscillations from the slices contribute constructively.  The resultant oscillation which has the frequency of $F=({\hbar c}/{2\pi e})(A_{\rm F})_{ext}$ is observed, where $(A_{\rm F})_{ext}$ is the extremal cross sectional area perpendicular to the applied magnetic field.  The cross section of the Fermi surface can be expanded around the extremal cross sectional area as,
\begin{eqnarray}
A_{\rm F}=(A_{\rm F})_{ext} \pm \frac{1}{2}\left|\frac{\partial^2 A_{\rm F}}{\partial k_H^2}\right|_{ext}{k_H}^2 \cdot \cdot \cdot\cdot.
\label{eq:A6}
\end{eqnarray}
Here + and - signs correspond to the minimum and maximum cross sectional areas, respectively.
The curvature factor $A^{\prime\prime}$ is defined by
\begin{eqnarray}
A^{\prime\prime}=\left|\frac{\partial^2 A_{\rm F}}{\partial k_H^2}\right|_{ext}.
\label{eq:A7}
\end{eqnarray}
When the curvature factor $A^{\prime\prime}$ is smaller, the phase difference of the oscillations from the slices near the extremal cross sectional area becomes smaller.  Then the amplitude of the resultant oscillation becomes larger when the curvature factor is smaller.

At a finite temperature, the distribution at the Fermi energy is broadened over the energy of the order of $k_{\rm B}T$ and we expect that the spread in the phase of the oscillation due to finite temperature is approximately proportional to $k_{\rm B}T$ and the phase smearing due to the finite temperature reduce the amplitude of the oscillation.  The phase of the oscillation changes by $2\pi$ when the energy changes by about $\hbar\omega_c$ and then the effect of the phase smearing due to temperature is expected to depend approximately on the ratio $k_{\rm B}T/\hbar\omega_c$.

In a real metal, there are always imperfections like impurities. These cause scattering of the conduction electrons. This effect may be taken into account by the broadening of the Landau level whose distribution is described by the Lorentzian form with width $\Gamma$.  By defining the Dingle temperature as
\begin{eqnarray}
T_{\rm D}=\Gamma/\pi k_B,	
\label{eq:A8}
\end{eqnarray}
$T_{\rm D}$ can be related with an electron life time $\tau$ as
\begin{eqnarray}
T_{\rm D} = \hbar/2\pi k_B<\tau>.	
\label{eq:A9}
\end{eqnarray}
Here $< ~ >$ denotes an average over the cyclotron orbit.  The broadening of the Landau level gives rise to phase smearing and reduces the oscillation amplitude.  The effect is expected to depend on the ratio $k_{\rm B}T_{\rm D}/\hbar\omega_c$.  In some cases, for example when the imperfections are dislocations or mosaic structures, the reduction of the signal may be better understood in terms of dephasing of the dHvA oscillations rather than scattering\cite{Shoenberg,Watt74}.   

As noted from the qualitative description for the reduction of the signal amplitude due to finite temperature and scattering,  each contribution to the reduction in the signal amplitude cannot be separated from the other unless we make assumptions for the field and temperature dependences of the scattering and effective mass as described later.  

The degeneracy due to the electron spin is removed under magnetic fields due to the Zeeman splitting and each Landau level splits into the two sublevels $\pm (1/2)g\mu_B H=\pm (1/2)(ge\hbar/2m_0c)H=\pm(1/2)\delta\epsilon$.  The energy spacing between the Landau levels of the up or down spin electrons does not change from $\hbar\omega_c$.  On the other hand, the magnetic field where the highest landau level matches the Fermi energy depends on the spin direction.  The phase difference between the two oscillations from the up and down spin electrons differ by $2\pi(\delta \epsilon/\hbar \omega_c)=2\pi(m_c g/2 m_0)$. Due to the phase difference the amplitude of the oscillation is reduced. The phase difference becomes $\pi$ when $m_c g/2 m_0=1/2$ and the oscillation amplitude happens to be zero.  This phenomena is called the spin splitting zero.  So far, the physical properties like the life time and the effective mass of the conduction electron is assumed to be spin independent. 

The explicit expression of the dHvA effect was given by Lifshitz and Kosevich\cite{Shoenberg,LK} as follows.  This formula is referred to as the LK formula.

\begin{equation}
\tilde{M} = -\left(\frac{e}{c\hbar}\right)^{3/2} \frac{\beta FH^{1/2}\Omega}{2^{1/2}\pi^{5/2}\left(A^{\prime \prime}\right)^{1/2}}
\sum_{p=1}^{\infty}\frac{R_{\rm{T}}R_{\rm{D}}R_{\rm{S}}}{p^{3/2}}\sin \left(2\pi p\left(\frac{F}{H}-\gamma\right)\pm \frac{\pi}{4}\right) 
\times (\mathbf{h}-\frac{1}{F}\frac{\partial F}{\partial \theta}\mathbf{h}_\theta -\frac{1}{F\sin\theta}\frac{\partial F}{\partial \phi}\mathbf{h}_\phi)
\label{eq:LK0}
\end{equation}

The sine terms come from the Fourier decomposition of the saw tooth form, although the saw tooth form is rounded by the reduction effect stated above.  The dHvA oscillation is composed of the fundamental frequency and its harmonic oscillations denoted by $p$.  Since the origin of the harmonic oscillations is the Fourier decomposition of a wave, the amplitude and phase of the harmonic oscillation has a definite relation to those of the fundamental oscillation as can be seen from eq.(\ref{eq:LK0}).  On the other hand, when the physical properties are spin dependent, such a definite relation does not hold.    
Here,  $\beta =(e\hbar)/(m_c c)$.  $\mathbf{h}$, $\mathbf{h}_{\theta}$ and $\mathbf{h}_{\phi}$ are the unit vectors given in the polar coordinates and is parallel to the field, $\theta$ and $\phi$ directions, respectively.  The $\pm$ sign in the sine term denotes that the oscillation arises from  the minimum extremal area ($+$) or the maximum extremal area ($-$). $R_{\rm{T}}$ denotes the effect of the finite temperature and is given by 
\begin{equation}
R_{\rm{T}}=\frac{2\pi^2pk_{\rm B}T/\beta H}{\sinh \left(2\pi^2pk_{\rm B}T/\beta H\right)}~.
\label{eq:RT1}
\end{equation}
$R_{\rm{D}}$ denotes the reduction effect due to the impurity scattering and is given by
\begin{equation}
R_{\rm {D}}=\exp (-2 \pi^2 p k_{\rm B}T_{\rm D}/\beta H)~.
\label{eq:RD1}
\end{equation}
\noindent $R_{\rm{S}}$ denotes the effect due to the spin splitting and is given by  
\begin{equation}
R_{\rm{S}}=\cos \left(\frac{1}{2}p\pi gm_c/m_0\right)~.
\label{eq:RS}
\end{equation}
It is difficult to judge whether the observed oscillation are composed of the oscillations of the two spin directions or comes from the one spin direction, unless we know the values of the other factors in the LK formula precisely as discussed in $\S$3.1.2.  As noted from the origin of the spin splitting phenomena, when the properties like effective mass and electron life time depend on the spin direction, the phase as well as the amplitude of the oscillation differ from those given by the LK formula.  The signal is ordinarily analyzed by using the Fourier analysis.   From the resultant power spectrum, we determine the amplitude and the frequency of the oscillation.  It is noted that the information of the phase of the oscillation is lost in the power spectrum, while the phase could contain important physical information in the case of the strongly correlated f electron system.

The LK formula was successfully applied to normal metals to reveal the electronic structures\cite{Shoenberg,Cracknell}.  Detailed studies to judge the validity of the equation, like the absolute measurements of the dHvA oscillation amplitude,  the evaluation of the value of the phase factor and so on have been also performed\cite{Shoenberg}.  

In most of the dHvA measurements, the field modulation method is used to detect the dHvA signal.  This method picks up the signal component parallel to the field direction, while the torque method does the perpendicular component.  The method applies a small ac field $h\sin(\omega t)$ in addition to the high DC field $H$.  For a small modulation field, the signal is given by
\begin{equation}
\tilde{M} (H+h\sin (\omega t))=\tilde{M}(H)+\frac{\partial \tilde{M}}{\partial H}h\sin (\omega t)+\frac{1}{2}\frac{\partial ^2\tilde{M}}{\partial H^2}h^2\sin ^2(\omega t) \cdot \cdot \cdot+\frac{1}{n!}\frac{\partial^n  \tilde{M}}{\partial H^n}h^n\sin ^n(\omega t)\cdot \cdot \cdot.
\label{eq:MdHvA}
\end{equation}
The induced voltage in a pickup coil is proportional to $\partial \tilde{M}/\partial t$ which is given by
\begin{equation}
\frac{\partial \tilde{M}}{\partial t}=\omega[\frac{\partial \tilde{M}}{\partial H}h\cos (\omega t)+\frac{1}{2}\frac{\partial ^2\tilde{M}}{\partial H^2}h^2\cos(2\omega t)\cdot \cdot \cdot + \frac{1}{2^{n-1}(n-1)!}\frac{\partial^n  \tilde{M}}{\partial H^n}h^n\cos(n \omega t) \cdot \cdot \cdot].
\label{eq:VdHvA}
\end{equation}
Here, only the lowest power of $h$ has been retained in the amplitude of each harmonic.  By using the phase sensitive detection technique, we pick up one of the components of $\cos(n\omega t)$.  With a large modulation field, the signal amplitude of the component  $\cos(n\omega t)$  is proportional to $J_n(2\pi F h/H^2)$.  Here $J_n$ is the $n$th Bessel function which shows an oscillatory behavior as a function of $Fh/H^2$.  To reach the first maximum of the function, we need a larger modulation field for a smaller frequency oscillation and vice versa.   This fact causes a heating problem to detect low frequency oscillations.  

The pressure effects on the electronic structures were also studied by using the dHvA effect\cite{Joss}. The effect was mostly studied on the frequency because the change in the effective mass is less than the experimental error in normal metals. The change in the frequency with pressure is very small in normal metals and therefore very accurate measurements were necessary.  The results are, so to say, not dramatic as compared with those of strongly correlated f electron systems which are demonstrated partially in the main text.   Since the pressure experiment of the strongly correlated f electron system uses the very low temperatures, the heating problem of the metallic pressure cell is more serious due to modulation field.  However, it turned out that the metallic pressure cell is practically compatible with the dilution refrigerator and the field modulation method, if we use a modulation field with low frequency and small amplitude\cite{Takashita97}.  But it is difficult to determine the effective mass of the low frequency oscillation accurately under pressure.  

\appendix\subsection{Phenomenon which are not described by the LK formula}
There are some other effects which are not described explicitly in the LK formula. In ferromagnetic metals, the observed oscillation is not periodic as a function of $1/H$ but periodic in the total induction $B=H+4\pi M$, because the relevant magnetic field for the quantization of an electron orbit is the total induction $B$ rather than the applied field $H$. Since we normally observe the oscillation as a function of applied field $H$, the observed frequency is modified from the true frequency.  The frequency of the dHvA oscillation in the ferromagnetic state could change with magnetic field, because the effective mass as well as the magnetization changes with field as described in $\S$4.2.  However, it is difficult to detect the small change in the frequency unless we take into account the change in the magnetization precisely.

When the magnetization changes rapidly with $H$, this effect sometimes affects the observed frequency significantly. Since the dHvA oscillation is the oscillation in magnetization, the electron motion is also influenced by the oscillation of magnetization $\tilde{M}$ and the observed behavior of the oscillation is modified largely from that predicted by the LK formula when the amplitude of $\tilde{M}$ is large.  The strength of the magnetic interaction is given by $8\pi |\tilde{M}F|/H^2$ which measures how rapidly the magnitude of the oscillatory magnetization changes with respect to the change in magnetic field.  As noted from this equation, the strength increases with decreasing temperature but has a maximum at a field as a function of magnetic field.   This effect does not seem to be large in the strongly correlated f systems because the oscillation amplitude is normally small.  The analysis of the effect was made semiclassically by replacing $H$ in eq.(\ref{eq:LK0}) by $B$ and by taking the phase smearing effect owing to inhomogeneity of the sample\cite{Shoenberg}.  Experimental as well as theoretical verification on this effect for the strongly correlated f electron system has not been made.  However, if the electron-electron interaction effect was suitably taken into account in $\tilde{M}$, the magnetic interaction effect could be treated semiclassically.  

When the frequency of the dHvA oscillation changes with magnetic field, it modifies the observed frequency as described in $\S$3.2.2.  However, it is noted from eq.(\ref{eq:LK0}) and also from eq.(\ref{eq:truefrequency}) that the observed frequency does not change with field, when the frequency or the Fermi surface cross section changes linearly with magnetic field.  When the Fermi surface splits into the up and down spin electron surfaces,  the observed frequency or the cross sectional area of the Fermi surface is the average of the up and down spin electron surface and does not change with field as long as the extremal cross sectional areas of the up and down spin electron Fermi surfaces change linearly with magnetic field.

We have not considered the many body interaction explicitly above. In normal metals the most significant many body interaction is the electron-phonon interaction.  The effective mass determined by the dHvA effect is found to agree well with the thermal effective mass.  The deviation from the LK formula has been found in mercury only for peculiar experimental conditions\cite{Khalid88}.

For the strongly correlated f electron system, the electron-electron interaction effect should be taken into account.    Most experimental results have been analyzed using the LK formula by assuming that the electron or quasiparticle with the enhanced effective mass makes a cyclotron motion under magnetic fields.  However, if the most significant many body effect is the Kondo effect, the assumptions that the Dingle temperature or effective mass does not depend on magnetic field and temperature may not be valid.  Even the Fermi surface or the electronic structure could be field and temperature dependent as described in the main text.  Efforts to derive a formula which incorporates the electron-electron interaction or the Kondo effect explicitly were made but they have not been widely used to analyze the experimental results\cite{Wasserman89,Wasserman96,Otsuki07}.   Although the LK formula may have to be reformulated,  we use the LK formula to analyze the experimental data and may obtain some results which are not consistent with the implicit assumptions of the LK formula.   However, as long as complimentary measurements like specific heat measurements indicate consistent results, we assume that the analyzed results like field dependent effective mass are physically useful and correct. 
 
To discuss the problems in the analysis of the dHvA effect measurements in the strongly correlated f electron system, we pick up the fundamental frequency component from the LK formula and describe it in a simplified form.  The dHvA effect signal amplitude $S_{amp}$ detected by the field modulation method is given by
\begin{equation}
S_{amp}=E(H)\cdot C \cdot \frac{\Omega\cdot F \cdot R^\prime_{\rm T} \cdot R^\prime_{\rm D}}{\sqrt{A^{\prime\prime}}\sqrt{H}}~.
\label{eq:Samp}
\end{equation}
Here, $E(H)$ is a factor which depends on experimental condition and is a function of $H$ for the field modulation method\cite{Shoenberg}.  $C$ is a constant.   ${R^\prime}_{\rm T}$ and ${R^\prime}_{\rm D}$ are given by 
\begin{equation}
R^\prime_{\rm T} = \frac{T}{\sinh(\lambda\mu T/H)}~,
\label{eq:RT2}
\end{equation}
and 
\begin{equation}
R^\prime_{\rm D} = \exp (-\lambda\mu T_{\rm D}/H)~.
\label{eq:RD2}
\end{equation}  
$\lambda$ is a constant given by 14.96 T/K and $\mu$ is the effective mass ratio $m^\ast/m_0$. 
From eq.(\ref{eq:Samp}) we obtain with $C \cdot \Omega \cdot F/\sqrt{A^{\prime\prime}}=C_0$, 
\begin{equation}
\ln \bigl( \frac{S_{amp}}{{R^\prime}_{\rm T}/\sqrt{H}}\bigr) = \ln (E(H) \cdot C_0) + \ln {R^\prime}_{\rm D}~.
\label{eq:lnSamp}
\end{equation} 
Here,
\begin{equation}
\ln {R^\prime}_{\rm D}= \frac{-\lambda\mu T_{\rm D}(T,H)}{H} ~.
\label{eq:lnRD}
\end{equation}  

In ordinary cases we plot the left hand side of eq.(\ref{eq:lnSamp}) against $1/H$.  When the field dependence of $E(H)$ comes only from the field modulation method and its effect is suitably corrected for the observed amplitude $S_{amp}$, we can fit a straight line to the data points. From the slope of the straight line, we obtain the Dingle temperature.  However, if the fitted line is not straight, we cannot determine the Dingle temperature because it is very difficult to determine the value of the first term in eq.(\ref{eq:lnSamp}) accurately. When the effective mass depends on magnetic field and temperature, we cannot determine the Dingle temperature unless we know the value of the effective mass as a function of magnetic field and temperature. On the other hand, we need to know the field and temperature dependence of the Dingle temperature to determine the effective mass.  These situations happen in the measurements of the dHvA effect on the strongly correlated f electron system.

However, in some cases the estimate of the Dingle temperature is possible.  For example, if $C_0$  and the effective mass do not depend on temperature, we can obtain 
the difference in the Dingle temperatures between two temperatures at a constant magnetic field 
from eqs.(\ref{eq:lnSamp}) and (\ref{eq:lnRD}). Since the temperature dependence of the 
frequency is ordinarily very tiny, $C_0$ can be assumed to be constant.  The Dingle temperature of the dilute Kondo alloy has been estimated and its temperature dependence is found to be similar to that of resistivity\cite{Matsumoto12}.  We may also evaluate the relative magnitude of reduction by measuring the amplitude of the oscillation keeping the experimental conditions including the volume and shape of the sample the same. 

The effective mass can be determined by analyzing the temperature dependence of the amplitude from eq.(\ref{eq:lnSamp}) when the Dingle temperature does not depend on temperature.  However, there are some cases where the temperature dependence of the signal amplitude cannot be described by the LK formula or in terms of the temperature independent scattering by the imperfections in the crystal.   For example, in the superconducting state the magnetic oscillations can be observed,  but the signal amplitude is affected by the flux lattice state which changes with magnetic field and temperature\cite{Isshiki08}.   When the second order phase transition temperature is low, the magnitude of the order parameter changes in the temperature range where the dHvA effect measurements are performed.  Consequently the change gives rise to the temperature dependence of the signal amplitude\cite{IsshikiD} which cannot be described by the LK formula.    
 
\appendix\subsection{dHvA effect measurements in dilute alloys and life time of quasiparticle}
In the present article, we describe the dHvA effect measurements on the alloys of the strongly correlated f electron systems.  We briefly review the results of the dHvA effect studies applied to very dilute alloys of normal metals\cite{Shoenberg,Springford}to discuss the origins of life time reduction in the strongly correlated f electron system. 

The experimental results in dilute alloys were analyzed based on the band structure calculation.  The change in the cross sectional area $\delta A$ by the alloyed impurity is given by 
\begin{equation}
\delta A = -\frac{2\pi}{\hbar^2}m_{\rm B} \langle {\rm Re} \Sigma(\mathbf{k},E_{\rm F} \rangle,
\label{eq:deltaA}
\end{equation}  
 and the Dingle temperature measured is given by
\begin{equation}
  T^\ast_{\rm D}=-{\bigl(}\frac{m_{\rm B}}{m^\ast\pi k_{\rm B}} \langle {\rm Im} \Sigma(\mathbf{k}, E_{\rm F})\rangle{\bigr)}.
\label{eq:DingleT}
\end{equation}  
\begin{equation}
\Sigma (\mathbf{k}, E_{\rm F})=c_0T_{\mathbf{k} \mathbf{k}}(E_{\rm F}).
\label{eq:Selfenergy}
\end{equation}  
Here, $m_{\rm B}$ is the band mass, $T$ is the T matrix, $\Sigma$ is the self energy and $c_0$ is the impurity concentration which is much less than 1 atomic \%.  In this calculation the Fermi energy is assumed to be unchanged.  This assumption may be valid because the perturbation of the potential is limited only to the region close to the impurity atom in a dilute alloy.  It is also noted that in an alloy the definition of the Fermi edge or the Fermi surface is ambiguous particularly when $|{\rm Im} \Sigma| >> |{\rm Re} \Sigma|$.  Even in that case, the theory assumes that the observed frequency change can be well expressed by eq.(\ref{eq:deltaA}).  The wave function of the host metal is expanded around the impurity site as 
\begin{equation}
\Psi _{\mathbf{k}}(\mathbf{r})=\sum_{l \Gamma \gamma} i^l a_{l \Gamma \gamma}(\mathbf{k}, E) R_l (r)  Y_{l \Gamma \gamma}(\theta,\phi).
\label{eq:Psi}
\end{equation}  
Here, muffin tin potentials are assumed for the host and impurity atoms.  $l$ is the azimuthal quantum number, $\Gamma$ is the irreducible representation corresponding to the symmetry of the lattice around the impurity, and $\gamma$ is one of the bases of the representation. $ Y_{l \Gamma \gamma}(\theta,\phi)$ is the cubic harmonics.  For simplicity, we may assume that the scattering in a metal is significant for $l \le 3$ and each irreducible representation does not appear more than once for $l \le 3$.  Then we denote ($l$, $\Gamma$) as $L$. 
T matrix is given by
\begin{equation}
T_{\mathbf{k} \mathbf{k}}=-\frac{\hbar^2}{2m_0\kappa}\sum_{L} P_{\rm L}(\mathbf{k},E)A_{\rm L}(E)\sin ( \delta^i_l-\delta^h_l ) e^{i ( \delta^i_l-\delta^h_l )} .
\label{eq:Tmatrix}
\end{equation}  
Here, $\hbar ^2\kappa ^2/2m_{\rm B}=E$.   $\delta ^i_l$ and $\delta ^h_l$ are the phase shifts corresponding to the impurity and host potential, respectively.  $P_{L}(\mathbf{k},E) $ is the amplitude of the wave function with the symmetry of $(l,\Gamma)$, i.e.
\begin{equation}
P_{L}(\mathbf{k},E)=\sum_\gamma |a_{\rm L \gamma}(\mathbf{k},E)|^2.
\label{eq:PL}
\end{equation}  
$A_{\rm L}$ denotes the effect of the repeated process that the scattered wave from the impurity is backscattered from the host atoms\cite{Holzwarth75}.  For a free electron like metal,  $A_{\rm L} \approx 1$  and  the effect does not need to be taken into account but may be significant for transition metals and probably for the strongly correlated f electron systems.  For free electron (or $\delta^h=0$) the phase shift of the impurity is related to the valence difference $\Delta z$ as 
\begin{equation}
 \frac{2}{\pi}\sum_{l} (2l+1)\delta^i_l=\Delta z.
\label{eq:FriedelSum}
\end{equation}  
The factor 2 comes from the spin degeneracy. 
Generally if we define the phase $\Delta\theta_L$ as
\begin{equation}
A_{\rm L}(E)=|A_{\rm L}| \exp (i\Delta\theta_{\rm L}),
\label{eq:Al}
\end{equation}  
the generalized Friedel sum rule is given by 
\begin{equation}
\frac{2}{\pi}\sum_{\rm L} g_{\rm L}((\delta^i_l-\delta^h_l)+\Delta\theta_{\rm L}) =\Delta z.
\label{eq:FriedelSum2}
\end{equation}  
Here,  $g_{\rm L}$ is the dimension of the irreducible representation $\Gamma$.
As noted from eqs.(\ref{eq:deltaA}), (\ref{eq:DingleT}), (\ref{eq:Tmatrix}), (\ref{eq:FriedelSum}) and (\ref{eq:FriedelSum2}), when the difference between the phase shifts due to the host and impurity potentials is large (small),  the changes in the frequency and the Dingle temperature will be normally large (small).  The changes also depend on the amplitude of the same symmetry component of the wave function with the potential.  The Friedel sum rule eq.(\ref{eq:FriedelSum}) or eq.(\ref{eq:FriedelSum2}) tells that the changes are normally large when the valence difference is large.  We also note that the rigid band model will be valid when each $\delta_l$ is small and has the same sign.  In the rigid band model, $\Delta{F}$ is given by\cite{Chollet68}
 \begin{equation}
\Delta{F}=\pm({\pi}^4/3)^{1/3}({\hbar}c/e)(m^{\ast}/m_{\rm th})({\Delta} n/n^{1/3}).
\label{eq:DeltaF}
\end{equation}  
Here, $m_{\rm th}$ is the thermal mass which is an average of $m^\ast$ over the Fermi surface, $n$ is the number of conduction electron per unit volume, and ${\Delta}n$ is the change in $n$ due to alloying.  $\pm$ corresponds to either electron (+) or hole (-) surface.  

The experimental studies on the dilute alloys of normal metals have difficulties owing to the fact that the signal amplitude is very much reduced with alloying a small amount of impurity atoms.  For example, the experiment on dilute alloys of Au-Ag reports that the Dingle temperature increases at a rate of about 9 K$/$at\% for the Belly orbit when Ag is dissolved in Au host metal. It is noted that the valence difference is zero and the potentials of Au and Ag as well as their lattice constants are quite similar to each other.  As can be seen from eqs.(\ref{eq:LK0}) and (\ref{eq:RD2}),  the value of the Dingle temperature indicates that the observed amplitude is reduced to 3 $\times 10^{-5}$ due to alloying 1 at \% impurity for the Belly oscillation whose effective mass is about 1 $m_0$.  In other alloys, the Dingle temperatures are considerably larger.  For example, in the case of Fe impurity in Au host,  $\Delta T_{\rm D}$ is 120 K/at.\%  and in the case of Sn impurity  $\Delta T_{\rm D}$ is 35 K/at.\%  \cite{Shoenberg,Springford,Lowndes73}.   These observations are strikingly different from those presented in this article.   Then alloys with only a tiny amount of the impurity could be studied by the dHvA effect.  Accordingly, the relative frequency change $\Delta F/F$ due to the impurity is very small. Normally $\Delta F/F$ of the order of $10^{-3}$ to $10^{-5}$ has to be measured.  Very accurate measurements were necessary to determine the frequency change.  The measurement of the Dingle temperature is also not easy but practicable in dilute alloys.  On the other hand,  reliable value of the effective mass can be obtained ordinarily only for two digits and the value becomes more unreliable when the signal amplitude is small.  Since the change in the effective mass is very small in the dilute alloys of normal metals with simple metal or transition metal element as an impurity, no studies on the effective mass have been reported in dilute alloys of normal metals except for a few semimetal alloys\cite{Springford} .

The dHvA effect measurements are also applied to dilute Kondo alloys with magnetic impurities of transition metal element.  Some of the results are successfully analyzed by the Friedel Anderson model\cite{Shiba73}.  One of the prominent effect in the Kondo alloy is the spin dependent scattering.  Since the amplitudes of the oscillations from the up and down spin electrons are different due to the spin dependent scattering,  the spin splitting zero phenomena observed at a particular field direction changes to the spin splitting minimum in a dilute Kondo alloy\cite{Coleridge70,Alles73}.    

Although it may not be apparent from the equations above, it is noted that the electron mean path $l$ is determined from an average distance between impurities and is given approximately by $v\tau$.  Here, $v$ is the velocity of the electron.  Then when the $v$ of an orbit is relatively large (small), then $\tau$ will be is small (large) or the Dingle temperature is large (small).  Ordinarily  if the band mass is large, the velocity of the conduction electron is small.  This relation can be found in a transition metal based alloy like Nb-Mo and other d band metal based alloys\cite{Aoki83}.   This relation is verified also in a heavy Fermion compound using two different orbits of CeIn$_3$ which have different effective masses\cite{Ebihara93} indicating the picture that the heavy quasiparticle makes a cyclotron motion under magnetic fields is valid.   Then, the Dingle temperature determined using the effective mass corresponds to the life time of the slowly moving heavy quasiparticle.

However, as demonstrated for the residual resistivity and the dHvA signal amplitude in the main text,  the scattering of the electron in the alloy of the strongly correlated f electron system depend not only on the concentration of the alloying element or the average distance between the impurities, but also on the many body effect.  There are also cases where the reduction of the signal amplitude cannot be described in terms of scattering,  but rather by many body fluctuations.  Examples are suggested in $\S$3.4 for the case of the oscillation amplitudes above $H_{\rm m}$.  It is also found in the low field state of  Ce$_x$La$_{1-x}$B$_ 6$ that the electron mean free path decreases with decreasing field towards the boundary between the Fermi liquid and non-Fermi liquid states\cite{Nakamura06,Endo06}.

\appendix\subsection{Measurement of frequency change by the phase shift of oscillation}
We describe how we determine the frequency change with temperature using the phase shift of the dHvA oscillation.  The same method can be applied also when the frequency is measured as a function of pressure\cite{Yamamizu04} or concentration of alloying element.

\begin{figure}[ht]
\begin{center}
\includegraphics[width=0.5\linewidth]{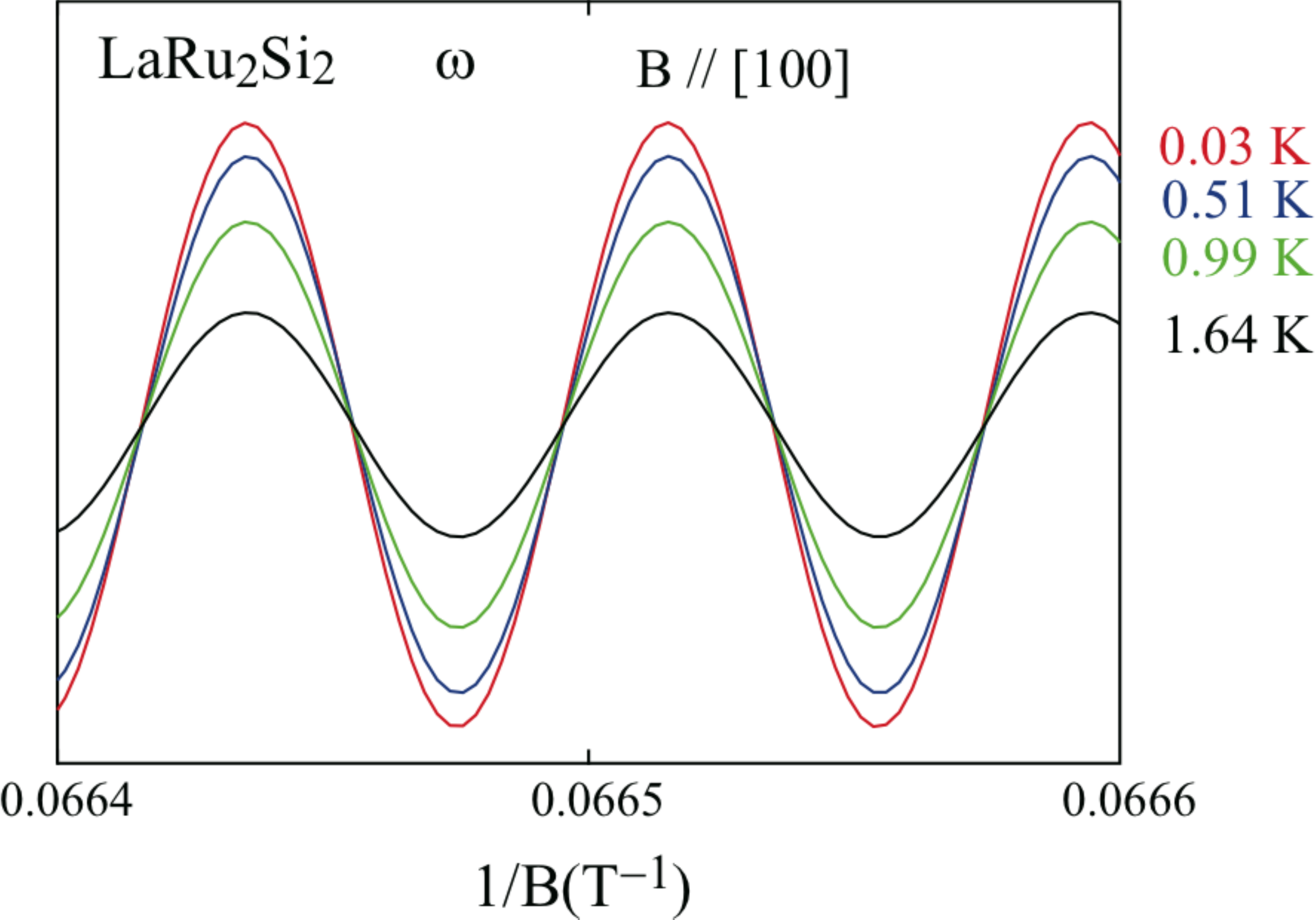}
\end{center}
\caption{(Color online) Traces of the $\omega$ oscillation in LaRu$_2$Si$_2$ at various temperatures as a function of inverse field.  The magnetic field is about 15 T and is applied parallel to the [100] direction.  }
\label{fig:90025FigA2}
\end{figure}

The $\omega$ oscillation of LaRu$_2$Si$_2$ at different temperatures are plotted as a function of magnetic field in Fig. \ref{fig:90025FigA2}. We assume that the dHvA oscillation has a definite phase at the infinite magnetic field strength.  Then we can assign the peak number from the infinite field to each maximum of the oscillation and identify the same peak number of the oscillations at different temperatures. The Fourier analysis is also employed for this identification.  When the magnetic fields for the positions of the same peak number are $H(T_2)$ at $T_2$ and $H(T_1)$ at $T_1$, then the relative change in the frequency between the frequencies $F(T_1)$ at $T_1$ and $F(T_2)$ at $T_2$ is given by\cite{Yamamizu04}, 

\begin{equation}
\frac{F(T_2)-F(T_1)}{F(T_1)} = \frac{H(T_2)-H(T_1)}{H(T_1)}  ~.
\label{eq:mDeltaF}
\end{equation}
 
This method can determine the relative change of the dHvA frequency very accurately, even if the absolute value of the magnetic field and accordingly the absolute value of the frequency cannot be determined accurately. The reliability of the measurement has been checked in several ways. For example, we confirmed that the phase shift of the $\omega$ oscillation with temperature in LaRu$_2$Si$_2$ cannot be detected as shown  in Fig. \ref{fig:90025FigA2}, while obvious changes can be found for the particular cases as demonstrated in $\S$4.1.1.  

\begin{figure}[ht]
\begin{center}
\includegraphics[width=0.5\linewidth]{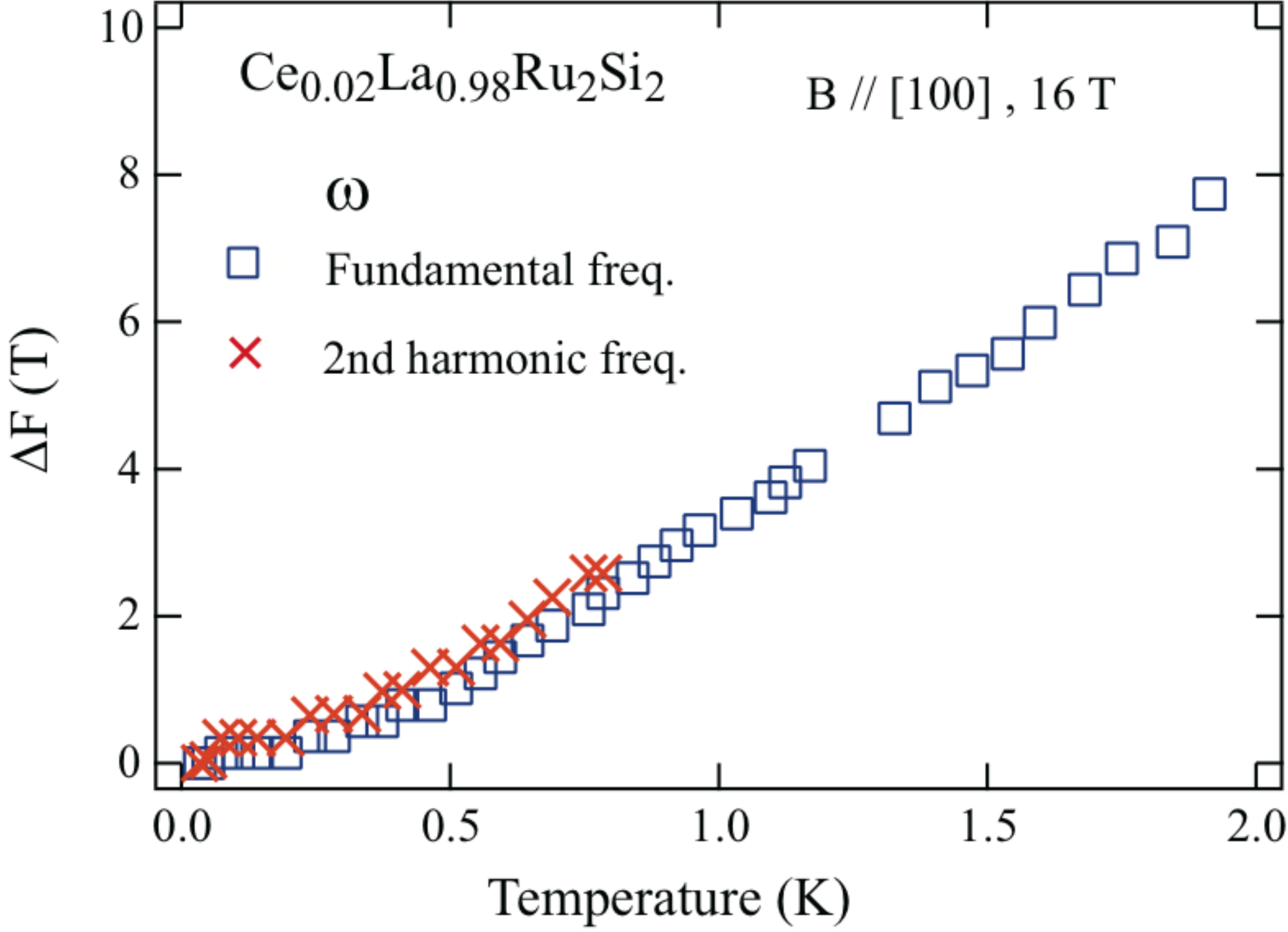}
\end{center}
\caption{(Color online) Frequency changes $\Delta F$ with temperature.  The magnetic field is about 16 T and applied parallel to the [100] direction.  The square and cross symbols denote the frequency changes derived from the phase shifts of the fundamental and second harmonic frequency oscillations, respectively. }
\label{fig:90025FigA3}
\end{figure}

The phase shift of the oscillation occurs in some cases from the spin dependent effective mass\cite{Endo06,Takashita96,Endo04,Endo05}.  Since the Ce concentration is small and the magnetic field is applied parallel to the hard axis of the magnetization for the case of $\S$4.1.1, this effect can be assumed to be negligible.  In fact, we do not find any characteristic feature of this effect by examining magnetic field and temperature dependences of the phase shift.  Particularly, if this effect is appreciably large, the behaviors of the phase shift are different for the fundamental and second harmonic frequency oscillations.  On the other hand, if the phase shift comes from the change in the frequency with temperature, the phase of the second harmonic frequency oscillation has a definite relation to the phase of the fundamental frequency oscillation, irrespective of temperature and magnetic field as described in $\S$A.1.  Therefore, the frequency change measured from the phase shift of the second harmonic frequency oscillation should be the same as that from the fundamental frequency oscillation.  In Fig. \ref{fig:90025FigA3} we show the changes in the frequency $\Delta F$ of the $\omega$ oscillation from the lowest temperatures which are obtained from the phase shifts of the fundamental and second harmonic frequency oscillations.  They agree very well with each other.

\begin{thebibliography}{9}
\bibitem{Matsumoto08}Y. Matsumoto, M. Sugi, N. Kimura, H. Aoki, I. Satoh, T. Terashima, and  S. Uji, J. Phys. Soc. Jpn. {\bf 77},  053703 (2008).
\bibitem{Harima11}H. Harima, News Letter of  Scientific Research on Innovative Area ``{\it Emergence of Heavy Electron and Their Ordering}''  vol.3, No1., p.34 (2011)  [in Japanese].
\bibitem{Flouquet95}J. Flouquet, S. Kambe, L. P. Regnault, P. Haen, J. P. Brison, F. Lapierre, and P. Lejey, Physica B \textbf{215}, 77 (1995) .
\bibitem{Flouquet02}J. Flouquet, P. Haen, S. Raymond, D. Aoki, and G. Knebel, Physica B \textbf{319}, 251 (2002).
\bibitem{Flouquet04}J. Flouquet, Y. Haga, P. Haen, D. Braithwaite, G. Knebel, S. Raymond, and S. Kambe, J. Mag. Mag. Mater. {\textbf{272-276}}, 27 (2004).
\bibitem{Flouquet05}J. Flouquet, A. Barla, R. Boursier, J. Derr, and G. Knebel, J. Phys. Soc. Jpn. \textbf{74}, 178 (2005).
\bibitem{Flouquet10}J. Flouquet, D. Aoki, W. Knafo, G. Knebel, T. D. Matsuda, S. Raymond, C. Proust, C. Paulsen, and P. Haen, J. Low Temp. Phys. \textbf{161}, 83 (2010).
\bibitem{Flouquet07}J. Flouquet, {\it On\ the\ Heavy\ Fermion\ Road} in Progress in Low Temperature Physics, vol. XV, ed. B. Halperin, (Elsevier B. V. 2007).
\bibitem{Yamagami92}H. Yamagami and A. Hasegawa, J. Phys. Soc. Jpn. {\bf 61},  2388 (1992).
\bibitem{Besnus85}M. J. Besnus, J. P. Kappler, P. Lehmann, and A. Meyer, Solid State Communi. {\bf 55}, 779 (1985).
\bibitem{Boucherle01}J.-X. Boucherle, F. Givord, S. Raymond, J. Schweizer, E. Leli\`ever-Berna, P. Lejay, and G. Fillion, J. Phys. Condes. Matter \textbf{13}, 10901 (2001).
\bibitem{Willers12}T. Willers, D. T. Adroja, B. D. Rainford, Z. Hu, N. Hollmann, P. O. K{\"o}rner, Y.-Y. Chin, D. Schmitz, H. H. Hsieh, H.-J. Lin, C. T. Chen, E. D. Bauer,  J. L. Sarrao, K. J. McClellan, D. Byler, C. Geibel, F. Steglich, H. Aoki, P. Lejay, A. Tanaka, L. H. Tjeng, and A. Severing, Phys. Rev. B {\bf 85}, 035117 (2012).
\bibitem{Amato94}A. Amato, R. Feyerherm, F. N. Gygax, A. Schenck, J. Flouquet, and P. Lejay, Phys. Rev. B \textbf{50}, 619(R) (1994).
\bibitem{Takahashi03}D. Takahashi, S. Abe, H. Mizuno, D. A. Tayurskii, K. Matsumoto, H. Suzuki, and Y. Onuki, Phys. Rev. B \textbf{67},  180407(R) (2003).
\bibitem{Mignod88}J. Rossat - Mignod, L. P. Regnault, J. L. Jacoud, C. Vettier, P. Lejey, J. Flouquet, E. Walker, D. Jaccard, and A. Amato, J. Mag. Mag. Mater. {\textbf{76\&77}}, 376 (1988).
\bibitem{Sato99}M. Sato, S. Kawarazaki, Y. Miyako, and H. Kadowaki, J. Phys. Chem. Solids \textbf{60}, 1203 (1999).
\bibitem{Kadowaki04}H. Kadowaki, M. Sato, and  S. Kawarazaki, Phys. Rev. Lett. \textbf{92}, 097204 (2004).
\bibitem{Fisher91} R. A. Fisher, C. Marcenat, N. E. Phillips, P. Haen, F. Lapierre, P. Lejay, J. Flouquet, and J. Voiron, J. Low Temp. Phys. {\textbf{84}}, 49 (1991).
\bibitem{Kadowaki86}K. Kadowaki and S. B. Woods, Solid State Communi. \textbf{58}, 507 (1986).
\bibitem{Tsujii05}N. Tsujii, H. Kontani, and Y. Yoshimura, Phys. Rev. Lett. \textbf{94}, 057201 (2005).
\bibitem{Haen87}P. Haen, J. Flouquet, F. Lapierre, P. Lejay, and G. Remenyi, J. Low Temp. Phys. {\bf 67}, 391 (1987).
\bibitem{Sakakibara95}T. Sakakibara, T. Tayama, K. Matsuhira, H. Mitamura, H. Amitsuka, K. Maezawa, and Y. Onuki, Phys. Rev. B {\bf 51}, 12030(R) (1995).
\bibitem{Paulsen90}C. Paulsen, A. Lacerda, L. Puech, P. Haen, P. Lejay, J. L. Tholence, and J. Flouquet, J. Low Temp. Phys. \textbf{81}, 317 (1990).
\bibitem{Lacerda89}A. Lacerda, A. de Visser, L. Puech, P. Lejay, P. Haen,  J. Flouquet, and A. de Visser, J. Appl. Phys. \textbf{67}, 5212 (1990).
\bibitem{Matsuda12a}Y. H. Matsuda, T. Nakamura, J. L. Her, S. Michimura, T. Inami, K. Kindo, and T. Ebihara, Phys. Rev. B \textbf{86}, 041109(R) (2012).
\bibitem{Rueff06}J.-P. Rueff, J.-P. Iti\'{e}, M. Taguchi, C. F. Hague, J.-M. Mariot, R. Delaunay, J.-P. Kappler, and N. Jaouen, Phys. Rev. Lett. \textbf{96}, 237403 (2006).
\bibitem{Matsuda07}Y. H. Matsuda, T. Inami, K. Ohwada, Y. Murata, H. Nojiri, Y. Murakami, H. Ohta, W. Zhang, and K. Yoshimura, J. Phys. Soc. Jpn. \textbf{76}, 034702 (2007).
\bibitem{Ohkawa89}F. J. Ohkawa, Solid State Communi. \textbf{71}, 907 (1989).
\bibitem{Ohkawa92}F. Ohkawa, Prog. Theor. Phys. Suppl. \textbf{108}, 209 (1992).
\bibitem{Ono98}Y. Ono, J. Phys. Soc. Jpn. \textbf{67}, 2197 (1998).
\bibitem{Ohara99}K. Ohara, K. Hanzawa, and K. Yoshida, J. Phys. Soc. Jpn. \textbf{68}, 521 (1999).
\bibitem{Watanabe00}S. Watanabe,  J. Phys. Soc. Jpn. \textbf{69}, 2947 (2000).
\bibitem{Satoh01}H. Satoh and F. J. Ohkawa, Phys. Rev. B \textbf{63}, 184401 (2001).
\bibitem{Miyake06}K. Miyake and H. Ikeda,  J. Phys. Soc. Jpn. \textbf{75}, 033704 (2006).
\bibitem{Bauer09}J. Bauer, Eur. Phys. J. B \textbf{68}, 201 (2009).
\bibitem{Ohara09}K. Ohara and  K. Hanzawa, J. Phys. Soc. Jpn. \textbf{78}, 044709 (2009).
\bibitem{Weickert10}F. Weickert, M. Brando, F. Steglich, P. Gegenwart and M. Garst, Phys. Rev. B {\bf81}, 134438 (2010).
\bibitem{Aoki11}D. Aoki, C. Paulsen, T. D. Matsuda, L. Malone, G. Knebel, P. Haen, P. Lejay, R. Settai, Y. Onuki, and J. Flouquet, J. Phys. Soc. Jpn. \textbf{80}, 053702 (2011).
\bibitem{Bercx12}M. Bercx and F. F. Assaad, Phys. Rev. B \textbf{86}, 075108 (2012).
\bibitem{Howczak12}O. Howczak and J. Spalek, J. Phys.: Condens. Matter \textbf{24}, 205602 (2012).
\bibitem{Kubo13a}K. Kubo, Phys. Status Solidi \textbf{10}, 544 (2013).
\bibitem{Kubo13b}K. Kubo, Phys. Rev. B \textbf{87}, 195127 (2013).
\bibitem{Matsumoto11}Y. Matsumoto, M. Sugi,  K. Aoki, Y. Shimizu,  N. Kimura, T. Komatsubara, H. Aoki, M. Kimata, T. Terashima, and S. Uji, J. Phys. Soc. Jpn. {\bf 80}, 074715 (2011).
\bibitem{Haen99} P. Haen, H. Bioud, and T. Fukuhara, Physica B {\textbf{259-261}}, 85 (1999).
\bibitem{Mignot90} J.-M. Mignot, J.-L. Jacoud, L.-P. Regnault, J. Rossat-Mignod, P. Haen, P. Lejay, Ph. Boutrouille, B. Hennion, and D. Petitgrand, Physica B {\textbf{163}}, 611 (1990).
\bibitem{Mignot91} J.-M. Mignot, L.-P. Regnault, J.-L. Jacoud, J. Rossat-Mignod, P. Haen, and P. Lejay, Physica B {\textbf{171}}, 357 (1991).
\bibitem{Wilhelm99}H. Wilhelm, K. Alami-Yadri, B. Revaz, and D. Jaccard, Phys. Rev. B {\bf59}, 3651 (1999).
\bibitem{Wilhelm04}H. Wilhelm and D. Jaccard, Phys. Rev. B \textbf{69}, 214408 (2004).
\bibitem{Runge95}E. K. R. Runge, R. C. Albers, N. E. Christensen, and G. E. Zwicknagl, Phys. Rev. B {\bf51}, 10375 (1995).
\bibitem{Mignot88} J.-M. Mignot, J. Flouquet, P. Haen, F. Lapierre, L. Puech, and J. Voiron,  J. Mag. Mag. Mater. \textbf{76\&77}, 97 (1988).
\bibitem{Shimizu12}Y. Shimizu, Y. Matsumoto, K. Aoki, N. Kimura, and H. Aoki, J. Phys. Soc. Jpn. \textbf{81}, 044707 (2012).
\bibitem{Pearson91}P. Villars, {\it Pearson's Handbook of Crystallogrphic Data for Intermetallic Phase}'' (ASM International 1991).
\bibitem{Severing89} A. Severing, E. Holland-Moritz, and B. Frick, Phys. Rev. B {\bf 39}, 4164 (1989).	
\bibitem{Hadzic86}M. Had\u{z}i\'{c}-Leroux, A. Hamzi\'{c}, A. Fert, P. Haen, F. Lapierre, and O. Laborde, Europhys. Lett. {\bf 1}, 579 (1986). 
\bibitem{Fert87}A. Fert and P. M. Levy, Phys. Rev. B \textbf{36}, 1907 (1987).
\bibitem{Lapierre87} F. Lapierre, P. Haen,  R. Briggs, A. Hamzi\'{c}, A. Fert, and J. P. Kappler, J. Mag. Mag. Mater. \textbf{63\&64}, 338 (1987).
\bibitem{Wilhelm01}H. Wilhelm, S. Raymond, D. Jaccard, O. Stockert, H v L$\ddot{\rm o}$hneysen, and A. Rosch, J. Phys: Condens. Matter {\bf 13}, L329 (2001).
\bibitem{Knebel01}G. Knebel, D. Braithwaite, G. Lapertot, P. C. Canfield, and J. Flouquet, J. Phys: Condens. Matter {\bf 13}, 10935 (2001).
\bibitem{Hattori10}K. Hattori and K. Miyake,  J. Phys. Soc. Jpn. \textbf{79}, 073702 (2010).
\bibitem{Sumiyama86}A. Sumiyama, Y. Oda, H. Nagano, Y. Onuki, K. Shibutani, and T. Komatsubara, J. Phys. Soc. Jpn. \textbf{55}, 1294 (1986).
\bibitem{Asano85}H. Asano, M. Umino, Y. Hataoka, Y. Shimizu, Y. Onuki, T. Komatsubara, and F. Izumi, J. Phys. Soc. Jpn. \textbf{54}, 3358 (1985). 
\bibitem{Knafo09}W. Knafo, S. Raymond, P. Lejay, and J. Flouquet, Nature Phys. {\bf 5}, 753 (2009).
\bibitem{Moriya95}T. Moriya and T. Takimoto, J. Phys. Soc. Jpn. \textbf{64}, 960 (1995).
\bibitem{Moriya03}T. Moriya and K. Ueda, Rep. Prog. Phys. \textbf{66}, 1299 (2003).
\bibitem{Kambe96}S. Kambe, J. Flouquet, P. Haen, and P. Lejay,  J. Low Temp. Phys. {\textbf{102}}, 477 (1996).
\bibitem{Lohneysen07}H. v. L\"{o}hneysen, A. Rosch,  M. Voyta, and P. W\"{o}lfle, Rev. Mod. Phys. \textbf{79}, 1015 (2007).
\bibitem{Sugi08}M. Sugi, Y. Matsumoto, N. Kimura, T. Komatsubara, H. Aoki, T. Terashima, and S. Uji, Phys. Rev. Lett. {\textbf{101}}, 056401 (2008).
\bibitem{Mignot89} J.-M. Mignot, A. Ponchet, P. Haen, F. Lapierre and J. Flouquet, Phys. Rev. B {\textbf{40}}, 10917 (1989).
\bibitem{Stryjewski}E. Stryjewski and N. Giordano, Adv. Phys. \textbf{26}, 487 (1977).
\bibitem{MatsumotoM}Y. Matsumoto, Master thesis, Graduate School of Science, Tohoku University, Sendai (2007).
\bibitem{Haen90}P. Haen, J. Voiron, F. Lapierre, J. Flouquet, and P. Lejay, Physica B \textbf{163}, 519 (1990).
\bibitem{Aoki12}D. Aoki, C. Paulsen, H. Kotegawa, F. Hardy, C. Meingast, P. Haen, M. Boukahil, W. Knafo, E. Ressouche, S. Raymond, and J. Flouquet, J. Phys. Soc. Jpn. \textbf{81}, 034711 (2012).
\bibitem{Holtmeier95}S. Holtmeier, P. Haen, A. Lacerda, P. Lejay, J. L. Tholence, J. Voiron, and J. Flouquet, Physica B \textbf{204}, 250 (1995).
\bibitem{Meulen91}H. P. van der Meulen,  A de Visser, T. T. J. M. Berendschot, J. J. M. Franse, J. A. A. J. Perenboom, D. Althof, and H. van Kempen, Phys. Rev. B \textbf{44}, 814 (1991).
\bibitem{YAoki98}Y. Aoki, T. D. Matsuda, H. Sugawara, H. Sato, H. Ohkuni, R. Settai, Y. Onuki, E. Yamamoto, Y. Haga, A. V. Andreev, V. Sechovsky, L. Havela, H. Ikeda, and K. Miyake, J. Magn. Magn. Mater. \textbf{177-181}, 271 (1998).
\bibitem{Ishida98}K. Ishida, Y. Kawasaki, Y. Kitaoka, K. Asayama, H. Nakayama, J. Floquet, Phys. Rev. B \textbf{57}, 11054(R) (1998).
\bibitem{Daou06} R. Daou, C. Bergemann, and S. R. Julian, Phys. Rev. Lett. {\bf 96}, 026401 (2006).
\bibitem{Continentino93}M. A. Continentino, Phys. Rev. B \textbf{47}, 11587(R) (1993).
\bibitem{Onuki87}Y. Onuki and T. Komatsubara, J. Magn. Magn. Mater. \textbf{63\&64}, 281 (1987).
\bibitem{Fukuhara96}T. Fukuhara, K. Maezawa, H. Ohkuni, J. Sakurai, H. Sato, H. Azuma, K. Sugiyama, Y. Onuki, and K. Kindo, J. Phys. Soc. Jpn. \textbf{65}, 1559 (1996).
\bibitem{Sugawara00}K. Sugawara, T. Namiki, S. Yasuda, T. D, Matsuda, Y. Aoki, H. Sato, N. Mushnikov, S. Hane, and T. Goto, Physica B \textbf{281\&282}, 69 (2000).
\bibitem{Aoki01}H. Aoki, M. Takashita, N. Kimura, T. Terashima, S. Uji, T. Matsumoto, and Y. Onuki, J. Phys. Soc. Jpn. {\bf 70}, 774 (2001).
\bibitem{Kitagawa11}S. Kitagawa, H. Ikeda, Y. Nakai, T. Hattori, K. Ishida, Y. Kamihira, M. Hirano, and H. Hosono, Phys. Rev. Lett. \textbf{107}, 277002 (2011).
\bibitem{Deppe12}M. Deppe, S. Lausberg, F. Weickert, M. Brando, Y. Skourski, N. Caroca-Canales, G. Geibel, and F. Steglich, Phys. Rev. B \textbf{85}, 060401(R) (2012).
\bibitem{Tsujii97b}N. Tsujii, J. He. F. Amita, K. Yoshimura, K. Kosuge, H. Michor, G. Hilscher, and T. Goto, Phys. Rev. B \textbf{56}, 8103 (1997).
\bibitem{Tsujii01} N. Tsujii, H. Mitamura, T. Goto, K. Yoshimura, K. Kosuge, T. Terashima, T. Takamasu, H. Kitazawa, S. Kato, and G. Kido, Physica B \textbf{294-295}, 284 (2001).
\bibitem{Mito12}T. Mito, T. Koyama, K. Nakagawara, T. Ishida, K. Ueda, T. Kohara, K. Matsubayashi, Y. Saiga, K. Munakata, Y. Uwatoko, M. Mizumaki, N. Kawamura, B. Idzikowski, and M. Reiffers, J. Phys. Soc. Jpn. \textbf{81}, 033706 (2012).
\bibitem{Torikachvili07}M. S. Torikachvili, S. Jia, E.D. Mun, S. T. Hannahs, R. C. Black, W. K. Neils, D. Martien, S. L. Bud'ko, and P. C. Canfield, Proc. Natl. Acad. Sci. U.S.A.  \textbf{104}, 9960 (2007).
\bibitem{Takeuchi10}T. Takeuchi, S. Yasui, M. Toda, M. Matsushita, S. Yoshiuchi, M. Ohya, K. Katayama, Y. Hirose, N. Yoshitani, F. Honda, K. Sugiyama, M. Hagiwara, K. Kindo, E. Yamamoto, Y. Haga, T. Tanaka, Y. Kubo, R. Settai, and Y. Onuki, J. Phys. Soc. Jpn. {\bf79}, 064609 (2010).
\bibitem{Ohya10}M. Ohya, M. Matsushita, S. Yoshiuchi, T. Takeuchi, F. Honda, R. Settai, T. Tanaka, Y. Kubo, and Y. Onuki, J. Phys. Soc. Jpn. \textbf{79}, 083601 (2010).
\bibitem{Onuki11}Y. Onuki. S. Yasui, M. Matsushita, S. Yoshiuchi, M. Ohya, Y. Hirose, N. D. Dung, F. Honda, T. Takeuchi, R. Settai, K. Sugiyama, E. Yamamoto, T. D. Matsuda, Y. Haga, T. Tanaka, Y. Kubo, H. Harima, J. Phys. Soc. Jpn. \textbf{80}, SA003 (2011).
\bibitem{Sugiyama99}K. Sugiyama, M. Nakashima, D. Aoki, K. Kindo, N. Kimura, H. Aoki, T. Komatsubara, S. Uji, E. Yamamoto, H. Harima, and Y. Onuki, Phys. Rev. B \textbf{60}, 9248 (1999).
\bibitem{Sugiyama99b}K. Sugiyama, M. Nakashima, H. Ohkuni, K. Kindo, Y. Haga, T. Honma, E. Yamamoto, and Y. Onuki, J. Phys. Soc. Jpn. \textbf{68}, 3394 (1999).
\bibitem{Sugiyama00}K. Sugiyama, M. Nakashima, M. Futoh, H. Ohkuni, T. Inoue, K. Kindo, N. Kimura, E. Yamamoto, Y. Haga, T. Honma, R. Settai, and Y. Onuki, Physica B \textbf{281\&282}, 244 (2000).
\bibitem{Yamada93}H. Yamada, Phys. Rev. B \textbf{47}, 11211 (1993). 
\bibitem{Sakakibara90}T. Sakakibara, T. Goto, K. Yoshimura, K. Murata, K. Fukamichi, J. Magn. Magn. Mater. \textbf{90 \&91}, 131 (1990).
\bibitem{Belitz05}D. Belitz, T. R. Kirkpatrick, and T. Vojta, Rev. Mod. Phys. \textbf{77}, 579 (2005).
\bibitem{Belitz05b}D. Belitz, T. R. Kirkpatrick, and J${\rm \ddot{o}}$rg Rollb${\rm \ddot{u}}$hler, Phys. Rev. Lett. \textbf{94}, 247205 (2005).
\bibitem{Kabeya12}N. Kabeya, H. Maekawa, K. Deguchi, N. Kimura, H. Aoki, and N. K. Sato,  J. Phys. Soc. Jpn. \textbf{81}, 073706 (2012).
\bibitem{Hoshino13}S. Hoshino and Y. Kuramoto, Phys. Rev. Lett. \textbf{111}, 026401 (2013).
\bibitem{Raymond98}S. Raymond, L. P. Regnault,  S. Kambe, J. Flouquet, and P. Lejay, J. Phys.: Condens. Matter {\textbf{10}}, 2363 (1998).
\bibitem{Sato04} M. Sato, Y. Koike, S. Katano, N. Metoki, H. Kadowaki, and S. Kawarazaki, J. Phys. Soc. Jpn. {\textbf{73}}, 3418 (2004).
\bibitem{Lonzarich88}G. G. Lonzarich, J. Magn. Magn. Mater. \textbf{76\&77}, 1 (1988).
\bibitem{Onuki92}Y. Onuki, I. Umehara, A. K. Albessard, T. Ebihara, and K. Satoh, J. Phys. Soc. Jpn. \textbf{61}, 960 (1992). 
\bibitem{Aoki92} H. Aoki, S. Uji, A. K. Albessard, and Y. Onuki, J. Phys. Soc. Jpn. {\bf 61}, 3457 (1992).
\bibitem{Aoki93a} H. Aoki, S. Uji, A. K. Albessard, and Y. Onuki, Phys. Rev. Lett. {\bf 71}, 2110 (1993).
\bibitem{Aoki93b} H. Aoki, S. Uji, A. K. Albessard, and Y. Onuki, J. Phys. Soc. Jpn. {\bf 62}, 3157 (1993).
\bibitem{Julian94}S. R. Julian, F. S. Tautz, G. J. McMullan, and G. G. Lonzarich, Physica B  \textbf{199\&200}, 63 (1994).
\bibitem{Tautz95} F. S. Tautz, S. R. Julian, G. J. McMullan, and G. G. Lonzarich, Physica B \textbf{206\&207}, 29 (1995).
\bibitem{Aoki95} H. Aoki, M. Takashita, S. Uji, T. Terashima, K. Maezawa, R. Settai, and Y. Onuki, Physica B \textbf{206\&207}, 26 (1995).
\bibitem{Takashita96}M. Takashita ,H. Aoki, T. Terashima, S. Uji, K. Maezawa, R. Settai, Y. Onuki, J. Phys. Soc. Jpn.  {\bf 65}, 515  (1996).
\bibitem{Denlinger01}J. D. Denlinger, G. -H. Gweon, J. W. Allen, C. G. Olson, M. B. Maple, J. L. Sarrao, P. E. Armstrong, Z. Fisk, and H. Yamagami,  J. Electron Spectrosc. Relat. Phenom. {\bf 117-118}, 347 (2001).
\bibitem{Yano08}M. Yano, A. Sekiyama, H. Fujiwara, Y. Amano, S. Imada, T. Muro, M. Yabashi, K. Tamasaku, A. Higashiya, T. Ishikawa, Y. Onuki, and S. Suga, Phys. Rev. B {\bf 77}, 035118 (2008).
\bibitem{Okane09}T. Okane, T. Ohkochi, Y. Takeda, S.-i. Fujimori, A. Yasui, Y. Saitoh, H. Yamagami, A. Fujimori, Y. Matsumoto, M. Sugi, N. Kimura, T. Komatsubara, and H. Aoki, Phys. Rev. Lett. {\bf 102}, 216401 (2009).
\bibitem{Koizumi11} A. Koizumi, G. Motoyama, Y. Kubo, T. Tanaka, M. Itou, and Y. Sakurai, Phys. Rev. Lett. {\bf 106}, 136401 (2011).
\bibitem{Yamagami93}H. Yamagami and A. Hasegawa, J. Phys. Soc. Jpn. {\bf 62}, 592 (1993).
\bibitem{Suzuki10}M.-T. Suzuki and H. Harima, J. Phys. Soc. Jpn. {\bf 79}, 024705 (2010).
\bibitem{Sakai13}O. Sakai, M.-T. Suzuki, H. Harima, and Y. Kaneta, JPS Conference Proceedings, in press.
\bibitem{Settai95} R. Settai, H. Ikezawa, H. Toshima, M. Takashita, T. Ebihara, H. Sugawara, T. Kimura, K. Motoki and Y. Onuki,  Physica B {\textbf{206\&207}}, 23 (1995).
\bibitem{Takashita98} M. Takashita, H. Aoki, C. J. Haworth, T. Matsumoto, T. Terashima, S. Uji, C. Terakura, T. Miura, K. Maezawa, R. Settai, and Y. Onuki, J. Magn. Magn. Mater. \textbf{177-181}, 417 (1998).
\bibitem{Shiba90}H. Shiba and P. Fazekas, Prog. Theor. Phys. Suppl. {\bf 101}, 403 (1990).
\bibitem{Shibata99}N. Shibata and K. Ueda, J. Phys.: Condens. Matter {\bf 11}, R1 (1999).
\bibitem{Oshikawa00}M. Oshikawa, Phys. Rev. Lett. {\bf 84}, 3370 (2000). 
\bibitem{Otsuki09}J. Otsuki, H. Kusunose, and Y. Kuramoto, Phys. Rev. Lett. {\bf102}, 017202 (2009).
\bibitem{Ikezawa97}H. Ikezawa, H. Aoki, M. Takashita, C. J. Haworth, S. Uji, T. Terashima, K. Maezawa, R. Settai, and Y. Onuki, Physica B {\bf 237-238}, 210 (1997).
\bibitem{King91}C. A. King and G. G. Lonzarich, Physica B {\textbf{171}}, 161 (1991).
\bibitem{IkezawaM}H. Ikezawa, Master thesis, Graduate School of Pure and Applied Sciences, University of Tsukuba, Tsukuba (1996). 
\bibitem{DoiM}Y. Doi, Master thesis, Graduate School of Science, Tohoku University (2012).
\bibitem{Felsch73}W. Felsch and K. Winzer, Solid State Communi. {\bf 13}, 569 (1973).
\bibitem{Costi00}T. A. Costi, Phys. Rev. Lett. {\bf 85}, 1504 (2000).
\bibitem{Coleridge70}P. T. Coleridge and I. M. Templeton, Phys. Revs. Lett. \textbf{24}, 108 (1970).
\bibitem{Alles73}H. Alles, R. J. Higgins, and D. H. Lowendes, Phys. Revs. Lett. \textbf{30}, 705 (1973).
\bibitem{Endo02} M. Endo, N. Kimura, H. Aoki, T. Terashima, S. Uji,  C. Terakura, and T. Matsumoto, J. Phys. Soc. Jpn. {\bf 71} Suppl., 127 (2002).
\bibitem{Endo03} M. Endo, N. Kimura, A. Ochiai, H. Aoki, T. Terashima, C. Terakura, S. Uji, T. Matsumoto, Acta Physica Polonica B {\bf 34}, 1031  (2003).
\bibitem{Sheikin03}I. Sheikin, A. Groger, S. Raymond, D. Jaccard, D. Aoki, H. Harima, J. Flouquet, Phys. Rev. B.  {\bf 67}, 094420 (2003). 
\bibitem{Nakayama04}M. Nakayama, N. Kimura, H. Aoki, A. Ochiai, C. Terakura, T. Terashima, and S. Uji, Phys. Rev. B {\bf 70}, 054421 (2004).
\bibitem{McCollam05}A. McCollam, S. R. Julian, P. M. C. Rourke, D. Aoki and J. Flouquet, Phys. Rev. Lett. {\bf 94}, 186401 (2005).
\bibitem{Daou06b}R. Daou, C. Bergemann, and S. R. Julian, Physica B \textbf{378}-\textbf{380}, 807 (2006).
\bibitem{Endo04} M. Endo, N. Kimura, H. Aoki, T. Terashima, S. Uji, T. Matsumoto, and T. Ebihara, Phys. Rev. Lett. {\bf 93}, 247003  (2004).
\bibitem{Endo05} M. Endo, N. Kimura, and H. Aoki, J. Phys. Soc. Jpn. \textbf{74}, 3295 (2005).
\bibitem{Endo06}M. Endo, S. Nakamura, T. Isshiki, N. Kimura, T. Nojima, H. Aoki, H. Harima, and S. Kunii, J. Phys. Soc. Jpn. {\bf 75}, 114704 (2006).
\bibitem{Spalek90}J. Spalek and P. Gopalan, Phys. Rev. Lett. \textbf{64}, 2823 (1990).
\bibitem{Spalek06}J. Spalek,  Phys. Stat. Sol. (b) {\bf 243}, 78 (2006).
\bibitem{Otsuki07}J. Otsuki J, H. Kusunose, and Y. Kuramoto,  J. Magn. Magn. Mater. \textbf{310}, 425 (2007).
\bibitem{Bauer07}J. Bauer and A. C. Hewson, Phys. Rev. B {\bf 76}, 035118 (2007).
\bibitem{Onari08}S. Onari, H. Kontani, and Y. Tanaka, J. Phsy. Soc. Jpn. {\bf 77}, 023703 (2008).
\bibitem{Iida11}H. Iida, Y. Kadota, M. Kogure, T. Sugawara, H. Aoki, and N. Kimura, J. Phys. Soc. Jpn. \textbf{80}, 083701 (2011).
\bibitem{Ogawa79}K. Ogawa, H. Aoki, and I. Nakatani, J. Phys. Chem. Solids.  {\bf 40}, 469 (1979). 
\bibitem{Ruitenbeck82}I. M. van Ruitenbeck, W. A. Verhoef, P. G. Mattocks, A. E. Dixon, A. P. J. van Deursen, and A. R. de Vroomen, J. Phys. F: Metal Phys. \textbf{12}, 2919 (1982).
\bibitem{Sigfusson84}T. I. Sigfusson, N. R. Bernhoet, and G. G. Lonzarich, J. Phys. F: Metal Phys. \textbf{14}, 2141 (1984).
\bibitem{Matsuhira97}K. Matsuhira, T. Sakakibara, H. Amitsuka, K. Tenya, K. Kamishima, T. Goto, and G. Kido, J. Phys. Soc. Jpn. \textbf{66}, 2851 (1997). 
\bibitem{Yoshiuchi09}S. Yoshiuchi, M. Toda, M. Matsushita, S. Yasui, Y. Hirose, M. Ohya, K. Katayama, F. Honda, K. Sugiyama, M. Hagiwara, K. Kindo, T. Takeuchi, E. Yamamoto, Y. Haga,  R. Settai, T. Tanaka, Y. Kubo, and Y. Onuki, J. Phys. Soc. Jpn. \textbf{78}, 123711 (2009).
\bibitem{Goodrich99}R. G. Goodrich, N. Harrison, A. Teklu, D. Young and Z. Fisk, Phys. Rev. Lett. {\bf 82}, 3669 (1999). 
\bibitem{Nakamura06}S. Nakamura, M. Endo,  H. Yamamoto, T. Isshiki, N. Kimura,  H. Aoki, T. Nojima, S. Otani and S. Kunii, J. Phys. Soc. Jpn. {\bf75}, 114 (2006).
\bibitem{Nakanishi01}Y. Nakanishi, T. Sakon, F. Takahashi, M. Motokawa, A. Uesawa, M. Kubota and T. Suzuki, Phys. Rev. B {\bf64}, 224402 (2001).
\bibitem{Okane12}T. Okane, Y. Takeda, H. Yamagami, A. Fujimori, Y. Matsumoto, N. Kimura, T. Komatsubara, and H. Aoki, Phys. Rev. B \textbf{86}, 125138 (2012).
\bibitem{Paschen04} S. Paschen, T. L\"{u}hmann, S. Wirth, P. Gegenwart, O. Trovarelli, C. Geibel, F. Steglich, P. Coleman, and Q. Si, Nature {\bf 432}, 881 (2004).
\bibitem{Gegenwart02} P. Gegenwart, J. Custers, C. Geibel, K. Neumaier, T. Tayama, K. Tenya,  O. Trovarelli, and F. Steglich,  Phys. Rev. Lett. {\bf 89}, 056402 (2002).
\bibitem{Terashima01}T. Terashima, T. Matsumoto, C. Terakura, S. Uji, N. Kimura. M. Endo, T. Komatsubara, and H. Aoki, Phys. Rev. Lett.  \textbf{87}, 166401 (2001).
\bibitem{Settai02}R. Settai, M. Nakashima, S. Araki, Y. Haga, T. C. Kobayashi, N. Tateiwa , and Y. Onuki, J. Phys. Condens. Matter \textbf{14}, L29 (2002).
\bibitem{Shishido05}H. Shishido, R. Settai, H. Harima, and Y. Onuki, J. Phys. Soc. Jpn. {\textbf{74}}, 1103 (2005).
\bibitem{Settai05}R. Settai, T. Kubo, T. Shiromoto, D. Honda, H. Shishido,  K. Sugiyama, Y. Haga, T. D. Matsuda, K. Betsuyaku, H. Harima,  T. C. Kobayashi,  and Y. Onuki, J. Phys. Soc. Jpn. \textbf{74}, 3016 (2005).
\bibitem{Terashima07}T. Terashima, Y. Takahide, T. Matsumoto, S. Uji, N. Kimura, H. Aoki, and H. Harima, Phys. Rev. B {\textbf{76}}, 054506 (2007). 
\bibitem{Goh08}S. K. Goh, J. Paglione, M. Sutherland, E. C. T. O'Farrell, C. Bergemann, T. A. Sayles, and M. B. Maple, Phys. Rev. Lett. {\textbf{101}}, 056402 (2008).
\bibitem{Si01}Q. Si, S. Rabello, K. Ingersent, and J. L. Smith, Nature \textbf{413}, 804 (2001).
\bibitem{Si10}Q. Si, Phys. Status Solidi B \textbf{247}, 476 (2010).
\bibitem{Vojta10}M. Vojta, J. Low Temp. Phys. \textbf{161}, 203 (2010).
\bibitem{Senthil03}T. Senthil, S. Sachdev, and M. Vojta, Phys. Rev. Lett. \textbf{90}, 216403 (2003).
\bibitem{Senthil04}T. Senthil, M. Vojta, and S. Sachdev, Phys. Rev. B \textbf{69}, 03511 (2004).
\bibitem{Senthil05}T. Senthil, S. Sachdev, and M. Vojta, Physica B \textbf{359-361}, 9 (2005).
\bibitem{Vojta08}M. Vojta, Phys. Rev. B \textbf{78}, 125109 (2008).
\bibitem{Watanabe07}H. Watanabe and M. Ogata, Phys. Rev. Lett. \textbf{99}, 136401 (2007).
\bibitem{Watanabe09}H. Watanabe and M. Ogata, J. Phys. Soc. Jpn. \textbf{78}, 024715 (2009).
\bibitem{Lanata08}N. Lanat$\grave{\rm a}$, P. Barone, and M. Fabrizio, Phys. Rev. B \textbf{78}, 155127 (2008).
\bibitem{Martin08}L. C. Martin and F. F. Assaad, Phys. Rev. Lett. \textbf{101}, 066404 (2008).
\bibitem{Martin10}L. C. Martin, M. Berex and F. F. Assaad, Phys. Rev. B \textbf{82}, 245105 (2010).
\bibitem{Leo08}L. De Leo, M. Civelli, and G. Kotliar, Phys. Rev. Lett. \textbf{101}, 256404 (2008).
\bibitem{Leo08b}L. De Leo, M. Civelli, and G. Kotliar, Phys. Rev. B. \textbf{77}, 075107 (2008).
\bibitem{Hoshino10}S. Hoshino, J. Otsuki, and Y. Kuramoto, Phys. Rev. B \textbf{81}, 113108 (2010).
\bibitem{Watanabe10}S. Watanabe and K. Miyake, J. Phys. Soc. Jpn. \textbf{79}, 033707 (2010).
\bibitem{Matsumoto12}Y. Matsumoto, T. Terashima, S. Uji, N. Kimura, and H. Aoki, J. Phys. Soc. Jpn. \textbf{81}, 054703 (2012).
\bibitem{Otsuki10}J. Otsuki, H. Kusunose, and Y. Kuramoto, J. Phys. Soc. Jpn. {\bf 79}, 114709 (2010).
\bibitem{Matsumoto10}Y. Matsumoto, N. Kimura, H. Aoki, M. Kimata, T. Terashima,  S. Uji, T. Okane, and H. Yamagami, J. Phys. Soc. Jpn. {\bf 79}, 083706 (2010).
\bibitem{Fujimori07} S.-I. Fujimori, Y. Satoh, T. Okane, A. Fujimori, H. Ymagami, Y. Haga, E. Yamamoto, and Y. Ounki,  Natutre Phys. {\bf 3}, 618 (2007).
\bibitem{Teklu00}A. A. Teklu, R. G. Goodrich, N. Harrison, D. Hall, Z. Fisk, and D. Young, Phys. Rev. B, {\bf 62}, 12875 (2000).
\bibitem{Sato85}N. Sato, A. Sumiyama, S. Kunii, H. Nagano, and T. Kasuya, J. Phys. Soc. Jpn. {\bf 54}, 1923 (1985).
\bibitem{Joss87}W. Joss, J. M. van Ruitenbeek, G. W. Crabtree, J. L. Tholence, A. P. J. van Deursen, and Z. Fisk, Phys. Rev. Lett. {\bf 59}, 1609 (1987).
\bibitem{Harrison04}N. Harrison, U. Alver, R. G. Goodrich, I. Vekhter, J. L. Sarrao, P. G. Pagliuso, N. O. Moreno, L. Balicas, Z. Fisk, D. Hall, R. T. Macaluso, and J. Y. Chan, Phys. Rev. Lett. {\bf93}, 186405  (2004).
\bibitem{Okane11}T. Okane, I. Kawasaki, A. Yasui, T. Ohkochi, Y. Takeda, S.-I. Fujimori, Y. Saitoh, H. Yamagami, A. Fujimori, Y. Matsumoto, N. Kimura, T. Komatsubara, and H. Aoki, J. Phys. Soc. Jpn. \textbf{80}, SA060 (2011).
\bibitem{Matsumoto12b}Y. Matsumoto, N. Kimura, T. Komatsubara, H. Aoki, N. Kurita, T. Terashima,  and S. Uji, J. Phys.: Conference Series {\bf 391}, 012042 (2012).
\bibitem{Raymond99}S. Raymond, P. Haen, R. Calemczuk, S. Kambe, B. F\r{a}k, P. Lejay, T. Fukuhara, and J. Flouquet,  J. Phys.: Condens. Matter {\textbf{11}}, 5547 (1999).
\bibitem{Besnus91}M. J. Besnus, A. Essaihi, N. Hamdaoui, G. Fischer, J. P. Kappler, A. Meyer, J. Pierre, P. Haen, and P. Lejay, Physica B \textbf{171}, 350 (1991).
\bibitem{Gold}A. V. Gold, {\it Electrons in Metals}, Solid State Physics,  ed. F. Cochran and R. R. Haering (Gordon and Breach, 1968) vol.1.
\bibitem{Shoenberg}D.Shoenberg, {\it Magnetic\ Oscillations\ in\ Metals} (Cambridge University Press, Cambridge, U.K., 1984).
\bibitem{Watt74}B. R. Watt, J. Phys. F: Metal Phys. \textbf{4}, 1371 (1974).
\bibitem{LK}I. M. Lifshitz and A. M. Kosevich, Zh. eksp. teor. fiz. \textbf{29},730 (1955). Soviet Phys. JETP \textbf{2}, 636 (1956).
\bibitem{Cracknell}A. P. Cracknell, {\it The\ Fermi\ Surfaces\ of\ Metals} (Taylor $\&$ Francis LTD 1971).
\bibitem{Joss}W. Joss, R. Griessen, and E. Fawcett,  {\it Electron states and Fermi surfaces of homogeneously strained metallic element}, Landolt - B\"{o}rnstein, New Series, Metals ; Phonon States, Electron states and Fermi surfaces,  ed. K. H. Hellwege and J. L. Olsen (Springer, Berlin, 1983) vol.III/13b.
\bibitem{Takashita97}M. Takashita, H. Aoki, T, Matsumoto, C. J. Hawaorth, T. Terashima, A. Uesawa,  and T. Suzuki, Phys. Rev. Lett. \textbf{78}, 1948 (1997).
\bibitem{Khalid88}M. A. Khalid,  P. H. P. Reinders, and M. Springford, J. Phys. : Met. Phys.  \textbf{18}, 1949 (1988).
\bibitem{Wasserman89}A. Wasserman, M. Springford, and A. C. Hewson, J. Phys. Condens. Matter \textbf{1}, 2669 (1989).
\bibitem{Wasserman96}A. Wasserman and M. Springford, Adv. Phys. \textbf{45}, 471 (1996).
\bibitem{Isshiki08}T. Isshiki, N. Kimura, H. Aoki, T. Terashima, S. Uji, K. Yamauchi, H. Harima, D. Jaiswal-Nagar, S. Ramakrishnan, and A. K. Grover, Phys. Rev. B \textbf{78}, 134528 (2008).
\bibitem{IsshikiD}T. Isshiki,  Doctor thesis, Graduate School of Science, Tohoku University, Sendai (2009).
\bibitem{Springford}  {\it Electrons\ at\ the\ Fermi\ Surface}, ed. M. Springford (Cambridge University Press, Cambridge, U.K., 1980).
\bibitem{Holzwarth75} N. A. Holzwarth, Phys. Rev. B \textbf{11}, 3718 (1975).
\bibitem{Chollet68} L.-F. Chollet and I. M. Templeton,  Phys. Rev. {\bf 170}, 656 (1968).
\bibitem{Lowndes73}D. H. Lowndes, K. Miller, R. G. Poulsen, and M. Springford,  Proc. Roy. Soc. \textbf{A 331}, 497 (1973).
\bibitem{Shiba73}H. Shiba, Prog. Theor. Phys. Jpn. {\bf 50}, 1797 (1973).
\bibitem{Aoki83}H. Aoki and K. Ogawa,  J. Phys. F: Met. Phys. \textbf{13}, 1821 (1983).
\bibitem{Ebihara93}T. Ebihara, I. Umehara, A. K. Albessard, K. Satoh, and Y. Onuki, J. Phys. Soc. Jpn. \textbf{61}, 1473 (1992).
\bibitem{Yamamizu04}T. Yamamizu, M. Endo, M. Nakayama, N. Kimura and H. Aoki, Phys. Rev. B {\bf 69}, 014423 (2004).

\end{thebibliography}
\end{document}